\documentclass{aa}

\usepackage{txfonts} 
\usepackage{graphicx}
\usepackage{natbib}
\usepackage{lscape}
\usepackage{array}
\usepackage{hyperref}

\newcommand{\hersc}{{\it Herschel}}

\newcommand{\spitz}{{\it Spitzer}}

\newcommand{\lsun}{$L_\odot$}
\newcommand{\msun}{$M_\odot$}
\newcommand{\zsun}{$Z_\odot$}
\newcommand{\mic}{$\mu$m}

\newcolumntype{R}[1]{>{\raggedleft\arraybackslash }b{#1}}
\newcolumntype{L}[1]{>{\raggedright\arraybackslash }b{#1}}
\newcolumntype{C}[1]{>{\centering\arraybackslash }b{#1}}

\newcommand{\bfaa}{}

\usepackage[normalem]{ulem}

\begin{document}

\title{DeGaS-MC: Dense Gas Survey in the Magellanic Clouds}
\vspace{4pt}
\subtitle{ I - An APEX survey of HCO$^+$ and HCN(2$-$1) toward the LMC and SMC}

\author{M. Galametz
\inst{1}\and
A. Schruba
\inst{2}\and
C. De Breuck
\inst{3}\and
K. Immer
\inst{4}\and
M. Chevance
\inst{5}\and
F. Galliano
\inst{1}\and
A. Gusdorf
\inst{6,7}\and
V. Lebouteiller
\inst{1}\and
M.Y. Lee
\inst{8}\and
S. C. Madden
\inst{1}\and
F. L. Polles
\inst{7}\and
T. A. van Kempen
\inst{9}
}

\institute{
AIM, CEA, CNRS, Universit\'{e} Paris-Saclay, Universit\'{e} Paris Diderot, Sorbonne Paris Cit\'{e}, F-91191 Gif-sur-Yvette, France,\\
\email{maud.galametz@cea.fr}
\and
Max-Planck-Institut f\"{u}r extraterrestrische Physik, Giessenbachstra{\ss}e 1, D-85748 Garching, Germany, 
\and
European Southern Observatory, Karl-Schwarzschild-Stras{\ss}e 2, D-85748 Garching-bei-M\"{u}nchen, Germany,
\and
Joint Institute for VLBI ERIC, Oude Hoogeveensedijk 4, 7991 PD Dwingeloo, The Netherlands, 
\and
Astronomisches Rechen-Institut, Zentrum f\"{u}r Astronomie der Universit\"{a}t Heidelberg, M\"{o}nchhofstra{\ss}e 12-14, 69120 Heidelberg, Germany, 
\and
Laboratoire de Physique de l'\'Ecole Normale Sup\'erieure, ENS, Universit\'e PSL, CNRS, Sorbonne Universit\'e, Universit\'e de Paris, Paris, France,
\and
LERMA, Observatoire de Paris, PLS research University, CNRS, Sorbonne Universit\'e, 75104 Paris, France,
\and
Korea Astronomy and Space Science Institute, 776 Daedeokdae-ro, 34055, Daejeon, Republic of Korea, 
\and
SRON Netherlands Institute for Space Research, Sorbonnelaan 2, 3584 CA Utrecht, The Netherlands.
}

\abstract
{Understanding the star-forming processes is key to understanding the evolution of galaxies. Investigating star formation 
requires precise knowledge of the properties of the dense molecular gas complexes where stars form and {{\bfaa a}} quantification of
how they are affected by the physical conditions
to which they 
are exposed. The proximity, low metallicity, and wide range of star 
formation activity of the Large and Small Magellanic Clouds (LMC and SMC) make them prime laboratories to study how local physical conditions 
impact the dense gas reservoirs and their star formation efficiency.}
{The aim of the Dense Gas Survey for the Magellanic Clouds (DeGaS-MC) project is to expand our knowledge of the relation between dense gas 
properties and star formation activity by targeting the LMC and SMC observed in the HCO$^+$(2$-$1) and HCN(2$-$1) transitions. }
{We carried out a {{\bfaa pointing}} survey targeting two lines toward $\sim$30 LMC and SMC molecular clouds 
using the SEPIA180 instrument installed on the Atacama Pathfinder EXperiment (APEX) telescope. We performed a follow-up mapping 
campaign of the emission in the same transition in 13 star-forming regions. This first paper provides line characteristic catalogs and integrated line-intensity maps of the sources.}
{HCO$^+$(2$-$1) is detected in 20 and HCN(2$-$1) in 8 of the 29 pointings observed. 
The dense gas velocity pattern follows the line-of-sight velocity field derived from the stellar population.
The three SMC sources targeted during the mapping campaign were unfortunately not detected in our mapping campaign but 
both lines are detected toward the LMC 30Dor, N44, N105, N113, N159W, N159E, and N214 regions.
The HCN emission is less extended than the HCO$^+$ emission and is restricted to the densest regions. 
The HCO$^+$(2$-$1)/HCN(2$-$1) brightness temperature ratios range from 1 to 7, which is consistent with the large ratios commonly observed in low-metallicity environments.
A larger number of {{\bfaa young stellar objects are}} found at high HCO$^+$ intensities and lower HCO$^+$/HCN flux ratios, and thus toward
denser lines of sight. The dense gas luminosities correlate with the star formation rate traced by {\bfaa the total infrared }luminosity over the 
two orders of magnitude covered by our observations, although substantial region-to-region variations are observed. 
}
{}


\authorrunning{Galametz M. et al}
\titlerunning{Dense Gas Survey in the Magellanic Clouds}
\maketitle


\section{Introduction}

The study of the star formation mechanisms is key to understanding the evolution of galaxies. 
To understand their star formation activity, it is crucial to have precise 
knowledge of the morphology and properties (mass, density, temperature, velocity) of their 
molecular gas complexes and how these are affected by the physical conditions to which they 
are exposed. H$_2$ is the most abundant molecular species in {\bfaa the interstellar medium (ISM)} but requires high 
temperatures to be excited, which makes it very difficult to observe in emission in cold 
clouds \citep{Shull1982,Wakelam2017}. Cold H$_2$ can be traced in absorption (UV radiation is 
absorbed by H$_2$ through the Lyman and Werner bands) but this technique can be 
limited by extinction along the line of sight \citep{Flower2012}.

CO is the most common molecule after H$_2$ and its observation has been extensively used to probe 
the molecular gas, from local star-forming \citep{Dickman1978,Bolatto2013} to high-redshift environments \citep{Carilli2013}. 
However, using CO luminosity as a tracer of H$_2$ has its limitations, as the CO emission does not solely arise from the 
densest phases of the cold gas where star formation occurs \citep{Liszt2010,Caldu2016} and can also be optically thick in the densest regions.
Additionally, the CO molecule can be subject to widespread photo-dissociation due to strong UV radiation fields and in porous ISM such as that encountered in low-metallicity environments \citep{Israel1997,Cormier2014,Chevance2016,Chevance2020}.
Moreover, its emission has a nonlinear relation with the H$_2$ column density it is supposed to trace \citep{Liszt2010}.

Many molecules have now been targeted to probe higher density regions in nearby galaxies, 
in particular molecules such as N$_2$H$^+$, HCN, HNC, HCO$^+$, CS, HC$_3$N, H$_2$CO 
\citep{Schlingman2011,Zhang2014,Kepley2018,Gallagher2018,JimenezDonaire2019}
and their isotopologs \citep[H$^{13}$CN, H$^{13}$NC, H$^{13}$CO$^+$;][]{Mauersberger1991_2,JimenezDonaire2017}.
Molecules like HCO$^+$ and HCN are commonly used to trace molecular gas that is more dense than detected in CO.
In Table~\ref{Lineused}, we report the dipole moments and critical densities for optically thin emission 
of the HCN and HCO$^+$ molecules for the J=2$-$1 transitions analyzed in the current study \citep{Schoier2005,Shirley2015}.
While substantial fractions of the CO emission have been argued to lie in extended translucent regions \citep{Pety2013,Leroy2017}, 
studies toward Galactic giant molecular clouds have suggested that HCN or HCO$^+$ emission could also arise from more extended diffuse areas
\citep{Nishimura2017,Pety2017,Shimajiri2017,Watanabe2017}, especially in the lowest J transitions. 
Reinforcing our analysis of the HCN or HCO$^+$ emissions in nearby environments is therefore crucial to better understanding 
their robustness as tracers of dense gas.

The HCO$^+$ and HCN luminosities have been shown to correlate over ten orders of magnitude with {\bfaa star formation rate (SFR)}
\citep{Gao_Solomon2004,Chen2015,Shimajiri2017,JimenezDonaire2019}.
Here again, a deeper knowledge of how the tracers evolve with the local conditions is crucial to be able to use 
them as {\bfaa star-forming} reservoirs at higher redshifts and in local environments as tools to investigate whether there is a universality in the dense 
gas properties that lead to star formation or if the star formation efficiency 
(SFE=$\Sigma_{SFR}$/$\Sigma_{\rm dense}$, i.e., the ratio of the {\bfaa SFR} surface 
density to the dense gas mass surface density) varies with the dense gas fraction or the 
gas physical conditions \citep{Garcia-Burillo2012,Usero2015,JimenezDonaire2019}. 
 
At a distance of $\sim$50-60 kpc, the Large and Small Magellanic Clouds (LMC and SMC)
are among the very few extragalactic sources in which we can resolve individual {\bfaa Giant Molecular Clouds (GMCs)} and study young 
stellar objects (YSOs), making them prime laboratories to compare Galactic star formation 
activity and efficiency with those of nearby galaxy populations.
The LMC and SMC host a wide range of star formation regimes, from quiescent clouds to very active star-forming 
regions, allowing us to study how the local physical conditions impact the dense gas reservoirs. 
Their low metallicities \citep[$\sim$1/2 and 1/5 \zsun, respectively;][]{Dufour1982,Russell1990} 
enable us to extend the parameter range to more primordial physical conditions. 
The elemental abundance of the LMC (respectively SMC) differs from that of more metal-rich environments, with an abundance in nitrogen
that is a factor of 10 (respectively 20) lower than that of the Milky Way but an abundance in oxygen only a factor of 3 (respectively 7)
lower \citep{Dufour1982}. This difference in the elemental abundance ratio is one of several hypotheses suggested to explain the 
variation of the HCO$^+$/HCN ratio with the galaxy host metallicity \citep{Bayet2012,Braine2017,Johnson2018,Kepley2018}.

Observations of HCO$^+$(1$-$0) and HCN(1$-$0) have been performed in several LMC star-forming complexes 
to study the impact of low-metallicity ISM and more extreme star-forming conditions on the dense gas tracers 
\citep{Chin1997,Heikkila1999,Wong2006,Wang2009,Seale2012,Anderson2014,Nishimura2016}, although in most 
cases, observations predominantly targeted very active star-forming regions. Additionally, very few 
positions have been targeted in the SMC in HCO$^+$ and HCN and our knowledge of its dense molecular gas content 
and its local variations is still very limited. Finally, the 1$-$0 transition is often the only transition observed:
to the best of our knowledge the 2$-$1 transition has never been observed in the Magellanic Clouds, despite its 
importance to complement the HCO$^+$ and HCN spectral line energy distributions and 
constrain dense gas properties such as excitation temperature or gas density.

The aim of the Dense Gas Survey in the Magellanic Clouds (DeGaS-MC; PI: Galametz) 
project is to expand our knowledge of the relation between the dense gas 
properties and star formation activity by targeting more diverse environments, 
from OB complexes to star-forming clouds of lower masses and less extreme UV radiation.
HCO$^+$(2$-$1) and HCN(2$-$1) observations were obtained toward $\sim$30 LMC and SMC molecular 
clouds using the SEPIA180 instrument installed on the Atacama Pathfinder Experiment 
(APEX) telescope at a resolution of 35\arcsec, thus $\sim$8-10 pc for the Magellanic Clouds.
A follow-up mapping campaign was then performed for a subset of these positions. \\

The goal of this first paper of the collaboration is to
{\it i)} present the survey characteristics, 
{\it ii)} provide catalogs and maps of the two targeted lines, and 
{\it iii)} to present the first results in order to showcase the dataset potential in terms of chemistry, dynamics, and dense gas fraction analysis. 
Future papers of the collaboration will present more detailed statistical analyses and modeling of the lines.
We describe the sample and observations in $\S$2. 
We present the detection statistics, the line properties, and the intensity maps in $\S$3.
We analyze the line characteristics (velocities, line widths, and line ratios) in $\S$4. 
We discuss the relation between the dense tracers and star formation activity as well as 
the dense gas fraction in $\S$5. We summarize our findings in $\S$6.

\begin{table}
\caption{Lines targeted and used in this analysis.}
\label{Lineused}
\centering
\begin{tabular}{ccccc}
\hline
\hline
                &\\
Molecule        & j $\rightarrow$ k     & Frequency     & n$_{crit}$$^a$        & $\mu_0$$^b$         \\
                &                               &       (GHz)   & (cm$^{-3}$)                   & (Debye)         \\
                &\\
\hline
                &\\
HCN             & 2$-$1 & 177.26        & 2.8 $\times$ 10$^6$   & 3.93 \\
HCO$^+$ & 2$-$1 & 178.38        & 4.2 $\times$ 10$^5$   & 2.98 \\
                &\\
\hline
                &\\
                CO              & 1$-$0 & 115.27        & 3.0  $\times$ 10$^3$  & 0.11 \\
                &\\
\hline
\end{tabular}
\begin{list}{}{}
\item[$^a$] Critical densities (optically thin approximation) are provided at 20~K for HCN and 
HCO$^+$ \citep{Shirley2015} and 100K for CO \citep{Lequeux2005}. We refer to \citet{Shirley2015} 
for more details on how radiative trapping can be taken  into account.
\item[$^a$] Dipole moments from \citet{Schoier2005}.
\end{list}
\end{table} 

\section{The DeGaS-MC survey}

The LMC is our nearest neighbor located at 50 kpc \citep{Schaefer2008}. It has
an almost face-on orientation \citep[23-37$\deg$][]{Subramanian2012} and a
half-solar metallicity \citep{Dufour1982}. In this paper, we assume a systemic velocity v$_{\rm sys}$=262.2 km~s$^{-1}$ 
for the LMC \citep{VanderMarel2002}.
The SMC is located at a distance of $\sim$60 kpc  \citep{Hilditch2005} and has a more metal-poor ISM
\citep[1/5-1/8\zsun][]{Russell1990}. We assume a systemic velocity v$_{\rm sys}$=158 km~s$^{-1}$ 
for the SMC \citep{Richter1987}. 

The SEPIA \citep[Swedish-European Southern Observatory PI receiver for APEX; ][159$-$211~GHz]{Belitsky2018} 
Band-5 instrument, now SEPIA180, is a dual polarization two-sideband receiver based on the ALMA Band-5 receivers. 
The central frequencies of the lower and upper sidebands (LSB and USB) are separated by 12~GHz. Each sideband 
is recorded by two XFFTS \citep[eXtended bandwidth Fast Fourier Transform Spectrometer;][]{Klein2012} units of 2.5~GHz 
width each, with a 1~GHz overlap. 
The backend provides 65536 channels for each of these 2.5~GHz sub-bands, which corresponds to a spectral resolution of 
0.065 km/s at 177~GHz. The beam size is close to 35\arcsec\ at this frequency, equivalent to $\sim$8-10~pc for the Magellanic Clouds. 
We refer to \citet{Billade2012}, \citet{Immer2016}, and \citet{Belitsky2018} for more details on the receivers and the offline 
data calibration. In order to observe both lines together, we choose a tuning frequency of 177.82~GHz in the USB, in the 
middle of the rest frequencies of the HCN(2$-$1) and HCO$^+$(2$-$1) lines (177.26 and 178.38~GHz, respectively).

\begin{figure*}
\centering
\hspace{-10pt}\includegraphics[width=17cm]{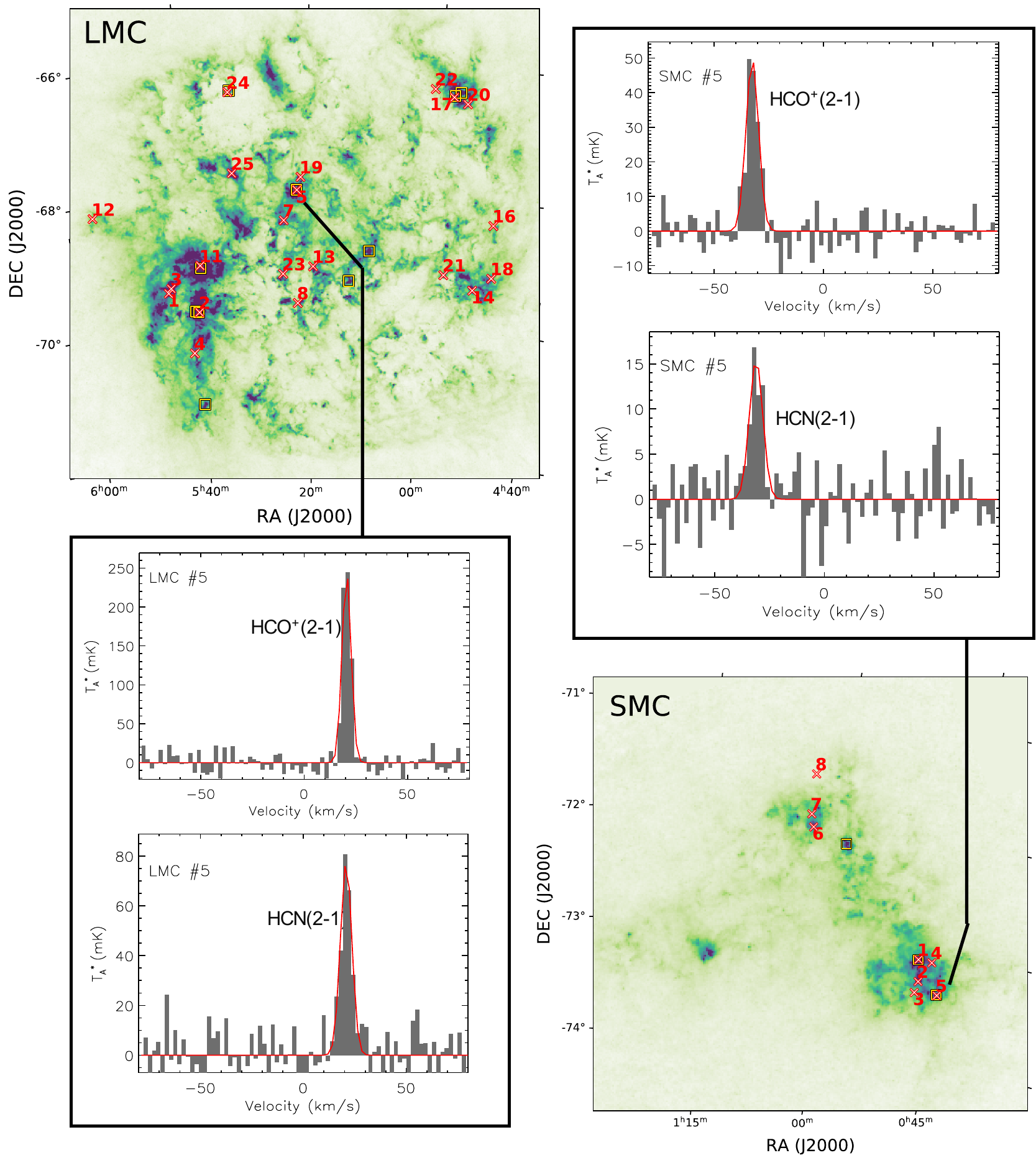}     
\caption{DeGaS-MC pointing campaign: the positions of our pointings are indicated by the red crosses
overlaid on the LMC and the SMC \hersc\ 500~\mic\ observations (FWHM = 36\arcsec).  
An example of the HCO$^+$(2$-$1) and HCN(2$-$1) spectra obtained towards regions LMC \#5 and SMC \#5 are shown 
(2 km~s$^{-1}$ spectral resolution). The x-axis is expressed in velocity with respect to the systemic velocities 
(v=0 corresponding to 262.2 km~s$^{-1}$ for the LMC and 158 km~s$^{-1}$ for the SMC). 
The catalogue of HCO$^+$(2$-$1) and HCN(2$-$1) spectra is shown in Figs.~\ref{LMCSMC_HCOp} and \ref{LMCSMC_HCN}. 
The yellow squares indicate the position of the sources targeted during the follow-up mapping campaign (see Sect.~\ref{section:mc}). 
Please note that the areas mapped are $\sim$4\arcmin$\times$4\arcmin, i.e., smaller than these squares.}
\label{Pointings} 
\end{figure*}

\subsection{The DeGaS-MC single pointing campaign}
\label{pc}

\subsubsection{Source selection}
 \citet{Wong2011} and \citet{Muller2010} provide catalogs of CO clouds in the LMC and SMC as part of the Magellanic Mopra Assessment (MAGMA)
project \footnote{\url{http://mmwave.astro.illinois.edu/magma/}}. We chose our pointing positions in the LMC to match the center of 
a selection of CO clumps provided by their catalogs, regularly ranging from CO luminosities of 8.9 $\times$ 10$^4$ down to 3.6 $\times$ 10$^3$~K~km~s$^{-1}$~pc$^{-2}$. 
We complemented the sample with targets from the SMC main star-forming body in order to probe a lower metallicity regime than the LMC.
From the originally proposed sample, 21 sources were observed in the LMC and 8 in the SMC.
The list of pointings is provided in Table~\ref{LineCharacteristics1} 
\footnote{Four additional sources were originally planned for the LMC but not observed within the allocated telescope time, 
hence the gaps in the source numbering of Table~\ref{LineCharacteristics1}} 
and their positions are overlaid on the \hersc\ 500~\mic\ emission map in Fig.~\ref{Pointings}. 
We note that since the SEPIA beam is small compared to the full extent of some of the 
CO complexes, some of our pointings might be missing the densest part of the targeted molecular complexes.

\vspace{5pt}
\subsubsection{Observations}
Observations were carried out in May and June 2016 (Project ID: 097.C-0758; PI: Galametz). Most of the regions were observed 
using the wobbler switch mode. Two LMC sources and one SMC source were observed using the total power mode. 
The system temperatures varied between 170 and 290 K. The precipitable water vapor was 0.9$<$pwv$<$2.0 mm. 
We selected off-positions for each pointing based on the LMC and SMC \hersc\ maps, typically a few degrees away 
from the observed region. The observations were carried out for a total time of 40h, with more than 16h on-source.

\begin{figure*}
\centering
\vspace{20pt}
\begin{tabular}{ccc}
\vspace{10pt}
{\bf \large 30Dor} & {\bf \large N159W} & {\bf \large N159E}\\
&\\
{\large HCO$^+$(2$-$1)} & 
{\large HCO$^+$(2$-$1)} & 
{\large HCO$^+$(2$-$1)} \\
\includegraphics[height=4.8cm]{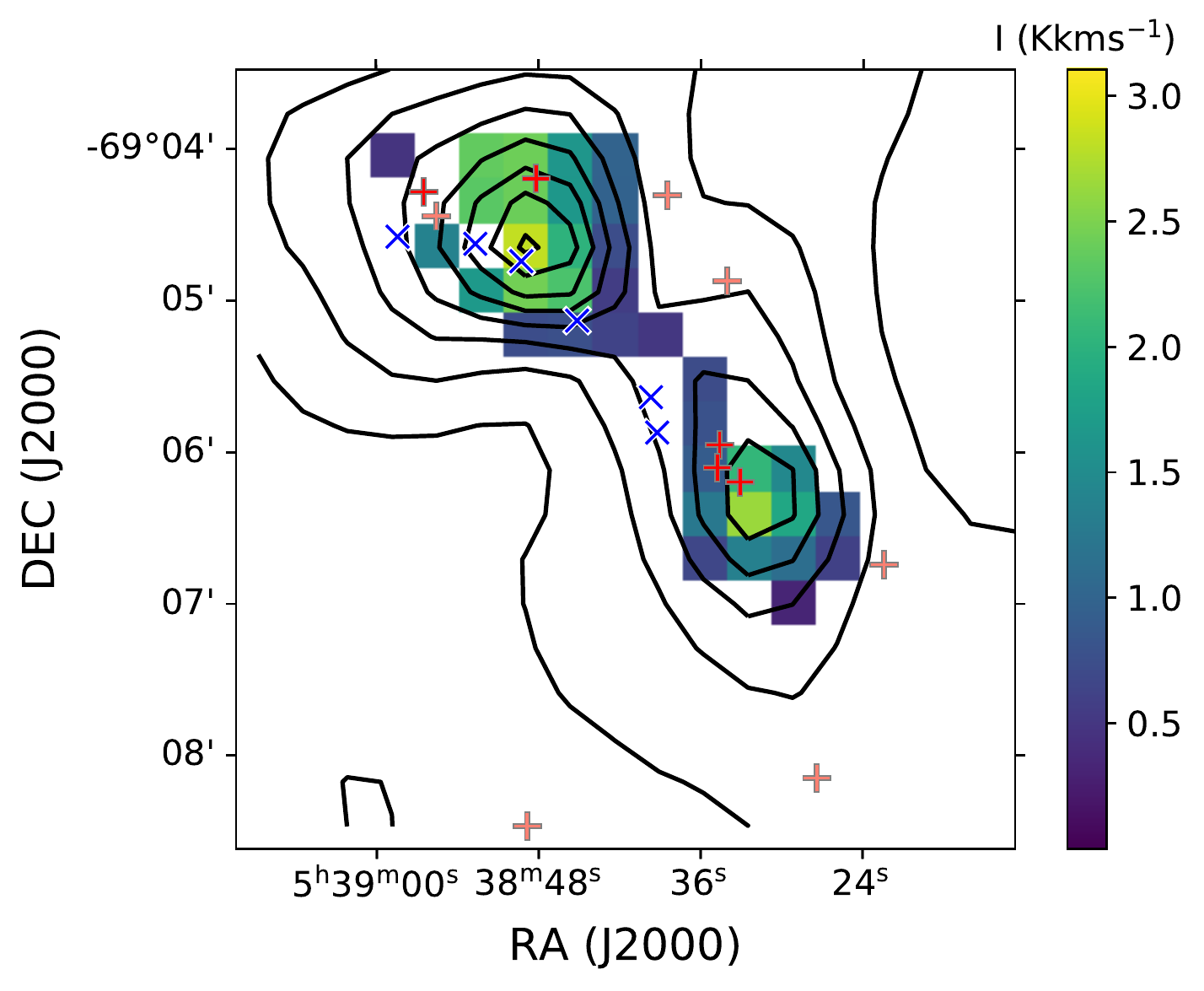}  &
\hspace{-0.5cm}\includegraphics[height=4.8cm]{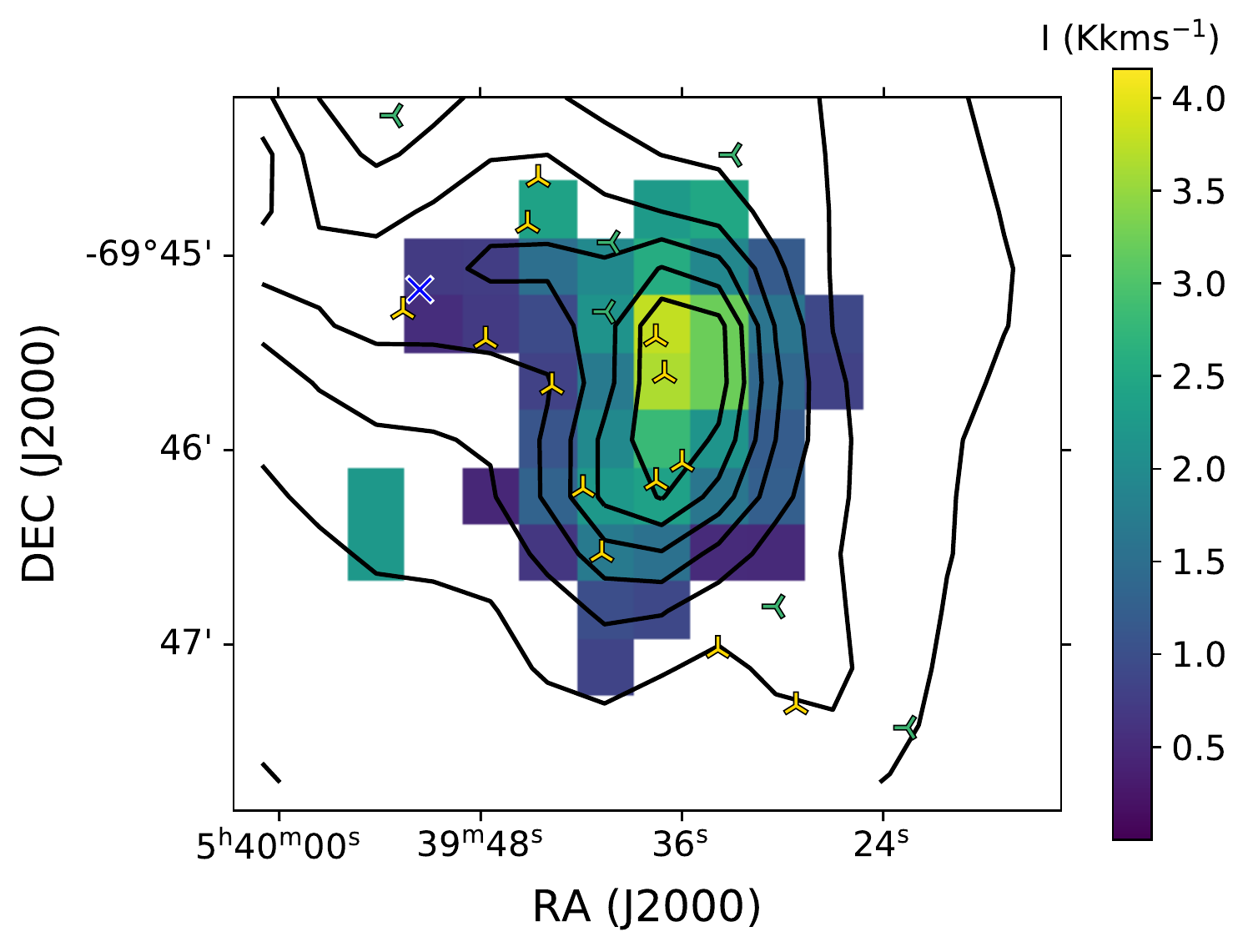} &
\hspace{-0.5cm}\includegraphics[height=4.8cm]{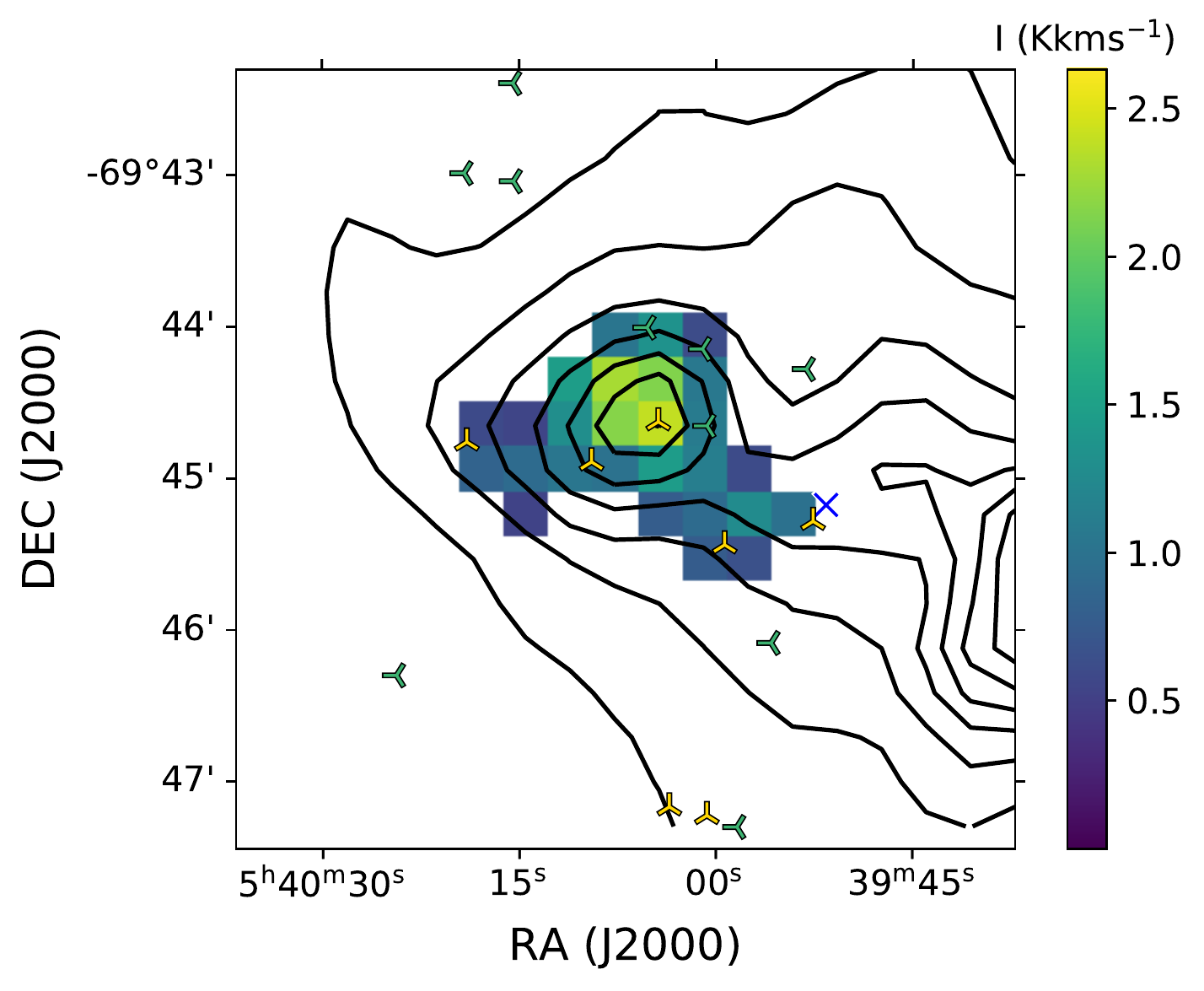}  \\
\includegraphics[height=4cm]{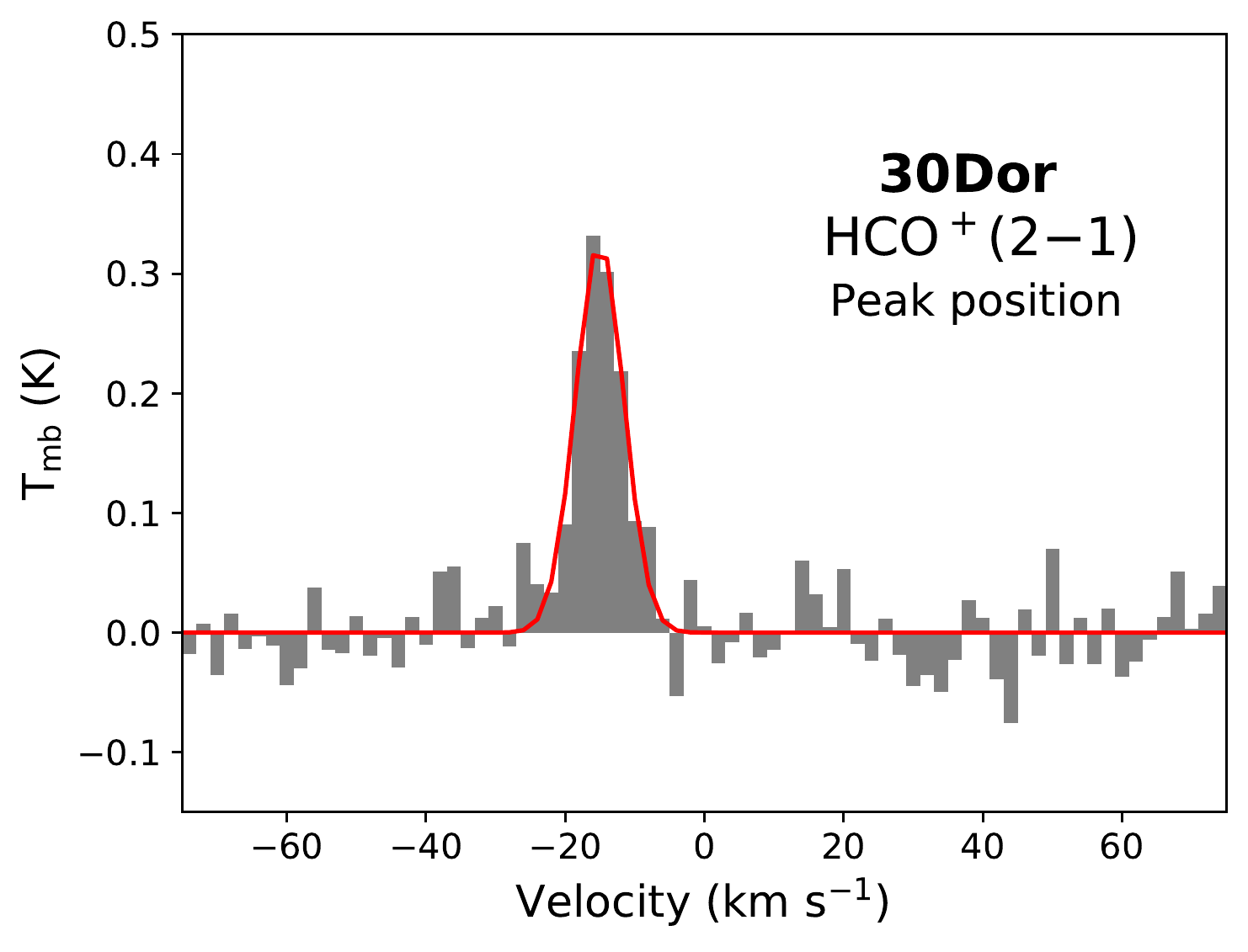}  &
\hspace{-0.5cm}\includegraphics[height=4cm]{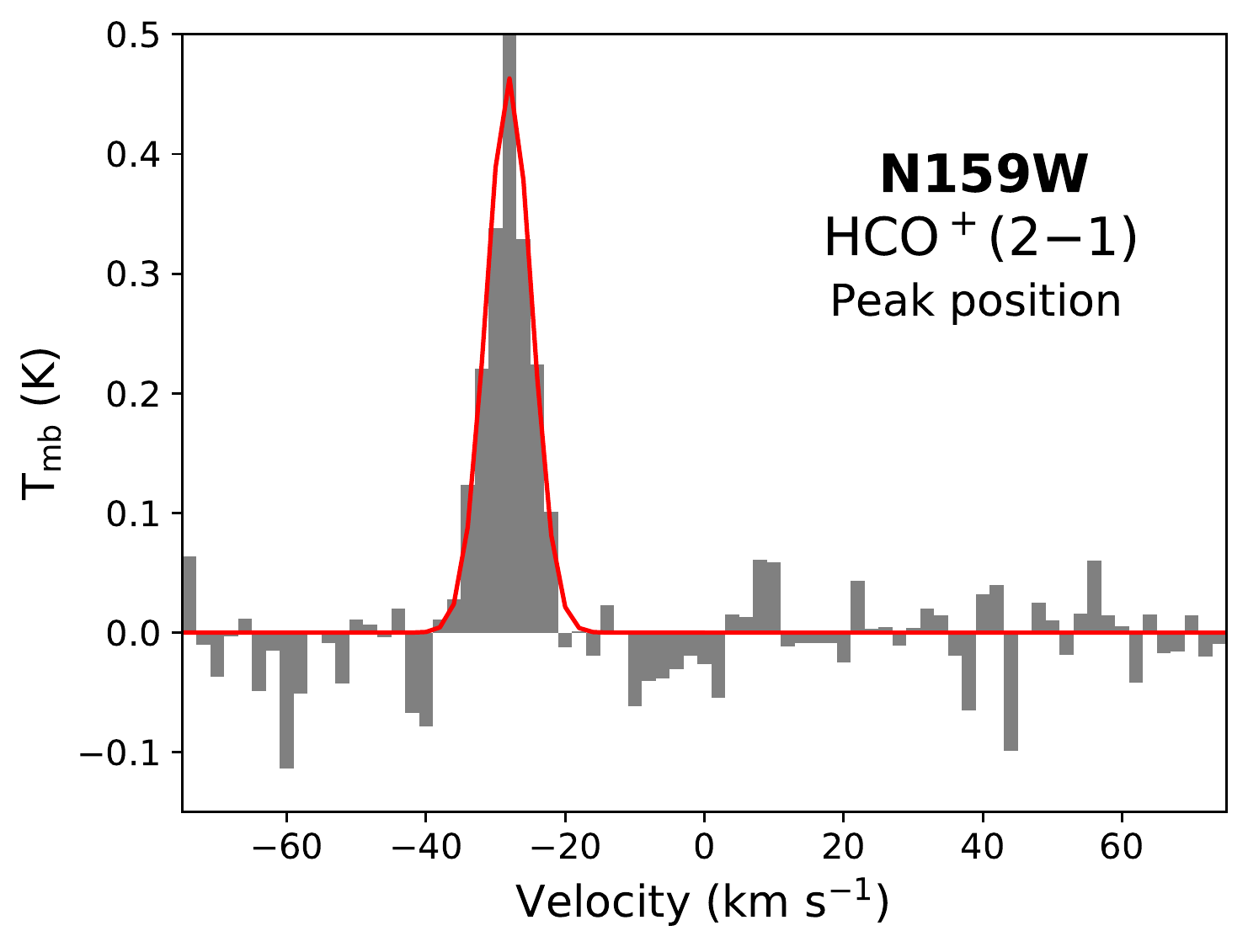} &
\hspace{-0.5cm}\includegraphics[height=4cm]{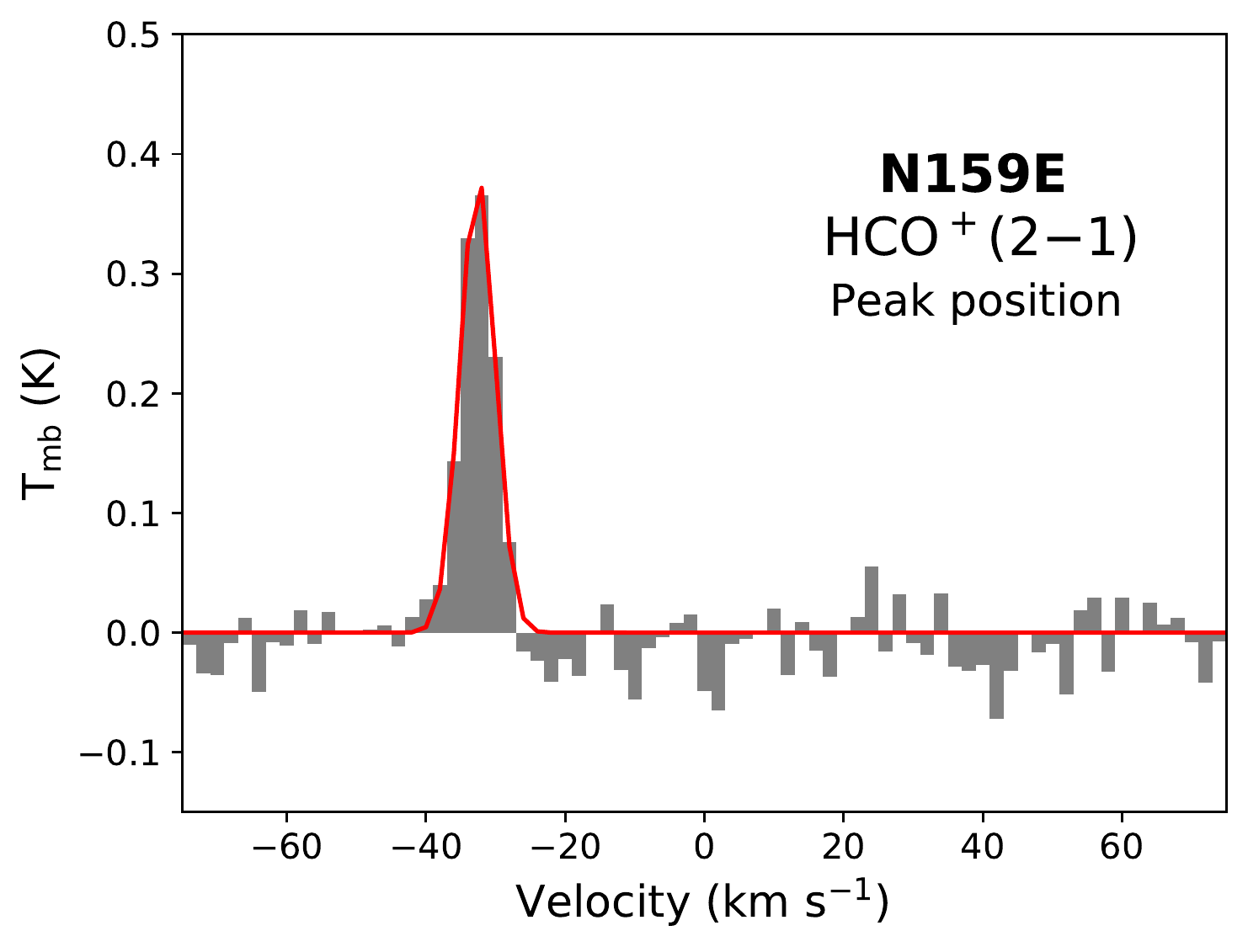} \\
&\\
{\large HCN(2$-$1)} & 
{\large HCN(2$-$1)} & 
{\large HCN(2$-$1)} \\
\includegraphics[height=4.8cm]{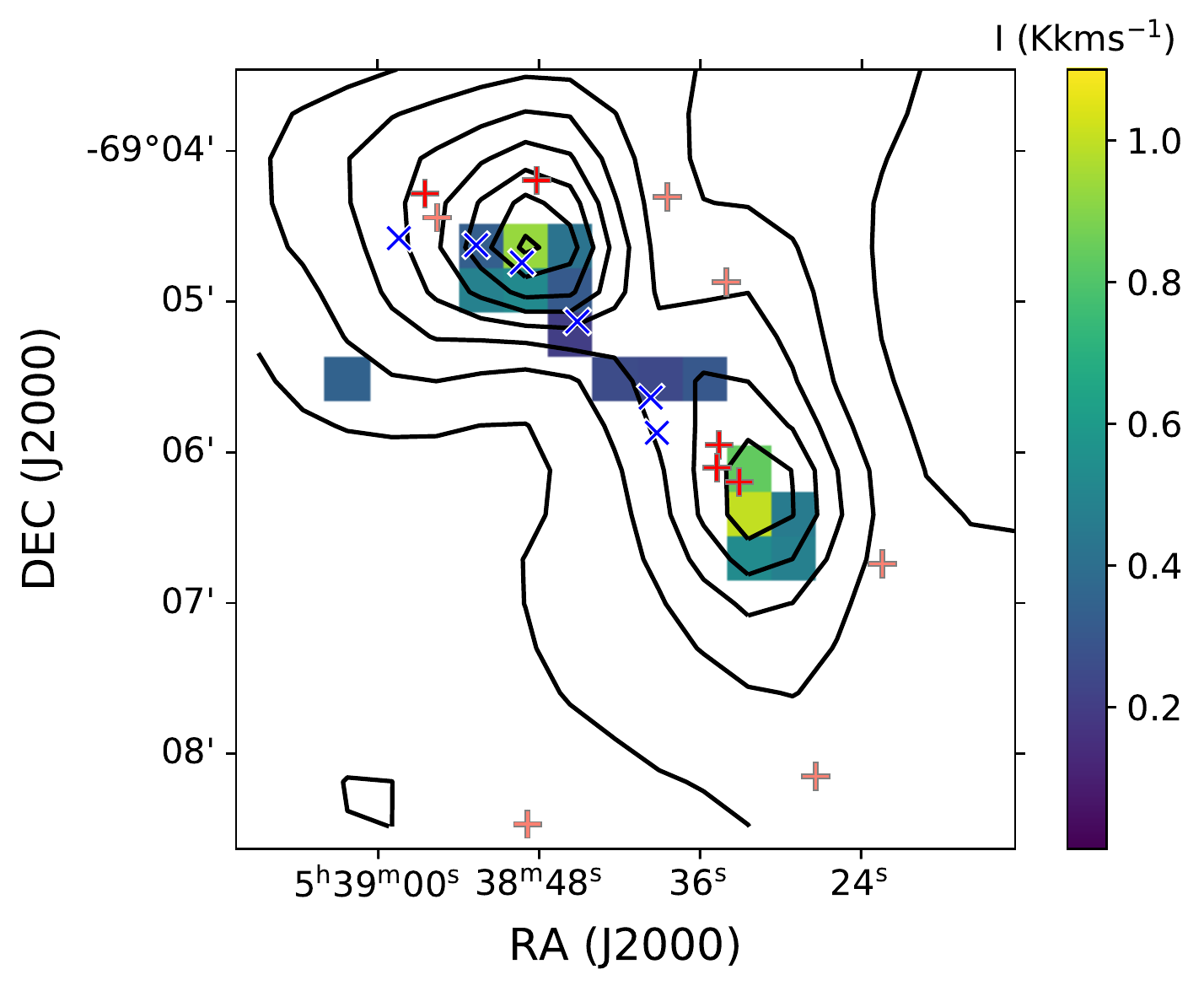}  &
\hspace{-0.5cm}\includegraphics[height=4.8cm]{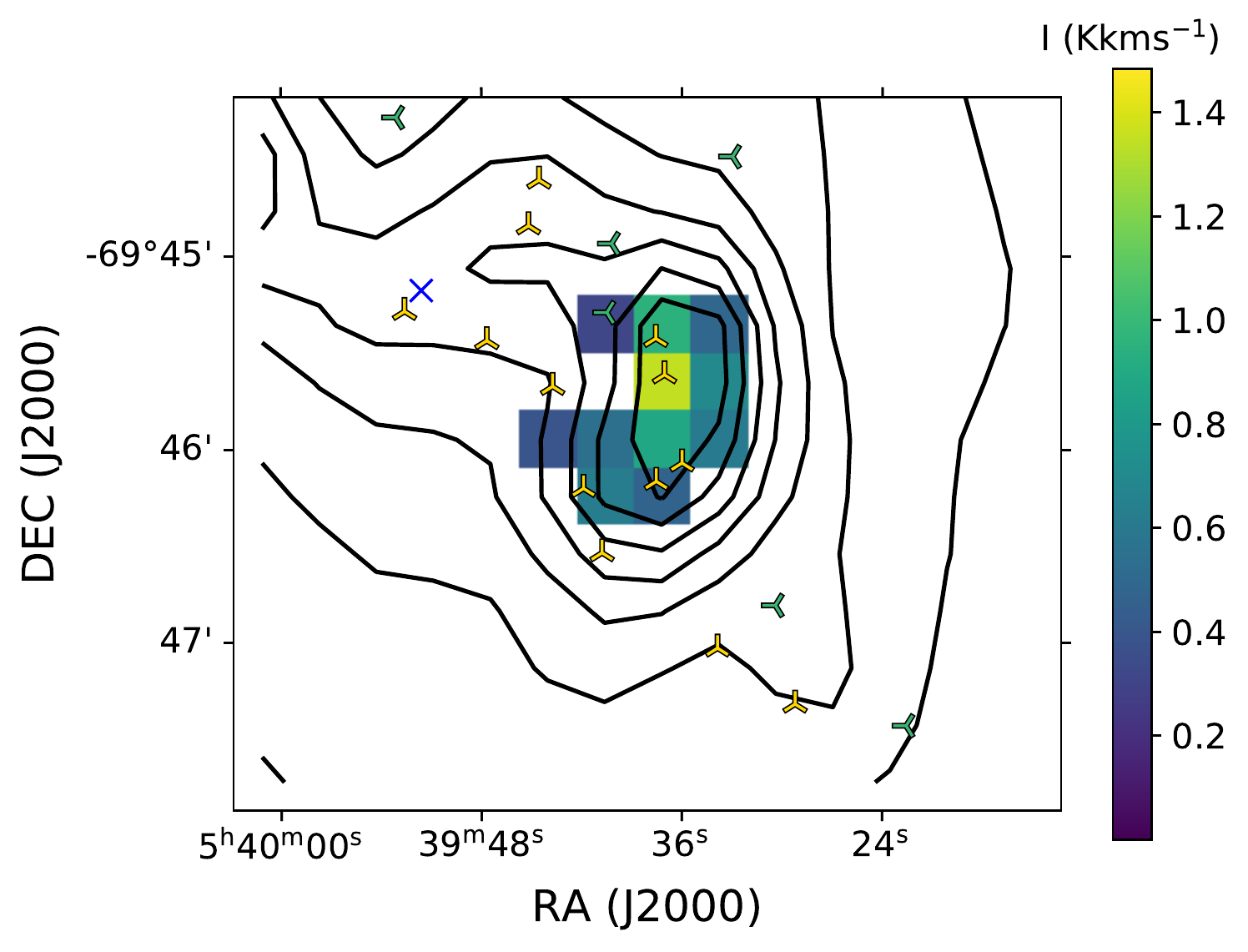} &
\hspace{-0.5cm}\includegraphics[height=4.8cm]{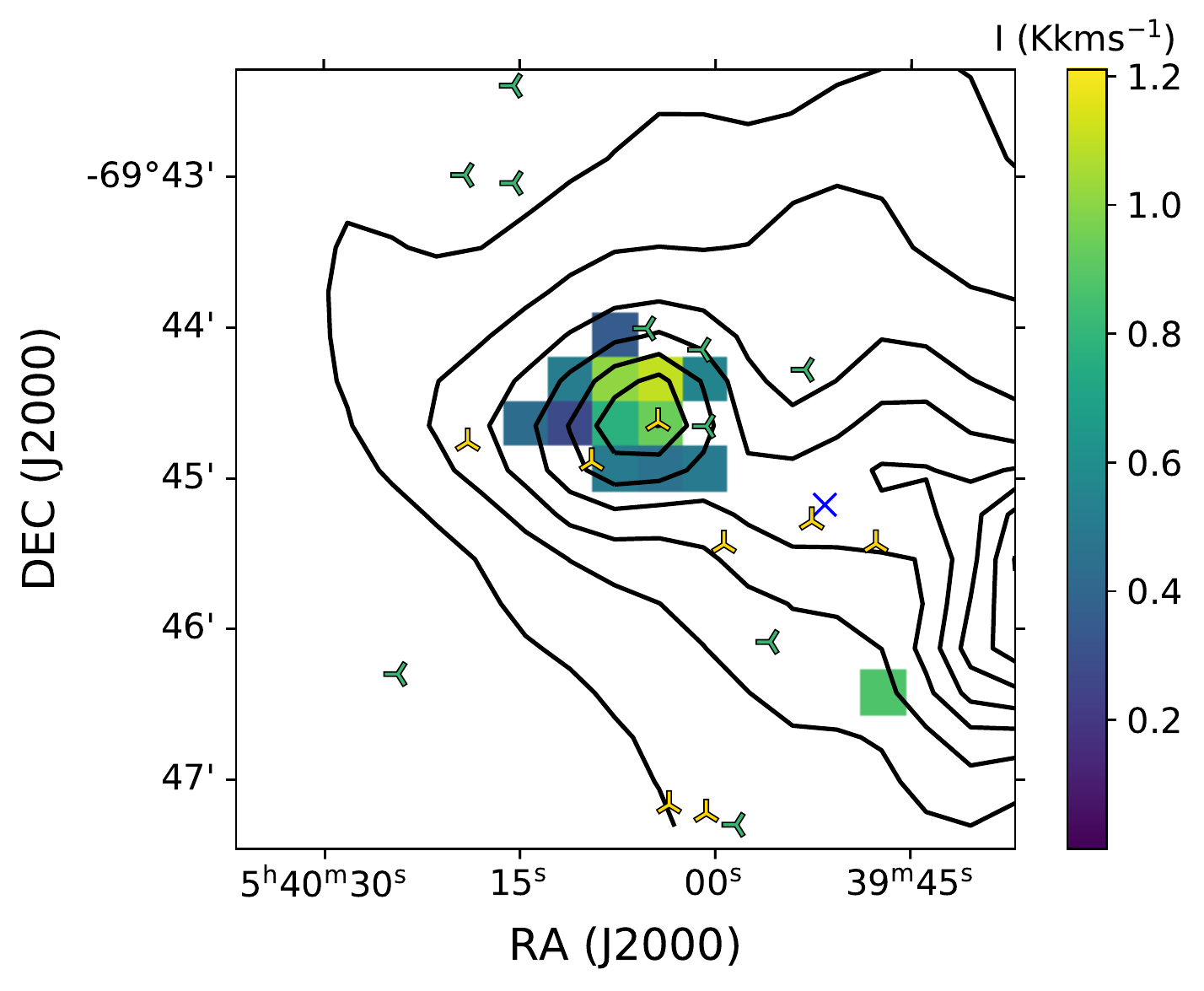}  \\
\includegraphics[height=4cm]{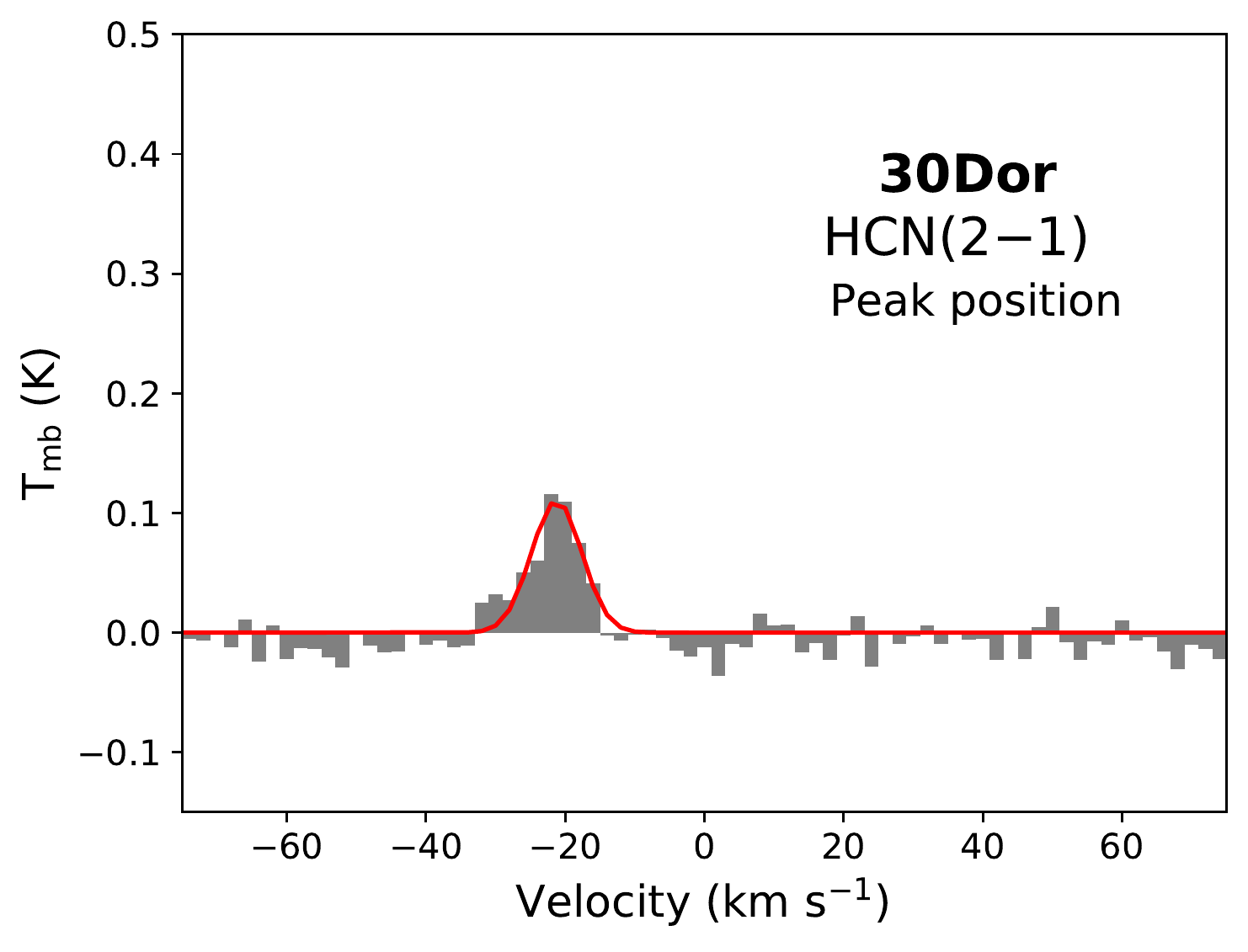}  &
\hspace{-0.5cm}\includegraphics[height=4cm]{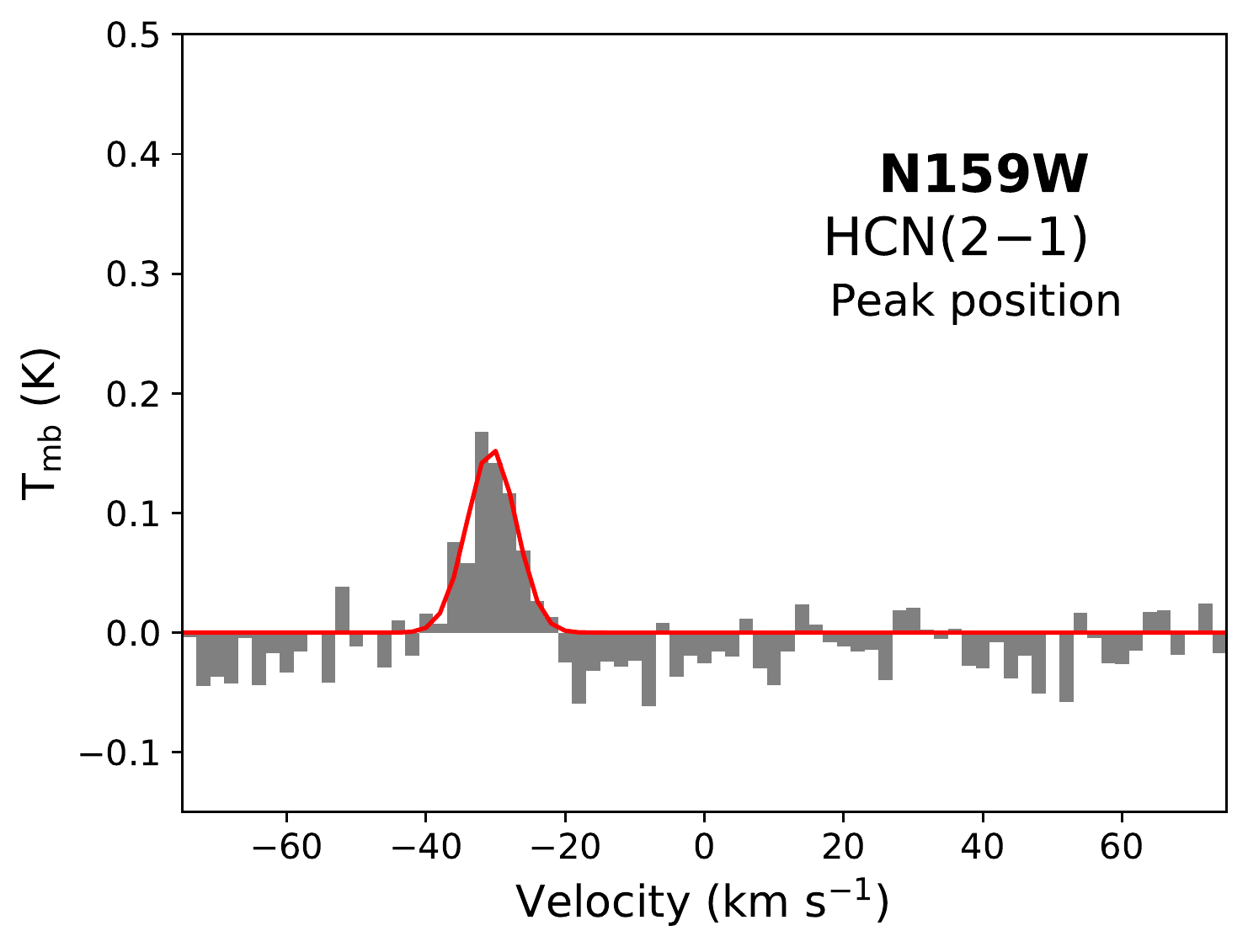} &
\hspace{-0.5cm}\includegraphics[height=4cm]{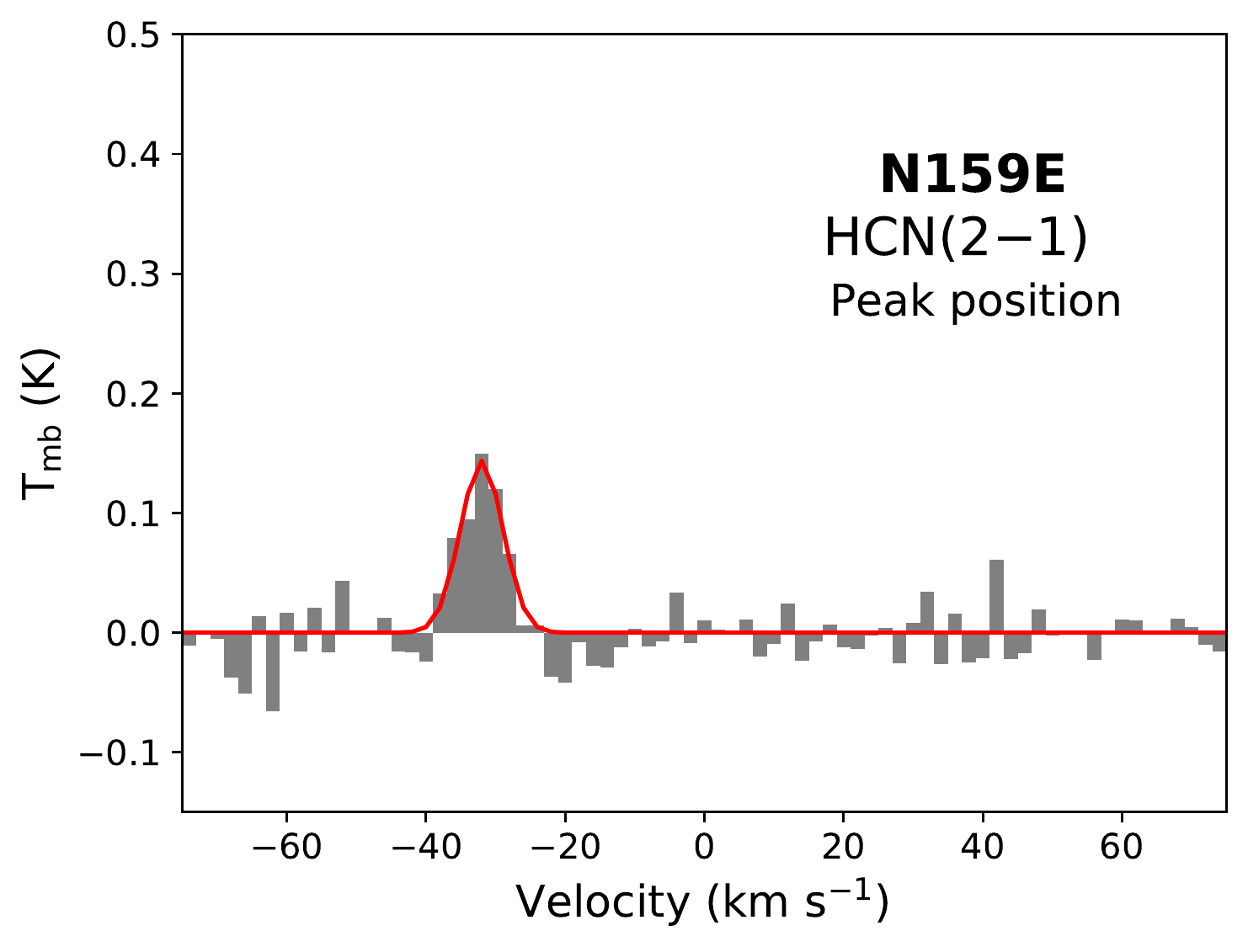}  \\
\vspace{10pt}
\end{tabular}
\vspace{10pt}
\caption{From top to bottom: HCO$^+$(2$-$1) and HCN(2$-$1) integrated intensity maps of 30Dor, N159W, and N159E, from left to right,
respectively (3-$\sigma$ detections). The cold dust emission as traced by \hersc\ PACS 160~\mic\ is overlaid as contours.
For each map, we show the corresponding spectrum at peak position (2 km~s$^{-1}$ spectral resolution) and overlay in red 
the Gaussian fit performed to derive the peak position line characteristics of Table~\ref{LineCharacteristics2}.
The bottom x-axis is expressed in velocity with respect to the systemic velocity (v=0 corresponding to 262.2 km~s$^{-1}$). 
Candidate YSOs retrieved from different catalogs are overlaid: 
purple down-tridents from \citet{Whitney2008}, 
blue crosses from \citet{Seale2012}, 
red and salmon plus symbols from \citet{Gruendl_Chu2009}, 
yellow up-tridents from \citet{Chen2009,Chen2010}, and
green left-tridents from \citet{Carlson2012}.
}
\label{IntensityMaps}
\end{figure*}


\begin{figure*}
\centering
\vspace{20pt}
\begin{tabular}{ccc}
\vspace{10pt}
{\bf \large N11B} & {\bf \large N11C} & {\bf \large N44}\\
&\\
{\large HCO$^+$(2$-$1)} & 
{\large HCO$^+$(2$-$1)} & 
{\large HCO$^+$(2$-$1)} \\
\includegraphics[height=4.8cm]{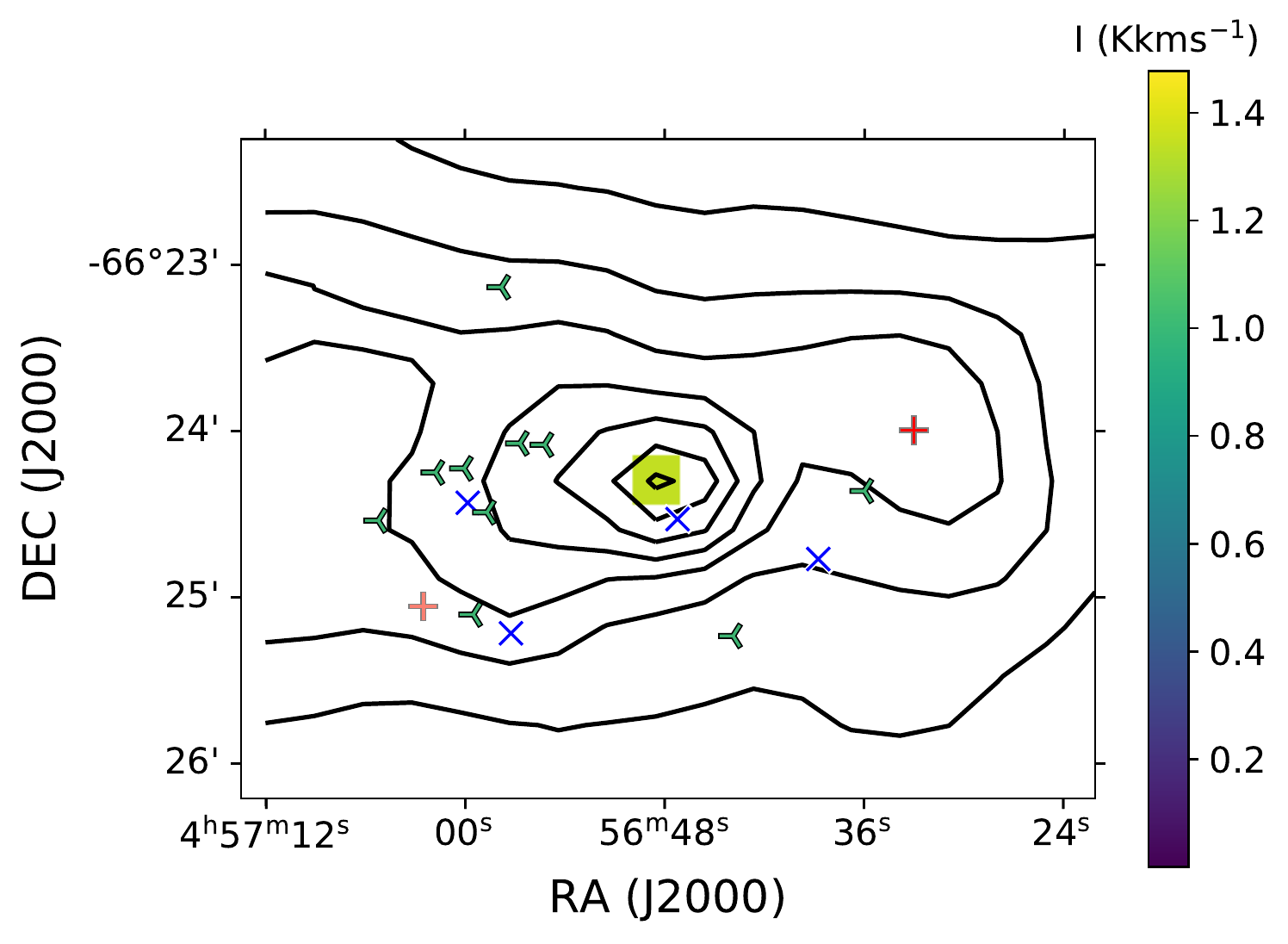}  &
\hspace{-0.5cm}\includegraphics[height=4.8cm]{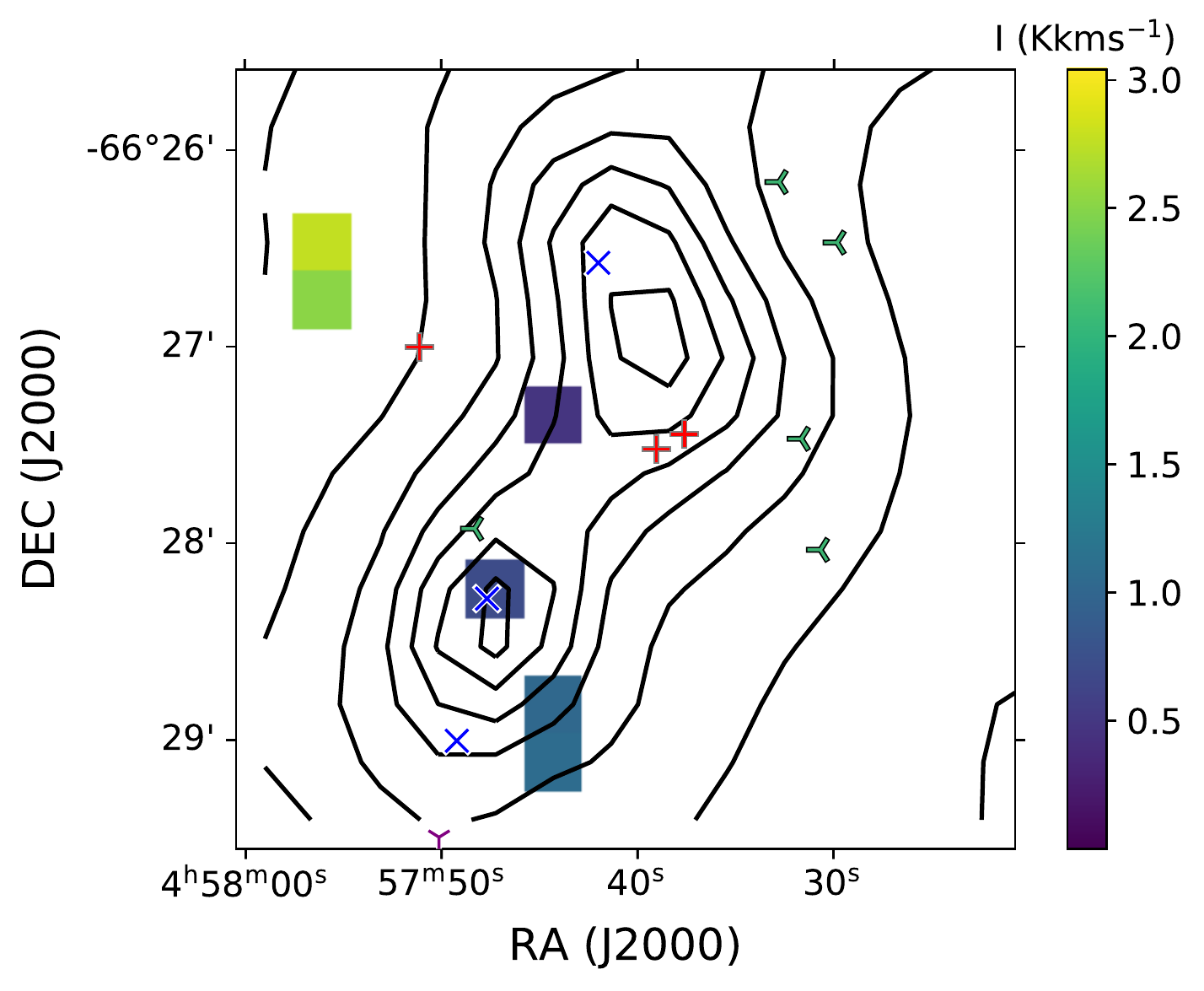} &
\hspace{-0.5cm}\includegraphics[height=4.8cm]{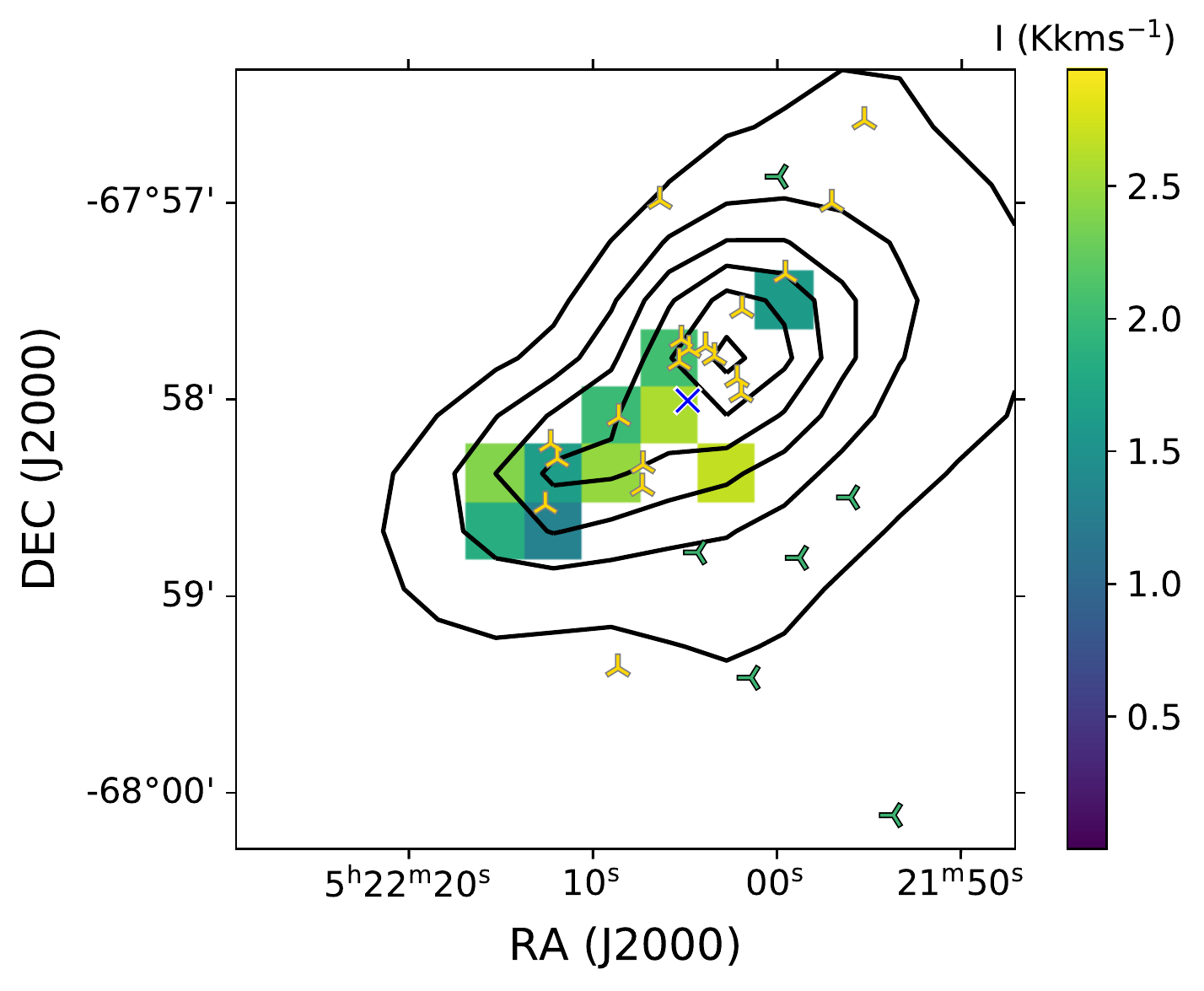}  \\
\includegraphics[height=4cm]{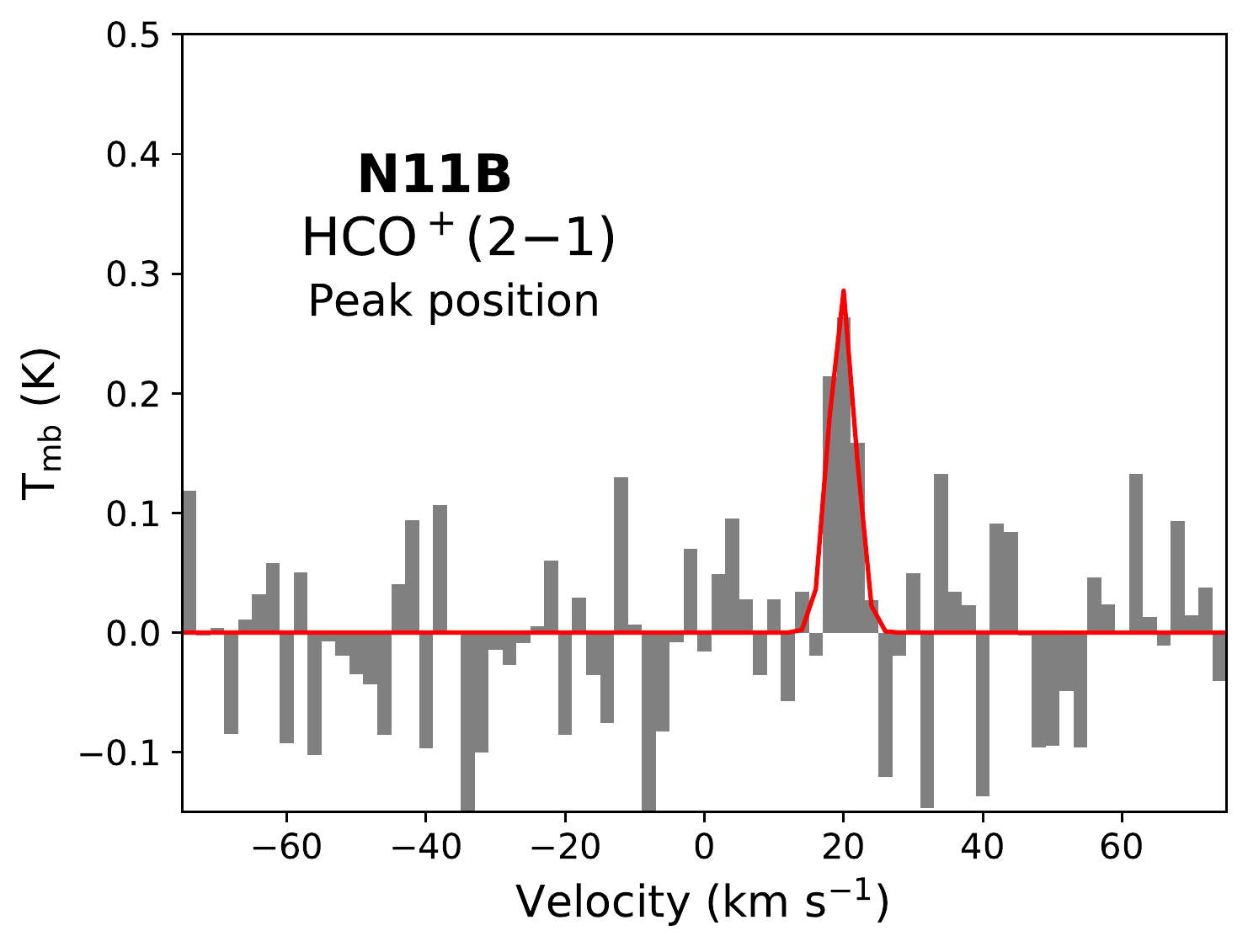}  &
\hspace{-0.5cm}\includegraphics[height=4cm]{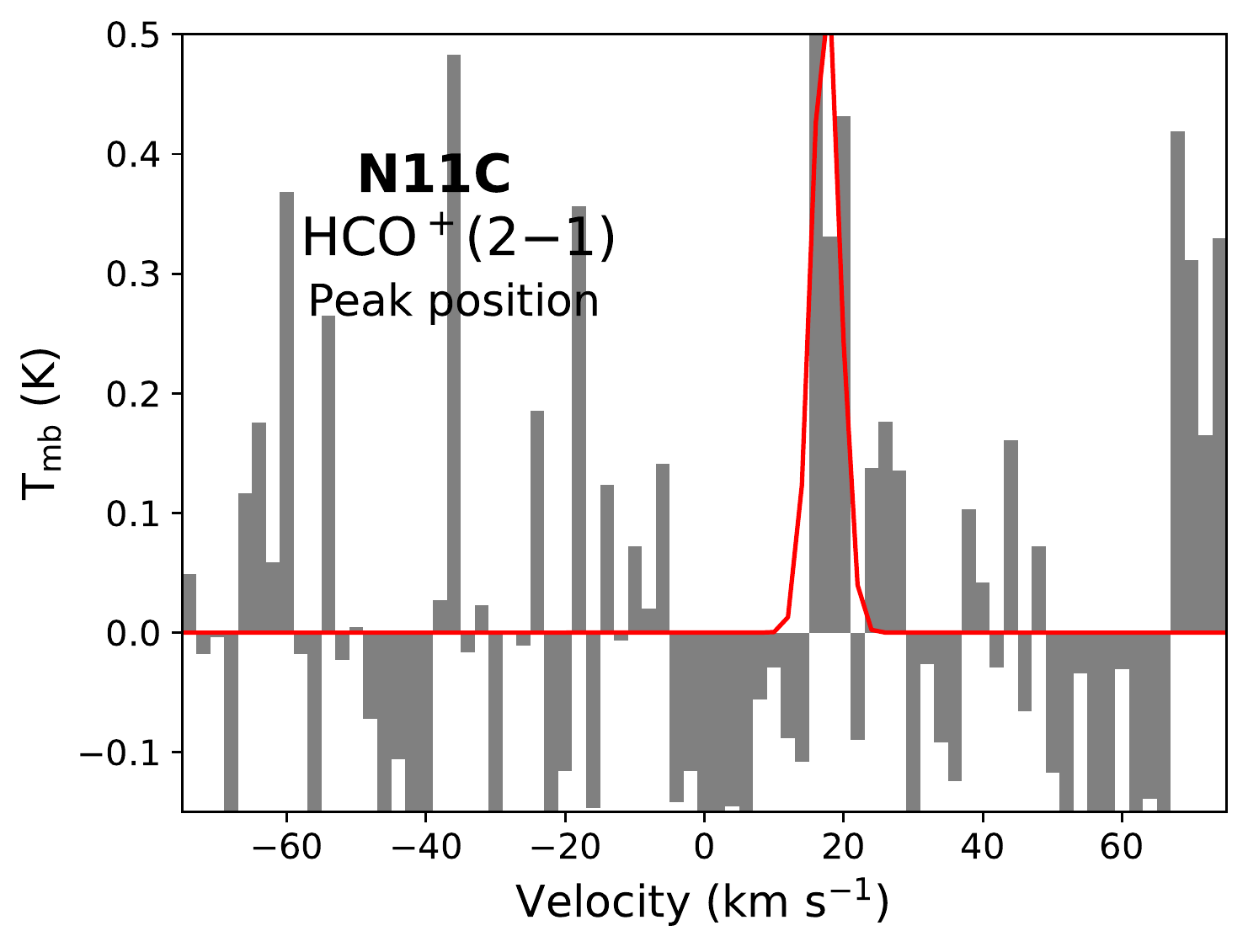} &
\hspace{-0.5cm}\includegraphics[height=4cm]{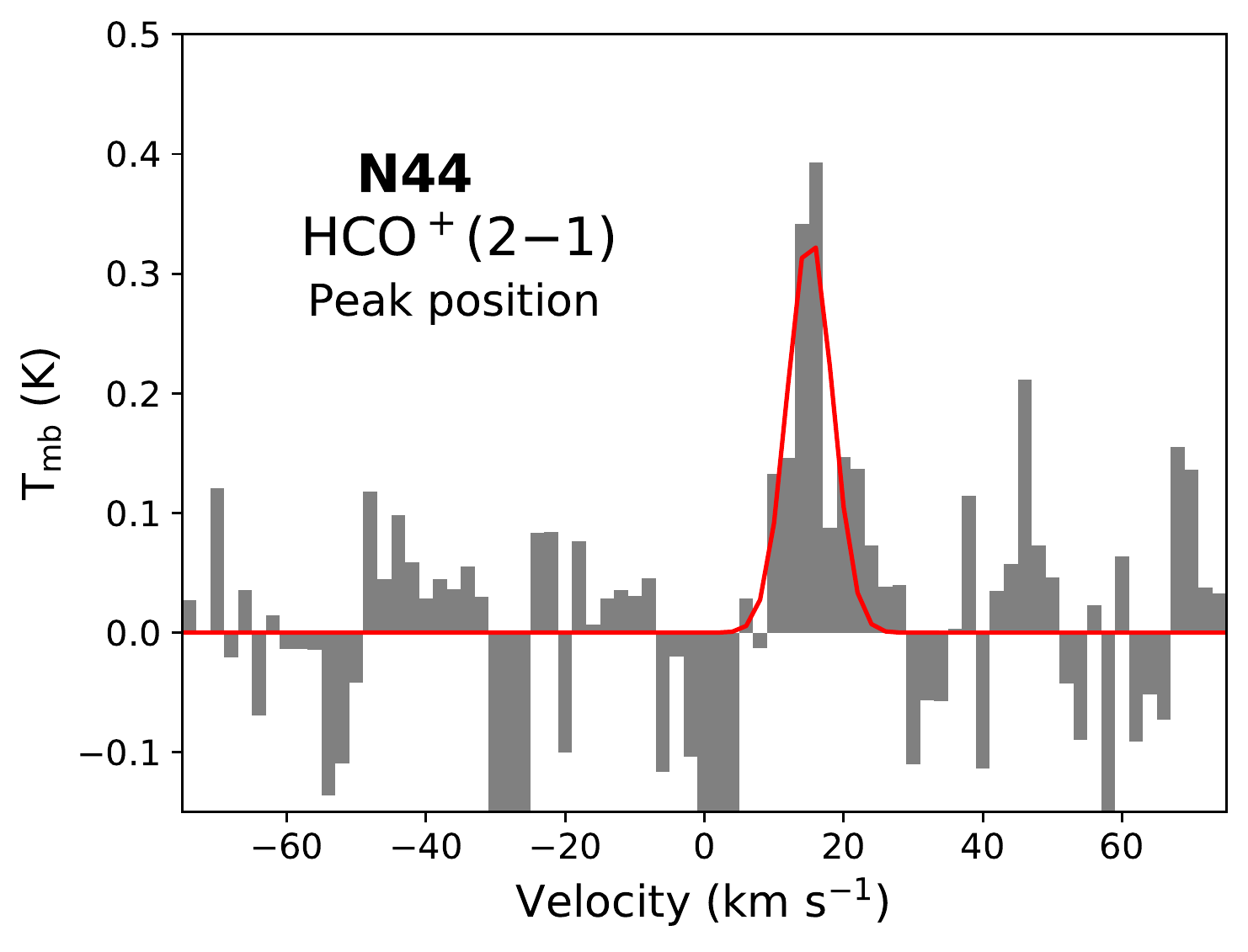} \\
&\\
{\large HCN(2$-$1)} & 
{\large HCN(2$-$1)} & 
{\large HCN(2$-$1)} \\
\includegraphics[height=4.8cm]{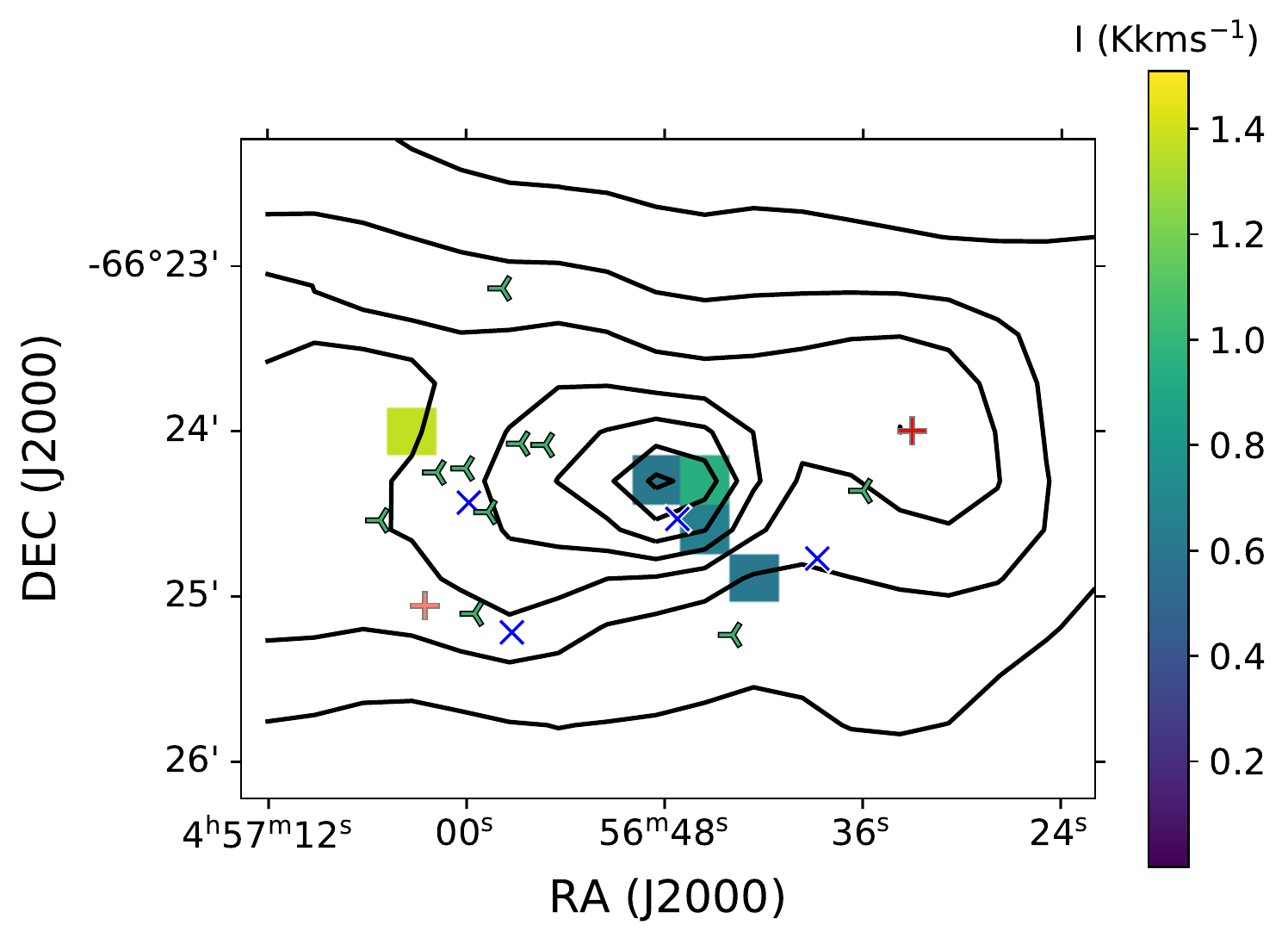}  &
\hspace{-0.5cm}\includegraphics[height=4.8cm]{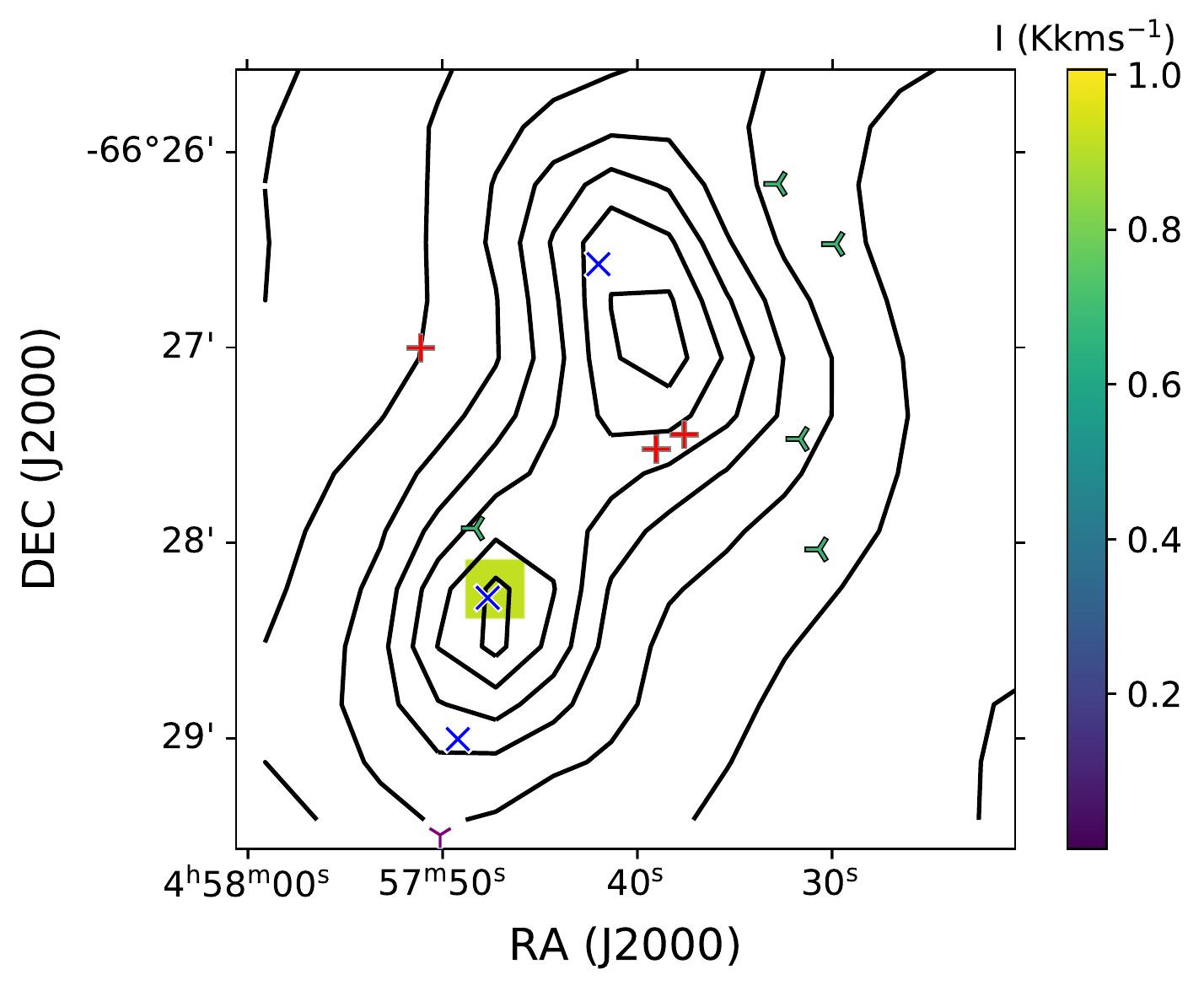} &
\hspace{-0.5cm}\includegraphics[height=4.8cm]{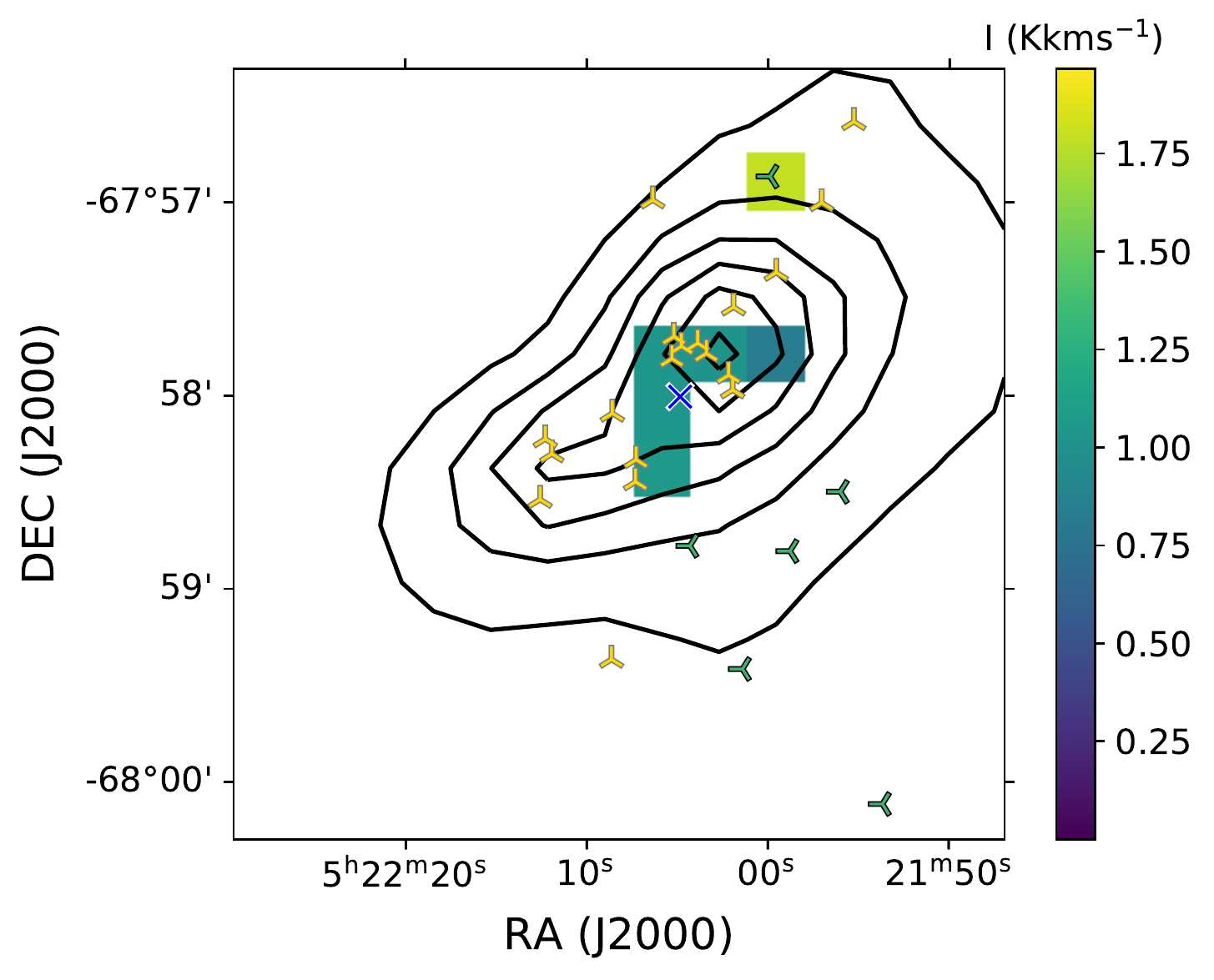}  \\
\includegraphics[height=4cm]{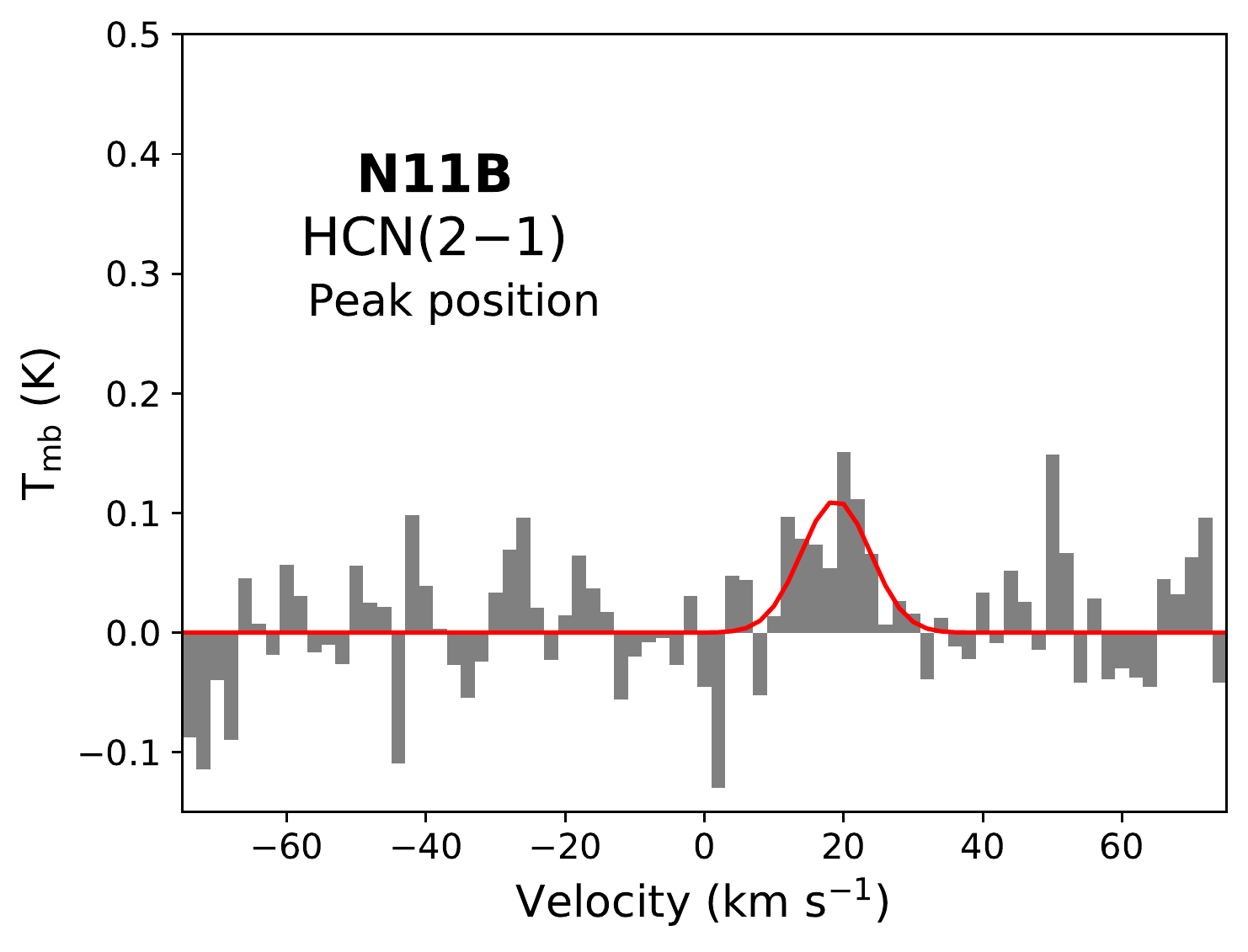}  &
\hspace{-0.5cm}\includegraphics[height=4cm]{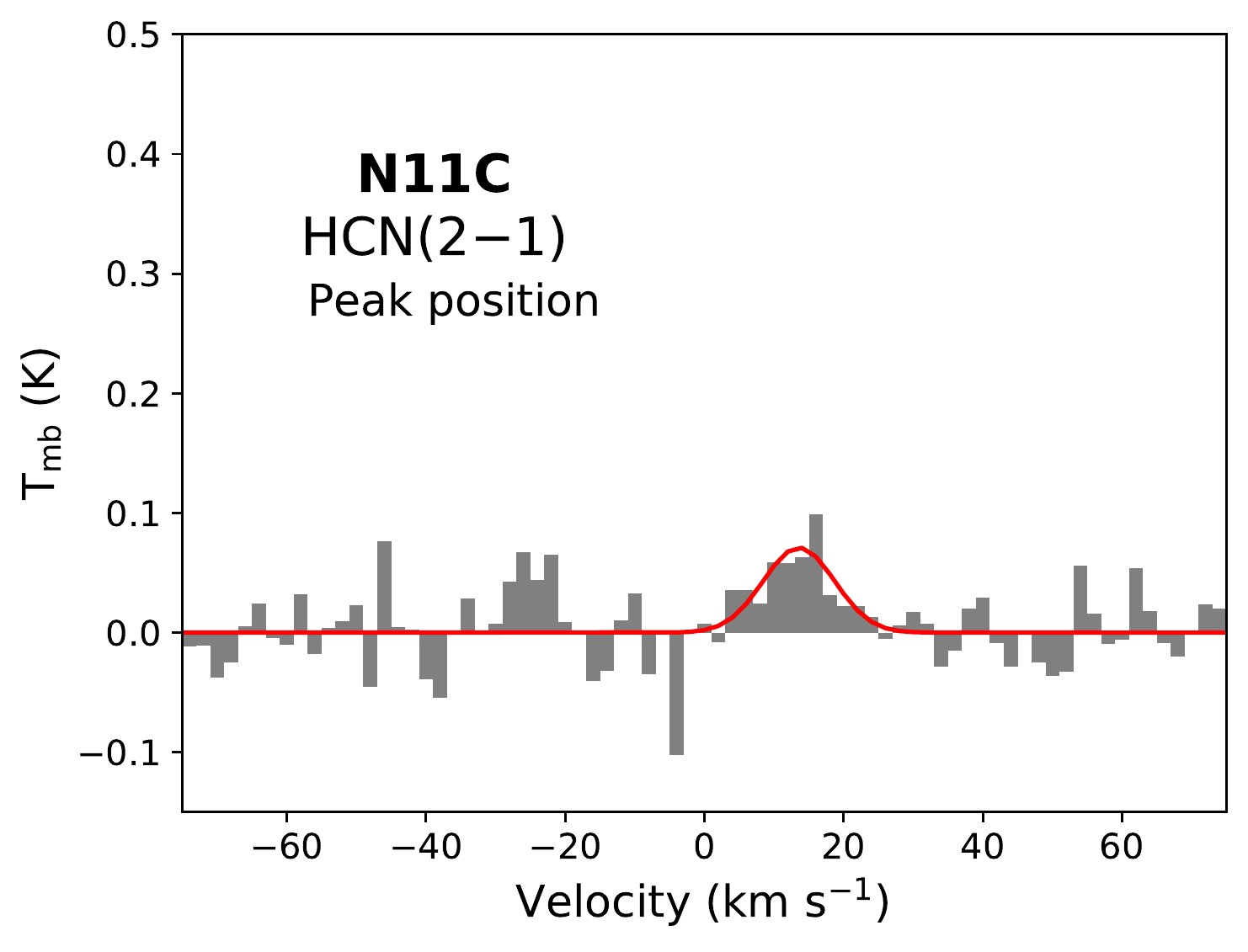} &
\hspace{-0.5cm}\includegraphics[height=4cm]{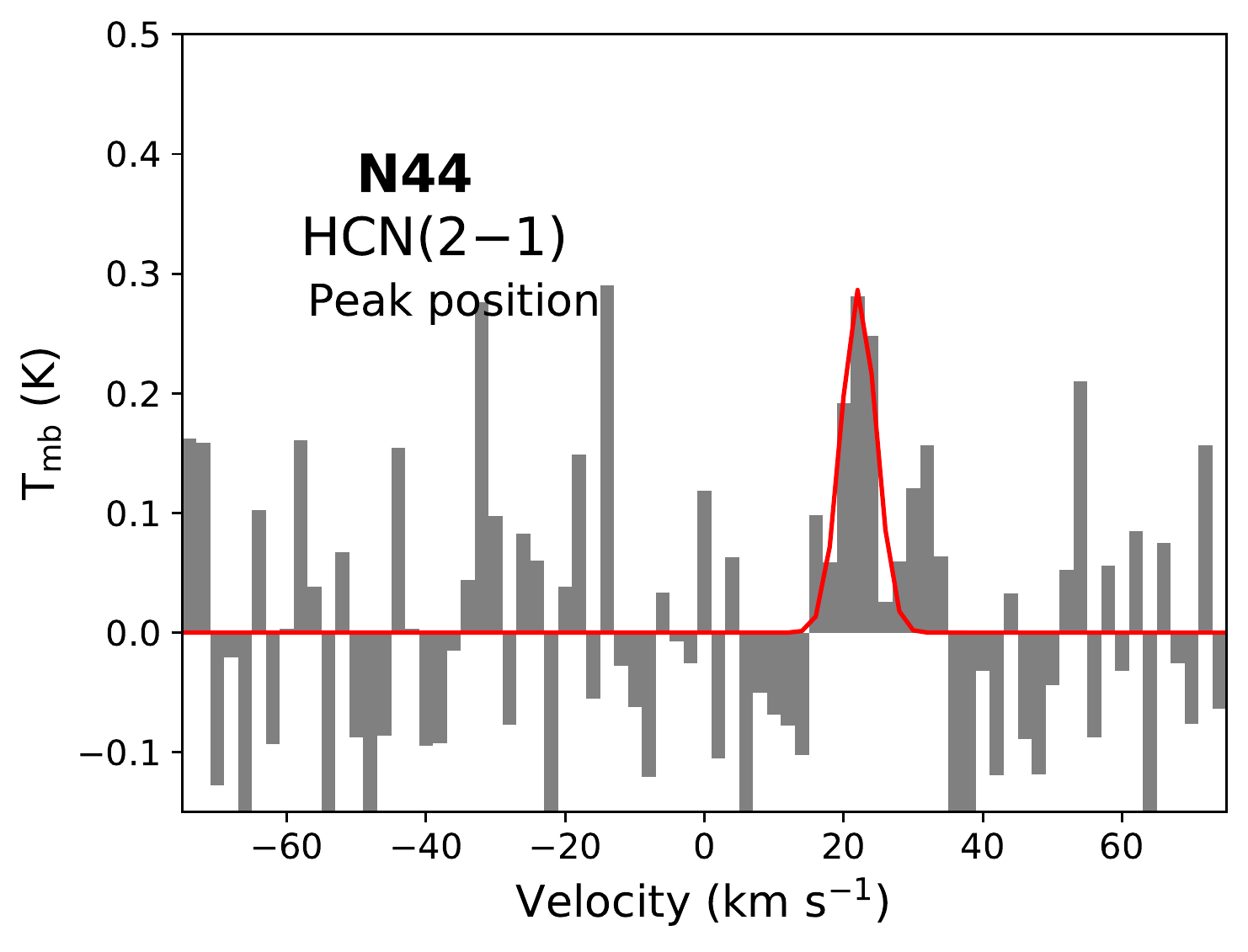}  \\
\end{tabular}
\vspace{10pt}
\caption{Same as Fig.~\ref{IntensityMaps}, but for N11B, N11C, and N44.}
\label{IntensityMaps2}
\end{figure*}


\begin{figure*}
\centering
\vspace{20pt}
\begin{tabular}{ccc}
\vspace{10pt}
 {\bf \large N55} & {\bf \large N105} & {\bf \large N113}\\
&\\
{\large HCO$^+$(2$-$1)} & 
{\large HCO$^+$(2$-$1)} & 
{\large HCO$^+$(2$-$1)} \\
\includegraphics[height=4.8cm]{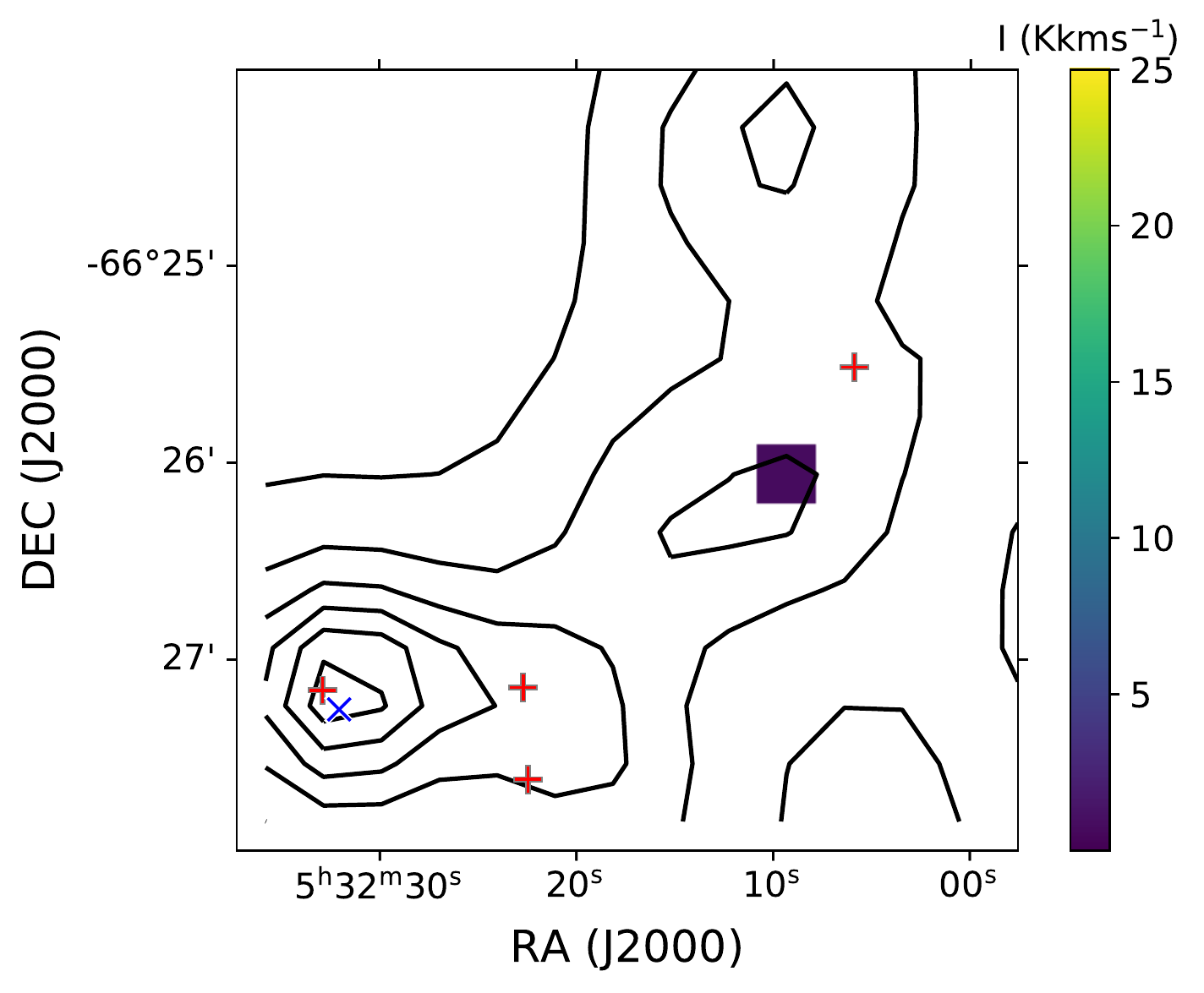}  &
\hspace{-0.5cm}\includegraphics[height=4.8cm]{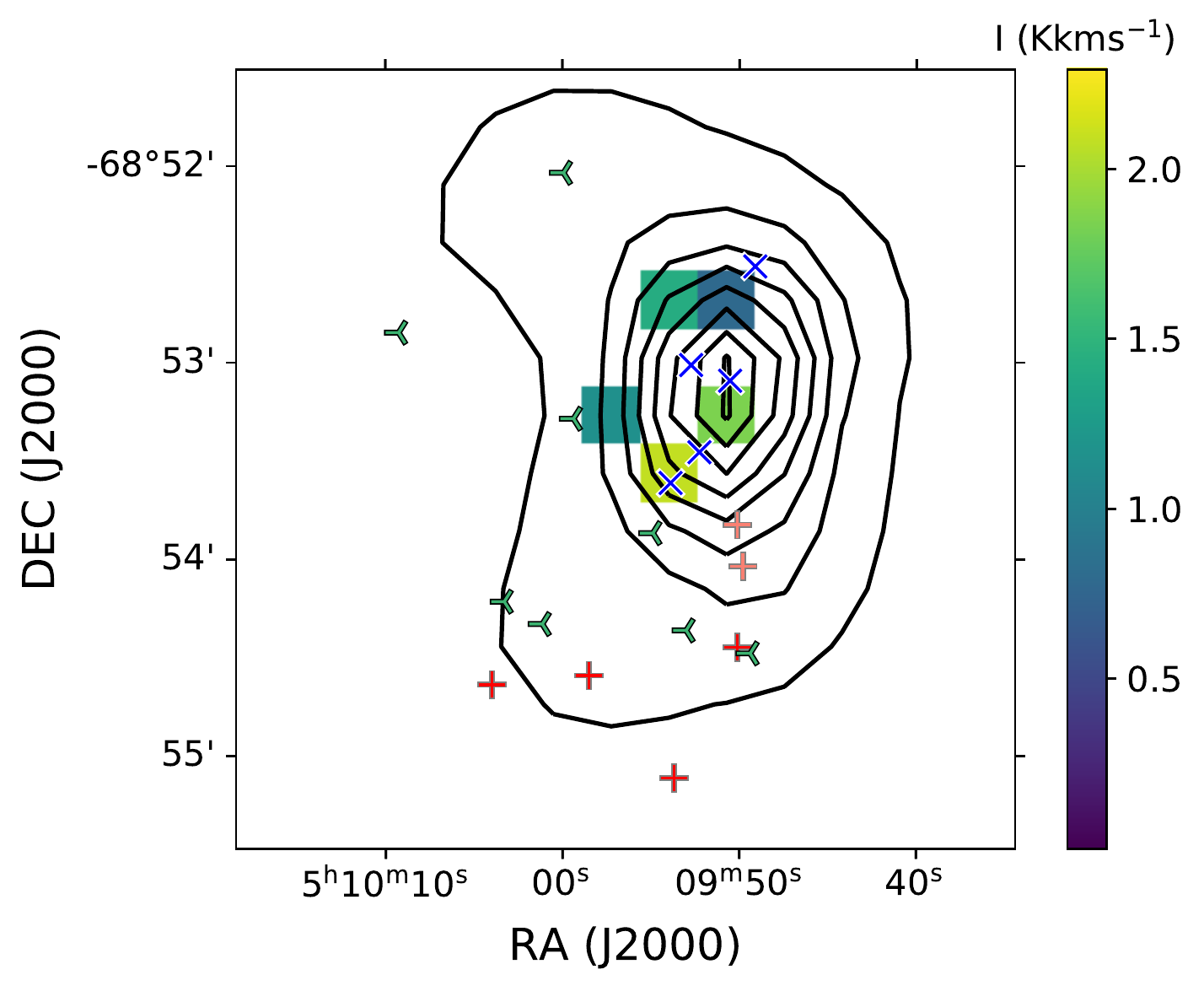} &
\hspace{-0.5cm}\includegraphics[height=4.8cm]{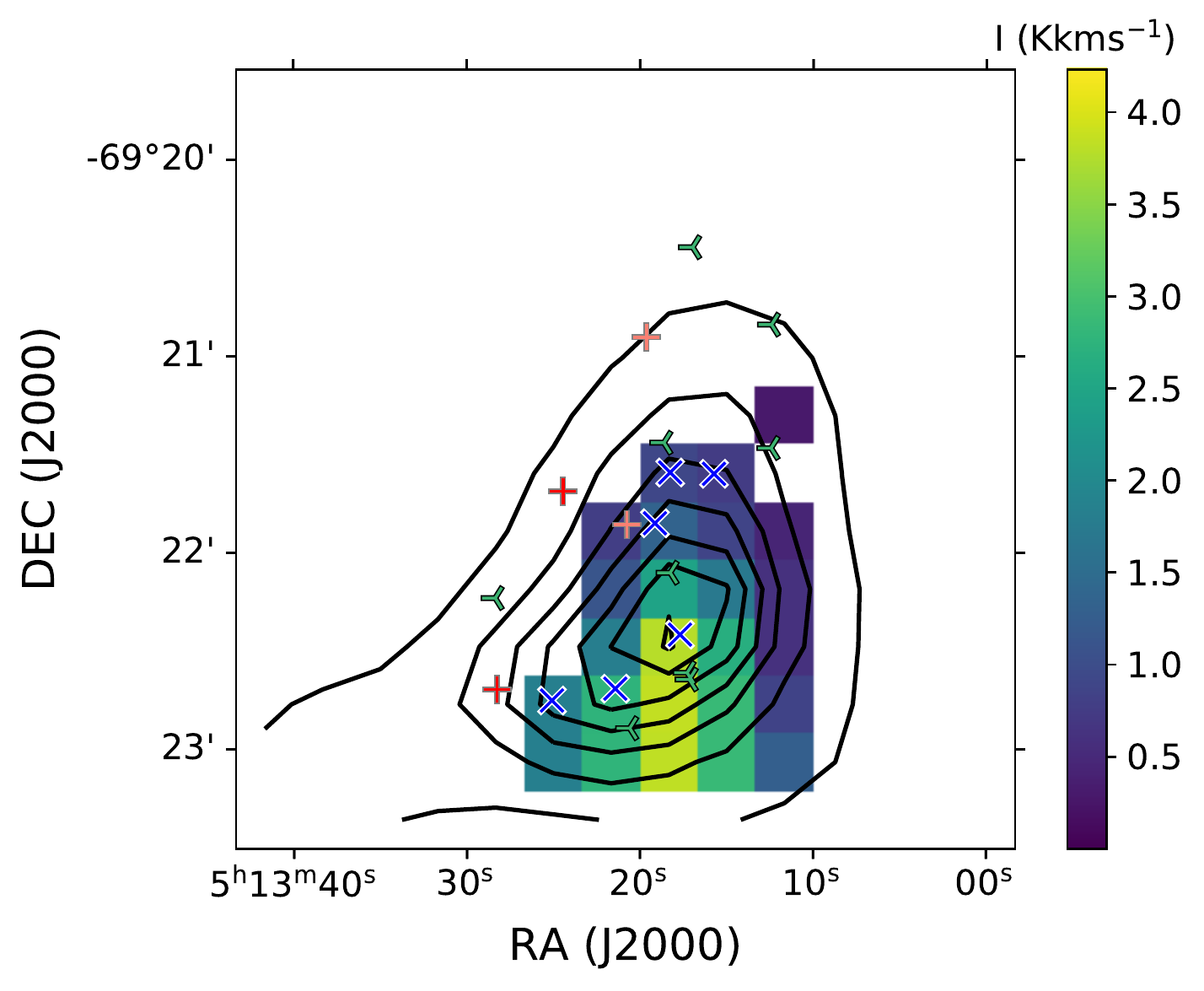}  \\

\includegraphics[height=4cm]{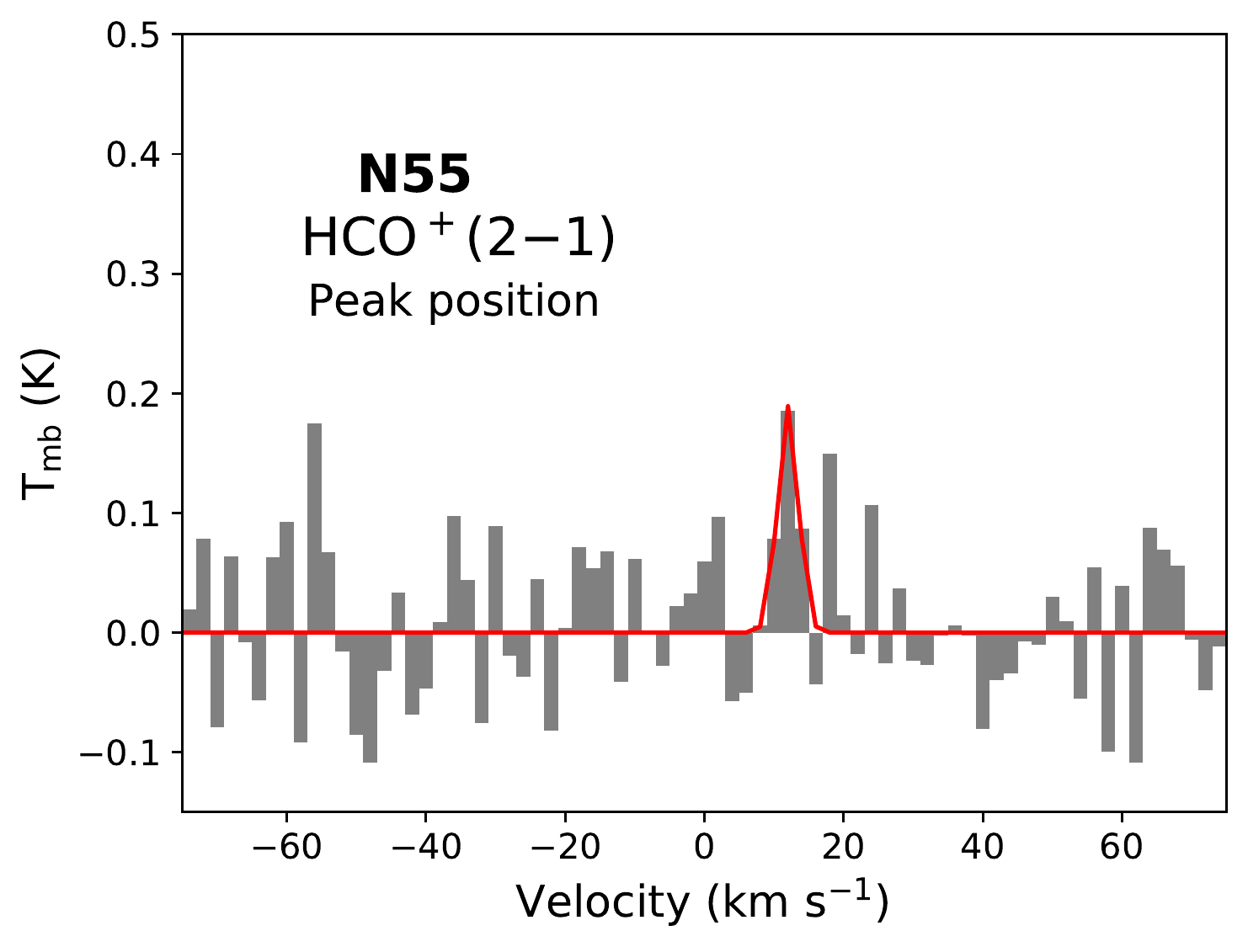}  &
\hspace{-0.5cm}\includegraphics[height=4cm]{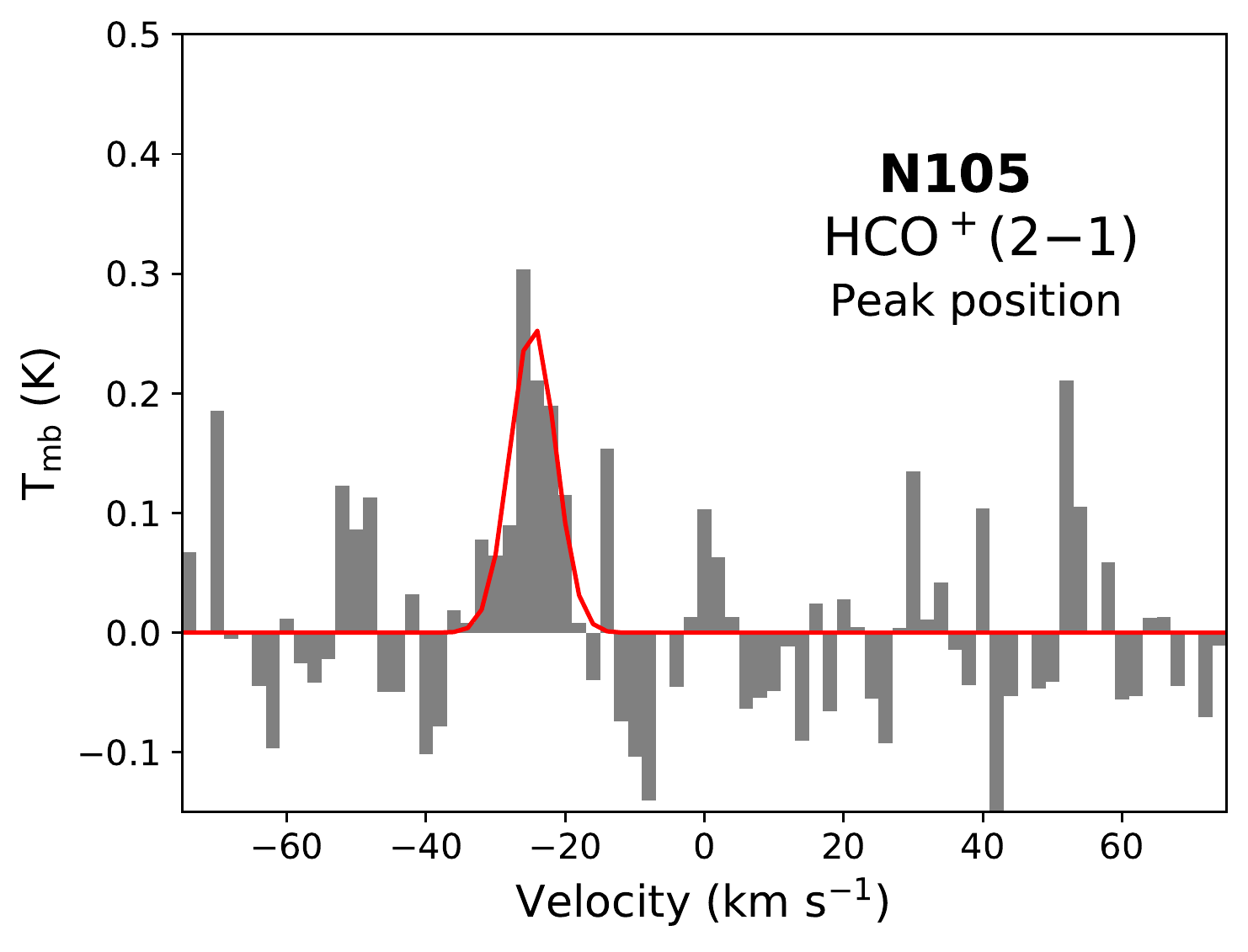} &
\hspace{-0.5cm}\includegraphics[height=4cm]{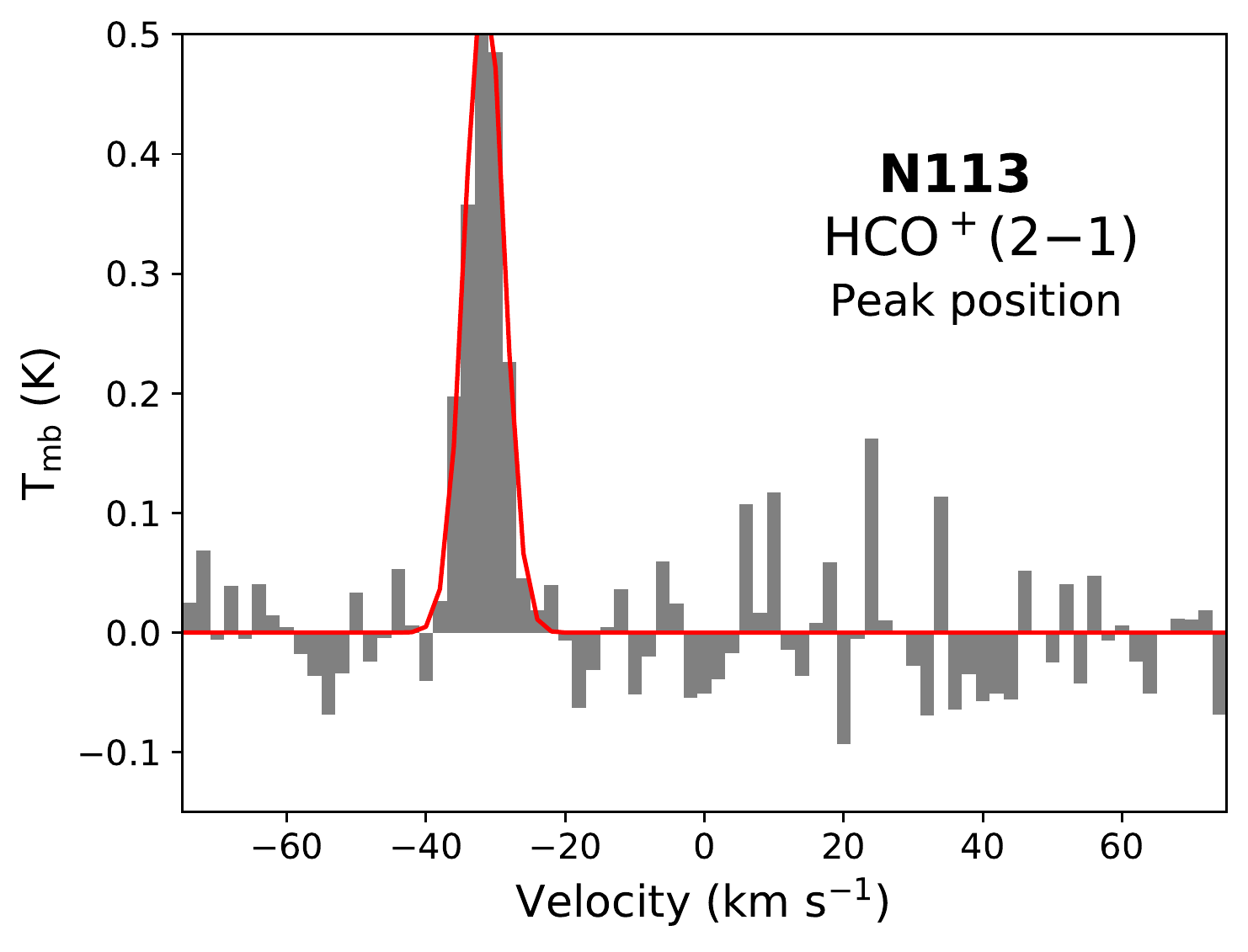}  \\
&\\
{\large HCN(2$-$1)} & 
{\large HCN(2$-$1)} & 
{\large HCN(2$-$1)} \\
\hspace{-0.5cm}\includegraphics[height=4.8cm]{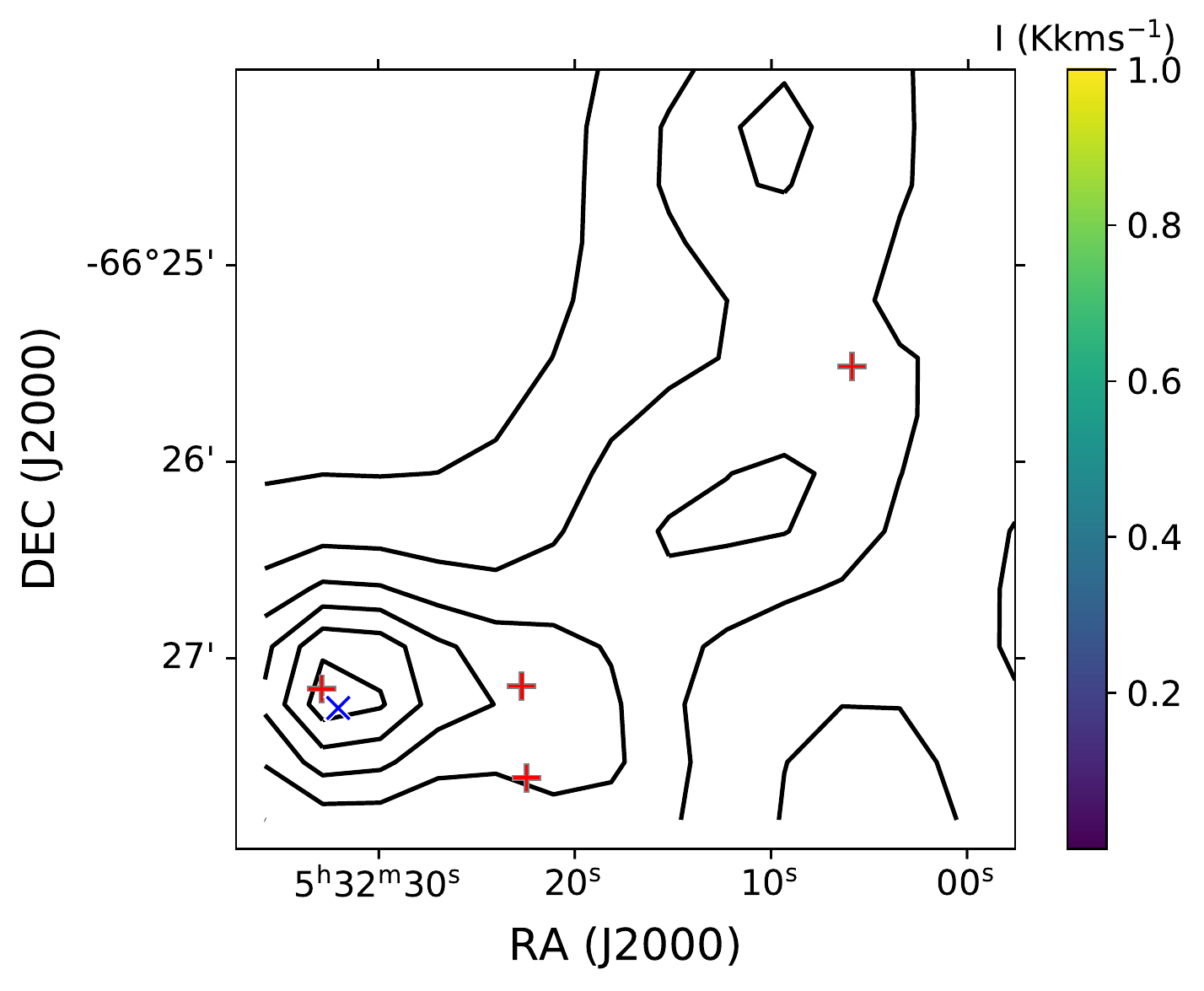}  &
\hspace{-0.5cm}\includegraphics[height=4.8cm]{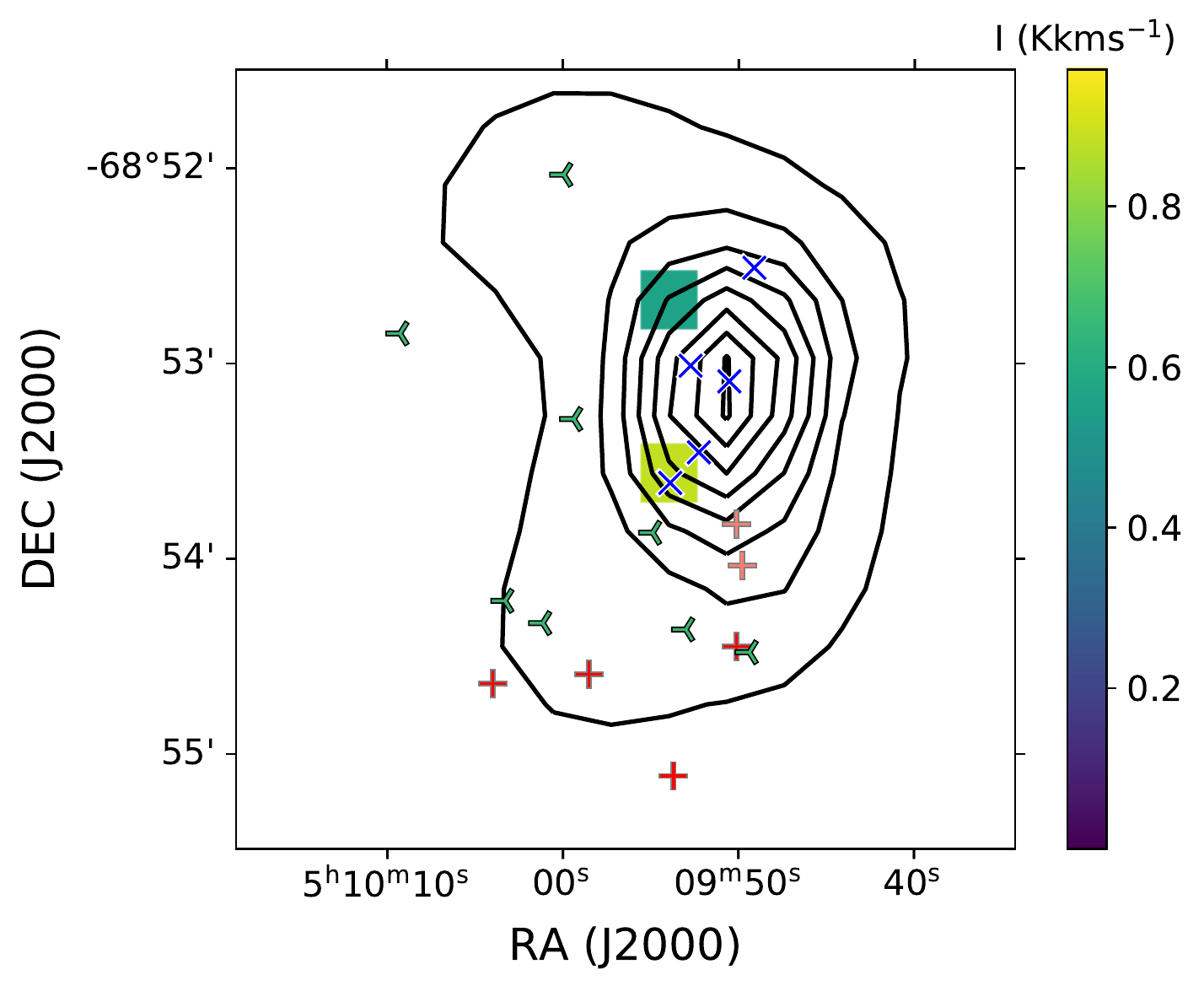} &
\includegraphics[height=4.8cm]{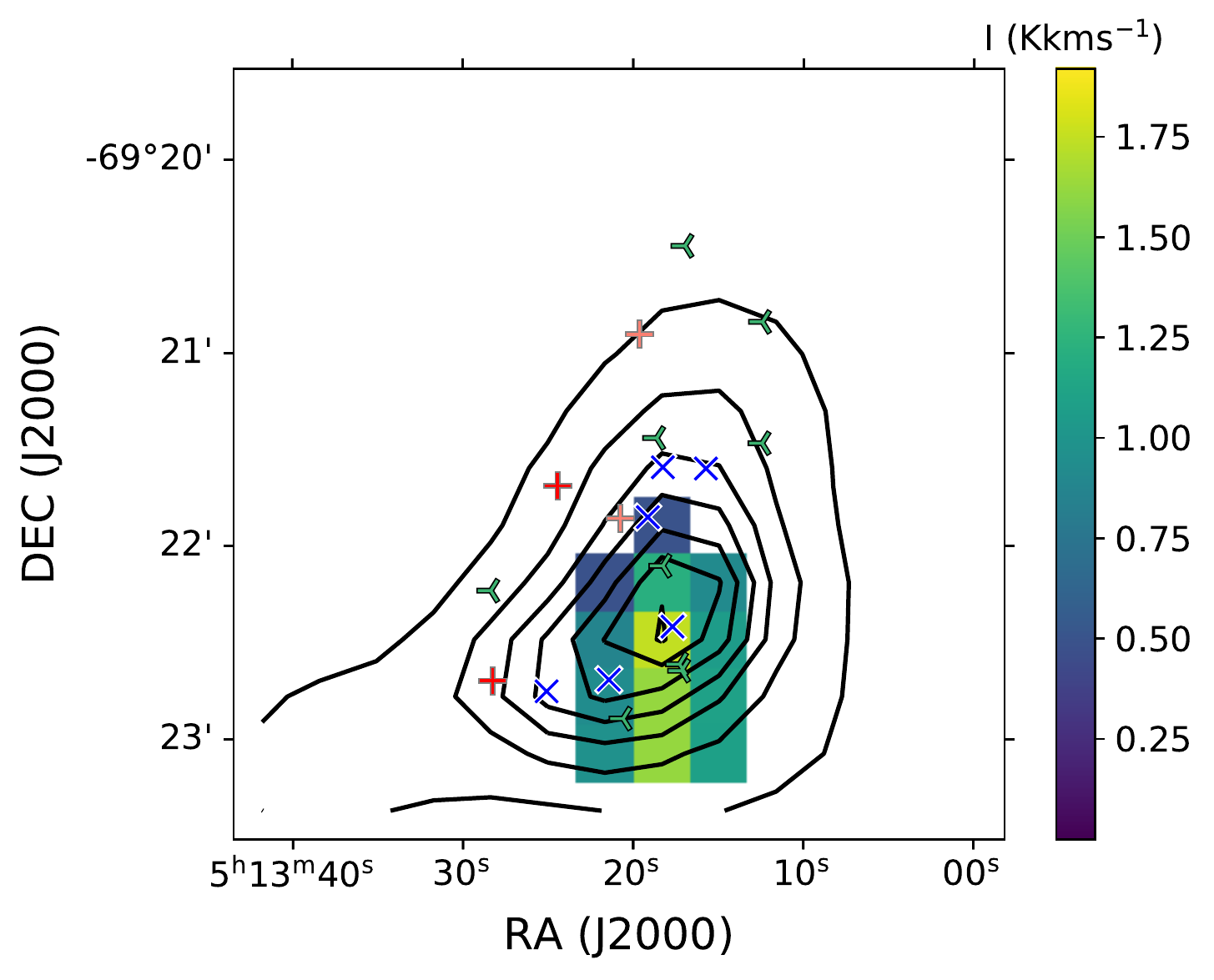}  \\
& \hspace{-0.5cm}\includegraphics[height=4cm]{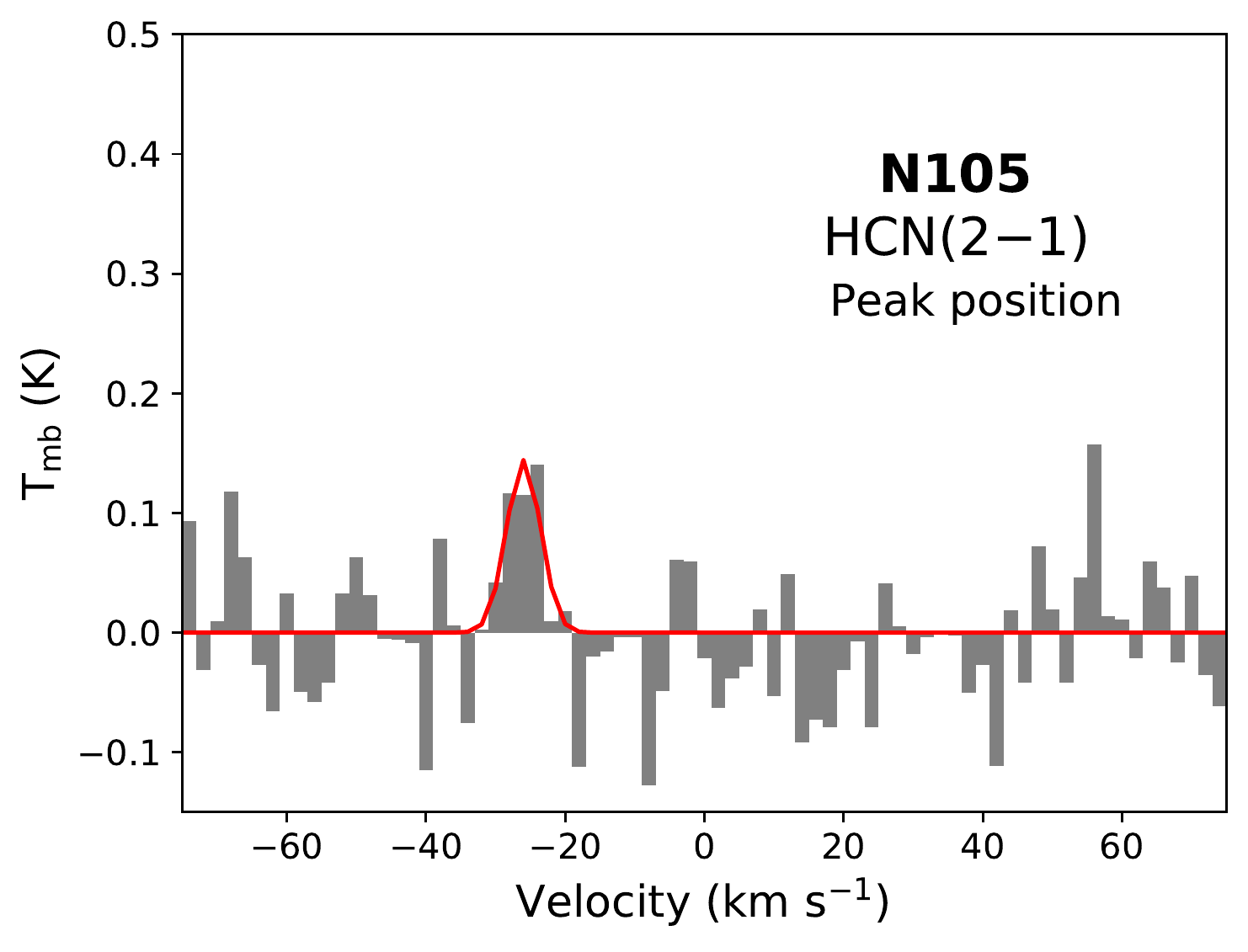} &
\hspace{-0.5cm}\includegraphics[height=4cm]{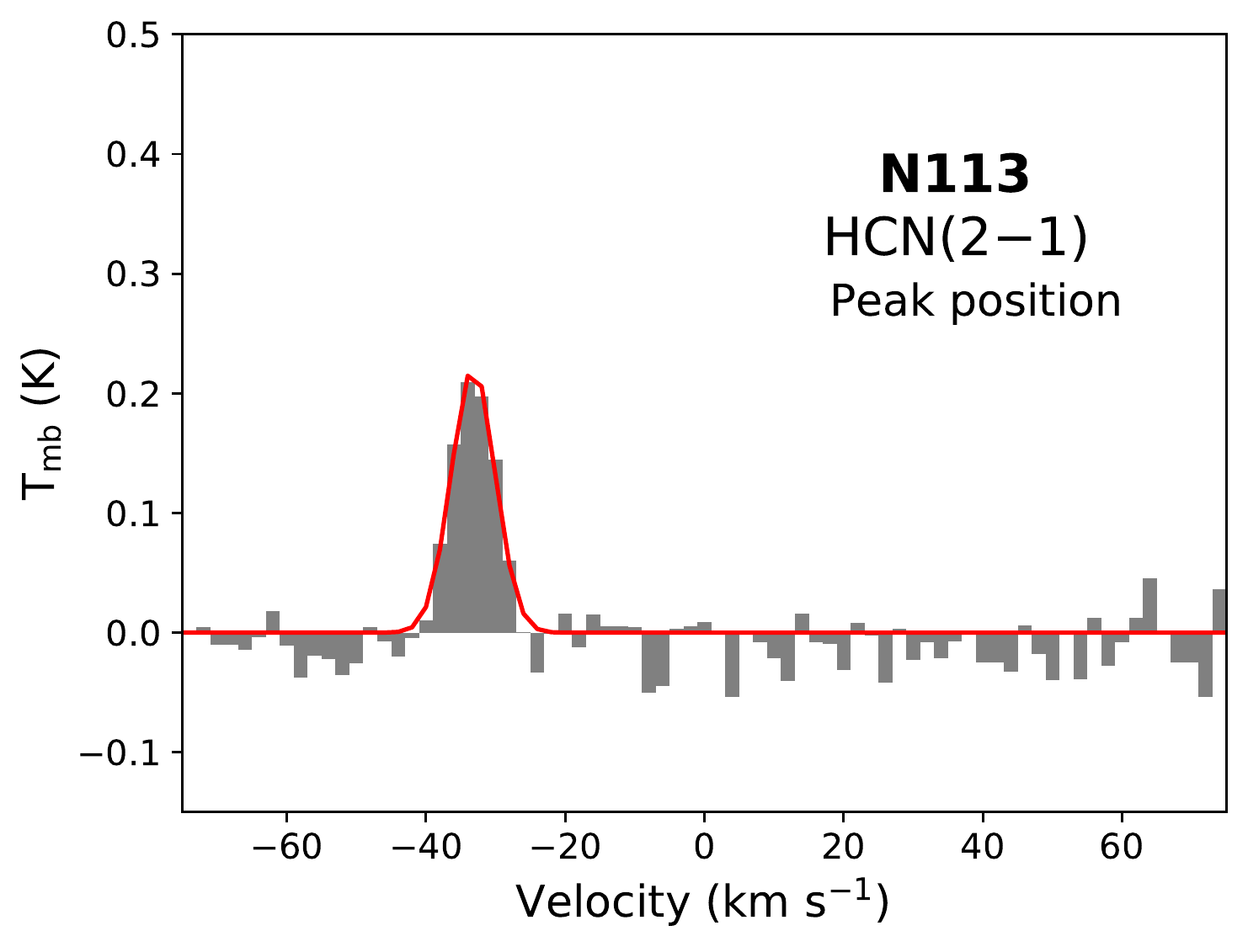}  \\
\end{tabular}
\vspace{10pt}
\caption{Same as Fig.~\ref{IntensityMaps}, but for N55, N105, and N113.}
\label{IntensityMaps3}
\end{figure*}


\begin{figure}
\centering
\vspace{20pt}
\begin{tabular}{ccc}
\vspace{10pt}
 {\bf \large N214}\\
\\
{\large HCO$^+$(2$-$1)} \\
\includegraphics[height=4.8cm]{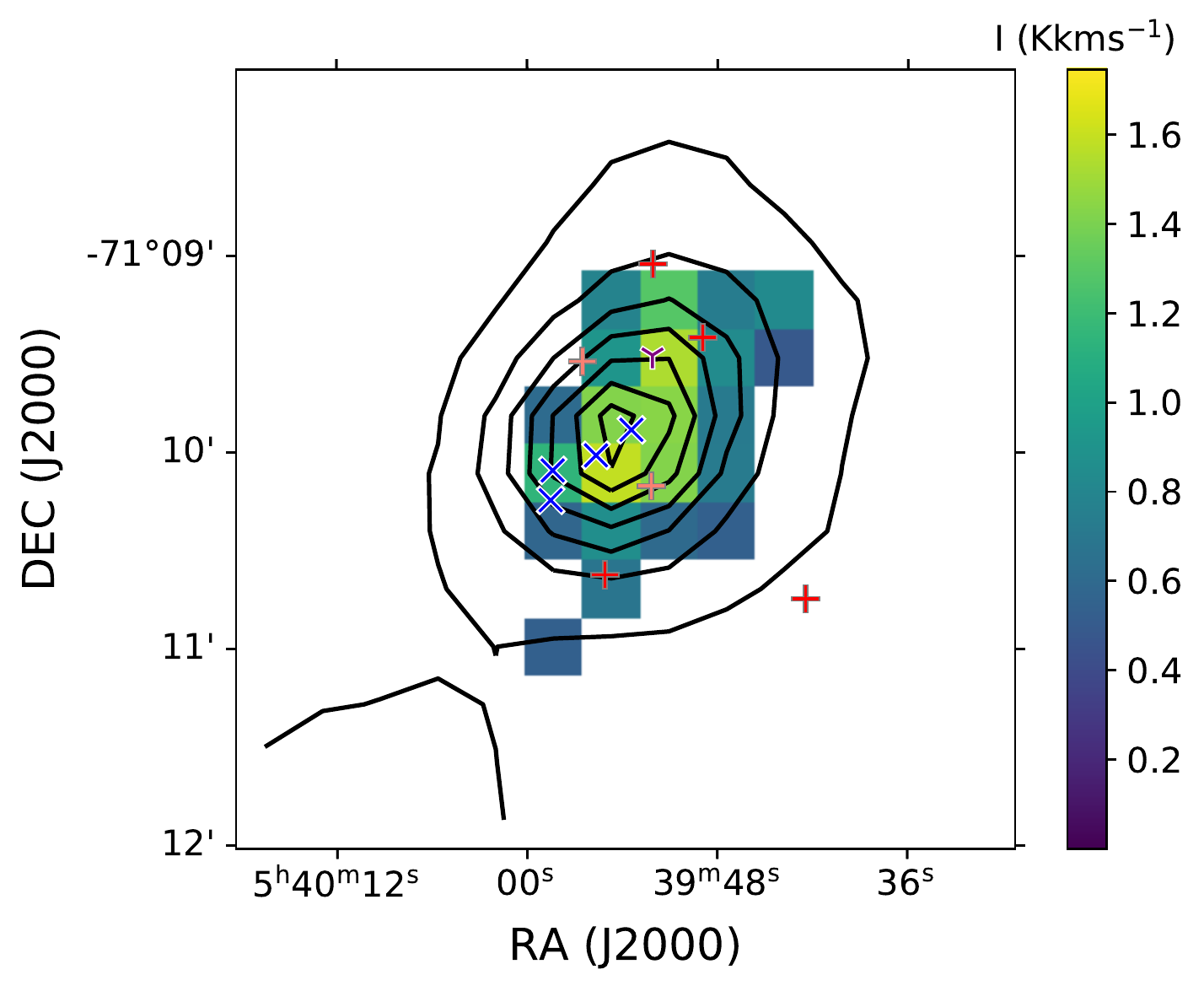}  \\
\includegraphics[height=4cm]{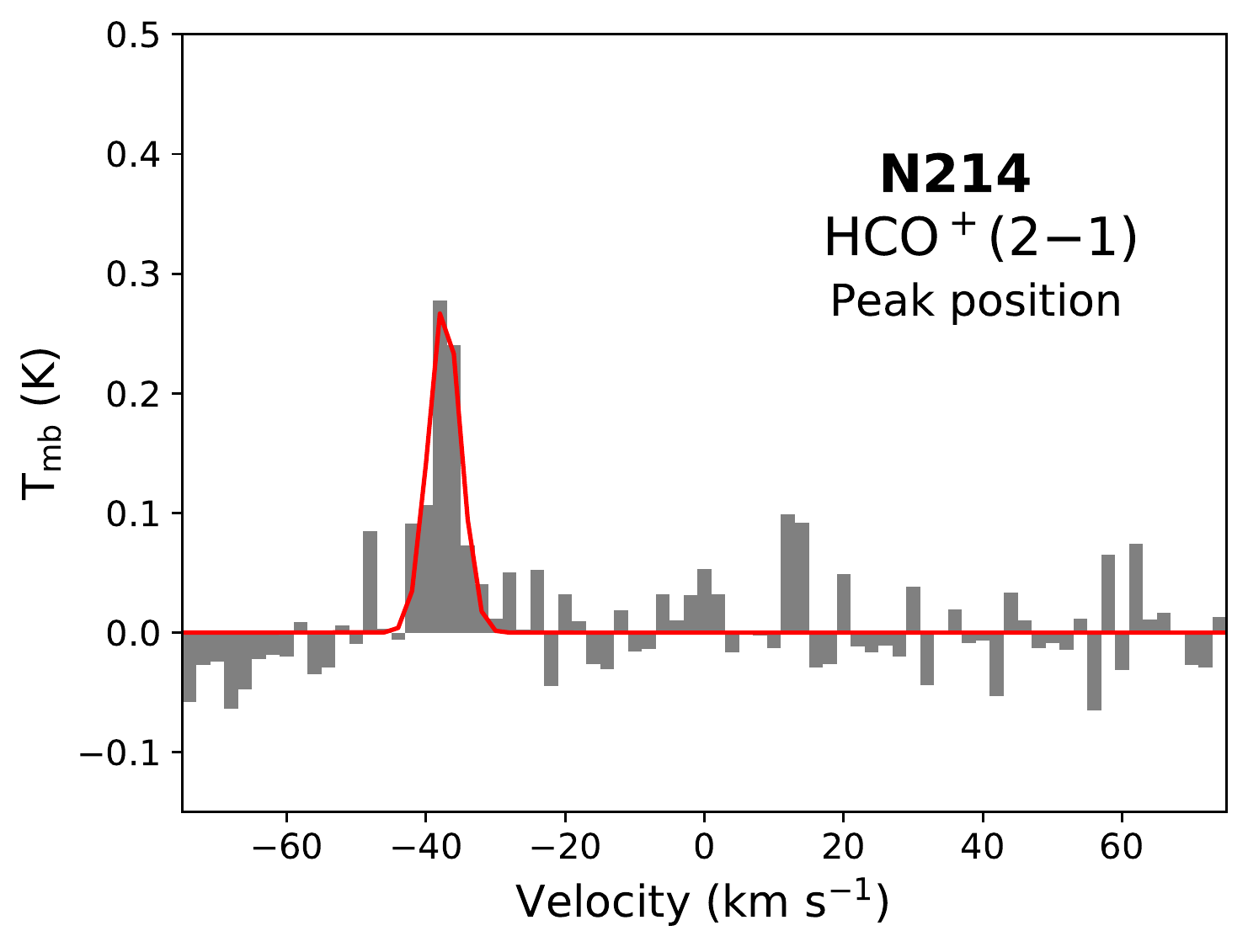}    \\
\\
{\large HCN(2$-$1)}  \\
\includegraphics[height=4.8cm]{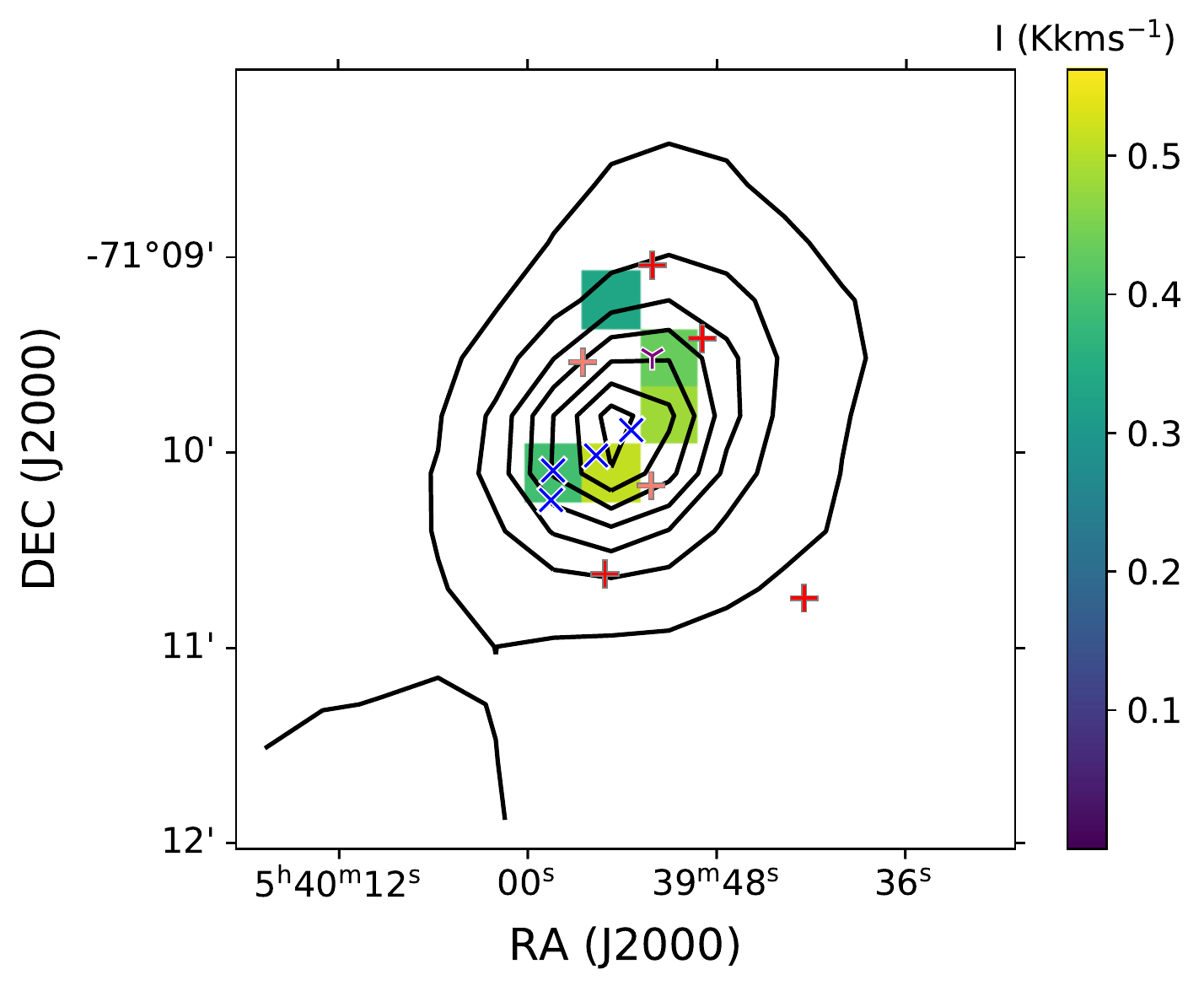}  \\
\includegraphics[height=4cm]{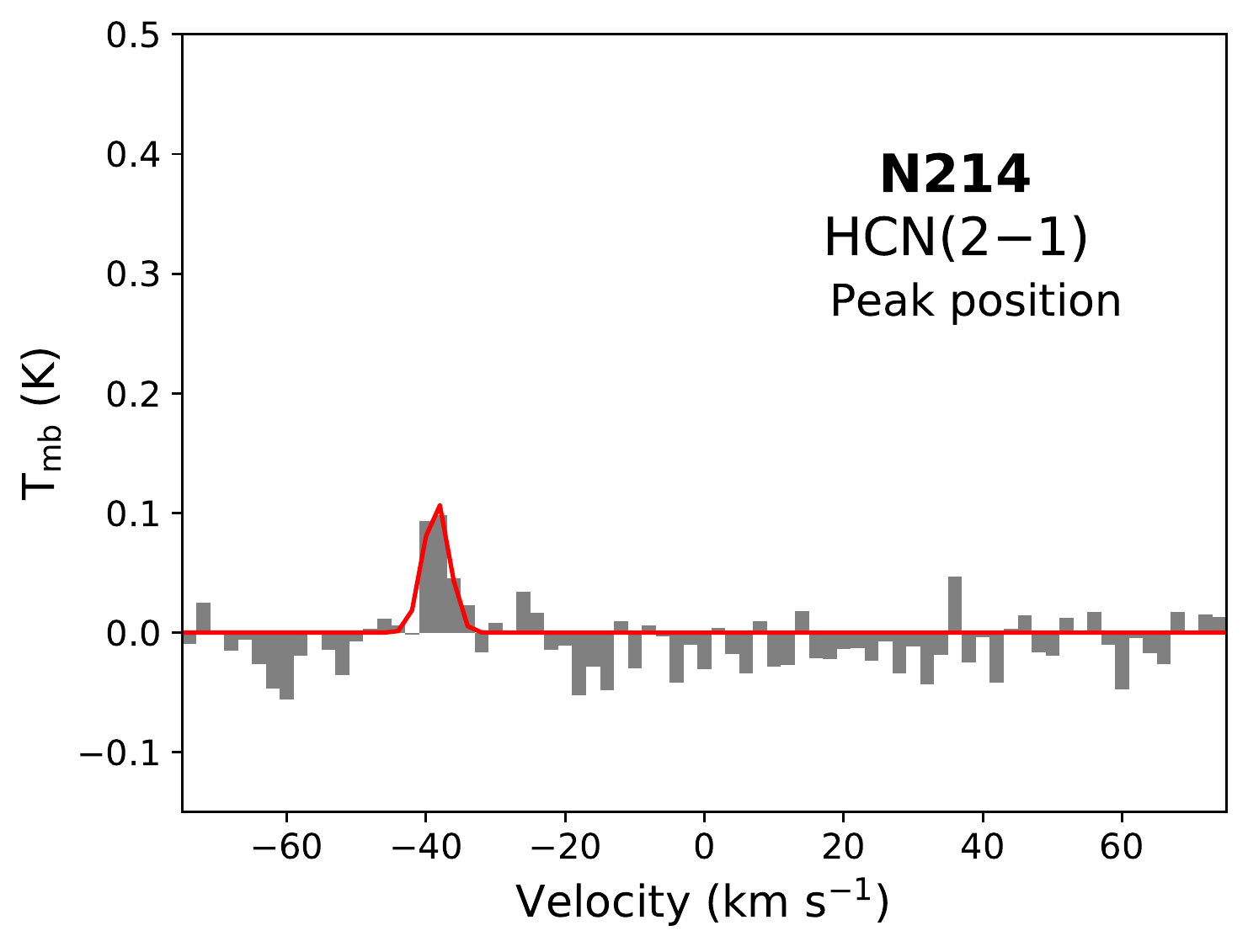}  \\
\end{tabular}
\vspace{10pt}
\caption{Same as Fig.~\ref{IntensityMaps}, but for N214.}
\label{IntensityMaps4}
\end{figure}

\subsection{The DeGaS-MC mapping campaign}
\label{section:mc}

\vspace{5pt}
\subsubsection{Source selection}
Following our successful detections of HCO$^+$(2$-$1) and HCN(2$-$1) in the LMC and SMC, we initiated 
a follow-up mapping campaign. The original selection of 28 sources included regions in which one or both tracers 
were detected during the pointing campaign. We also added regions that were not targeted during the pointing campaign, 
including known bright star-forming regions (i.e., N11B, N105, N113 in the LMC, N66 in the SMC) or 
sources with previously published HCO$^+$(1$-$0) and HCN(1$-$0) data (i.e., N214). Regions where deep YSO analyses had been carried out \citep{Chen2009,Chen2010,Carlson2012,Oliveira2013} were targeted in priority. 
Thirteen sources were finally mapped during the allocated time: ten in the LMC and
three in the SMC. The observed sources are listed in Table~\ref{LineCharacteristics2} and the mapping positions overlaid 
with yellow squares on Fig.~\ref{Pointings}.

\vspace{5pt}
\subsubsection{Observations}
The mapping campaign benefited from 6h of Director's Discretionary Time (Project ID: E-297.C-5052A-2016; PI: Galametz) 
and 140 h of {\bfaa allocated} time in Period 99 (OSO and Max-Planck APEX time; Project ID: O-099.F-9301A-2017; PIs: Galametz \& Schruba). 
The DDT observations, carried out in August 2016, were the first to demonstrate the feasibility of mapping extragalactic 
dense molecular clouds with the SEPIA180 receiver using on-the-fly mode. 
The rest of the observations were carried out from April to June 2017, with 0.8$<$pwv$<$2.5 mm for most of the data 
(pwv$\sim$4 on May 20 2017). The size of the area covered around the sources is of about 50-70pc.

\subsection{Data reduction}

Data of the pointing and mapping campaigns were reduced using CLASS/GILDAS \footnote{Gildas Continuum and Line Analysis Single-dish 
Software, \url{http://www.iram.fr/IRAMFR/GILDAS}} \citep{Pety2005}. For the pointing campaign data, we removed a linear baseline from each individual 
spectrum before averaging the data. The final spectra were smoothed to a velocity resolution of 2 km~s$^{-1}$. 
Following \citet{Belitsky2018}, we assume an antenna forward efficiency $\eta_{f}$ of 0.95, and a beam efficiency $\eta_{mb}$ of 0.67 
to convert the antenna temperature T$^*_A$ into $\mathrm{T_{mb}}$ \footnote{\url{http://www.eso.org/sci/activities/apexsv/sepia/sepia-band-5.html}}. 
The rms reached during the pointing campaign ranges (in T$^{mb}$) from 7.8 to 17.6 mK for both Clouds. The exact rms value for each line of sight is indicated in Table~\ref{LineCharacteristics1}. 
For the mapping campaign data, two linear baseline subtractions were performed before and after a resampling of the data to 
a resolution of 2 km~s$^{-1}$. The final pixel size of the spectral cubes is half a beam size (18\arcsec). The rms reached throughout the 
maps (estimated with {\it go noise} in CLASS) are provided in Table~\ref{LineCharacteristics2}.

\subsection{Ancillary data available}

\subsubsection{CO observations}

A $^{12}$CO(1$-$0) mapping of the LMC and SMC has been performed using NANTEN \citep{Fukui2008}
(2.6\arcmin\ resolution).
A follow-up survey was carried out with the Mopra MAGMA project \citep[][]{Hughes2010,Muller2010,Wong2011},
allowing us to trace the structures of GMCs at a resolution of 45\arcsec. 
A cloud decomposition has been performed using the CPROPS decomposition technique \citep{Rosolowsky2006} and catalogs of the 
CO cloud properties are available on the VizieR database (\url{https://vizier.u-strasbg.fr/viz-bin/VizieR?-source=J/ApJS/197/16}).
Observations of $^{12}$CO(1$-$0) in the SMC were conducted as part of the MAGMA-SMC program \citep{Muller2013}
but maps are not released at the time of writing.
Observations of the $^{12}$CO(2$-$1) transition in the SMC were also performed using the SHeFI/APEX-1 receiver
as part of the CO Survey of the Small Magellanic Cloud (COSSA) large programme (ESO periods P92-95) by 
T. van Kempen et al. (in prep) at a 28\arcsec\ resolution.

\subsubsection{Spitzer and Herschel maps}
\label{section:derivingLTIR}

The dust emission in the LMC and SMC has been extensively observed in the IR and submillimeter(submm) regimes. Both Magellanic Clouds were fully 
mapped with \spitz\ as part of the Surveying the Agents of a Galaxy's Evolution project  \citep[SAGE; ][]{Meixner2006}. 
Observations are available at 3.6, 4.5, 5.8, and 8 \mic\ (IRAC) and 24, 70, and 160 \mic\ (MIPS). We refer to \citet{Meixner2006} and \citet{Gordon2014} for a 
detailed description of the various steps of the data reduction. 
Both Clouds were also observed with {\it Herschel} as part of the successor project of SAGE, the HERschel Inventory of The Agents of Galaxy's Evolution project
\citep[HERITAGE; ][]{Meixner2010,Meixner2013}. 
They were mapped at 100 and 160 \mic\ (PACS) and 250, 350, and 500 \mic\ (SPIRE). Details of the \hersc\ data reduction and image treatment 
can be found in \citet{Meixner2013} and \citet{Gordon2014}. 
The coarsest resolution of this IR-to-submm dataset is 36\arcsec, thus equal to the beam of our SEPIA180 observations. 

We convolve all the IR-submm maps to this limiting resolution to be able to locally compare the dense gas tracers with the dust emission and 
the total IR luminosity L$_{TIR}$, an extensively used proxy for the SFR in extragalactic studies \citep{Kennicutt1998,Murphy2011}. 
To estimate L$_{TIR}$, we use the 100-160-250 L$_{TIR}$ empirical calibration derived for nearby galaxies from the \hersc/KINGFISH project by \citet{Galametz2013b}. 
For the DeGaS-MC pointings, we produce a full LMC and SMC L$_{TIR}$ map and the L$_{TIR}$ per pointing using a photometric aperture of the 
size of the SEPIA beam.
For the DeGaS-MC maps, the \hersc\ PACS 100 and 160 \mic\ and SPIRE 250~\mic\ maps are first cut and regridded to our intensity map pixel grids for each region, and were then combined to produce an individual L$_{TIR}$ map for each region targeted. 
Uncertainties on the L$_{TIR}$ are estimated using a Monte-Carlo technique, varying  the \hersc\ fluxes and calibration coefficients 5000 times 
within their error bars following Gaussian distributions. These uncertainties are dominated by the calibration uncertainties of the PACS and 
SPIRE instruments \citep[10\% at 100~\mic\ and 250~\mic, 20\% at 160~\mic;][]{Poglitsch2010,Griffin2010}.

\section{Detection statistics and line characteristics}

\subsection{Pointing campaign}

The spectra obtained during the pointing campaign are provided in Figs~\ref{LMCSMC_HCOp} (HCO$^+$ observations) 
and \ref{LMCSMC_HCN} (HCN observations). We fit the spectra with Gaussian line profiles. 
We thus assume single-component symmetric profiles (we discuss this hypothesis in Section~\ref{section:linewidth}).
We consider the line as detected when the peak brightness is 3-$\sigma$ above the S/N of the Gaussian fit.
HCO$^+$(2$-$1) is detected above 3-$\sigma$ in 15 out of the 21 pointings observed in the LMC, HCN(2$-$1) in six of them.
HCO$^+$(2$-$1) is detected in 5 out of the 8 regions observed in the SMC, HCN(2$-$1) in two of them.
For both Magellanic Clouds, HCN(2$-$1) is only detected in the brightest HCO$^+$ sources, namely 
those whose HCO$^+$(2$-$1) integrated intensity is $>$0.35 K~km~s$^{-1}$. 
The brightest detections in the LMC are the LMC pointings \#2, \#5, \#17, \#24 and \#25 located in the 
star-forming regions N159W, N44, N11C, N55, and N57, respectively \footnote{Correspondence between
the pointing number and the name can be found in Table~\ref{LineCharacteristics1}}. All these regions are known as 
complexes where large numbers of stars are formed and which host large CO reservoirs \citep{Israel2003_2}.
The brightest detections of the targeted SMC sources are found for pointings SMC \#1 and \#5 in the N27 and N13 star-forming regions, respectively.
N27 is also the brightest CO source observed in the SMC  \citep[][T.~van Kempen et al. in prep]{Israel2003_2}.

From the Gaussian fit, we derive the velocity offset from the systemic velocity, line widths, and integrated 
intensities and report the line characteristics in Table~\ref{LineCharacteristics1}.
Table~\ref{LineCharacteristics1bis} lists the corresponding HCO$^+$(2$-$1)/HCN(2$-$1) ratios along with the L$_{TIR}$
and corresponding SFR for each pointing.

\begin{figure}
\centering
\begin{tabular}{c}
 \includegraphics[width=8.8cm]{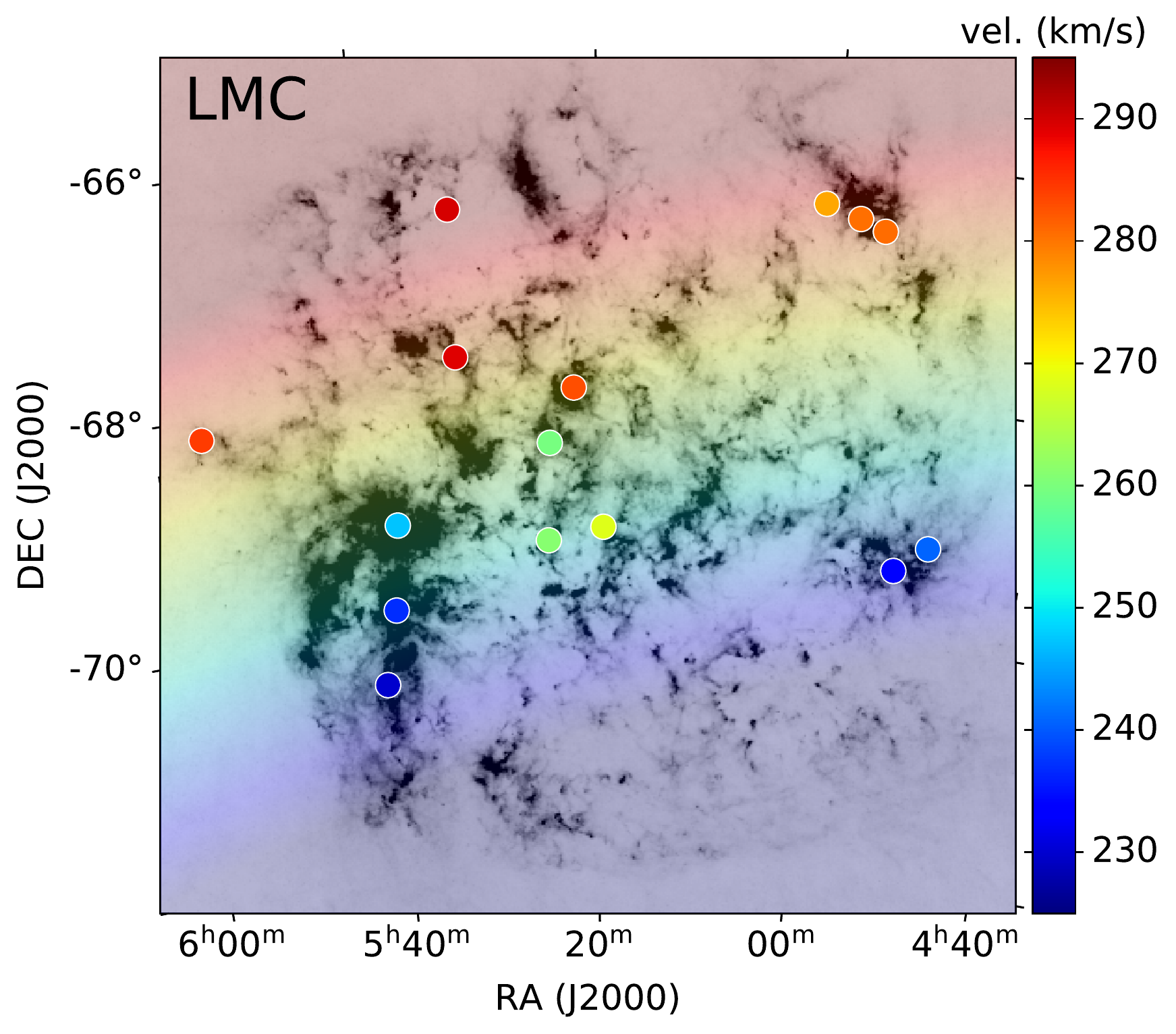}\\
 \includegraphics[width=8.8cm]{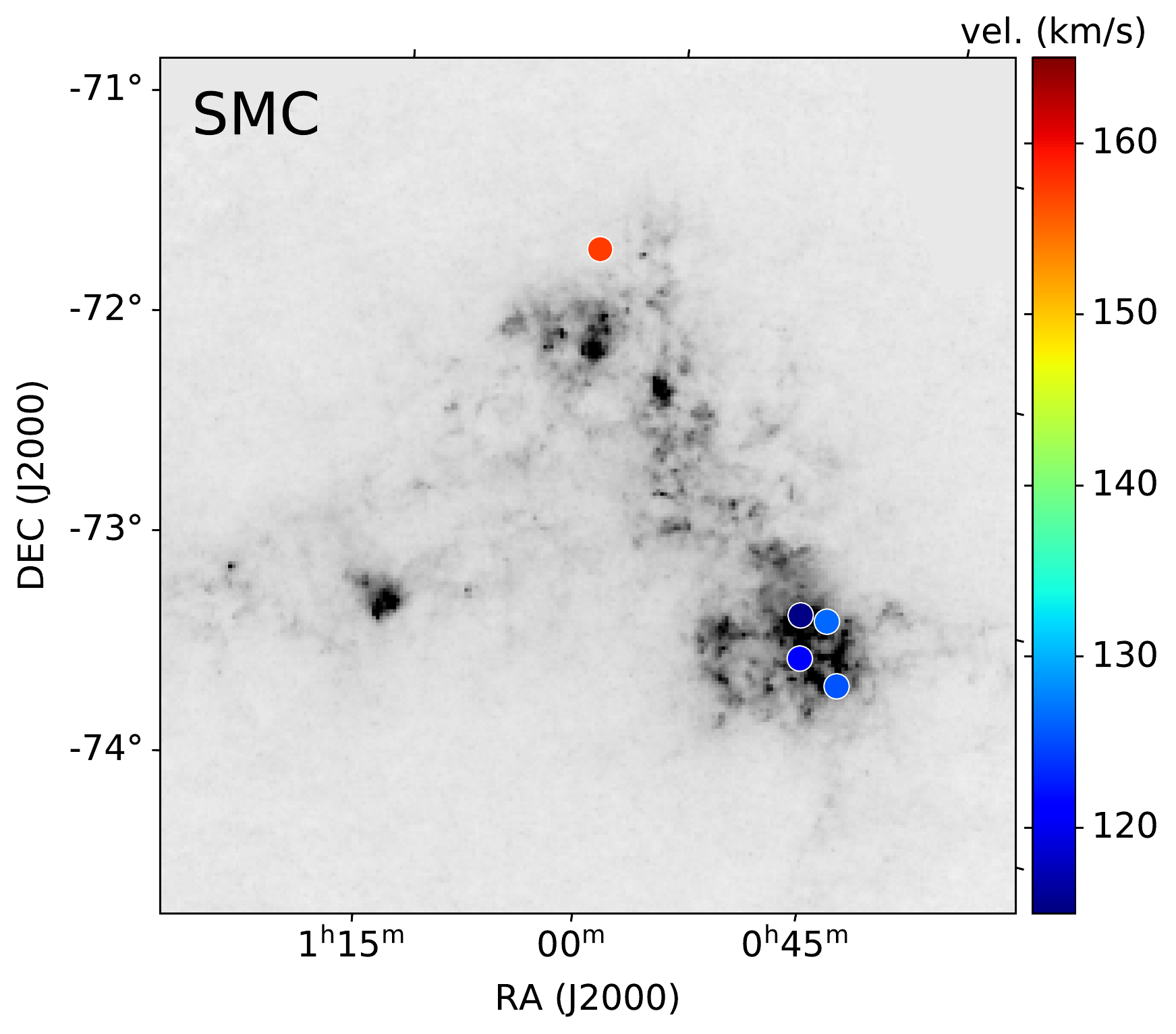} \\    
\end{tabular}
   \caption{LMC and SMC {\bfaa LOS} velocity fields as traced by the HCO$^+$ line. The pointings are
    represented by circles coded by their associated velocity (in units of km~s$^{-1}$) according to the provided color scales.
    For the LMC, we derive a velocity gradient of 0.31$\pm$0.1 km~s~$^{-1}$~kpc$^{-1}$ {\bfaa (P.A. = 5$^{\circ}$; overlaid in rainbow)}.
}
   \label{LOSvelocities}
\end{figure}

\begin{figure*}
\centering
\begin{tabular}{c}
\hspace{-5pt} \includegraphics[width=15cm]{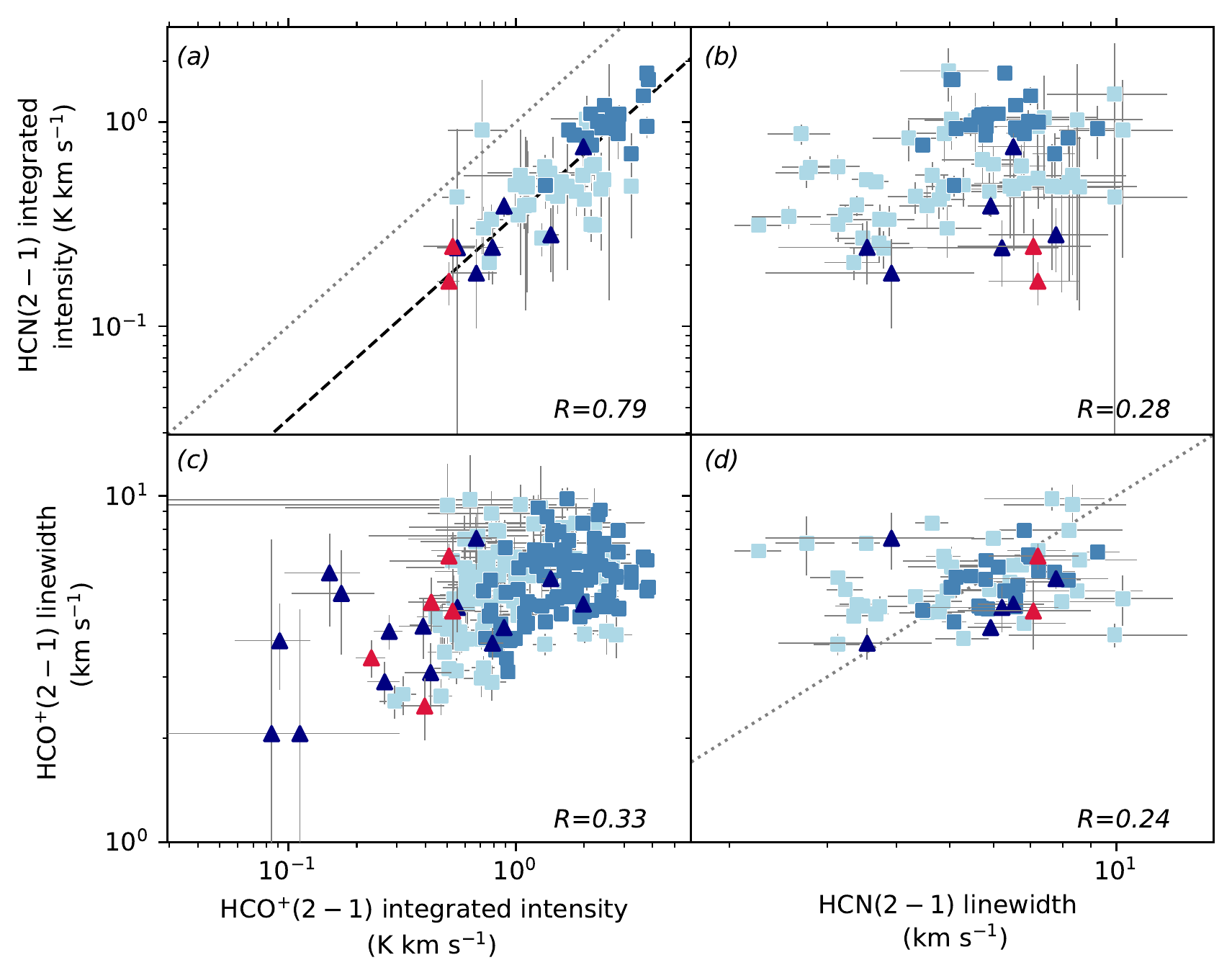}\\
\end{tabular}
   \caption{Relations between the HCO$^+$(2$-$1) and HCN(2$-$1) integrated intensities and the line widths (FWHM) of the lines.
   The LMC pointings appear as blue triangles and the SMC pointings as red triangles. 
   The light and dark blue squares indicate the individual detections from the 
    LMC spectral maps (18\arcsec\ pixels) detected at a 3 and 5-$\sigma$ level, respectively. 
    In panel (a), the dashed black line indicates the best linear fit to all the data.
    In panels (a) and (c), the dotted gray line is the 1:1 relation. }
   \label{HCO+HCNrelation}
\end{figure*}

\subsection{Mapping campaign}

From the reduced spectral cubes (mapping campaign), we derive integrated 
line-intensity maps by performing a Gaussian fit to each spectrum.
We restrict our calculation to pixels whose peak brightness is 3-$\sigma$ above the {\bfaa S/N} of the Gaussian fit. 
The HCO$^+$(2$-$1) and HCN(2$-$1) intensity maps of the LMC regions are shown in Figs.~\ref{IntensityMaps}, 
~\ref{IntensityMaps2}, ~\ref{IntensityMaps3}, and ~\ref{IntensityMaps4}. 
The final pixel size is 17\farcs6, thus half the beam size.
The \hersc\ 160~\mic\ contours (convolved at a 36\arcsec\ resolution) delineate 
the dust reservoirs observed at similar resolution to the SEPIA observation.
The 160~\mic\ is also a good proxy of the TIR luminosity in the region, tracing the SFR pockets.
We also show in the bottom panels the HCO$^+$(2$-$1) and HCN(2$-$1) spectra associated with the peak of their respective 
emission. The line characteristics (position, velocity offsets, linewidths and integrated line intensities) at the peak position 
for both lines are provided in Table~\ref{LineCharacteristics2} along with the corresponding SFR at this position.\\

Both lines are well detected in LMC 30Dor, N44, N105, N113, N159W, N159E and 
N214 while very few detections are obtained in N11B and N11C. For most of the 
sources, the location of the HCN and HCO$^+$ emission peaks correspond to 
peaks in the dust emission, except for N113 whose molecular line peaks seem 
to be shifted 35\arcsec\ south from the dust peak, as already noted in \citet{Wong2006}.
The HCO$^+$ emission is more extended than the HCN emission (except in N11B), with HCN
restricted to the densest regions of the targeted clouds. This more extended morphology is consistent 
with previous 3mm HCO$^+$ and HCN observations obtained in the 
LMC at higher resolution with the ATCA by \citet{Seale2012} in N105, N113, N159, and N44. We note that
our SEPIA observations trace the larger scale emission in the clouds surrounding the 
individual clumps that the ATCA and now ALMA observations are able to separate.
The HCN and HCO$^+$ difference in spatial extent is also consistent with observations of 
Galactic GMCs \citep{Pety2017}.
Part of the difference can be linked with the fact that HCN(2$-$1) probes higher densities 
\citep[n$_{crit}$ = 2.8 $\times$ 10$^6$~cm$^{-3}$ at 20~K) than HCO$^+$ 
(n$_{crit}$ = 4.2 $\times$ 10$^5$~cm$^{-3}$ at 20~K; ][]{Shirley2015}.
The (sub-thermal) excitation of both lines in the lower density parts of the clouds could also contribute to the
spatial difference we observe, along with a lower relative abundance of HCN in the outer part of the clouds.

We do not detect HCN toward N55. This is surprising, as recent observations with the Atacama Large Millimeter Array (ALMA) 
have resolved HCN clumps in the southern region of N55. 
The ALMA observations are centered at RA=05$^h$32$^m$31.50$^s$ DEC=$-$66$^{\circ}$26\arcmin22.5\arcsec, which is at
 the border of our map (A.~J.~Nayana et al., submitted to ApJ). 
The ALMA map beam size is 4\arcsec\ $\times$ 3\arcsec: the absence of detection could suggest that 
the emission detected with ALMA is smeared out in an APEX beam.
For N11B and N55, the spectral cubes have among the highest rms values of the sample: deeper integrations 
toward these two regions might reveal the weak HCN emission in these clouds. We finally note that we do not obtain any 
detection of HCO$^+$ or HCN in our SMC maps, here again because the mapping observations were unfortunately not sufficiently deep.


\section{Analysis}

\subsection{Velocity gradient across the Clouds}

The velocity gradients in the LMC have been studied using stellar emission \citep{Feitzinger1977,Schommer1992} and
atomic gas \citep[][]{Kim2003}. Our observations enable us to compare the {\bfaa line-of-sight (LOS)} velocity gradient of 
 the dense gas motions with these previous estimates. The correlation between the velocity offsets of the HCO$^+$ and 
 the HCN lines is extremely good (Pearson correlation coefficient R=0.99). We therefore use the velocity offsets of the 
 HCO$^+$ lines for the velocity field analysis.

We derive the individual LOS velocities of the dense gas in each pointing where the line is detected. 
We overlay the respective velocities as colored circles on the LMC map in Fig.~\ref{LOSvelocities}. 
Velocities in the LMC range from 229.7 km~s$^{-1}$ in the southern part of the galaxy (LMC \#4) to 289.6 km~s$^{-1}$ in 
the north (LMC \#24). This velocity pattern follows the LOS velocity field which was derived using 7000 carbon stars velocities by \citet{VanderMarel2014}.
It also corresponds to the velocity range where H{\sc i} peak column densities of 7 $\times$ 10$^{21}$ cm $^{-2}$ are found by \citet{Kim2003} (225-310~km~s$^{-1}$). 
More quantitatively, we estimate velocity gradient via a least-square minimization, with v$_{grad}$ = v$_{sys}$ + a$\Delta\alpha$ + b$\Delta\beta$, where $\Delta\alpha$ and $\Delta\beta$ are the offsets with respect to the center of mass taken at 5$^h$27.6$^m$, -69.87$^{\circ}$ \citep[as determined by][]{VanderMarel2002}. We find a velocity gradient of 0.31$\pm$0.1 km~s~$^{-1}$~kpc$^{-1}$ with a position angle of 5$^{\circ}$. We overlay the fitted velocity gradient in Fig.~\ref{LOSvelocities}.

The velocity field of the SMC is far less constrained by our pointing observations, as we only have a few detections of HCO$^{+}$ 
mostly in the southern part of the bar, with velocities ranging from 115 to 127 km~s$^{-1}$. The velocity 
in the northern part of the bar is constrained by the detection of HCO$^+$ in SMC \#8 and is equal to 157 km~s$^{-1}$.
The overall velocity field is consistent with that derived from the H{\sc i} observations by \citet{Stanimirovic2004}.
These latter authors find a large-scale gradient of about 70 km~s$^{-1}$ from the southwest to the northeast, and of about 
40 km~s$^{-1}$ between the two large supershells 37A and 304A corresponding to our SW and NE detections, respectively.

\subsection{HCO$^{+}$ and HCN intensities}

Previous surveys have been performed to derive HCO$^+$(1$-$0)/HCN(1$-$0) ratios toward six of our targeted 
regions. Estimates taken from the literature are summarized in Table~\ref{Literature10} for reference, with 
studies performed for the 1$-$0 transition toward H{\sc ii} regions at cloud scale \citep[for instance by][]{Chin1997} 
down to clump scales \citep{Seale2012,Anderson2014}. Observations of the 4$-$3 transition in N159 are reported 
by \citet{Paron2016} (with a HCO$^+$/HCN ratio of 8.75). 
To quantitatively compare the two molecular line emissions in the 2$-$1 transition, we show in 
Fig.~\ref{HCO+HCNrelation} (a) how the HCN(2$-$1) integrated intensity relates to HCO$^+$(2$-$1) integrated intensity.
We observe that both tracers scale relatively tightly with each other (Pearson correlation coefficient R=0.79). 
The dashed line indicates the linear fit to the data with the intercept fixed to 0. {\bfaa The relation is}:

\begin{equation}
\frac{I_{HCN(2-1)}}{K~km~s^{-1}} ~=~ 0.35 ~~ \frac{I_{HCO^+(2-1)}}{K~km~s^{-1}}
\end{equation}

Our targets are systematically brighter in HCO$^+$(2$-$1) 
than in HCN(2$-$1), with a mean HCO$^+$/HCN integrated intensity ratio (calculated with intensities in K~km~s${-1}$) 
of approximately 3 for the LMC pointings and 2.5 for the SMC pointings. 
The ratio is always above unity in both Magellanic Clouds (except one LOS in N11C) contrary to the Galaxy 
\citep{Pety2017,Shimajiri2017} or more metal-rich nearby galaxies\footnote{HCN intensities observed toward the
center of metal-rich active galaxies should however be taken with caution as they could be enhanced via IR 
pumping or affected by shocks.} 
\citep{Nguyen1992,Martin2015,Knudsen2007}. 
The highest HCO$^+$(2$-$1)/HCN(2$-$1) ratio ($\sim$7) is found in the luminous 30Dor region.
Overall, the HCO$^+$(2$-$1)/HCN(2$-$1) observed ratios are consistent with the fact that high HCO$^+$-to-HCN values 
are expected in the LMC and SMC, as the ratio is known to increase with decreasing metallicity \citep{Braine2017,Johnson2018,Kepley2018}.

The use of RADEX or LVG methods have shown that low-J HCN lines usually become thermalized at higher densities than 
the HCO$^+$ lines \citep{Muller2009,Mills2013}. Therefore, variations in the HCO$^+$/HCN ratio could be 
driven by local variations in density. Combining our dataset with observations of lower and higher excitation lines would be 
needed to properly quantify these densities. Variations in the HCO$^+$/HCN ratio could also be driven by local variations in the 
species ratio itself \citep{Nishimura2016}. Finally, higher HCO$^+$ intensities are also often attributed to the variations in the N/O abundance ratio with metallicity
\citep{Nicholls2017,Vincenzo2018,Berg2019}, in particular the elemental deficiency of nitrogen leading 
to a deficiency in N-bearing molecules \citep[HCN but also HNC; see][]{Nishimura2016}. 
The two Magellanic Clouds also have different metallicities, but unfortunately HCO$^+$ and HCN are only detected simultaneously in
two LOSs for the SMC, which does not allow us to robustly analyze metallicity effects between the two Clouds.
For these two SMC LOSs, the dense gas tracer intensities follow the overall relation of Fig.~\ref{HCO+HCNrelation} (a).

\begin{table}
\caption{HCO$^+$(1$-$0)/HCN(1$-$0) $^a$ estimates from literature.}
\label{Literature10}
\centering
\begin{tabular}{ccccc}
\hline
\hline
                &\\
Region  && HCO$^+$(1$-$0) & Reference & Facility            \\
\cline{3-3}
                && HCN(1$-$0) &  \\ 
                &&\\
\hline
                &&\\
30Dor   && 1.9-2.7      & \citet{Heikkila1999}  & SEST \\
                && 3.4-9.4      & \citet{Anderson2014}  & ATCA \\
N11B    && 2.3          & \citet{Nishimura2016} & MOPRA \\
N44             && 2.3          & \citet{Chin1997}              & SEST \\       
                && 2.8          & \citet{Nishimura2016} & MOPRA \\
                && 2.9-7.5      & S12, A14 $^b$                 & ATCA \\
N105    && 2.2-4.8      & S12, A14              & ATCA \\ 
N113            && 1.5          & \citet{Wang2009}              & SEST \\
                && 1.7-2.0      & \citet{Wong2006}              & MOPRA \\
                && 2.3          & \citet{Nishimura2016} & MOPRA \\
                && 1.9-9.8      & S12, A14                      & ATCA \\ 
N159    && 2.2          & \citet{Chin1997}              & SEST \\
                && 3.3          & \citet{Nishimura2016} & MOPRA \\
                && 2.4-9.3      & S12, A14                      & ATCA \\
N214    && 2.1          & \citet{Chin1997}              & SEST \\
                &\\
\hline
\end{tabular}
\begin{list}{}{}
\item $^a$ expressed as integrated intensity (in K~km~s$^{-1}$) ratios  
\item $^b$ \citet{Seale2012,Anderson2014}: the values reported are the ratio range covered by the resolved clumps.
\end{list}
\end{table} 

\subsection{HCO$^{+}$ and HCN line widths}
\label{section:linewidth}

On the scales traced by our survey, the line widths provide information on the GMC dynamics.
Giant molecular clouds are known to follow scaling relations between their size, luminosity, and line width \citep{Larson1981},
suggesting that a relation also exists between the mass of a star-forming cores and its dynamical status.
These scaling relations have been studied in CO for extragalactic GMCs \citep[e.g. by][]{Bolatto2008,Schruba2019,Sun2020},
including the LMC where line widths have been found to be narrower \citep{Hughes2010} or wider \citep{Indebetouw2013} than
the overall size--line width relation, depending on the region probed. 
As far as dense gas tracers are concerned, in the Galaxy, HCO$^+$ line widths of Galactic dense clouds in the plane or at high-latitude 
are in the [2-5] km~s$^{-1}$ range \citep{Pirogov1995,Schlingman2011} while this value raises to 6 km~s$^{-1}$
toward Galactic high-mass star-forming cores \citep{Shirley2003}. Other parameters than mass alone can 
also affect the line width, such as inter-clump pressure or the cloud turbulence \citep{Heyer2009,Schruba2017,Schruba2019}.

Figure~\ref{HCO+HCNrelation}d shows the relation between the HCO$^+$(2$-$1) and HCN(2$-$1) (FWHM) line widths
while the side panels (b) and (c) show how the line width varies as a function of the 
line integrated intensity. Part of the relationship between line width 
and its integrated intensity could be due to the definition of the line integrated intensity 
itself~\footnote{The integrated intensity directly depends on the FWHM and the peak intensity.}, 
although three sources (LMC \#4, \#7 and \#13) seem to be slightly above the overall positive correlation in panel 
(c). These three regions have no HCN detection, which prevents us from confirming the deviation from this line.
The HCN and HCO$^+$ line widths and line profiles are relatively similar, both covering line width
values from 2 to 10 km~s$^{-1}$. For the LMC pointing sample campaign (blue triangles), the mean HCO$^+$(2$-$1) line width is 
4.3 km~s$^{-1}$ (5.6 km~s$^{-1}$ for HCN). The mean line width in the SMC is similar for HCO$^{+}$ (4.4 km~s$^{-1}$) 
but larger for HCN (7.1 km~s$^{-1}$) whose detection remains limited to the densest central, and therefore probably more turbulent, parts of the clouds. 
We observe that our pointing campaign, which targeted lower cloud masses, includes LOSs with lower HCO$^+$ line widths 
$<$3 km~s$^{-1}$ in LMC\#12, \#22, and \#23. The largest line widths are found toward the 30Dor (9.8 km~s$^{-1}$), 
N159 (9.4 km~s$^{-1}$), and N44 (8.1 km~s$^{-1}$) star-forming regions, a signature that 
these regions could host more massive clumps than the other regions targeted. 


We note a larger dispersion in the HCN(2$-$1) line widths compared to HCO$^+$(2$-$1) line widths
(see Fig.~\ref{HCO+HCNrelation} d). This dispersion might be linked with the fact that we model the 
HCN(2$-$1) line as a single component in spite of the line having a hyperfine structure of a few km/s, which is
not resolved at our working spectral resolution of 2km/s.

\begin{figure*}
\centering
\begin{tabular}{c}
 \includegraphics[width=13cm]{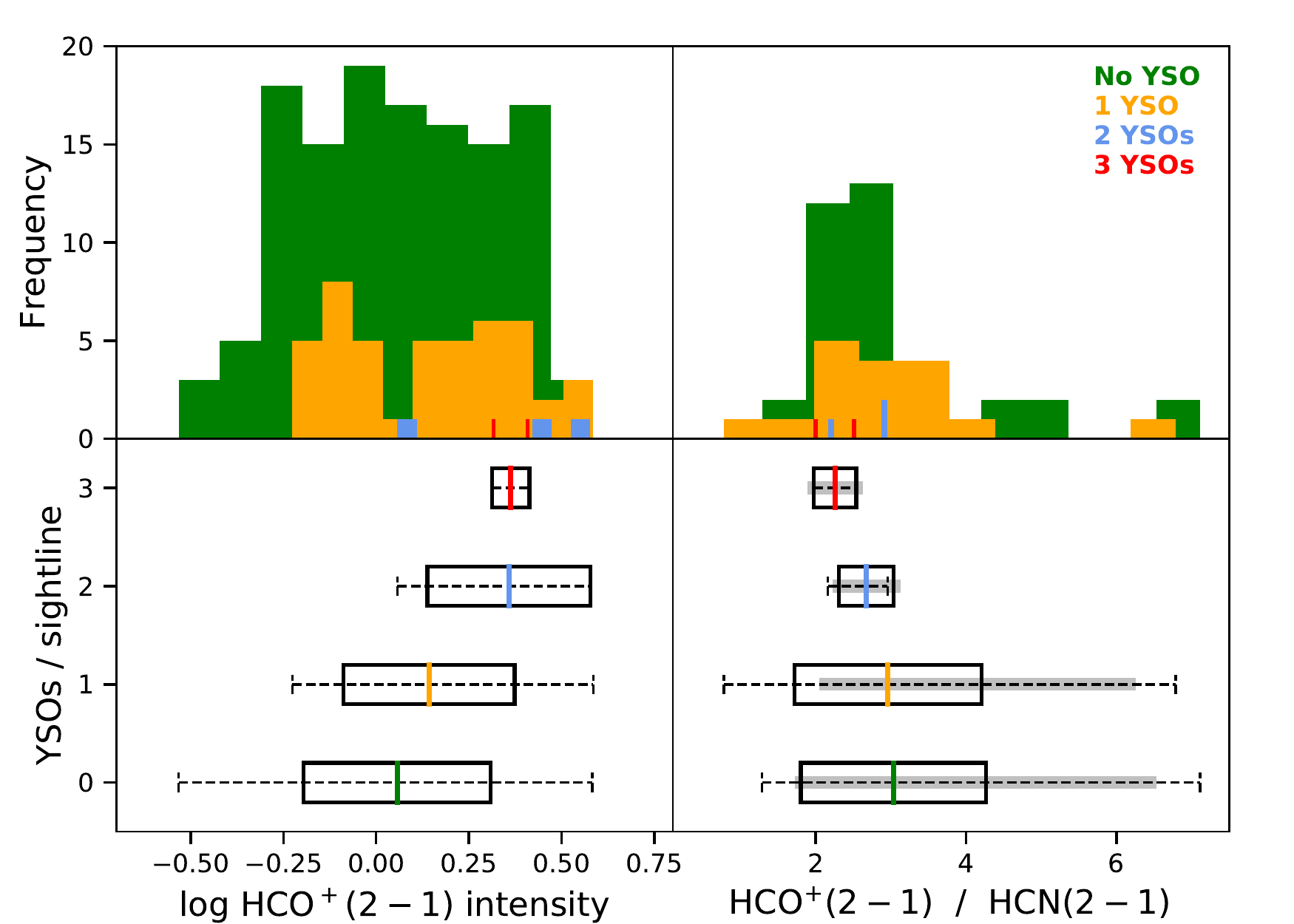}\\
\end{tabular}
   \caption{{\it Top:} Histograms of the HCO$^+$(2$-$1) integrated intensities in K~km~s$^{-1}$ (left) 
   and HCO$^+$(2$-$1)/HCN(2$-$1) integrated intensity ratios (right) 
   of all the LOSs observed. LOSs with 0, 1, 2, and 3 YSOs are shown in green, orange, blue, and red, respectively.
   {\it Bottom:} Characteristics of each of these distributions. The black box indicates the standard deviation, the 
   colored line the mean value, and the dashed line indicates the minimum and maximum HCO$^+$(2$-$1) 
   intensity (left) or HCO$^+$(2$-$1)/HCN(2$-$1) (right) for each distribution. The gray thick line indicates 
   the standard deviation of the distribution when LOSs that have HCO$^+$ detection and HCN upper limits 
   are included.}
   \label{YSOdistribution}
\end{figure*}

\begin{figure*}
\centering
\begin{tabular}{c}
 \includegraphics[width=14cm]{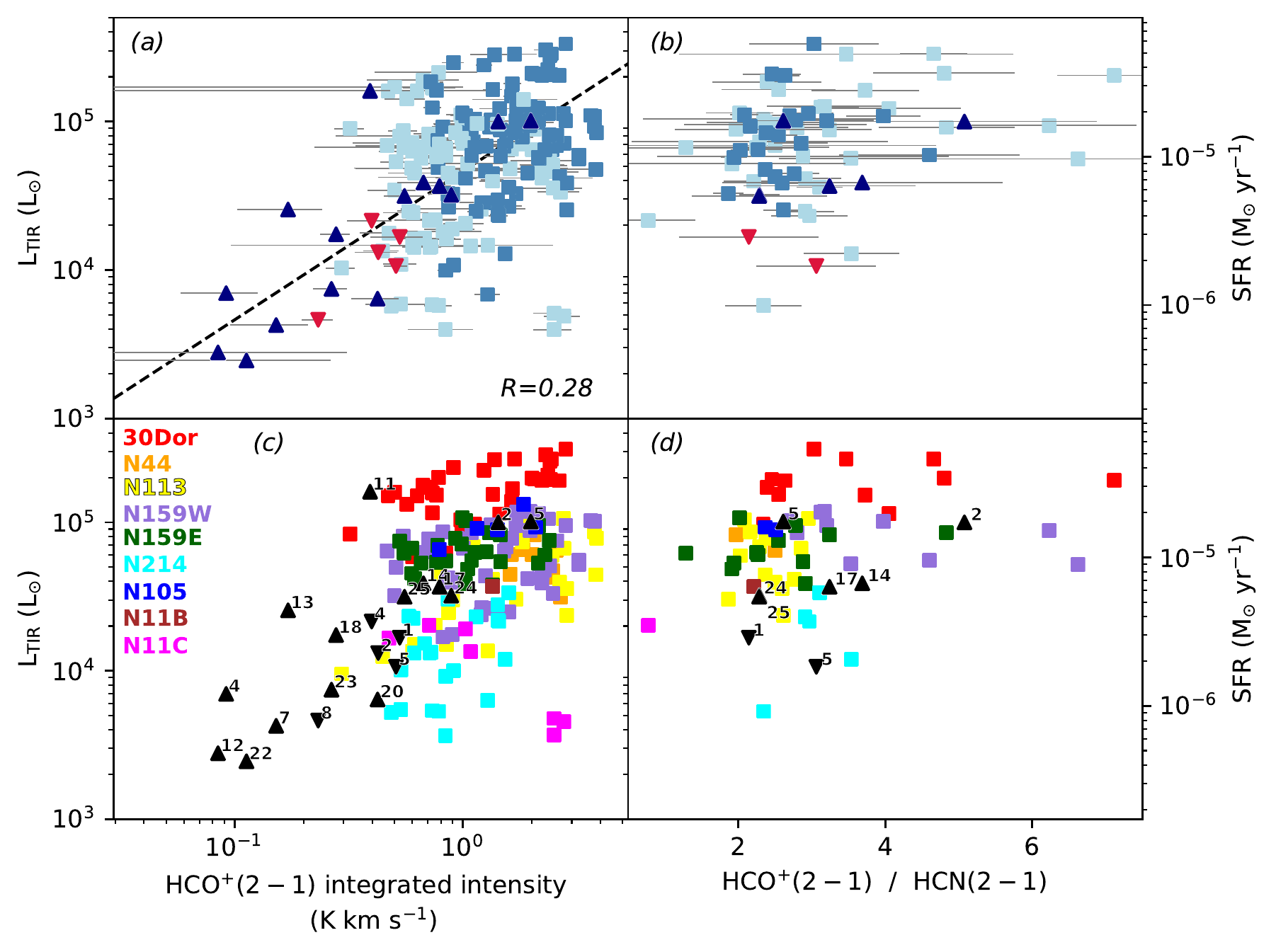}\\
\end{tabular}
\vspace{-10pt}
   \caption{{\it Panels (a) $and$ (b):} Variation of the HCO$^+$(2$-$1) intensity and the HCO$^+$(2$-$1)/HCN(2$-$1) ratio
    as a function of the SFR traced by L$_{TIR}$ for the LMC (blue upward triangles) 
    and SMC (red downward triangles) pointings. Intensities are in units of K~km~s$^{-1}$.
    The light and dark blue squares indicate the individual detections from the 
    LMC spectral maps (18\arcsec\ pixels) detected at a 3 and 5-$\sigma$ level, respectively. 
    In panel (a), the dashed line indicates the linear fit to the data. The Pearson's correlation coefficient is indicated
    {\it Panels (c) and  (d):} Same but with colors indicating the star-forming region of each resolved element. 
    Black triangles indicate LMC and SMC pointings shown with their corresponding DeGaS-MC numbers. 
    }
   \label{HCO+HCN_vs_LTIR}
\end{figure*}


\section{Discussion}

\subsection{Relation with the YSO populations}

Most of the sources targeted during the mapping campaign benefit from intensive identification campaigns of their population
of YSOs \citep{Whitney2008,Gruendl_Chu2009,Chen2009,Chen2010,Seale2012,Carlson2012}. We retrieved these
catalogs from the VizieR database and overlay the YSO candidates on the HCO$^+$ and HCN maps (Fig.~\ref{IntensityMaps}). If sources 
are listed in several catalogs, they only appear once in the maps and in our analysis.
We note the double color-code for the \citet{Gruendl_Chu2009} catalog: red plus symbols on our maps indicate sources classified as YSOs while salmon-colored 
plus symbols indicate sources that have been primarily classified as YSOs but show other characteristics that could 
suggest another nature (star or evolved star, diffuse source, background galaxies; classified as `CA',`CS',`CG' or `CD' in the catalog).
From their spatial distribution, we produce maps of YSO candidates per LOS for each targeted SF region.
The total number of YSOs in each mapped field is reported in Table~\ref{LineCharacteristics2}.

We plot the distribution of the number of YSO candidates per LOS as a function of the HCO$^+$(2$-$1) integrated intensity 
in Fig.~\ref{YSOdistribution} (top left). The bottom panels highlight the characteristics of each distribution: the black boxes indicate the 
standard deviation for each distribution and the colored line their mean value. We observe a higher number of YSOs at high HCO$^+$ or HCN intensities,
which is consistent with the expectations that stars should primarily form in high-density gas \citep[][among many others]{Myers1987,Yun1999,Sadavoy2014}.
The highest number of YSOs per LOS is found toward N44 (3 YSOs/pixel) and N159W (2 YSOs/pixel).
However, we note that an extensive and targeted analysis was performed for these two sources by \citet{Chen2009,Chen2010}, 
which might introduce {\bfaa biases}. Deeper analysis of YSOs on the other regions or at higher resolution (with JWST 
for instance) might also reveal more YSOs clustered toward the central regions, in particular in crowded environments such as 30Dor. 

Figure~\ref{YSOdistribution} (right) shows the distribution of the number of YSO candidates per LOS, this time as a function of the
HCO$^+$(2$-$1)/HCN(2$-$1) ratio. To reinforce the statistics, we also indicate with the gray thick line (bottom right panel) how the standard 
deviation of each distribution evolves if we also consider the HCN upper limits combined with the HCO$^+$ detections. In both cases, 
more candidates seem to be found toward LOSs with lower HCO$^+$/HCN ratios, 
which is consistent with the previous findings of \citet{Seale2012} at clump scales. However, our statistics are sparse for the $>$1 YSO / LOS bins. 
If real, such a correlation of the YSO number with the line ratio would not be easy to interpret, as both lines can be 
independently affected by local conditions  (temperature, radiation field, shocks). Nevertheless, one explanation could be that given the 
difference in thermalization of the two lines, lower HCO$^+$/HCN ratios could be found toward LOSs containing gas at larger densities, 
hence where YSOs could preferentially form. Part of the correlation could also be driven by chemistry effects; for instance, HCN and 
HCO$^+$ have been shown to trace different ionization conditions in YSO outflows \citep[]{Tappe2012}.

\subsection{Scaling relations with the star formation rate}

The correlation between HCO$^+$ and HCN luminosities and SFR has been studied
over ten orders of magnitude, from individual Galactic clouds to full galaxies \citep{Gao_Solomon2004,Chen2015,Usero2015,JimenezDonaire2019}.
Our observations help us investigate the relation on GMC scales, thus populating the bridge between these two scales 
in a low-metallicity environment whose ISM structure and abundance might affect the relation. 
Given the larger detection rate of HCO$^+$(2$-$1), we selected this line for the HCO$^+$(2$-$1) versus
SFR comparison in panels (a) and (c) of Fig.~\ref{HCO+HCN_vs_LTIR}. The IR luminosity L$_{TIR}$, which is a proxy of the recent star formation, 
is derived as described in Section~\ref{section:derivingLTIR}. The L$_{TIR}$ are then converted in SFR using the calibration from \citet{Murphy2011}. 
Both L$_{TIR}$ and SFR are indicated on the left and right y-axes, respectively. The uncertainties on L$_{TIR}$, although
plotted in Fig.~\ref{HCO+HCN_vs_LTIR}, are smaller than the symbol size. L$_{TIR}$ and/or SFR for the pointings and toward the peak positions of
the maps are indicated in Tables~\ref{LineCharacteristics1bis} and ~\ref{LineCharacteristics2}.

Our DeGaS-MC pointings nearly cover two orders of magnitude in terms of SFR and the HCO$^+$(2$-$1) 
luminosity correlates well with this SFR (Pearson's R=0.6) throughout the LMC and the SMC (triangles). The dispersion around the 
relation increases when the full dataset is taken into account (Pearson's R=0.28). The relationship between HCO$^+$(2-1) integrated 
intensity and SFR (dashed line in Fig.~\ref{HCO+HCN_vs_LTIR}a) is as follows:

\begin{equation}
\label{HCOSFR}
\frac{SFR}{M_\odot~yr^{-1}} ~~=~~ 7.7 \times 10^{-6} ~~\frac{I_{HCO^+(2-1)}}{K~km~s^{-1}}
.\end{equation}

The correlation confirms that in the Magellanic Clouds, as in nearby spirals \citep[e.g.][]{Bigiel2015} up to 
high-redshifts starbursts \citep[e.g.,][]{Oteo2017}, the {\bfaa SFR} is closely related with denser regions (we show
in $\S$\ref{section:DG} that the correlation is more scattered with the total molecular gas traced by the CO(1$-$0) line).
Similarly to what is observed in other nearby galaxies \citep{Querejeta2019}, we observe strong variations in the dense-gas-luminosity--SFR relation from one SF region to another and within the regions themselves. While N159E, N159W, N113, and N44 follow 
the global trend, 30Dor and N214 are diverging from the overall relation with opposite behaviors.
30Dor systematically presents SFRs larger than predictions from Eq.~\ref{HCOSFR}, which translates into a higher SF efficiency, 
while N214 has a more quiescent SFR than predicted.\\

Several studies have also suggested that the HCO$^+$/HCN ratio varies with star formation activity in molecular clouds.
In M31 for instance, \citet{Brouillet2005} show that the HCO$^+$ emission is slightly stronger than the HCN emission toward GMCs with a high star formation activity (presence of an associated large H{\sc ii} region) compared to more quiescent clouds. 
Panels (b) and (d) of Fig.~\ref{HCO+HCN_vs_LTIR}  show the relation between the HCO$^+$(2$-$1)/HCN(2$-$1) ratio and the SFR. These results suggest 
that the local relation between the two parameters is actually not straightforward when probed on GMC scales. For our selection of sources, the 
HCO$^+$(2$-$1)/HCN(2$-$1) ratio and the SFR are indeed only weakly correlated, with most of our {\bfaa LOSs} having ratios of around
two to three and a strong dispersion at SFR $>$ 5 $\times$ 10$^{-6}$ \msun~yr$^{-1}$. However, we note that the largest HCO$^+$/HCN ratios are 
preferentially found towards more actively star-forming LOSs in N159W and 30Dor. 

As will be shown in Fig.~\ref{DenseGasFraction2} (bottom), the HCO$^+$(2$-$1)/HCN(2$-$1) ratio does not depend strongly
on the local CO intensity toward the same LOS. No correlation is found either with the local temperature (that we traced 
via the 100 \mic~/~350 \mic\ luminosity ratio). For instance, all LOSs in 30Dor are warmer than the other regions, but the 
majority of these LOSs cover similar HCO$^+$(2$-$1)/HCN(2$-$1) ratios (between 2 to 4) to the other regions.

\begin{figure}
\centering
\begin{tabular}{c}
\hspace{-10pt} \includegraphics[width=9cm]{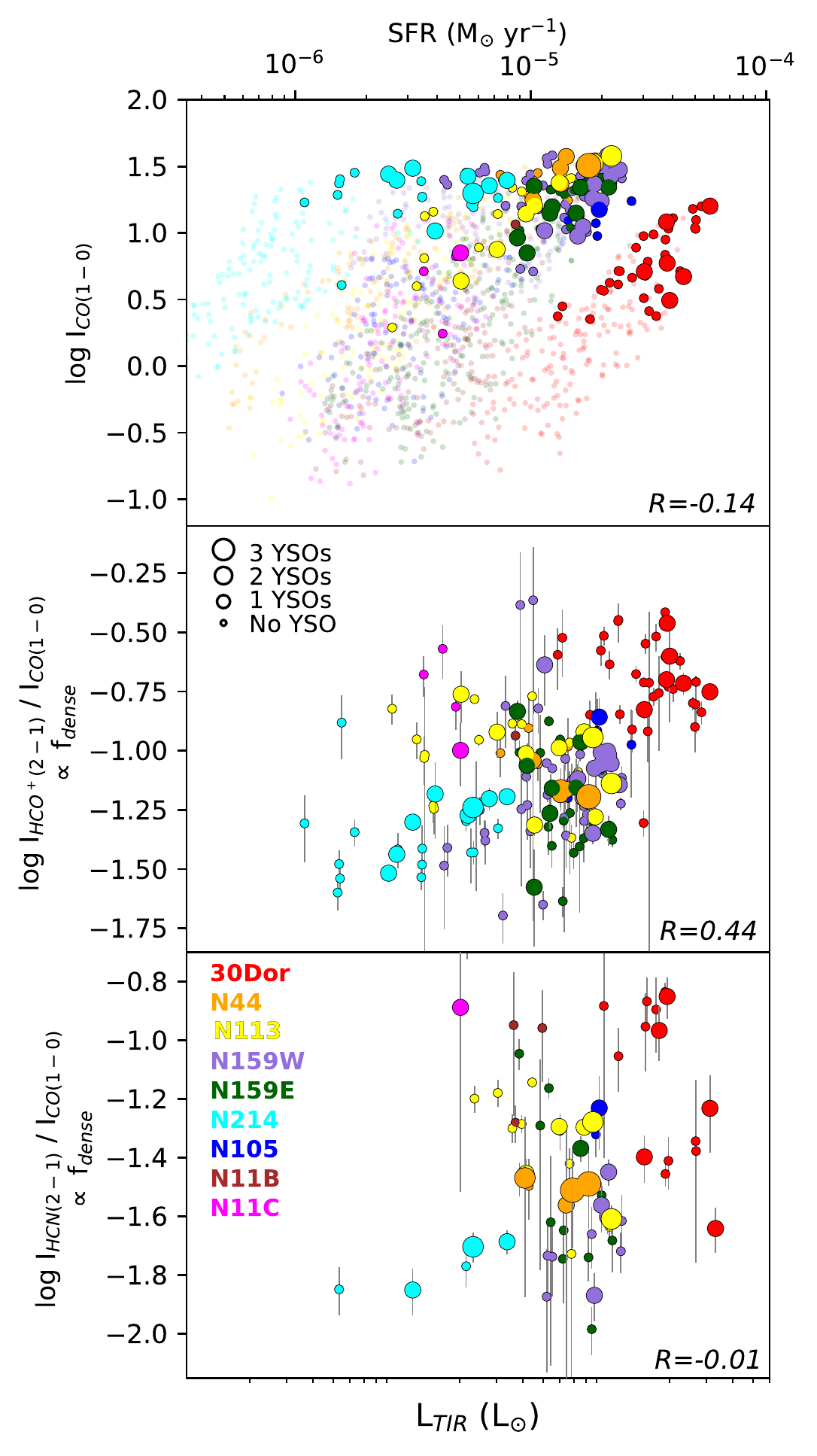}\\
\end{tabular}
\vspace{-10pt}
   \caption{{\it Top - } Relation between the CO(1$-$0) intensity (in K~km~s$^{-1}$) and the local L$_{TIR}$ 
   for {\bfaa LOSs} where HCO$^+$(2$-$1) has been detected with DeGaS-MC (and associated Pearson correlation coefficient R). 
   The shaded points complement the relation for all {\bfaa LOSs} where CO(1$-$0) has been detected.
   The colors separate the various star-forming regions and the increasing size of the symbols indicates an increase in the YSO 
   number (0, 1, 2 or 3) for each LOS. The corresponding local SFRs are indicated on the upper x-axis.
   {\it Middle and bottom -} Variation of the HCO$^+$(2$-$1)/CO(1$-$0) (middle), a proxy for the dense gas fraction, 
   and of the HCN(2$-$1)/CO(1$-$0) (bottom) ratio (with intensities in K~km~s$^{-1}$)
   as a function of the local L$_{TIR}$. }
   \label{DenseGasFraction1}
\end{figure}

\begin{figure}
\centering
\begin{tabular}{c}
\hspace{-10pt} \includegraphics[width=9cm]{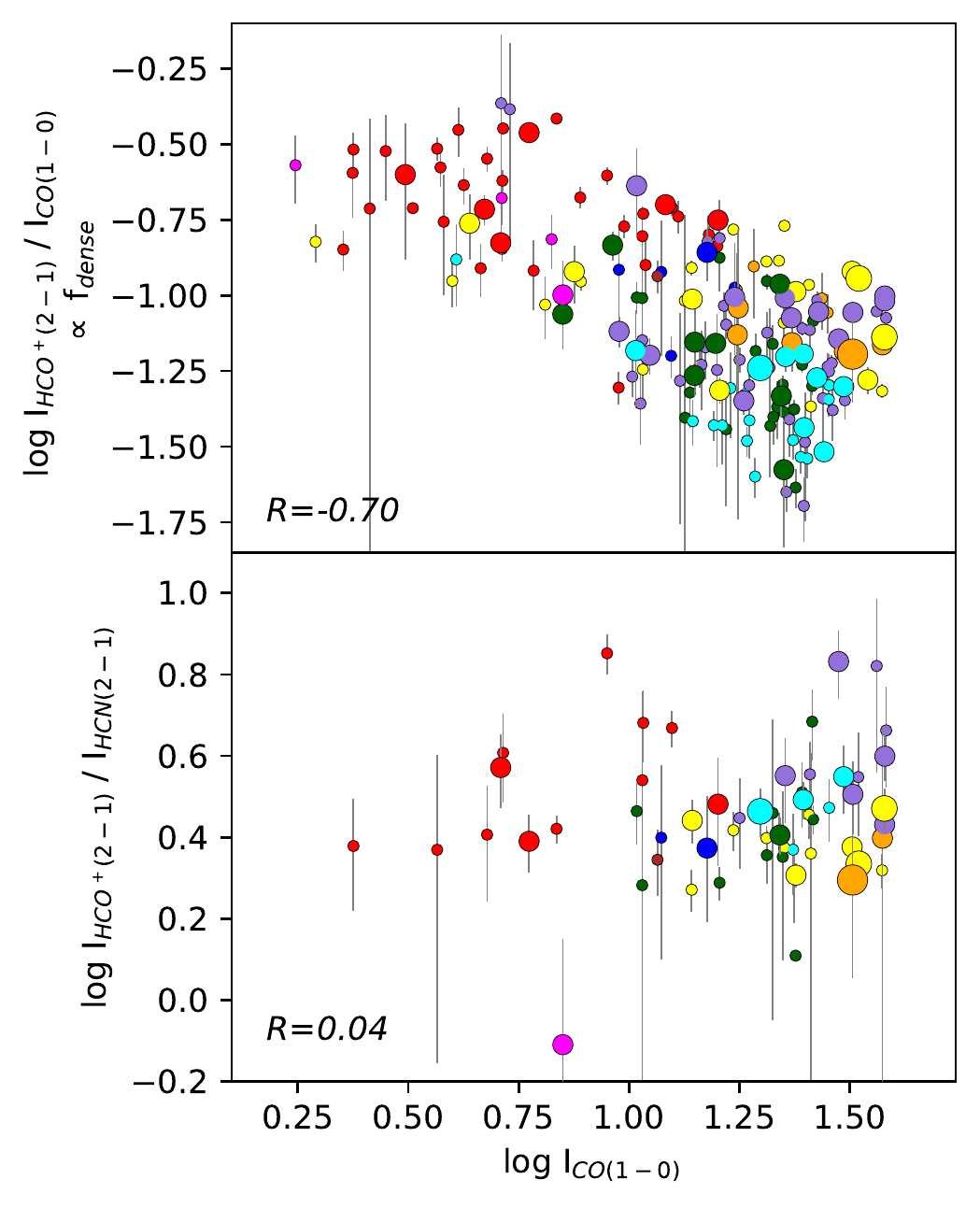}\\
\end{tabular}
\vspace{-10pt}
   \caption{Variation of the HCO$^+$(2$-$1)/CO(1$-$0) (top), a proxy for the dense gas fraction, 
   and of the HCO$^+$/HCN(2$-$1) (bottom) ratio
   as a function of CO luminosity (all intensities in K~km~s$^{-1}$). 
   For symbol sizes and colors: see convention in Fig.~\ref{DenseGasFraction1}. }
   \label{DenseGasFraction2}
\end{figure}

\subsection{Dense gas fraction}
\label{section:DG}

The HCN/CO or HCO$^+$/CO flux ratios are commonly used as tracers of the dense gas fractions \citep{Bigiel2015,Usero2015}. 
The HCO$^+$(1$-$0)/CO(1$-$0) ratio has been derived toward a few LMC molecular clouds \citep{Johansson1998,Heikkila1999}.
Targeting dense clumps within 30Dor, \citet{Anderson2014} also find relatively 
low (10\%) dense gas fractions in the region. As a comparison, Milky Way estimates of the dense gas fraction in 
Galactic center clouds have led to values of about 15\% \citep[][]{Mills2018}.
To analyze how this ratio varies throughout the LMC, we use the CO(1$-$0) map from the MAGMA survey (second data release). 
A map is provided at a resolution of 45\arcsec, close to that of our analysis (35\arcsec). As this difference in resolution is 
minor and the beam profile shapes are rather uncertain, we do not attempt to convolve the data and directly regrid the CO intensity map to 
the pixel grids of our individual intensity maps. The CO, HCN, and HCO$^+$ data are kept in K~km~s$^{-1}$. We assume that 
the CO(1$-$0) line is not saturated.

Figure~\ref{DenseGasFraction1} (top) first presents the relation between the CO(1$-$0) intensity with the SFR 
(as traced by the TIR luminosity) for {\bfaa LOSs} where HCO$^+$(2$-$1) has been detected during our DeGaS-MC campaign. The shaded points 
complement the relation for all {\bfaa LOSs} where CO(1$-$0) has been detected in our fields of view. The very weak correlation
between the two parameters (whether  or not we include the {\bfaa LOSs} where HCO$^+$ is not detected) 
reinforces the numerous previous claims that the HCO$^+$ line emission is a far better tracer of the gas that forms stars than the CO(1$-$0) line.
Part of the lack of correlation could be linked with saturation effects affecting the CO emission.
Figure~\ref{DenseGasFraction1} (middle and bottom) highlights the substantial variations of the HCO$^+$(2$-$1)/CO(1$-$0) and
HCN(2$-$1)/CO(1$-$0) ratios, and thus of the dense gas fraction from one LMC region to another. The highest 
dense gas fractions are found in 30Dor while the lowest are found in N214. Figure~\ref{DenseGasFraction1}
also provides information on how the dense gas fractions evolve as a function of L$_{TIR}$: the 30Dor and N214 regions
are also the two regions with respectively highest and lowest SFR on average. The positive correlation between L$_{TIR}$ and the
HCO$^+$(2$-$1)/CO(1$-$0) is therefore in line with the expected correlation between the SFR and the dense gas fraction, 
though the relation presents significant scatter (Pearson's R=0.37). \\

The size of the circles in Fig.~\ref{DenseGasFraction1} is proportional to the number of YSOs per {\bfaa LOS}. 
The number of YSOs does not seem to correlate with the HCO$^+$/CO ratio, which indicates that YSOs are not preferentially 
found where the dense gas fraction is the highest. We note however that the lack of a strong correlation could also be linked with the complexity 
of detecting YSOs toward crowded regions or the lower mass YSO populations, which could lead to an underestimation of the number 
of YSOs toward certain LOSs.\\

Finally, in \citet{Gallagher2018_2}, a positive correlation is found between the dense gas fraction (traced by HCN(1$-$0)/CO(1$-$0)) 
and the cloud-scale molecular gas surface density (traced by CO(2$-$1)) at kiloparsec-scales. However, substantial 
galaxy-to-galaxy variations are observed and the authors predict that a study at cloud scales could reveal intrinsic variations
from region to region \citep[e.g., as found by][for individual molecular clouds in M51]{Querejeta2019}. 
Figure~\ref{DenseGasFraction2} (top) shows how the dense gas fraction (here traced via HCO$^+$(2$-$1)/CO(1$-$0) ratio)
scales with the CO(1$-$0) integrated intensity in our survey. Substantial scatter is observed from one star-forming region to 
another. Contrary to the findings of \citet{Gallagher2018_2}, the two tracers seem to be anti-correlated. We note that the DeGaS-MC trend is derived 
for a particular cloud regime: low-Z, massive, star-forming clouds. Observations of more quiescent clouds would be necessary 
to construct the full LMC-scale relation as in \citet{Gallagher2018_2}. The scatter in Fig.~\ref{DenseGasFraction2} is mostly driven 
by the shifted 30Dor region. The CO emission in this region could be ``underluminous" because of photo-dissociation by the strong 
star formation and stellar feedback, while HCO$^+$ would be better shielded \citep{Schruba2017}.
As mentioned previously, Fig.~\ref{DenseGasFraction2} (bottom) shows that the HCO$^+$(2$-$1)/HCN(2$-$1) ratio does not 
depend strongly on the local CO intensity toward the same {\bfaa LOS}.

\subsection{Potential impact of X-rays on the chemistry?}

As agents of ionization, X-rays can lead to an enhancement of the ionization rate. 
High ionization rates have a direct impact on chemistry; they are suspected 
to decrease the HCO$^+$ abundance \citep{Krolik1983,Maloney1996} and have been found to
boost the abundance of free electrons. A as electrons might be dominant for the excitation of the 
HCN molecule over HCO$^+$ \citep{GoldsmithKauffmann2017},
high ionization rates will also result in an enhancement of the excitation of HCN, although how 
these various chemical processes dominate in the studied regions remains unclear. 

In our survey, the lowest HCO$^+$(2$-$1)/HCN(2$-$1) ratio is found in the N11C 
region ($<$1), in the southern part of the complex. This southern region is irradiated 
by the LH~13 massive cluster and hosts various compact stellar clusters (Sk$-$66$^{\circ}$41 and HNT) 
as well as extremely luminous stars, such as Wo~599, which has been  suggested to be the main ionizing source of N11C
\citep{Heydari1987,Heydari2000}. X-ray emission has been associated to LH~13 in particular \citep{Naze2004} and could 
be responsible for the lower HCO$^+$(2$-$1)/HCN(2$-$1) observed in N11C south. 
The other region observed toward the N11 bubble, N11B, also contains an embedded extremely 
young OB association (with multiple O and other massive stars). 
Its associated diffuse X-ray emission exceeds that expected from the stars and 
could be potentially powered by stellar winds \citep{MacLow1998}.
Here again, the presence of ionization enhanced by X-rays could partly explain the 
very weak HCO$^+$ detected toward N11B.
X-rays probably affect the HCO$^+$(2$-$1) and HCN(2$-$1) emission
as well in 30Dor. However, most of the X-ray emission observed by ROSAT \citep{Kennicutt1994_2}
or CHANDRA \citep{Townsley2006} toward the sub-region targeted by our DeGaS-MC survey 
arises from the massive stellar cluster R136 located at RA=5$^h$38$^m$42.4$^s$ 
DEC=$-$69$^{\circ}$06\arcmin03.4\arcsec, which is in-between the two HCO$^+$(2$-$1)
and HCN(2$-$1) peaks, a region devoid of molecular emission in our survey.

\subsection{Future work}

This first paper of this series highlights the potential of the DeGaS-MC survey to characterize the dense molecular gas 
at GMC scales and show how it evolves within a low-metallicity ISM. 
Future papers of the collaboration will provide a more detailed analysis of the line profiles and 
velocity information and how it translates in terms of dense molecular gas dynamics. 
Combined with other transitions, we will perform a modeling of the various regions with 
photo-dissociation region or X-ray illuminated cloud models using the CLOUDY model 
by \citet{Ferland2017} or multi-structural, thermo-chemical 
modeling by \citet[and Bisbas, Schruba \& van Dishoeck in prep]{Bisbas2019}. {\bfaa This will 
allow us to} derive constraints on the density structure or kinetic temperature 
of the clouds and quantify the impact of the local physical conditions
(ambient gas density, turbulence, stellar population, impact of X-rays on 
the ionization, supernova shocks) on the dense gas content.

In this analysis, we mention that no correlation is found between the 
HCO$^+$(2$-$1)/HCN(2$-$1) ratio and the local temperature traced via the 
100 \mic~/~350 \mic\ luminosity ratio, but a more thorough dust SED modeling 
of the various regions using the {\it Herschel} measurements would allow us to properly quantify 
the dust temperature, derive dust (and gas) column densities, and compare them with the molecular intensities. 
The modeling would also help us to probe whether the source-to-source dispersion we observe in 
Figs.~\ref{DenseGasFraction1} and ~\ref{DenseGasFraction2} could be partly caused by the differences in the 
temperature structure of the different sources.
As most of the line mapping survey targets solar-metallicity environments, these analyses would allow us to 
extend studies of ISM chemical composition at different dust (or gas) columns 
to the low-metallicity ISM conditions.

We finally note that this survey provides crucial constraints on the fluxes expected 
for follow-up studies of larger areas and/or at higher resolution with ALMA, but could also, via stacking 
analysis of the diffuse emission, provide constraints for unresolved studies of 
high-redshift galaxies.


\section{Summary}

We report new observations of HCO$^+$ and HCN molecules in the 2$-$1 transition 
obtained with the APEX SEPIA180 instrument. We performed a pointing campaign toward 
30 LMC and SMC regions followed up by a mapping campaign of bright star-forming regions in the Clouds.

\vspace{5pt}
\noindent {\it (i)} HCO$^+$(2$-$1) is detected in two thirds of the regions observed in the LMC and SMC and 
HCN(2$-$1) in about one-quarter of the 29 DeGaS-MC pointings observed, with HCN(2$-$1) detected 
only toward the brightest HCO$^+$ sources. 
The three targeted SMC sources were unfortunately not detected in our mapping campaign but 
both lines are detected toward the LMC 30Dor, N44, N105, N113, N159W, N159E, and N214 regions.
For most of the sources targeted during the mapping campaign, the HCN(2$-$1) and HCO$^+$(2$-$1) peaks 
correspond to peaks in the molecular gas and dust emission. The HCN emission is generally less extended than the 
HCO$^+$ emission and is restricted to the densest regions. The largest line widths are found toward 30Dor, N159, and 
N44 star-forming regions and could be a sign that these regions host more massive, high-column-density clumps. 

\vspace{5pt}
\noindent {\it (ii)} The HCO$^+$(2$-$1)/HCN(2$-$1) brightness temperature ratio is above unity in both Magellanic Clouds 
and can reach values up to 7. These high values are consistent with the generally high ratios observed in
low-metallicity environments. The ratio does not seem to strongly correlate with the gas surface density or temperature 
and only moderately correlates with the SFR, with the relation showing strong source-to-source dispersion. 

\vspace{5pt}
\noindent {\it (iii)} The highest number of YSOs per LOS is found at higher HCO$^+$ intensities, 
consistent with the fact that YSOs are generally expected toward denser LOSs. More YSO candidates 
are preferentially found at lower HCO$^+$/HCN flux ratios, although more statistics would be needed to 
confirm and interpret the trend.

\vspace{5pt}
\noindent {\it (iv)} The highest dense gas fractions are found in 30Dor and the lowest are found in N214. 
The dense gas fraction correlates with the SFR, although most of the relation is driven by these two 
more active or more quiescent regions. Finally, substantial region-to-region variations are observed in the 
dense gas fraction versus the ({\bfaa LOS}) molecular surface density.


\section*{Acknowledgments}
We would like to first thank the referee for his/her thorough reading and very helpful feedback.
This publication is based on data acquired with the Atacama Pathfinder Experiment. 
We thank the APEX observing staff for completing this large program as well as their constant 
motivation and dedication to the community.
APEX is a collaboration between the Max-Planck-Institut fur Radioastronomie, the European 
Southern Observatory (ESO), and the Onsala Space Observatory. 
SEPIA is a collaboration between Sweden and ESO. 
MG has received funding from the European Research Council (ERC) under the
European Union Horizon 2020 research and innovation programme (MagneticYSOs project, grant agreement
No 679937, PI: Maury).
MC gratefully acknowledges funding from the 
Deutsche Forschungsgemeinschaft (DFG, German Research Foundation) through an Emmy Noether 
Research Group (grant number KR4801/1-1) and the DFG Sachbeihilfe (grant number KR4801/2-1).
FLP acknowledges funding from the ANR grant LYRICS (ANR-16-CE31-0011). 
This work was partly supported by the Programme National ``Physique et Chimie du Milieu Interstellaire" (PCMI) 
of CNRS/INSU with INC/INP co-funded by CEA and CNES. 


\bibliographystyle{aa}
\bibliography{mybiblio.bib}

\appendix

\section{Pointing campaign}

\begin{landscape}
\begin{table}
\caption{Characteristics of the pointing campaign sample.}
\label{LineCharacteristics1}
\centering
\begin{tabular}{ccccccccccccccc}
\hline
\hline
\vspace{-5pt}
&\\
Name    & Region $^a$           & $\alpha$ \& $\delta$ (J2000) &MAGMA $^b$ & COSSA $^c$   & rms$^d$ && \multicolumn{3}{c}{HCO$^+$(2$-$1) $^e$}   && \multicolumn{3}{c}{HCN(2$-$1)}              \\
\vspace{-5pt}
&\\\cline{8-10}\cline{12-14}
\vspace{-5pt}
&\\
&& ($^h$,$^m$,$^s$)~~~($^{\circ}$,$\arcmin$,$\arcsec$)& L$_{CO(1-0)}$                       &L$_{CO(2-1)}$& (mK) && Vel. Offset $^f$ & FWHM            & $\mathrm{\int T_{mb}~dv}$~$^g$        && Vel. Offset       & FWHM             & $\mathrm{\int T_{mb}~dv}$    \\
&&     &{\tiny(K~km/s pc$^2$)} & {\tiny(K~km/s pc$^2$)}&&& {\tiny(km~s$^{-1}$)}   & {\tiny(km~s$^{-1}$)} & {\tiny(K~km~s$^{-1}$)}       && {\tiny(km~s$^{-1}$)} & {\tiny(km~s$^{-1}$)}  & {\tiny(K~km~s$^{-1}$)}            \\
\vspace{-5pt}
&\\
\hline
\vspace{-5pt}
&\\
LMC \\
\#1  &                          &       05:44:34.32~~-69:26:13.2        & 8.9$\times$10$^4$~{\tiny(511)}          &-& 8.9                 
& & -                   & -                     & $<$0.05                              & & -                     & - & $<$0.05  \\
\#2  & {\tiny N159W}            &       05:39:31.20~~-69:45:57.6        & 4.8$\times$10$^4$~{\tiny(416)}          &-& 17.6                
& & -25.4$\pm$0.2       & 5.8$\pm$0.4 & 1.43$\pm$0.12   & & -23.9$\pm$0.9         & 7.8$\pm$1.9 & 0.284$\pm$0.097 \\
\#3  &                          &       05:44:02.88~~-69:22:08.4 & 4.5$\times$10$^4$~{\tiny(504)}                 &-& 6.7 
& & -                   & -                     & $<$ 0.04                             & & -                     & - &  $<$0.04  \\
\#4  & {\tiny N219}             &       05:40:59.76~~-70:22:48.0 & 3.4$\times$10$^4$ ~{\tiny(471)}           &-& 7.2 
& & -32.5$\pm$0.6       & 3.8$\pm$1.1 & 0.091$\pm$0.034                 & & -                     & - & $<$0.04  \\
\#5  & {\tiny N44}              &       05:22:06.00~~-67:58:30.0        & 2.8$\times$10$^4$~{\tiny(229)}          &-& 16.6                
& & 20.7$\pm$0.1        & 4.9$\pm$0.2 & 1.98$\pm$0.10   & & 21.0$\pm$0.3         & 6.5$\pm$0.9 & 0.758$\pm$0.128 \\
\#7  &                                  &       05:24:20.64~~-68:26:02.4        & 2.5$\times$10$^4$~{\tiny(275)}          &-& 6.5         
& & -2.5$\pm$0.7        & 6.0$\pm$1.8 & 0.151$\pm$0.055                 & & -                     & - & $<$0.03  \\
\#8  &                          &       05:22:11.28~~-69:41:45.6        & 2.5$\times$10$^4$~{\tiny(233)}          &-& 7.2                 
& & -                   & -                     & $<$0.04                            & & -                     & - & $<$0.04  \\
\#11  & {\tiny 30Dor}           &       05:38:44.88~~-69:03:32.4        & 1.7$\times$10$^4$~{\tiny(398)}          &-& 8.9         
& & -14.8$\pm$0.3       & 4.2$\pm$0.8 & 0.39$\pm$0.08   & & -                   & - & $<$0.05 \\
\#12  &                         &       05:55:42.48~~-68:09:57.6        & 1.5$\times$10$^4$~{\tiny(543)}          &-& 5.7         
& & 21.9$\pm$0.2        & 2.1$\pm$5.4 & 0.085$\pm$0.22          & & -                   & - & $<$0.03  \\
\#13  &                         &       05:19:29.04~~-69:08:27.6        & 1.4$\times$10$^4$~{\tiny(196)}          &-& 8.1         
& & 6.2$\pm$0.9         & 6.1$\pm$1.7 & 0.173$\pm$0.068                 & & -                     & - & $<$0.04   \\
\#14  & {\tiny N79S}            &       04:51:54.96~~-69:22:51.6        & 1.3$\times$10$^4$~{\tiny(34)}           &-& 14.6                
& & -29.3$\pm$0.5       & 7.5$\pm$1.4 & 0.675$\pm$0.153         & & -28.4$\pm$0.5         & 3.9$\pm$1.6 & 0.182$\pm$0.084 \\      
\#16  & {\tiny N76}             &       04:49:36.00~~-68:22:04.8        & 1.1$\times$10$^4$~{\tiny(25)}           &-& 9.9         
& & -                   & -                     & $<$0.05                              & & -                     & - & $<$0.05  \\
\#17  & {\tiny N11C}            &       04:57:48.24~~-66:28:51.6        & 9.2$\times$10$^3$~{\tiny(78)}           &-& 15.3                
& & 18.2$\pm$0.1        & 3.8$\pm$0.4 & 0.788$\pm$0.091         & & 19.2$\pm$0.4         & 4.4$\pm$1.1 & 0.243$\pm$0.083 \\
\#18  & {\tiny N79W}            &       04:48:53.76~~-69:10:04.8        & 8.6$\times$10$^3$~{\tiny(13)}           &-& 7.9         
& & -21.5$\pm$0.3       & 4.1$\pm$0.5 &  0.277$\pm$0.040                & & -                     & - & $<$0.04  \\
\#19  &                                 &       05:21:28.32~~-67:46:40.8        & 7.5$\times$10$^3$~{\tiny(221)}                  &-& 8.3         
& & -                   & -                     & $<$0.04                             & & -                     & - & $<$0.04 \\
\#20  & {\tiny N11I}            &       04:55:38.64~~-66:34:08.4        & 7.1$\times$10$^3$~{\tiny(63)}           &-& 13.2                
& & 18.5$\pm$0.2        & 3.1$\pm$0.7 & 0.424$\pm$0.098         & & -                   & - & $<$0.07 \\
\#21  &                                 &       04:57:06.48~~-69:11:24.0        & 6.6$\times$10$^3$~{\tiny(70)}           &-& 7.1         
& & -                   & -                     & $<$ 0.04                            & & -                     & - & $<$0.04 \\
\#22  & {\tiny N15}             &       05:00:44.64~~-66:22:37.2        & 5.6$\times$10$^3$~{\tiny(98)}           &-& 6.8         
& & 14.0$\pm$0.1        & 2.1$\pm$2.6 & 0.115$\pm$0.151                 & & -                     & - & $<$0.04  \\
\#23  &                         &       05:24:38.64~~-69:14:52.8        & 4.7$\times$10$^3$~{\tiny(278)}                  &-& 10.1                
& & -1.2$\pm$0.2        & 2.9$\pm$0.4 & 0.266$\pm$0.043                 & & -                     & - & $<$0.05  \\
\#24  & {\tiny N55}             &       05:32:30.00~~-66:27:10.8        & 4.1$\times$10$^3$~{\tiny(338)}                  &-& 5.5         
& & 27.4$\pm$0.1        & 4.2$\pm$0.1 & 0.888$\pm$0.035         & & 27.7$\pm$0.2         & 5.9$\pm$0.5 & 0.390$\pm$0.045 \\
\#25  & {\tiny N57}             &       05:32:30.72~~-67:41:20.4        & 3.6$\times$10$^3$~{\tiny(339)}                  &-& 16.0                
& & 26.9$\pm$0.3        & 4.8$\pm$0.7 & 0.554$\pm$0.103         & & 26.1$\pm$0.8        & 6.2$\pm$1.7 & 0.243$\pm$0.086  \\
\vspace{-5pt}
&\\
\hline
\vspace{-5pt}
&\\
SMC \\
\#1     & {\tiny N27}   &       00 48 21.79~~-73 05 45.5        & -             &2.8$\times$10$^4$& 17.6    
& &  -42.6$\pm$0.4      & 4.6$\pm$1.0   & 0.517$\pm$0.133 & & -41.1$\pm$0.9 & 7.1$\pm$1.9 & 8.4$\pm$3.1 \\
\#2     & {\tiny N23}   &       00:47:55.30~~-73:17:21.6        & -             &1.7$\times$10$^4$& 11.1    
& &  -36.9$\pm$0.3      & 4.9$\pm$0.9   & 0.428$\pm$0.090 & & - & - & $<$0.06  \\
\#3     &                       &       00:48:08.50~~-73:23:31.5                & -               &-& 6.4         
& &  -                          & -                     & $<$0.03                                 & & - & - & $<$0.03  \\
\#4     & {\tiny N12A}  &       00:46:41.13~~-73:06:07.0        & -             &1.2$\times$10$^4$& 12.6    
& &  -31.5$\pm$0.2      & 2.5$\pm$0.5   & 0.401$\pm$0.086   & & - & - & $<$0.07  \\
\#5     & {\tiny N13}   &       00:45:18.46~~-73:22:52.3        & -             &5.9$\times$10$^3$& 6.9     
& &  -32.5$\pm$0.2      & 6.7$\pm$0.6   & 0.513$\pm$0.056       & & -31.2$\pm$0.6 & 7.2$\pm$1.3 & 5.6$\pm$1.4 \\
\#6     & {\tiny N76}   &       01:03:8.533~~-72:03:49.6        & -             &3.4$\times$10$^2$& 10.4    
& &  -                          & -                     & $<$0.06                          & & - & - & $<$0.05  \\
\#7     &                       &       01:03:31.89~~-71:56:56.2        & -               &2.8$\times$10$^3$& 6.2         
& &  -                          & -                     & $<$0.03                          & & - & - & $<$0.03  \\
\#8     & {\tiny N71}   &       01:00:53.81~~-71:35:21.5        & -             &2.3$\times$10$^3$& 7.4     
& &  -0.8$\pm$0.2       & 3.4$\pm$0.4   & 0.232$\pm$0.033        & & - & - & $<$0.04  \\
&\\
\hline
\end{tabular}
\begin{list}{}{}
\item[$^a$] Name of the closest H{\sc ii} region as identified by \citet{Henize1956}.
\item[$^b$] CO luminosity of the closest CO(1$-$0) cloud in the MAGMA physical decomposition catalogue. L$_{CO}$ is derived using the CPROPS method. The hosting cloud ID number in the catalogue is indicated in parenthesis.
\item[$^c$] CO luminosity of the associated CO(2$-$1) cloud taken as part of the COSSA project (PI: van Kempen, private communication). 
The extraction is also performed with CPROPS. 
\item[$^d$] The rms is estimated from the HCO$^+$(2$-$1) spectra (in T$_{mb}$) with 2 km~s$^{-1}$ channel width. 
\item[$^e$] The various line characteristics are derived via a Gaussian fit to the line. Values are provided for 3-$\sigma$ detections. 
\item[$^f$] We use a systemic velocity v$_{\rm sys}$=262.2 km~s$^{-1}$ for the LMC and 158 km~s$^{-1}$ for the SMC. 
\item[$^g$] Upper limits are estimated from the rms and assuming a linewidth of 5~km~s$^{-1}$.
\end{list}
\end{table} 
\end{landscape}

\begin{table}
\vspace{30pt}
\caption{Pointing campaign: Line ratios and SFR}
\label{LineCharacteristics1bis}
\centering
\begin{tabular}{cccc}
\hline
\hline
\vspace{-5pt}
&\\
Name    & HCO$^+$/ HCN      & L$_{TIR}$ (\lsun)  & SFR (\msun~yr$^{-1}$)    \\
& \\
\hline
\vspace{-5pt}
&\\
LMC \\
\#1     & -                             & 3.4$\times$10$^3$ &5.9$\times$10$^{-7}$ \\
\#2     & 5.02$\pm$1.78 & 1.0$\times$10$^5$ &1.7$\times$10$^{-5}$ \\
\#3     & -                             & 1.2$\times$10$^4$ &2.0$\times$10$^{-6}$ \\
\#4     & -                             & 7.0$\times$10$^3$ &1.2$\times$10$^{-6}$ \\
\#5     & 2.62$\pm$0.45 & 1.0$\times$10$^5$ &1.7$\times$10$^{-5}$ \\
\#7     & -                             & 4.3$\times$10$^3$ &7.3$\times$10$^{-7}$ \\
\#8     & -                             & 2.0$\times$10$^4$ &3.4$\times$10$^{-6}$ \\
\#11    & -                             & 1.6$\times$10$^5$ &2.7$\times$10$^{-5}$ \\
\#12    & -                             & 2.8$\times$10$^3$ &4.8$\times$10$^{-7}$ \\
\#13    & -                             & 2.6$\times$10$^4$ &4.4$\times$10$^{-6}$ \\
\#14    & 3.69$\pm$1.90 & 3.9$\times$10$^4$ &6.7$\times$10$^{-6}$ \\
\#16    & -                             & 1.6$\times$10$^3$ &2.7$\times$10$^{-7}$ \\
\#17    & 3.21$\pm$1.14 & 3.7$\times$10$^4$ &6.3$\times$10$^{-6}$ \\
\#18    & -                             & 1.7$\times$10$^4$ &3.0$\times$10$^{-6}$ \\
\#19    & -                             & 2.7$\times$10$^3$ &4.6$\times$10$^{-7}$ \\
\#20    & -                             & 6.4$\times$10$^3$ &1.1$\times$10$^{-6}$ \\
\#21    & -                             & 5.2$\times$10$^3$ &9.0$\times$10$^{-7}$ \\
\#22    & -                             & 2.5$\times$10$^3$ &4.2$\times$10$^{-7}$ \\
\#23    &-                              & 7.5$\times$10$^3$ &1.3$\times$10$^{-6}$ \\
\#24    & 2.28$\pm$0.28 & 3.2$\times$10$^4$ &5.5$\times$10$^{-6}$ \\
\#25    & 2.28$\pm$0.92 & 3.2$\times$10$^4$ &5.4$\times$10$^{-6}$\\
\vspace{-5pt}
&\\
\hline
\vspace{-5pt}
&\\
SMC \\
\#1 & 2.13$\pm$0.96 &   1.7$\times$10$^4$ &     2.9$\times$10$^{-6}$ \\
\#2 & -                          &      1.3$\times$10$^4$ &     2.3$\times$10$^{-6}$ \\
\#3 & -                          &      3.0$\times$10$^3$ &     5.2$\times$10$^{-7}$ \\
\#4 & -                          &      2.1$\times$10$^4$ &     3.7$\times$10$^{-6}$ \\
\#5 & 3.05$\pm$0.80      &      1.1$\times$10$^4$ &     1.8$\times$10$^{-6}$ \\
\#6 & -                          &      9.1$\times$10$^3$ &     1.6$\times$10$^{-6}$ \\
\#7 & -                          &      3.4$\times$10$^3$ &     5.9$\times$10$^{-7}$ \\
\#8 & -                          &      4.6$\times$10$^3$ &     8.0$\times$10$^{-7}$ \\
&\\
\hline
\end{tabular}
\end{table}

\begin{figure*}
\centering
\vspace{30pt}
\begin{tabular}{ccc}
\includegraphics[height=4cm]{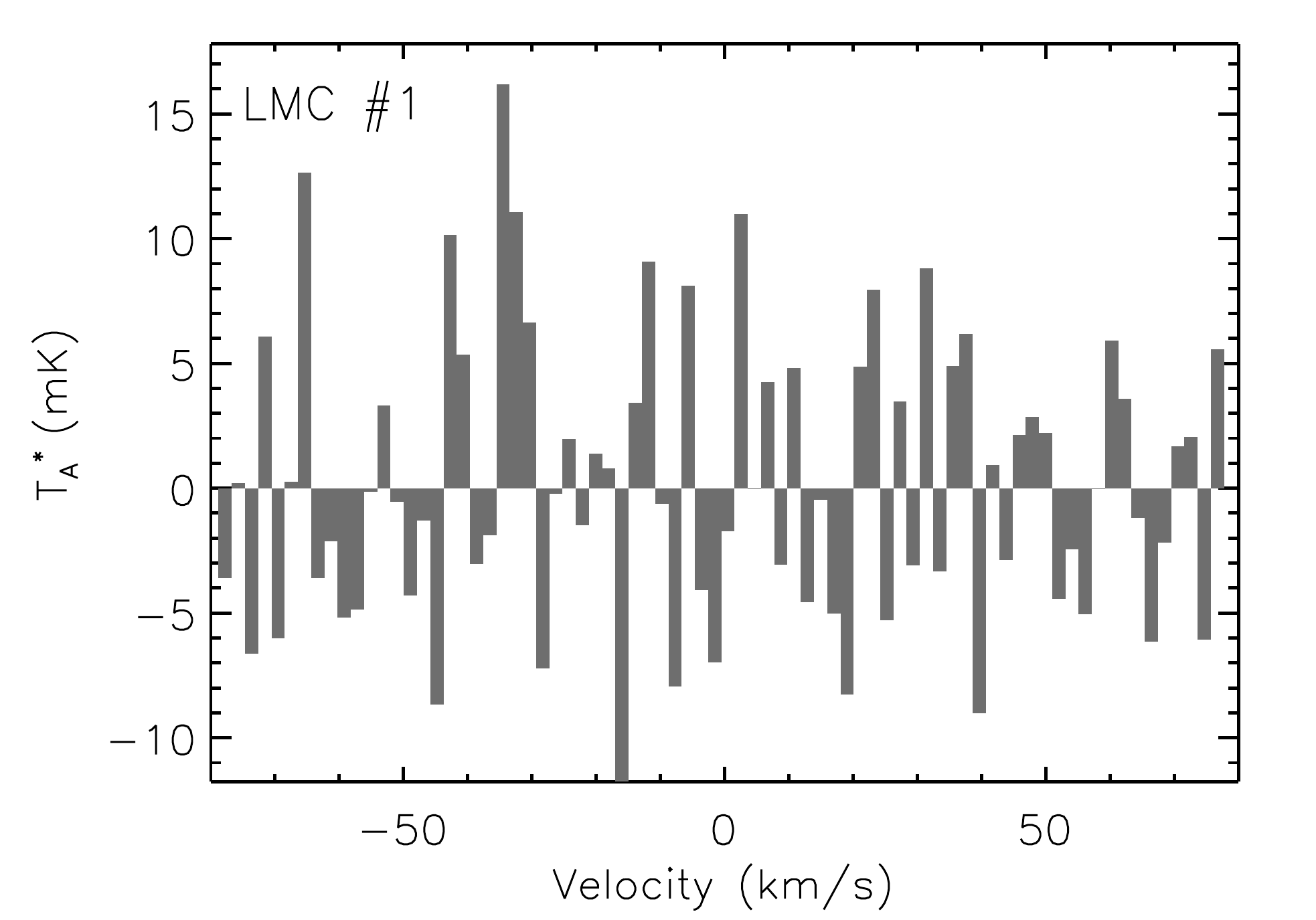}  & 
\hspace{-0.4cm}\includegraphics[height=4cm]{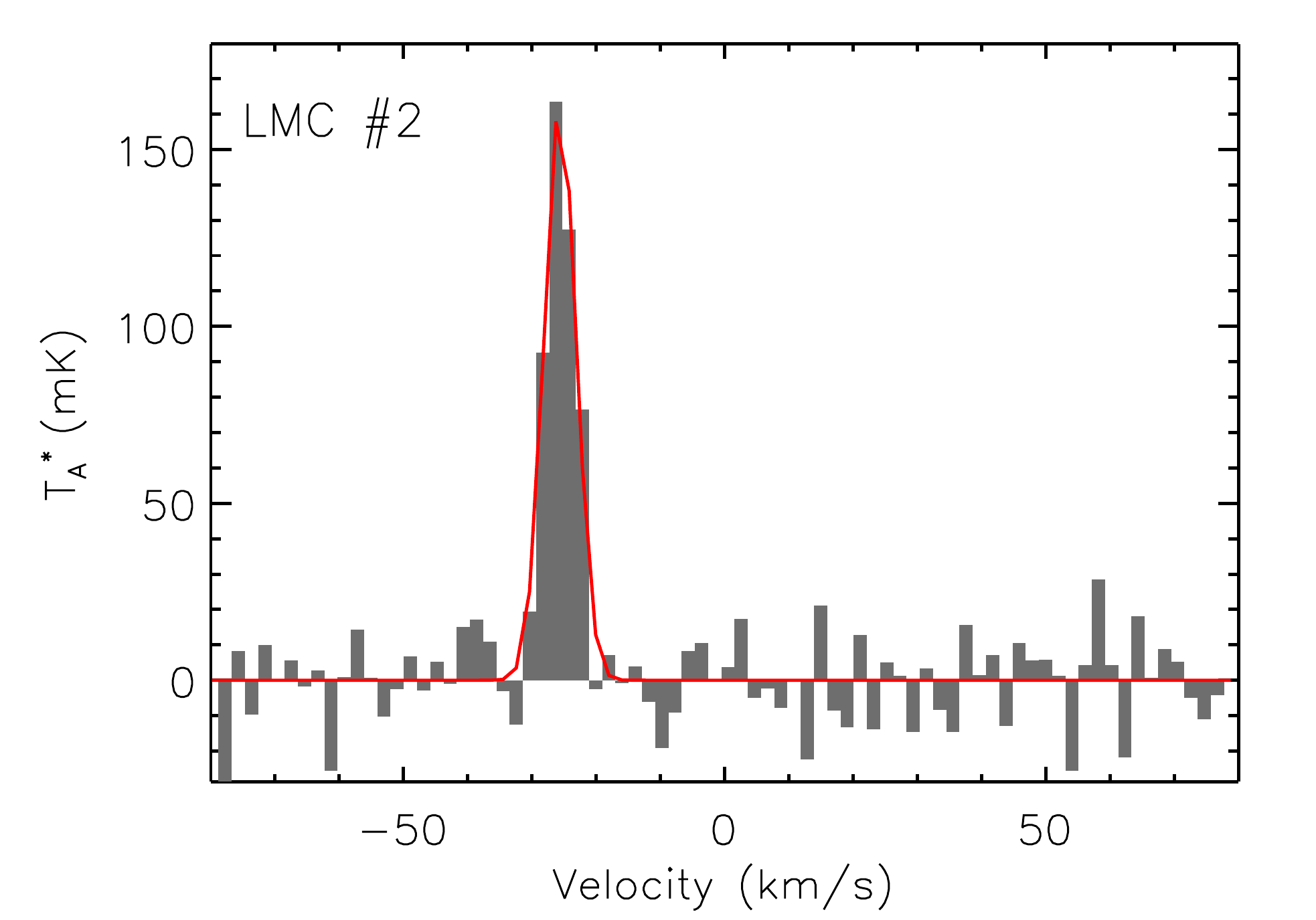}  &
\hspace{-0.4cm}   \includegraphics[height=4cm]{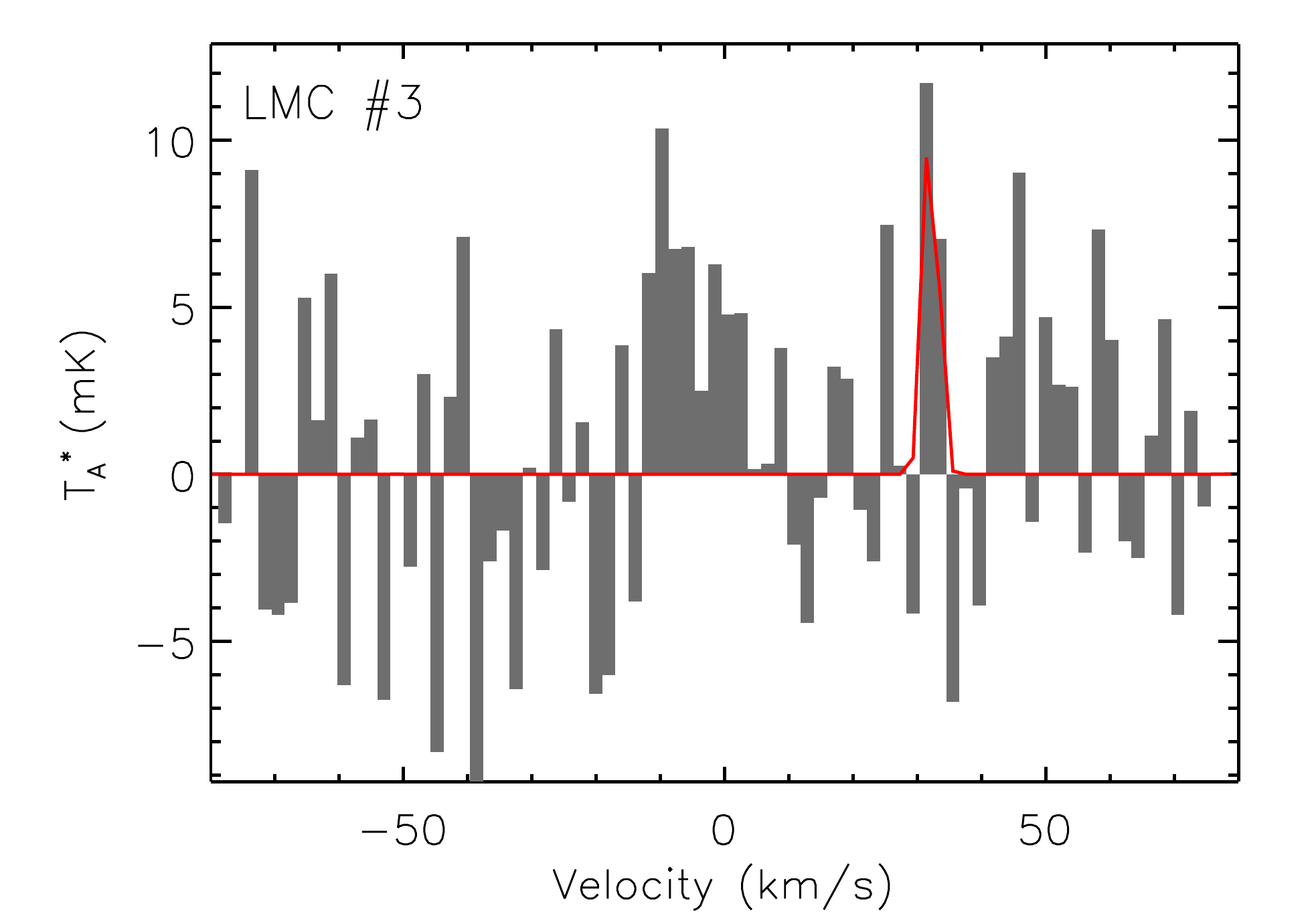} \\
\includegraphics[height=4cm]{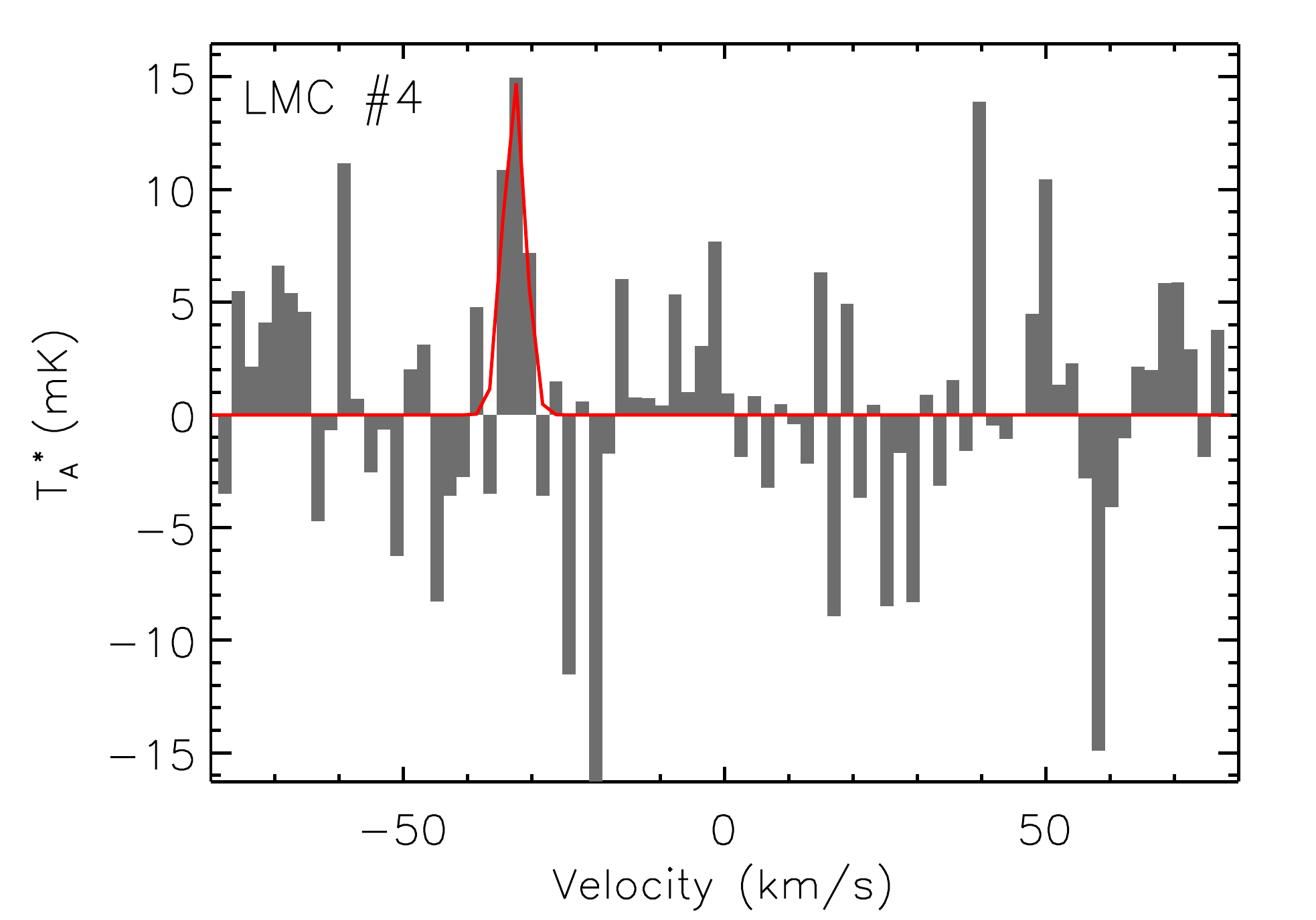}  & 
\hspace{-0.4cm}\includegraphics[height=4cm]{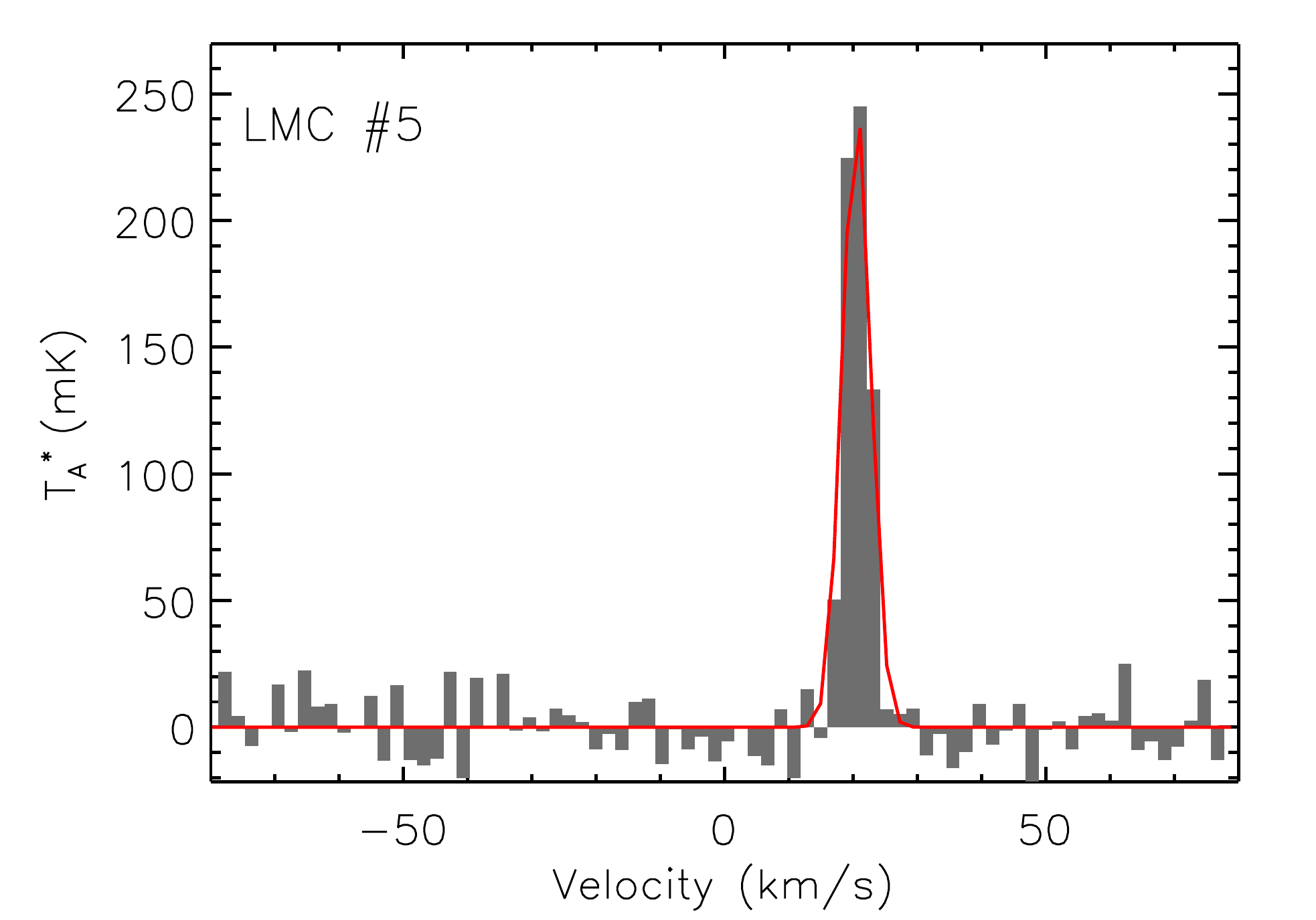}  &
\hspace{-0.4cm}  \includegraphics[height=4cm]{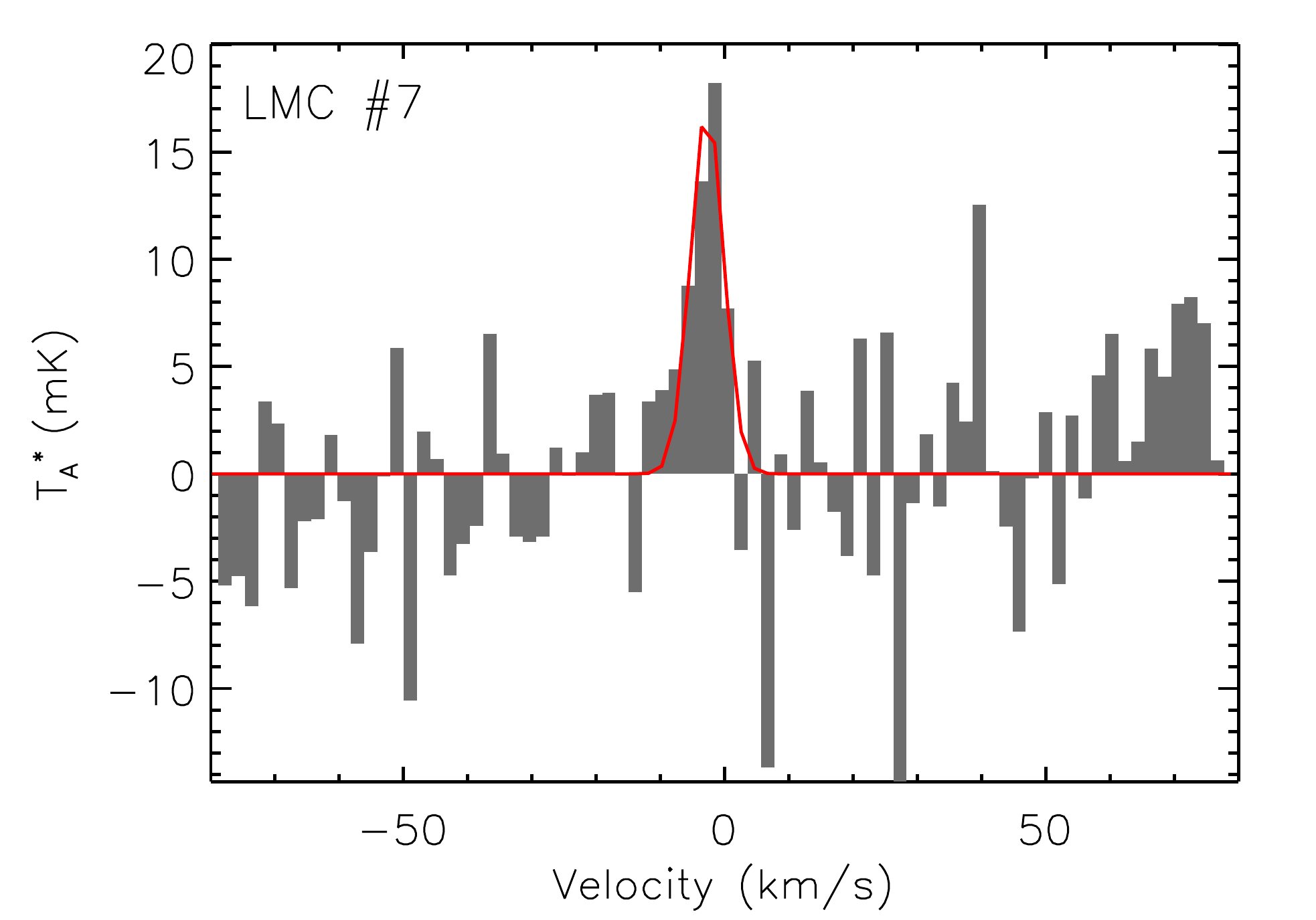} \\
\includegraphics[height=4cm]{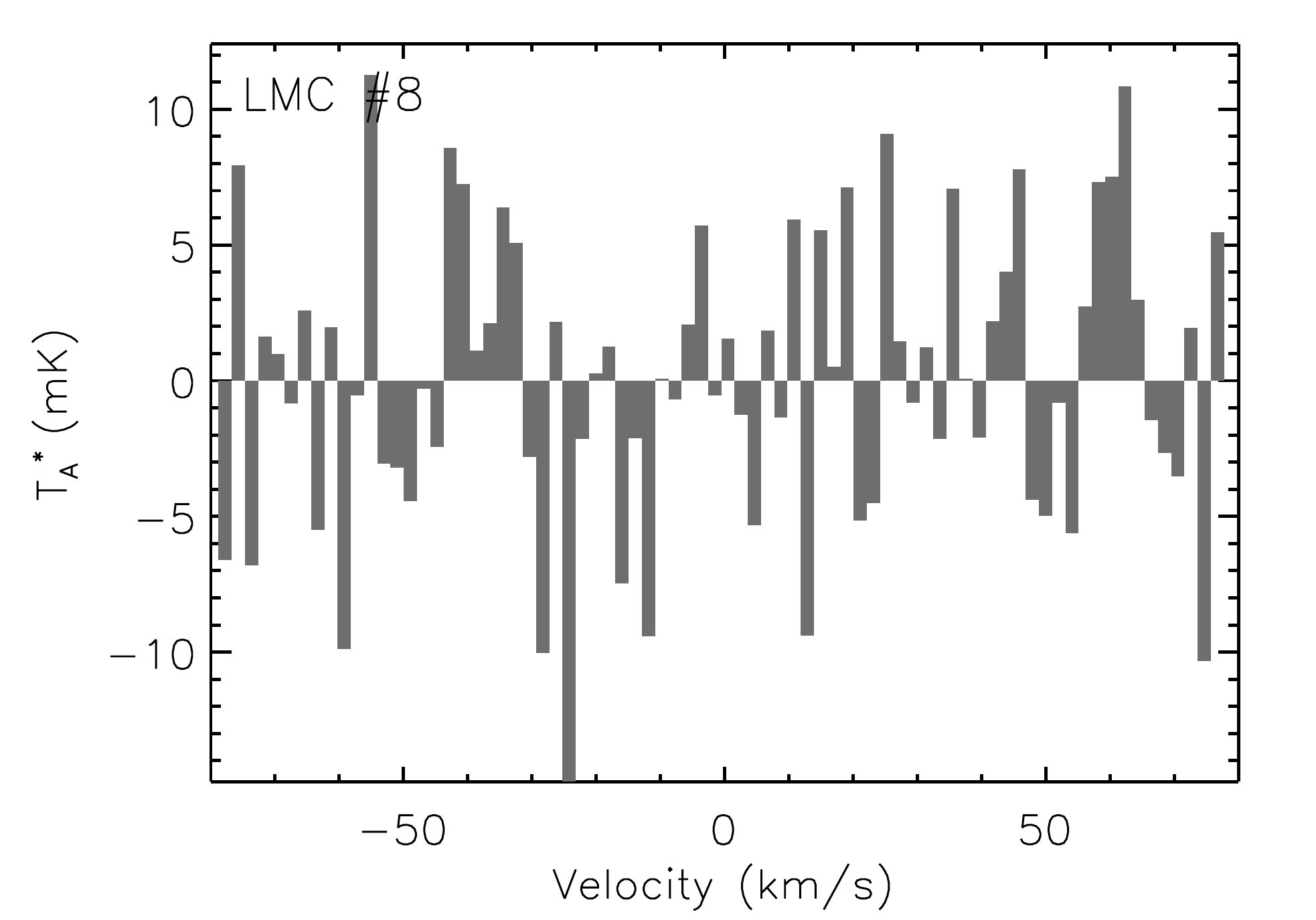}  & 
\hspace{-0.4cm}\includegraphics[height=4cm]{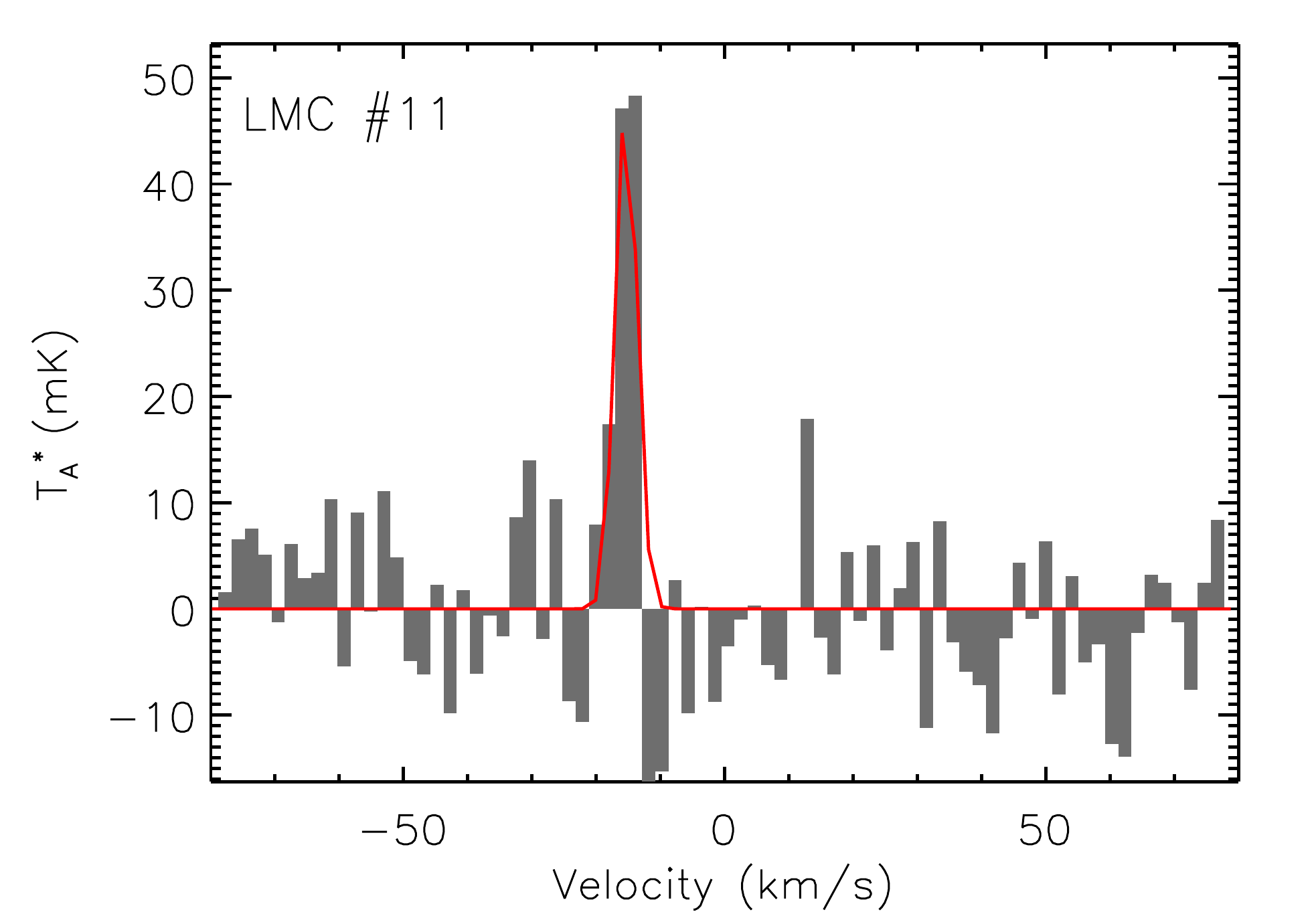}  &
\hspace{-0.4cm}  \includegraphics[height=4cm]{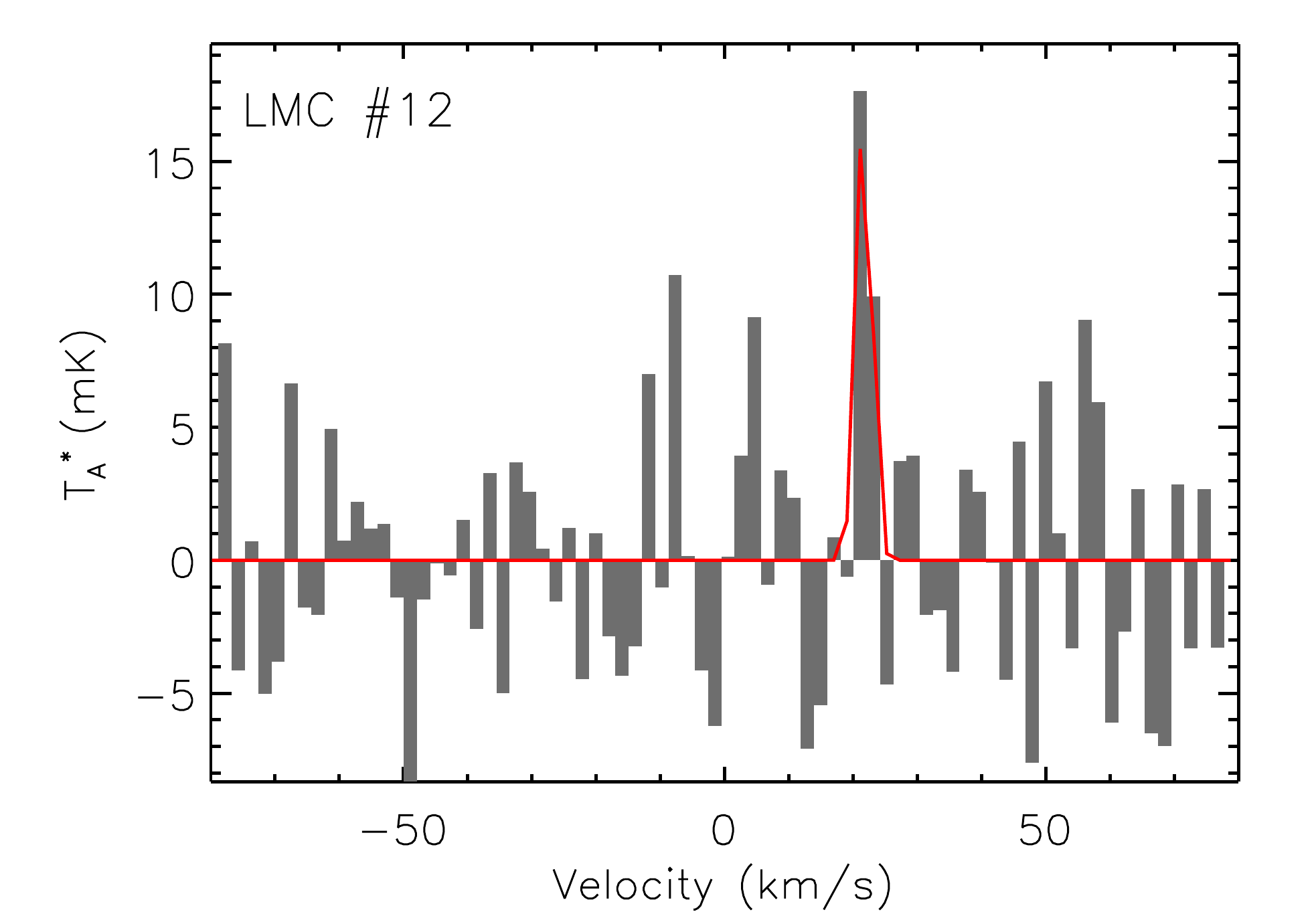} \\ 
\includegraphics[height=4cm]{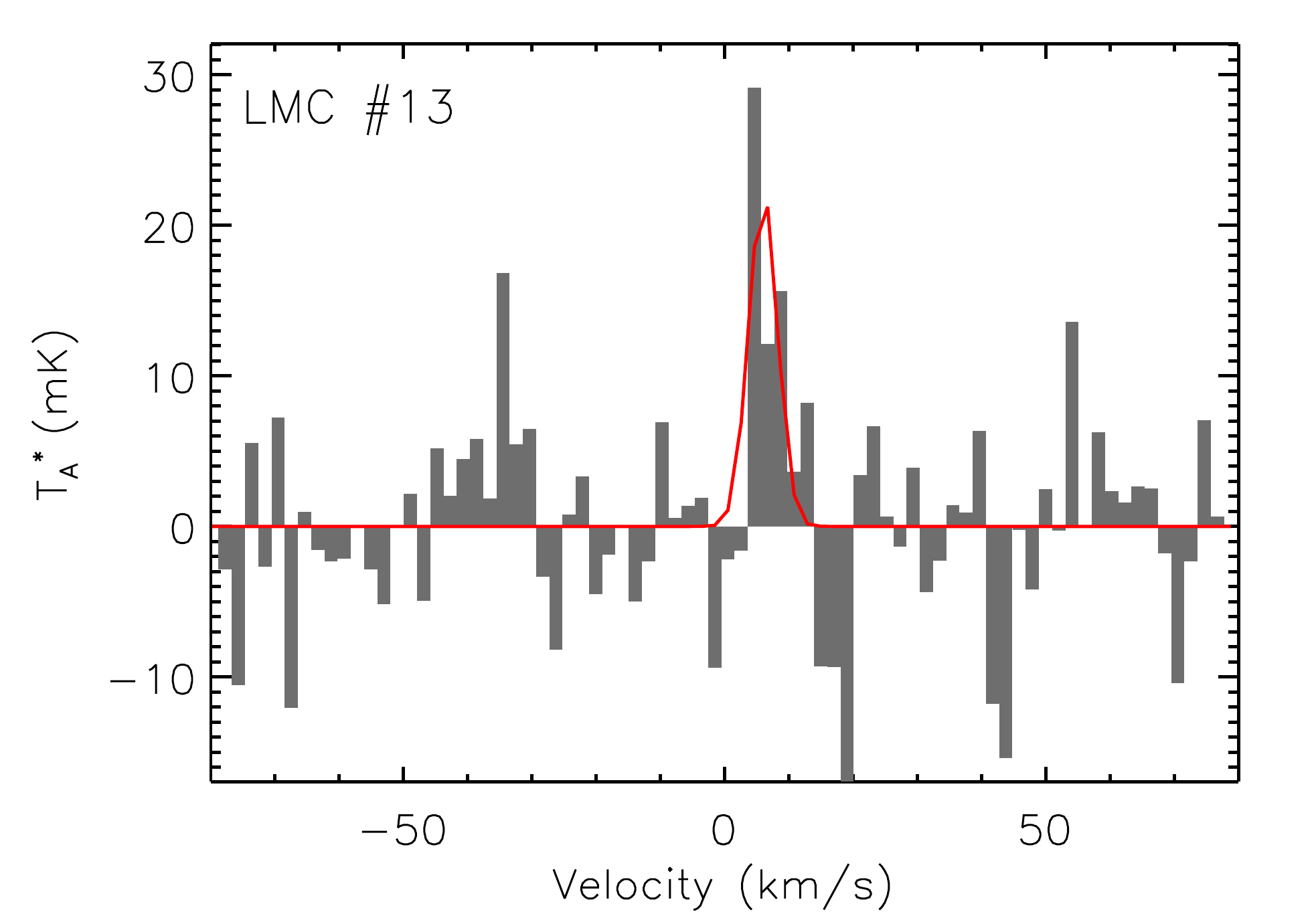}  & 
\hspace{-0.4cm}\includegraphics[height=4cm]{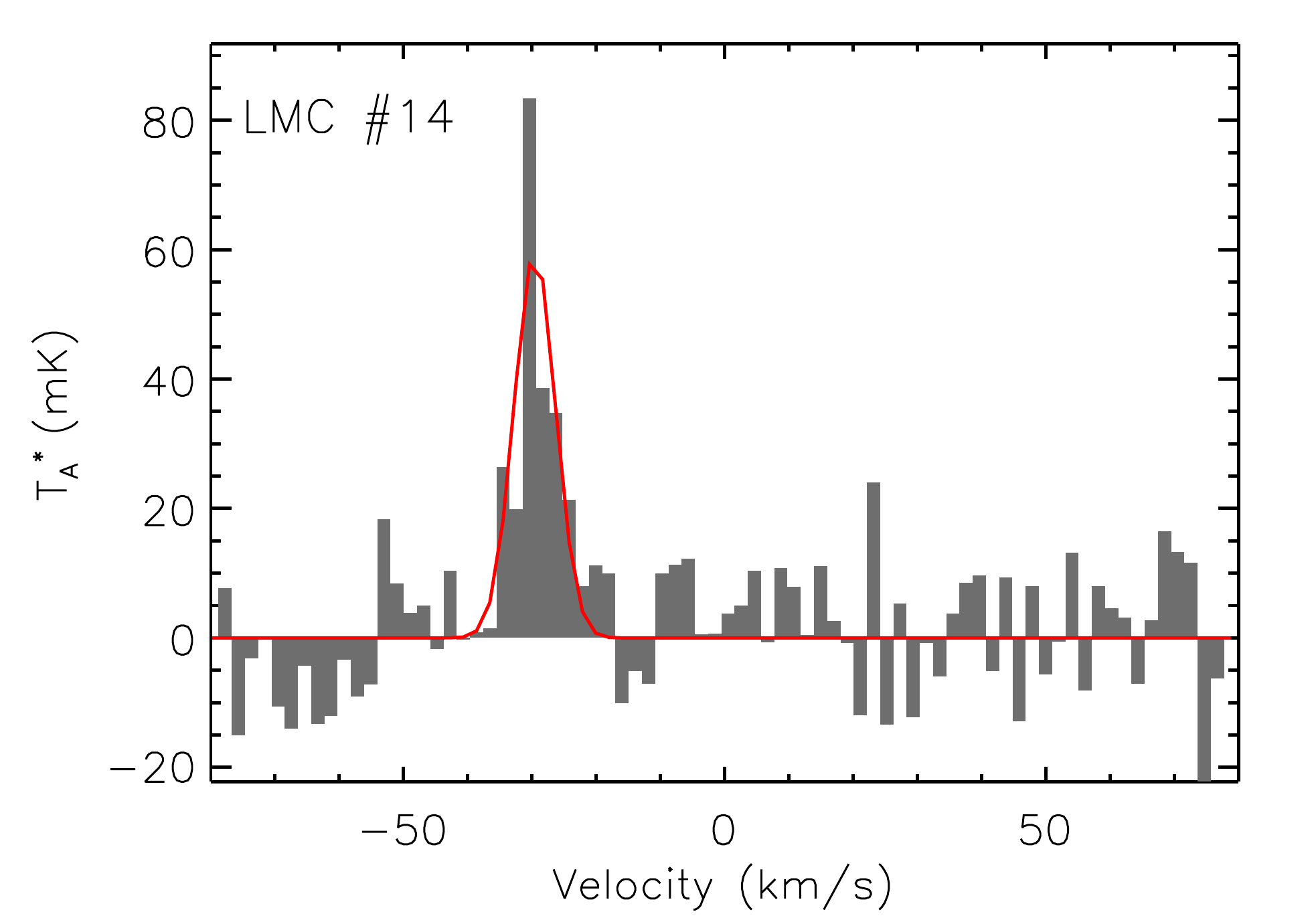}  &
\hspace{-0.4cm}  \includegraphics[height=4cm]{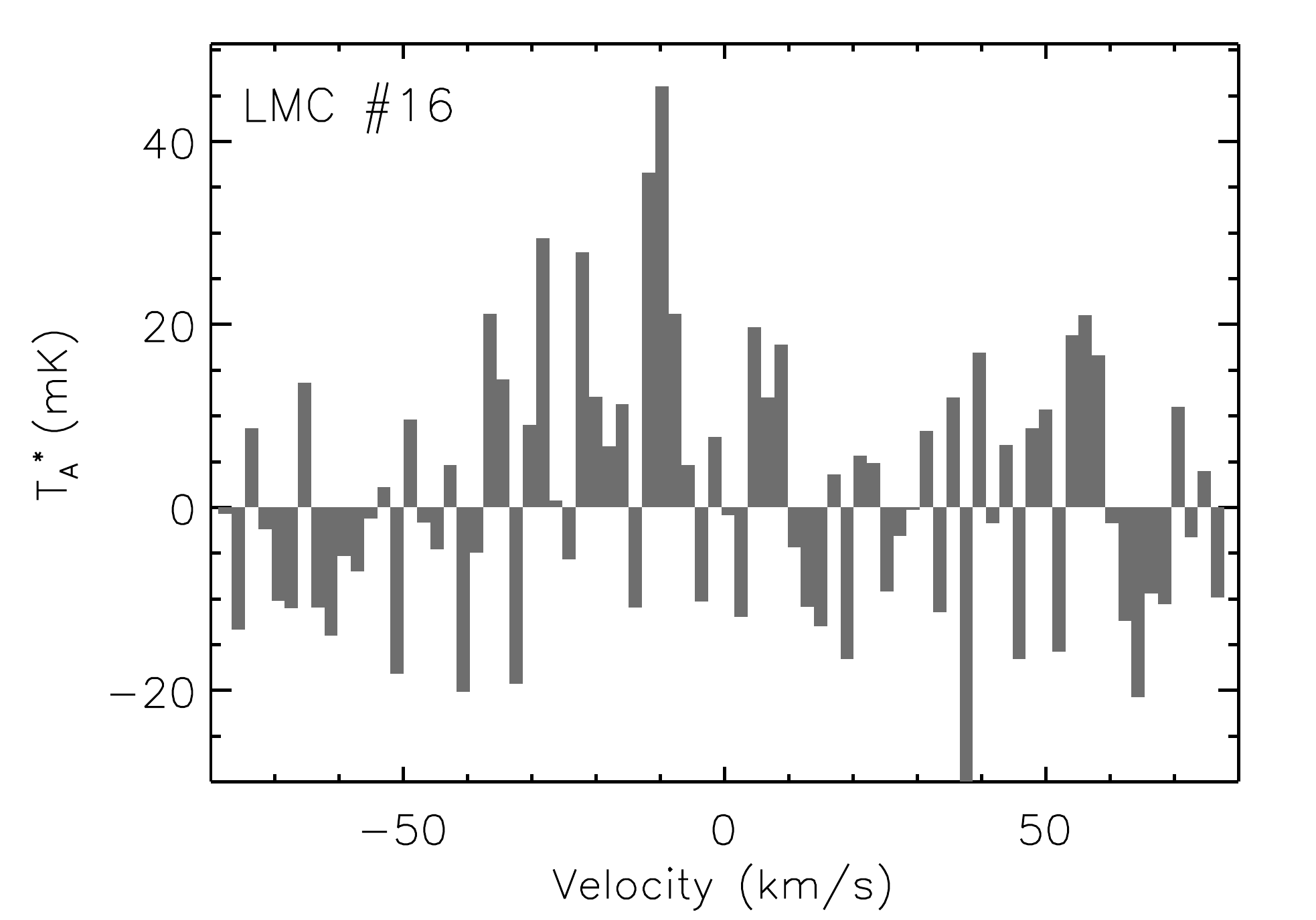} \\ 
\includegraphics[height=4cm]{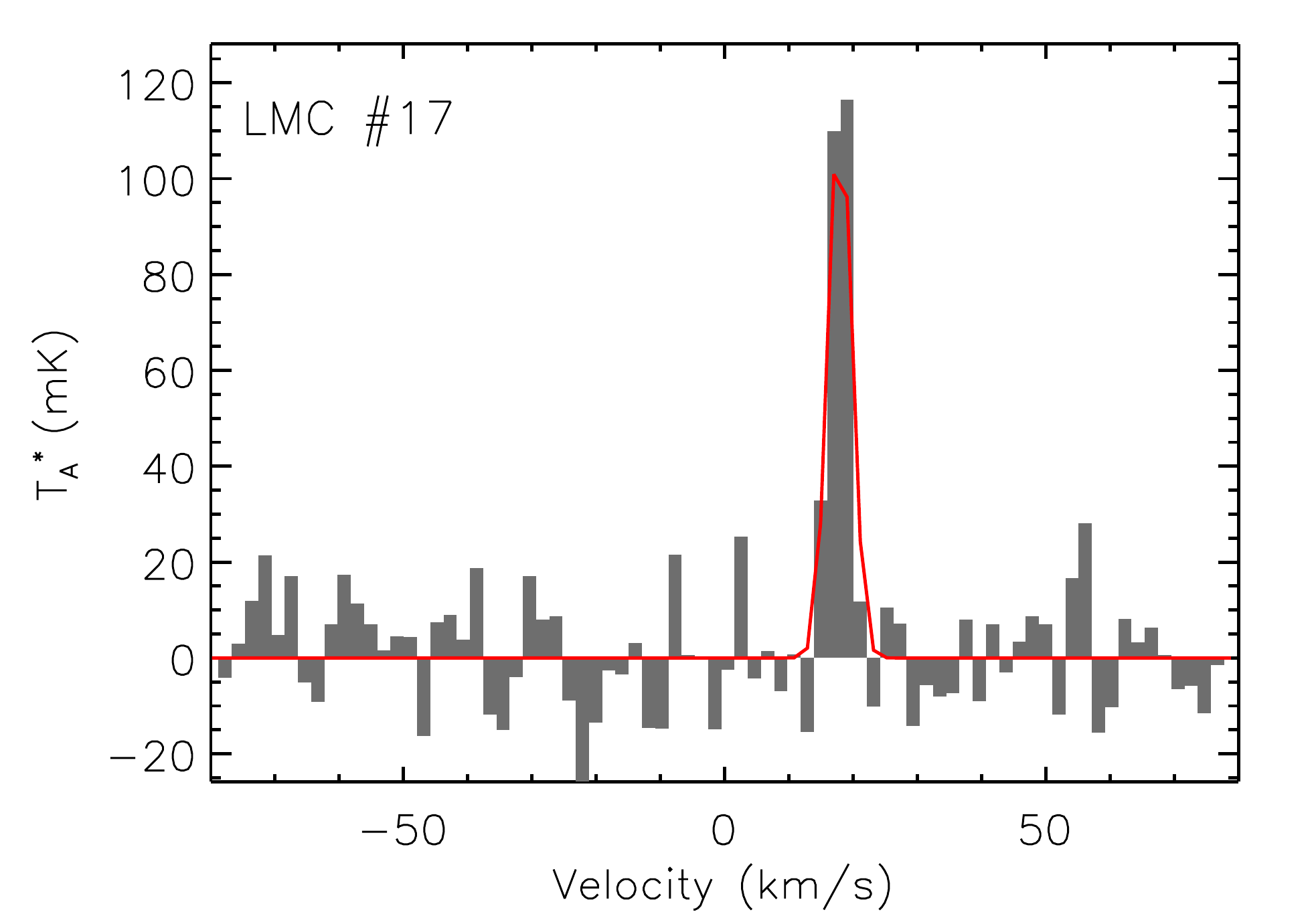}  & 
\hspace{-0.4cm}\includegraphics[height=4cm]{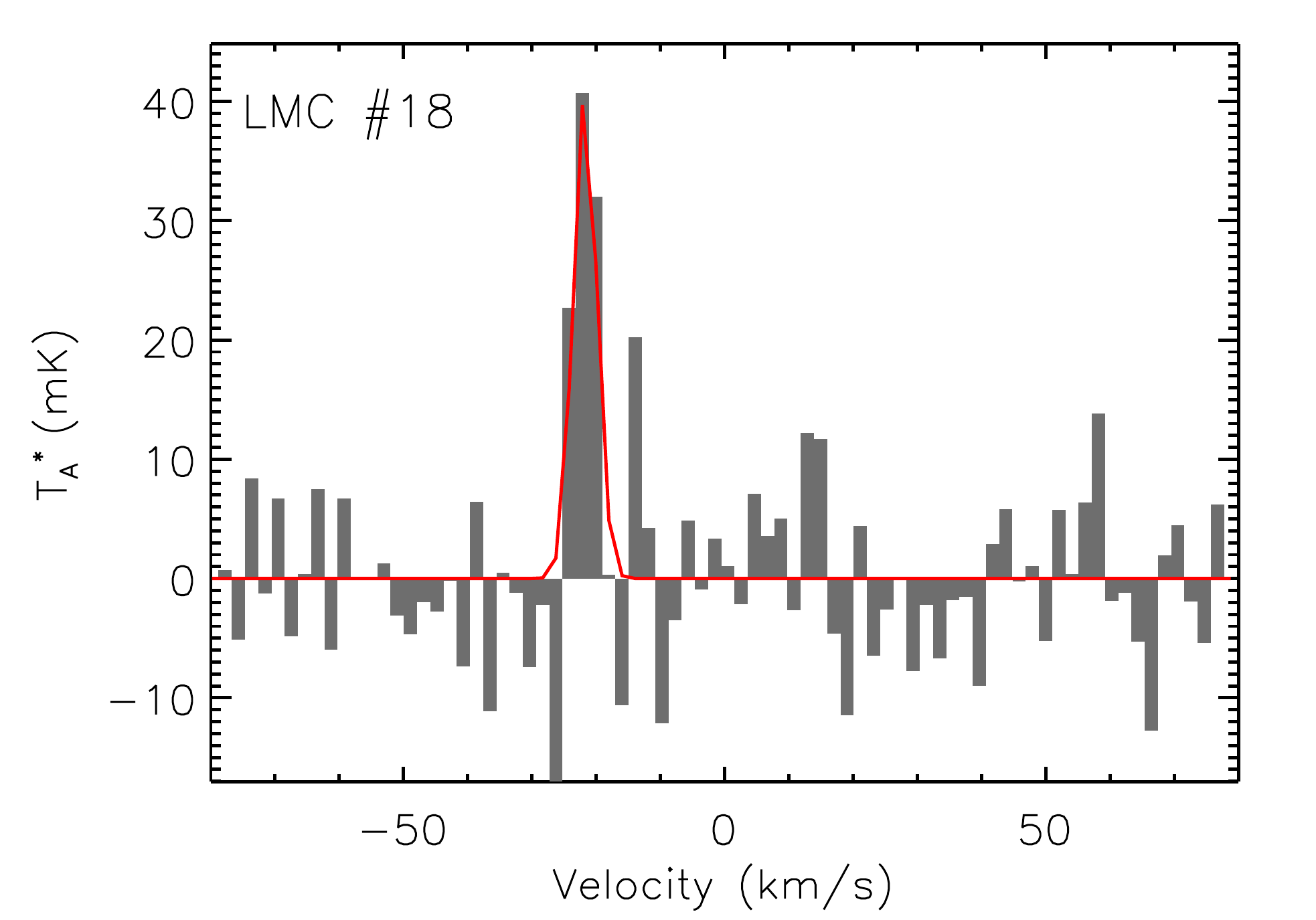}  &
\hspace{-0.4cm}   \includegraphics[height=4cm]{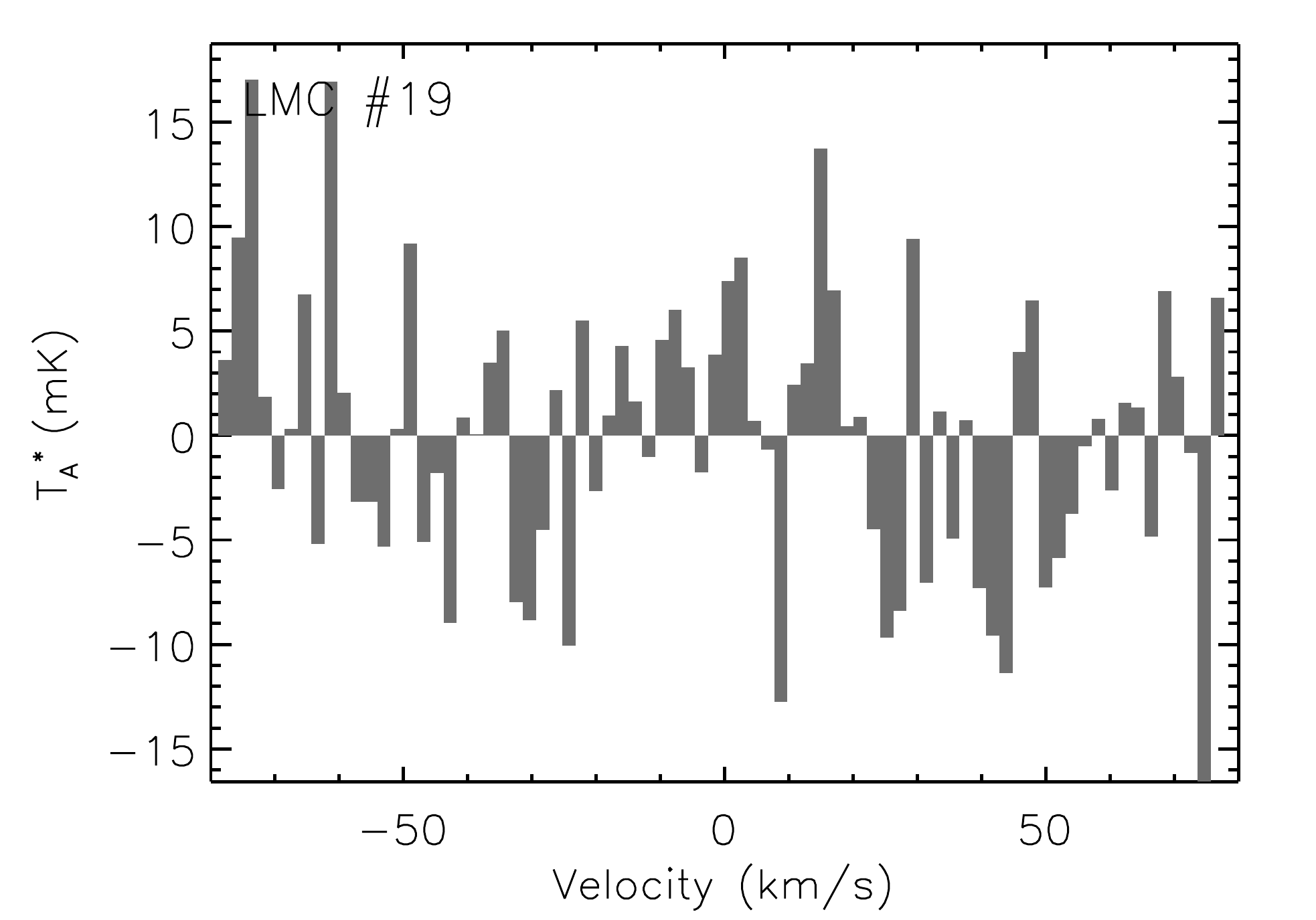}\\ 
\end{tabular}
   \caption{HCO$^+$(2$-$1) observations toward the LMC and SMC (2 km~s$^{-1}$ spectral resolution). 
   The bottom x-axis is expressed in velocity with respect to the systemic velocity (v=0 corresponding to 
   262.2 km~s$^{-1}$ for the LMC and 158 km~s$^{-1}$ for the SMC). We indicate with a red line the Gaussian
   fit on the lines detected at a 3-$\sigma$ level.}
   \label{LMCSMC_HCOp}
\end{figure*}
\addtocounter {figure}{-1}

\begin{figure*}
\centering
\vspace{30pt}
\begin{tabular}{ccc}
\includegraphics[height=4cm]{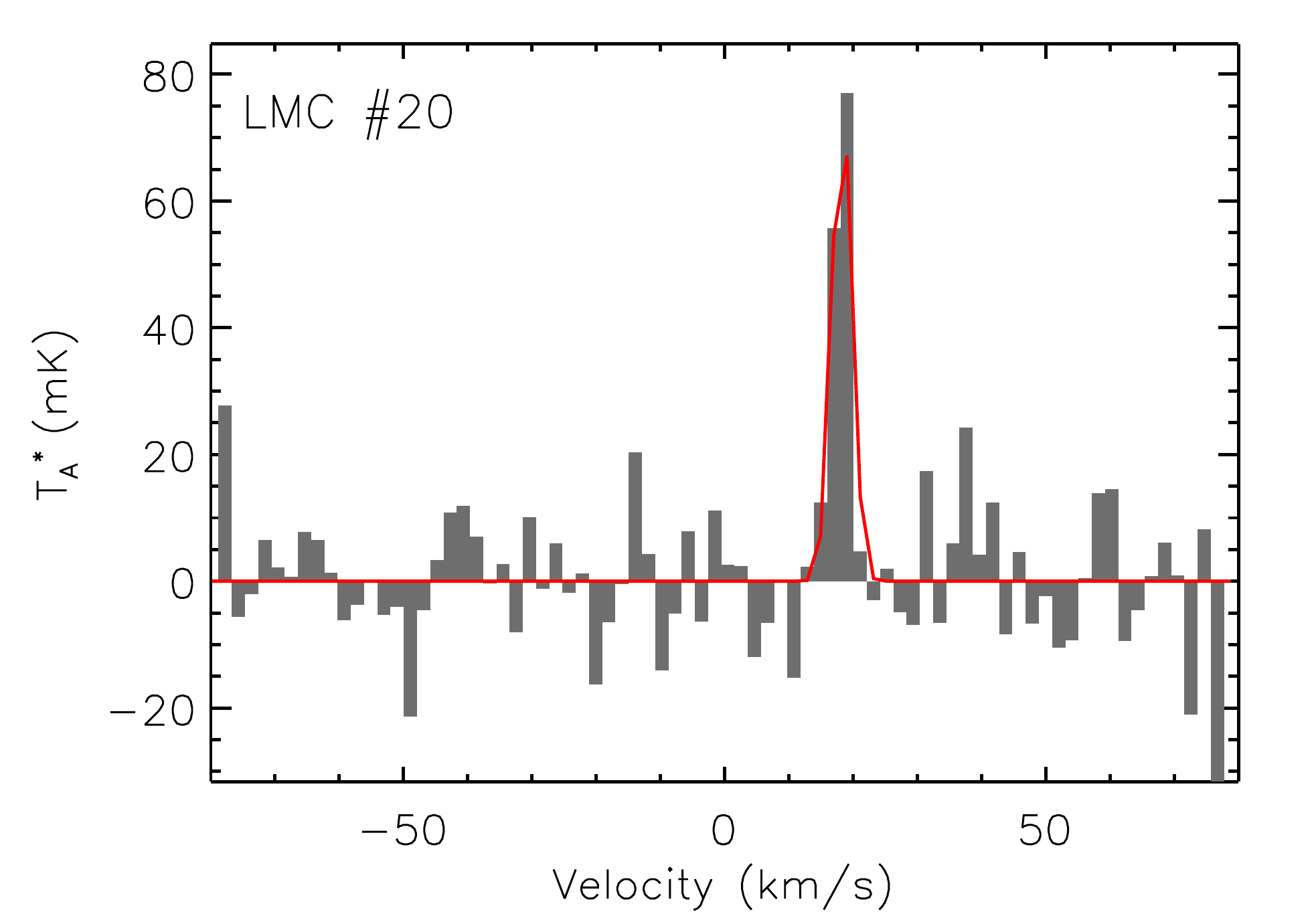}  & 
\hspace{-0.4cm}\includegraphics[height=4cm]{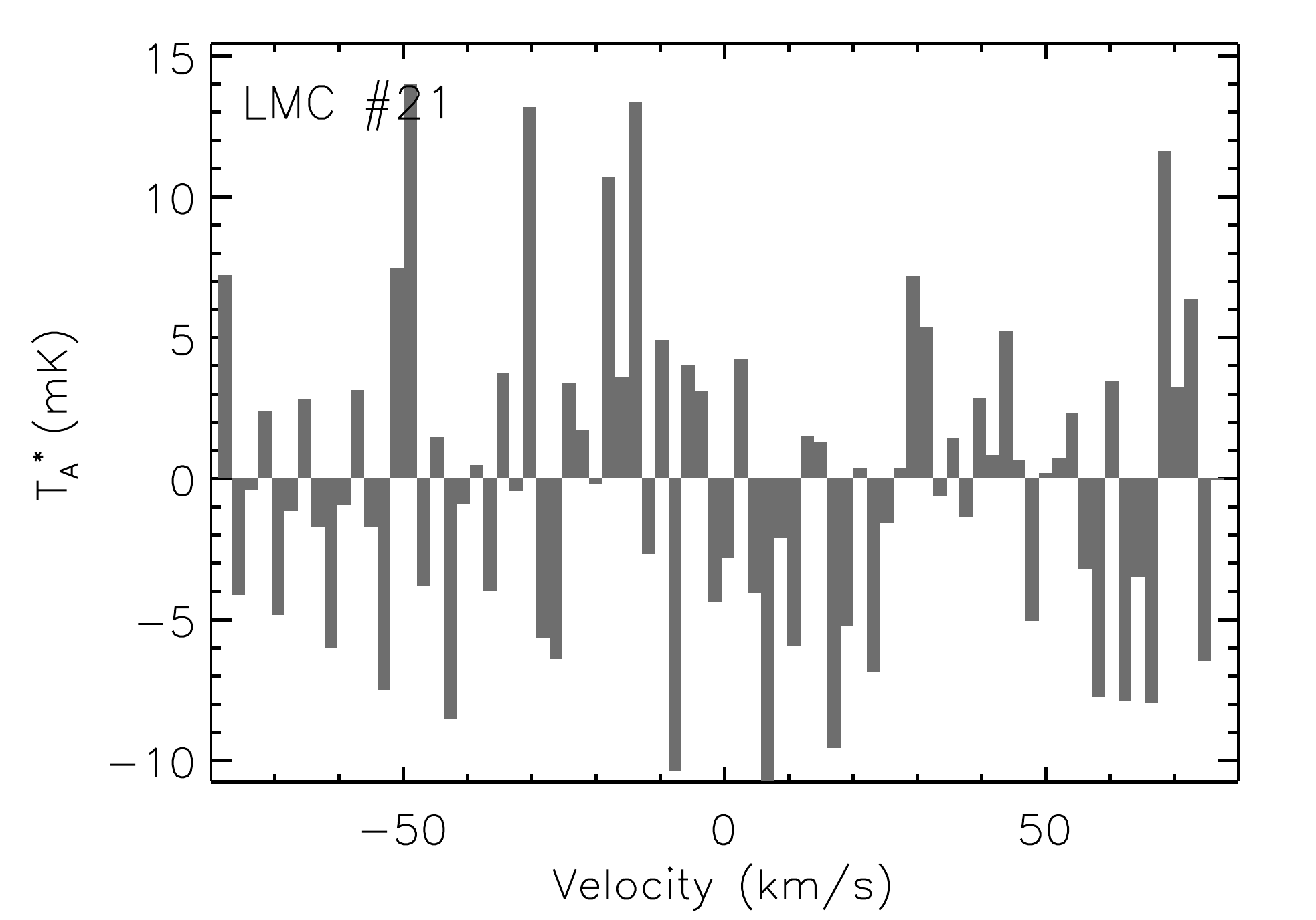}  &
\hspace{-0.4cm}   \includegraphics[height=4cm]{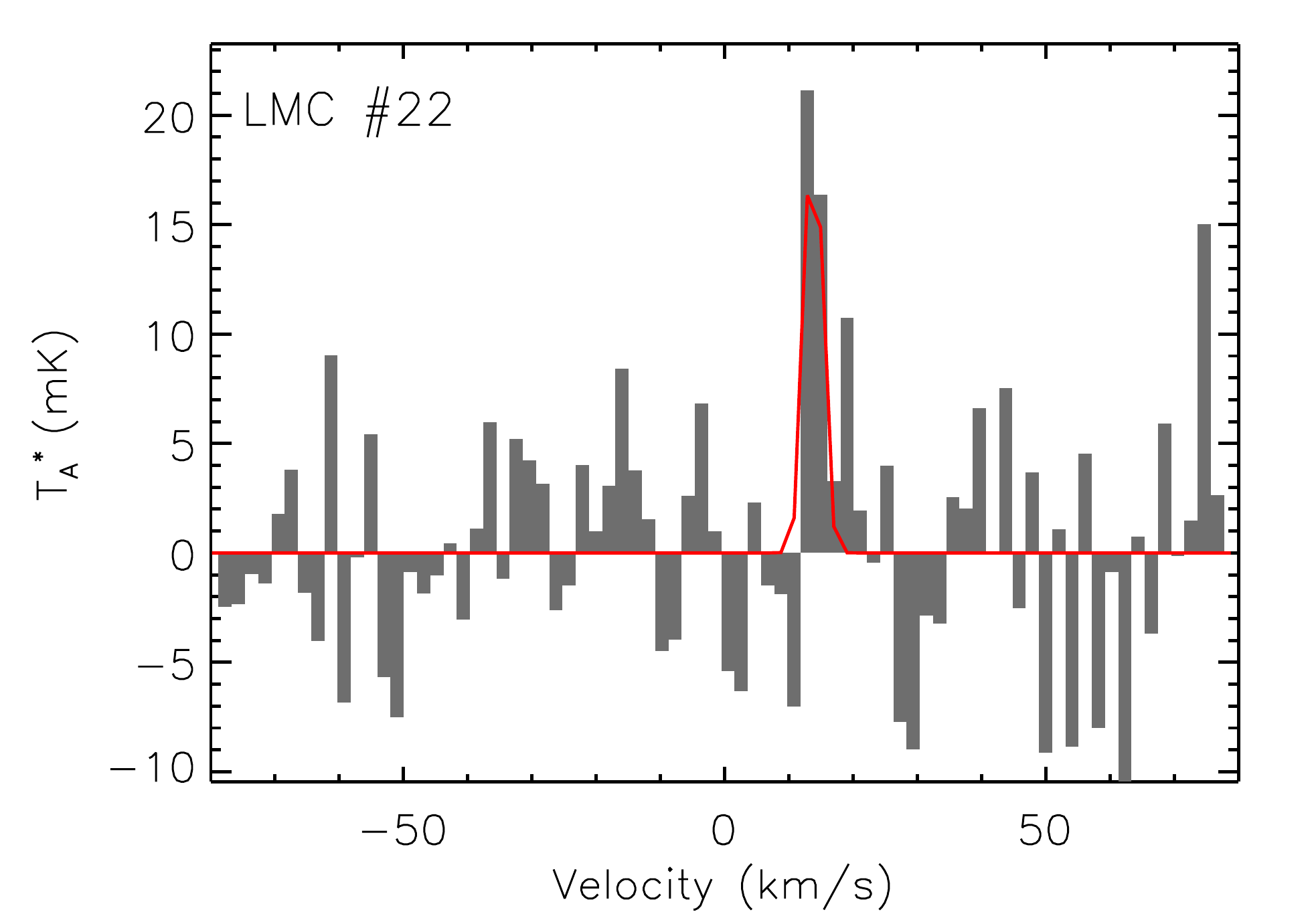} \\
\includegraphics[height=4cm]{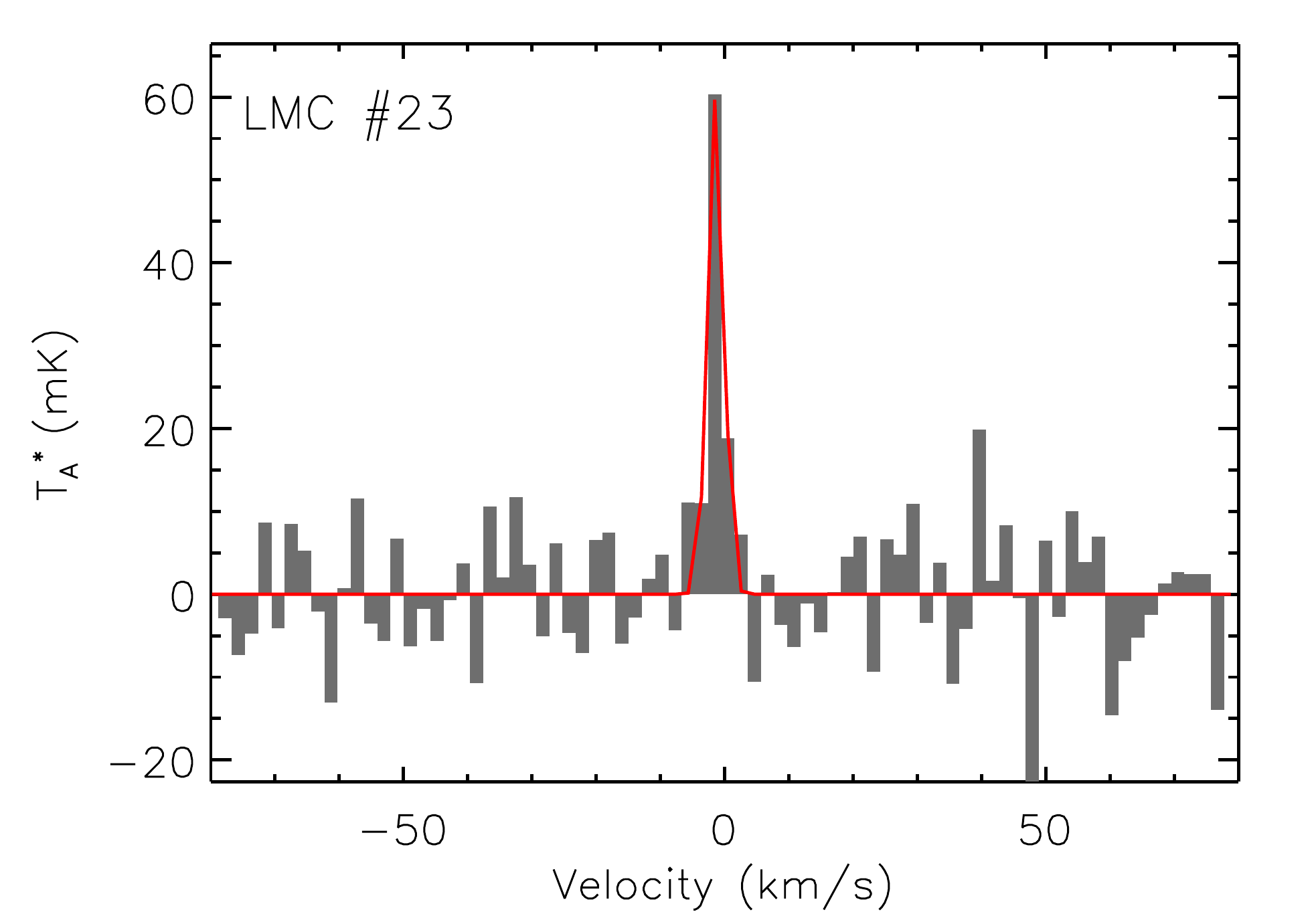}  & 
\hspace{-0.4cm}\includegraphics[height=4cm]{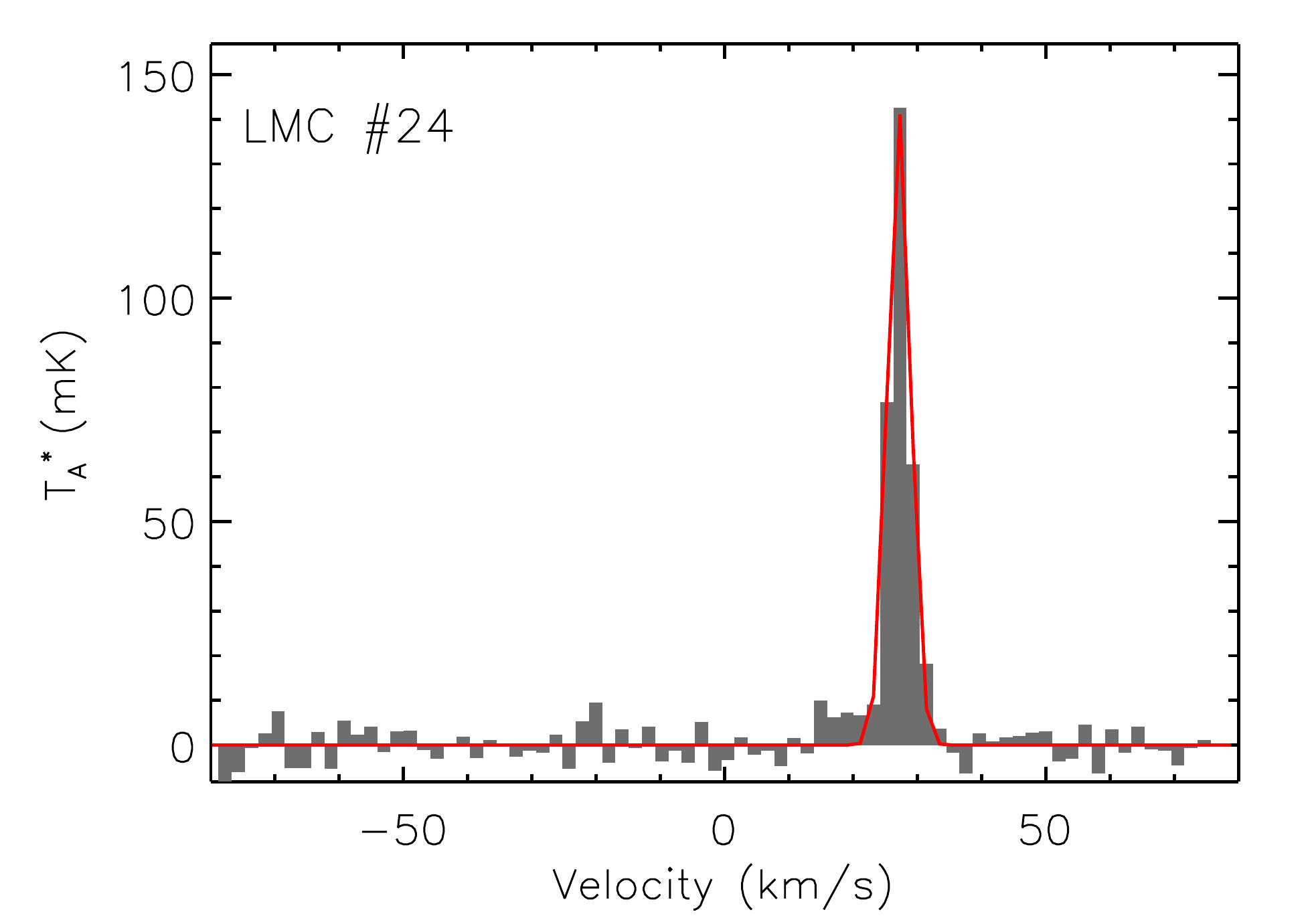}  &
\hspace{-0.4cm}  \includegraphics[height=4cm]{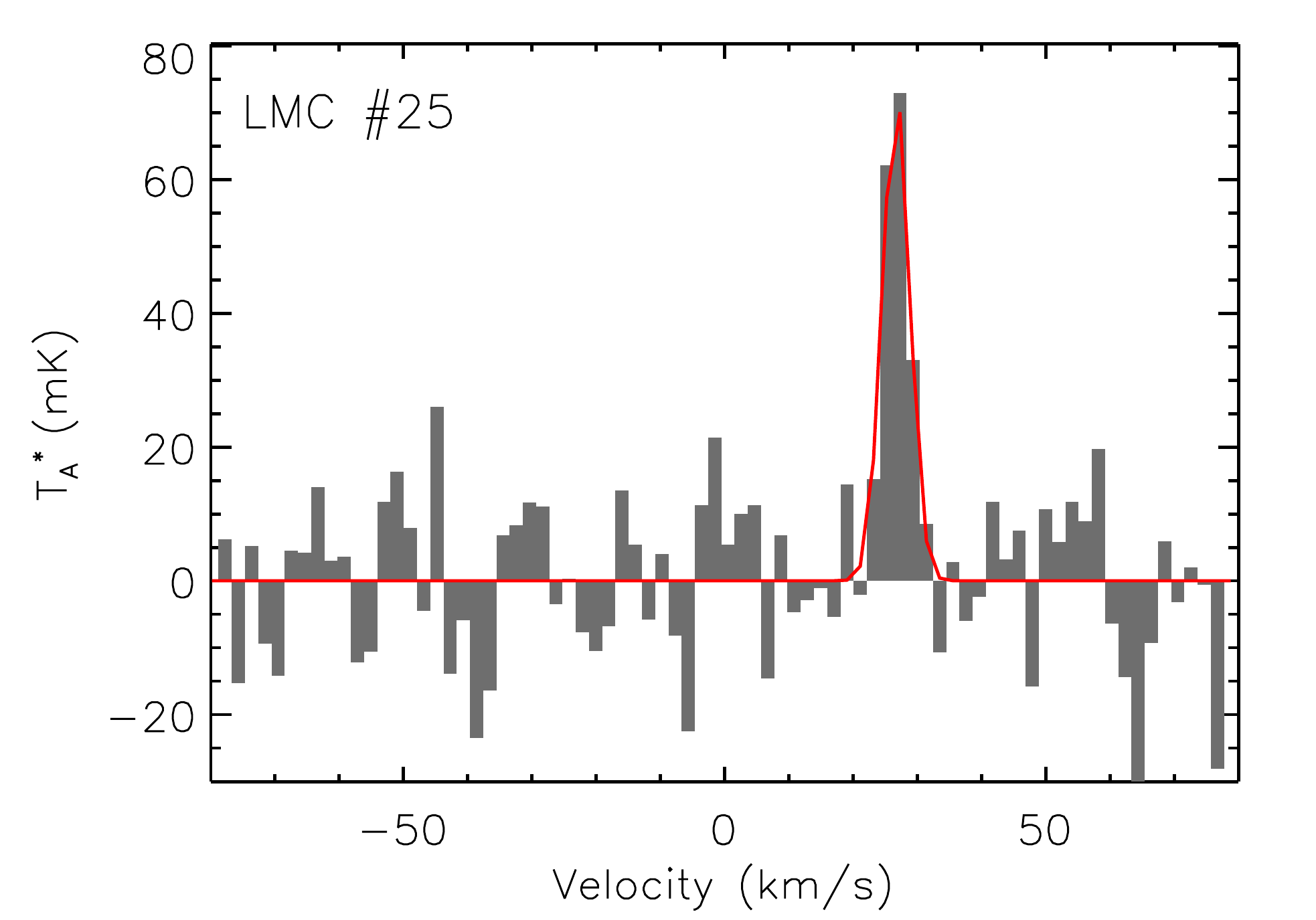} \\
\includegraphics[height=4cm]{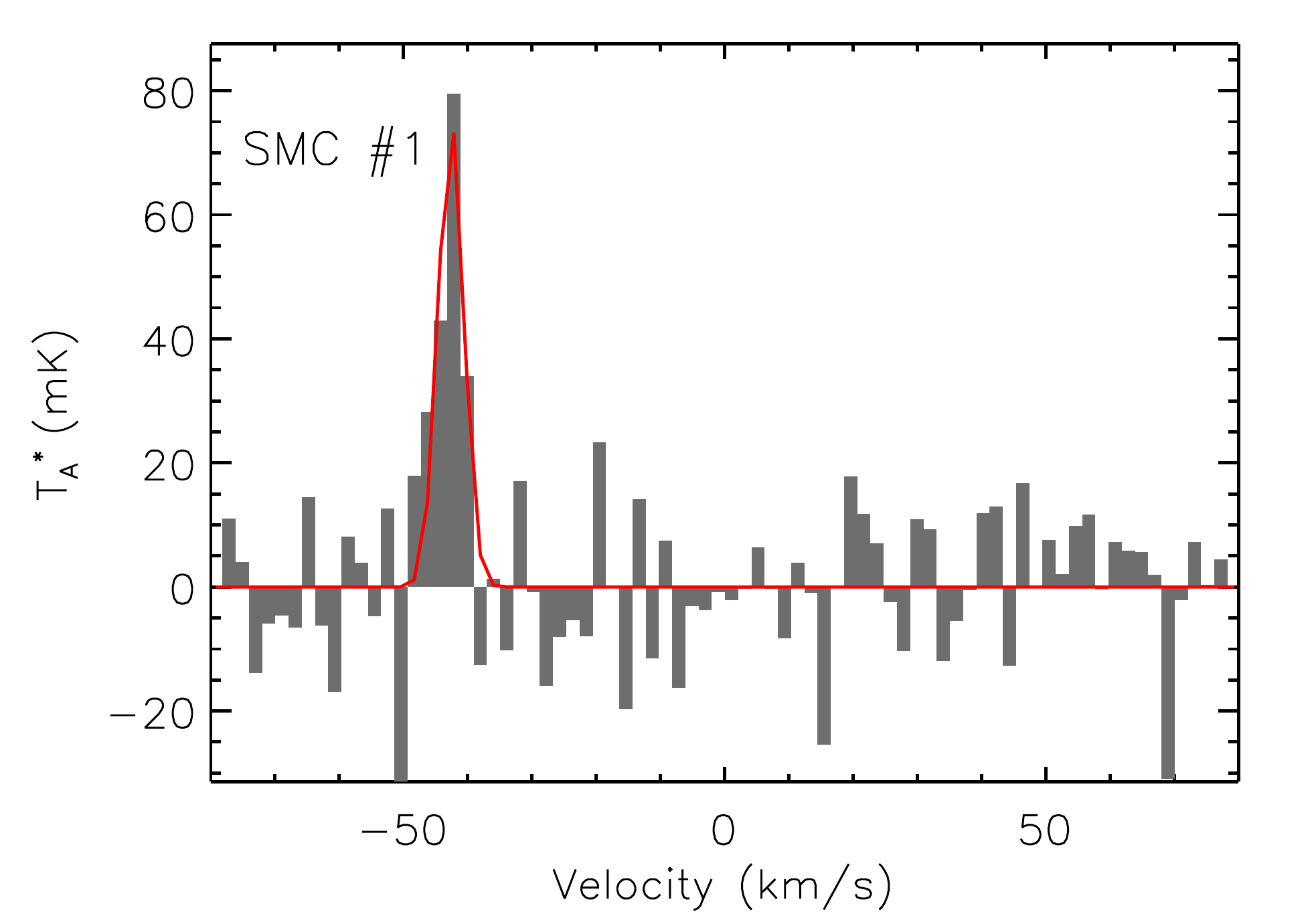}  & 
\hspace{-0.4cm}\includegraphics[height=4cm]{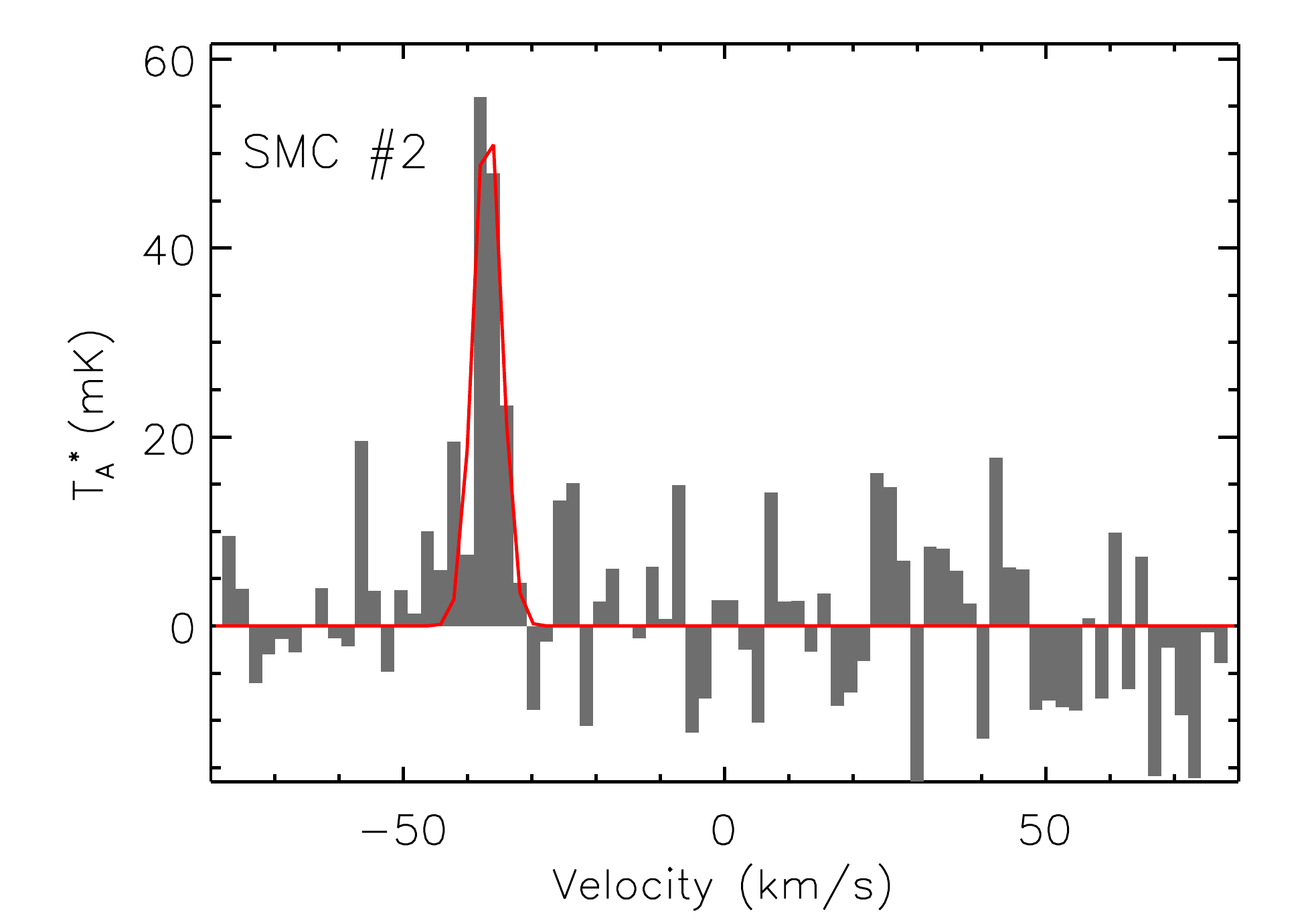}  & 
\hspace{-0.4cm}\includegraphics[height=4cm]{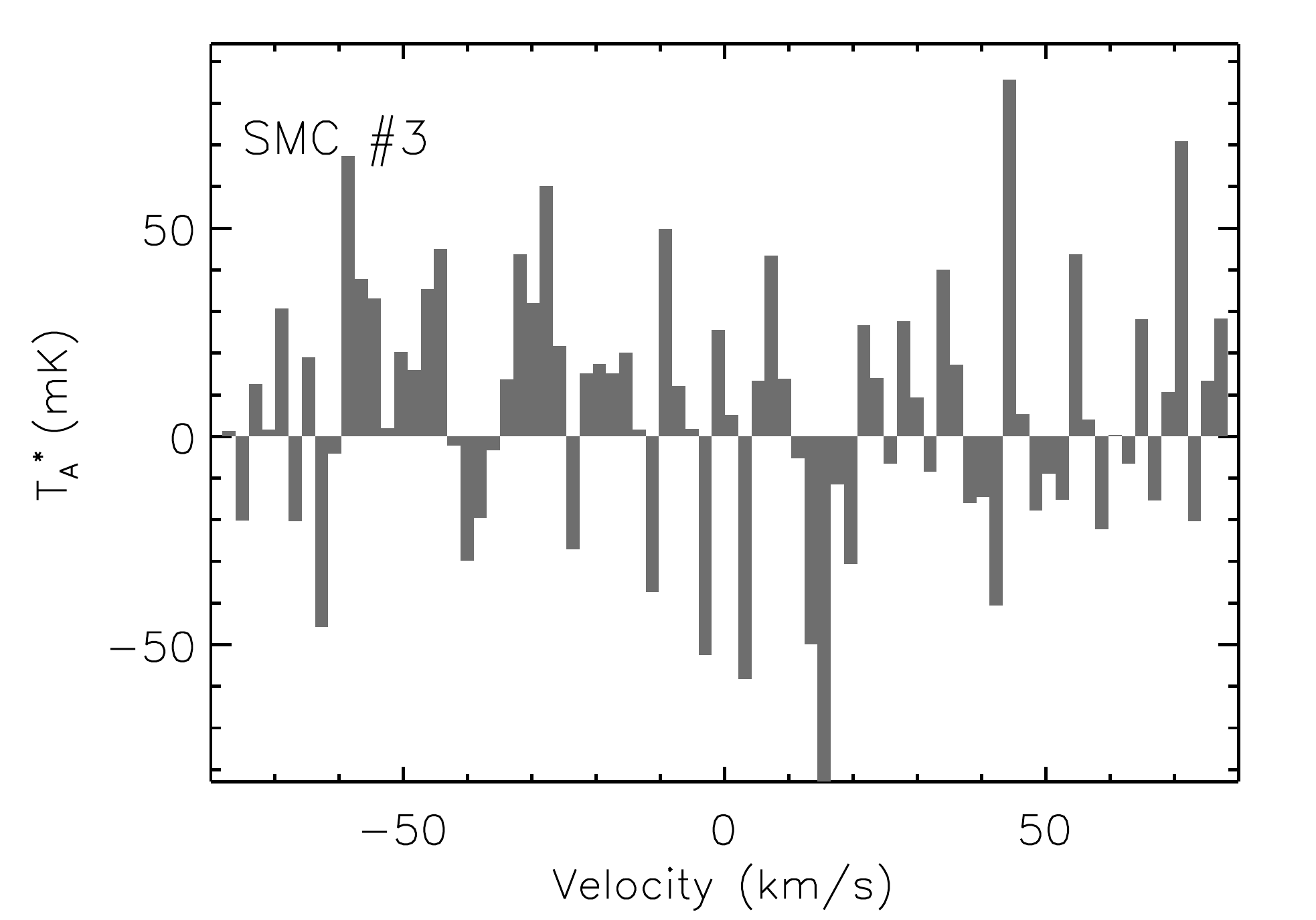} \\
\includegraphics[height=4cm]{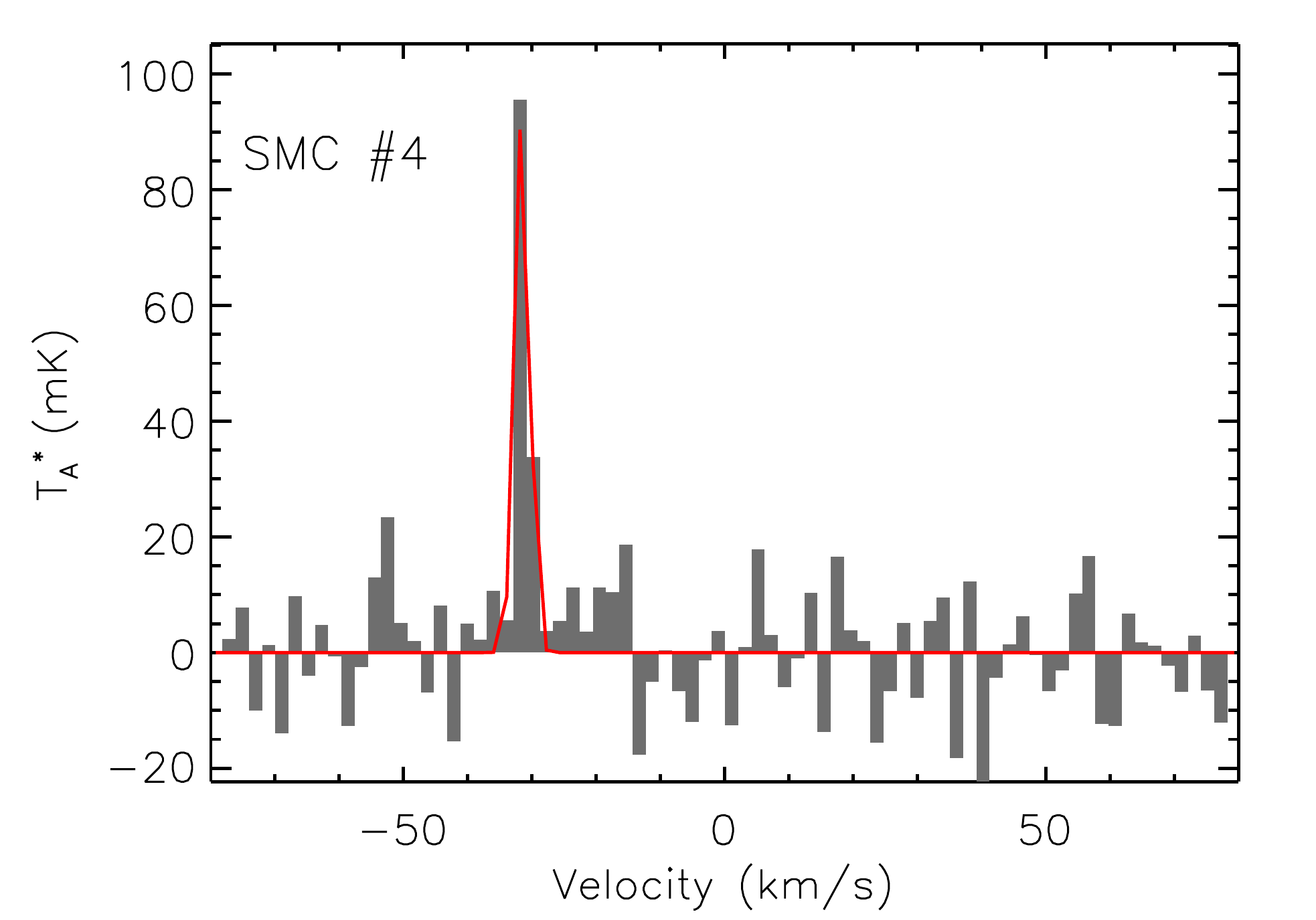}  & 
\hspace{-0.4cm}\includegraphics[height=4cm]{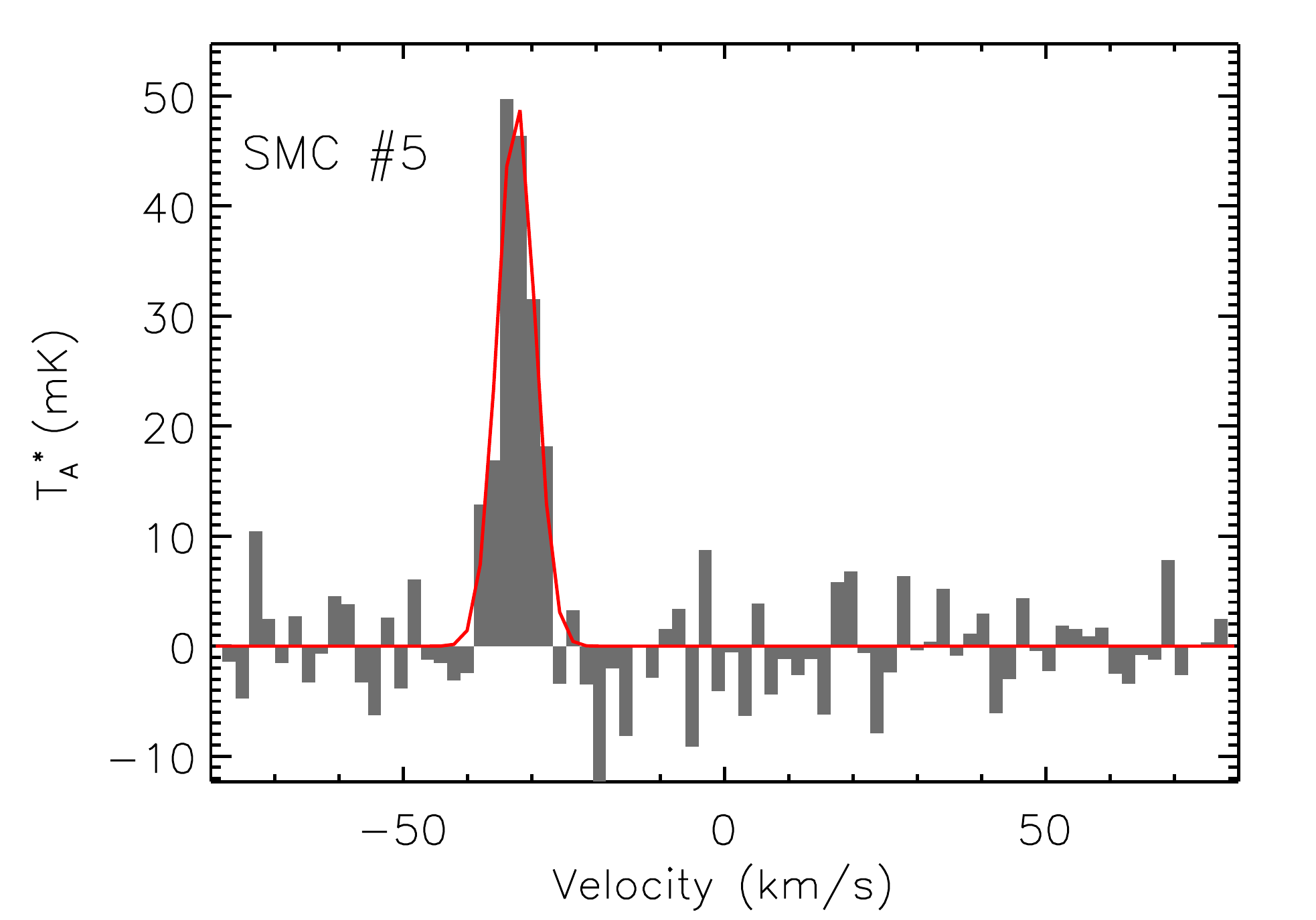}  & 
\hspace{-0.4cm}\includegraphics[height=4cm]{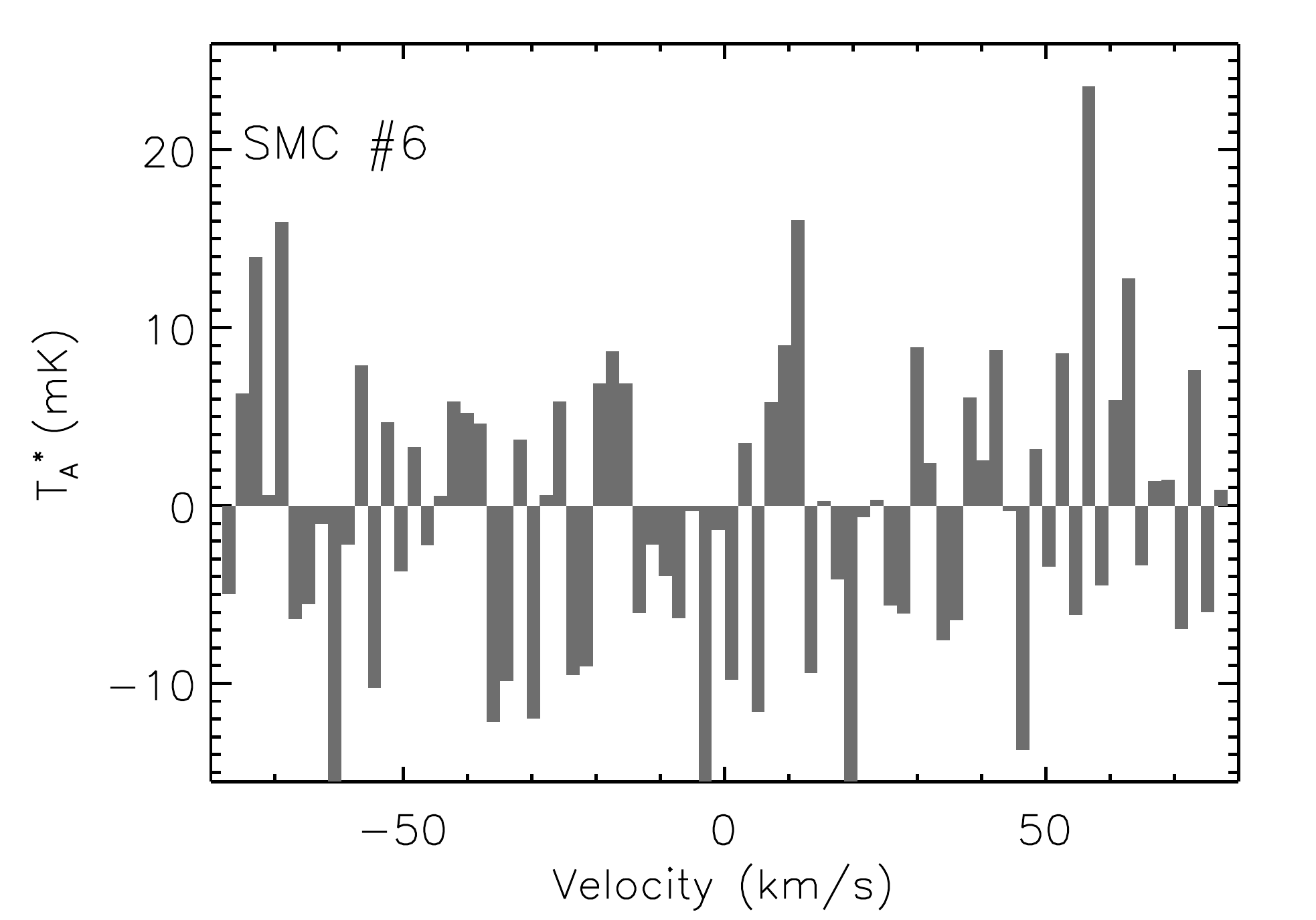} \\
\end{tabular}
\begin{tabular}{cc}
\includegraphics[height=4cm]{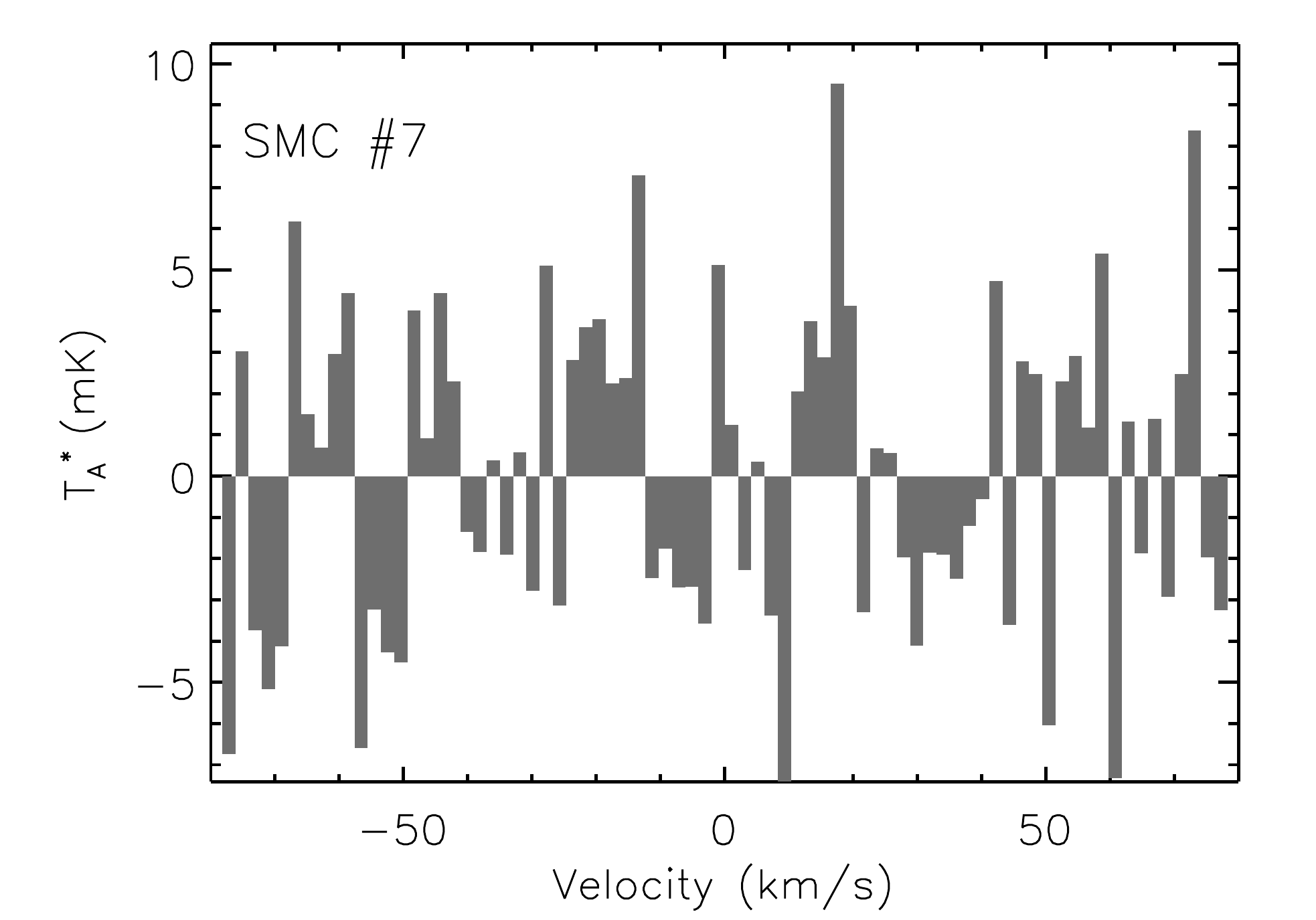}  &
\includegraphics[height=4cm]{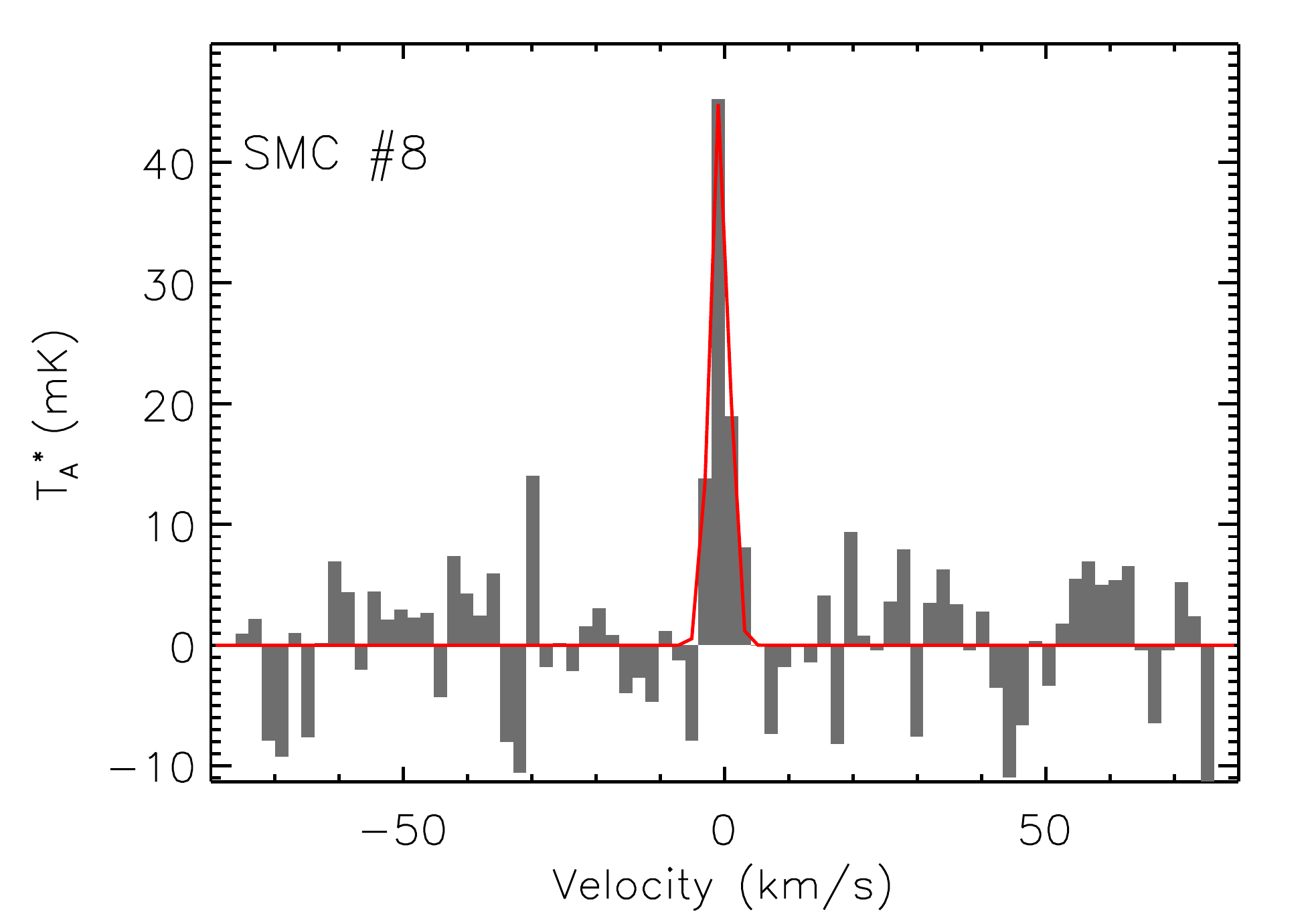}  \\
\end{tabular}
\caption{continued.}
\end{figure*}

\begin{figure*}
\centering
\begin{tabular}{ccc}
\includegraphics[height=4cm]{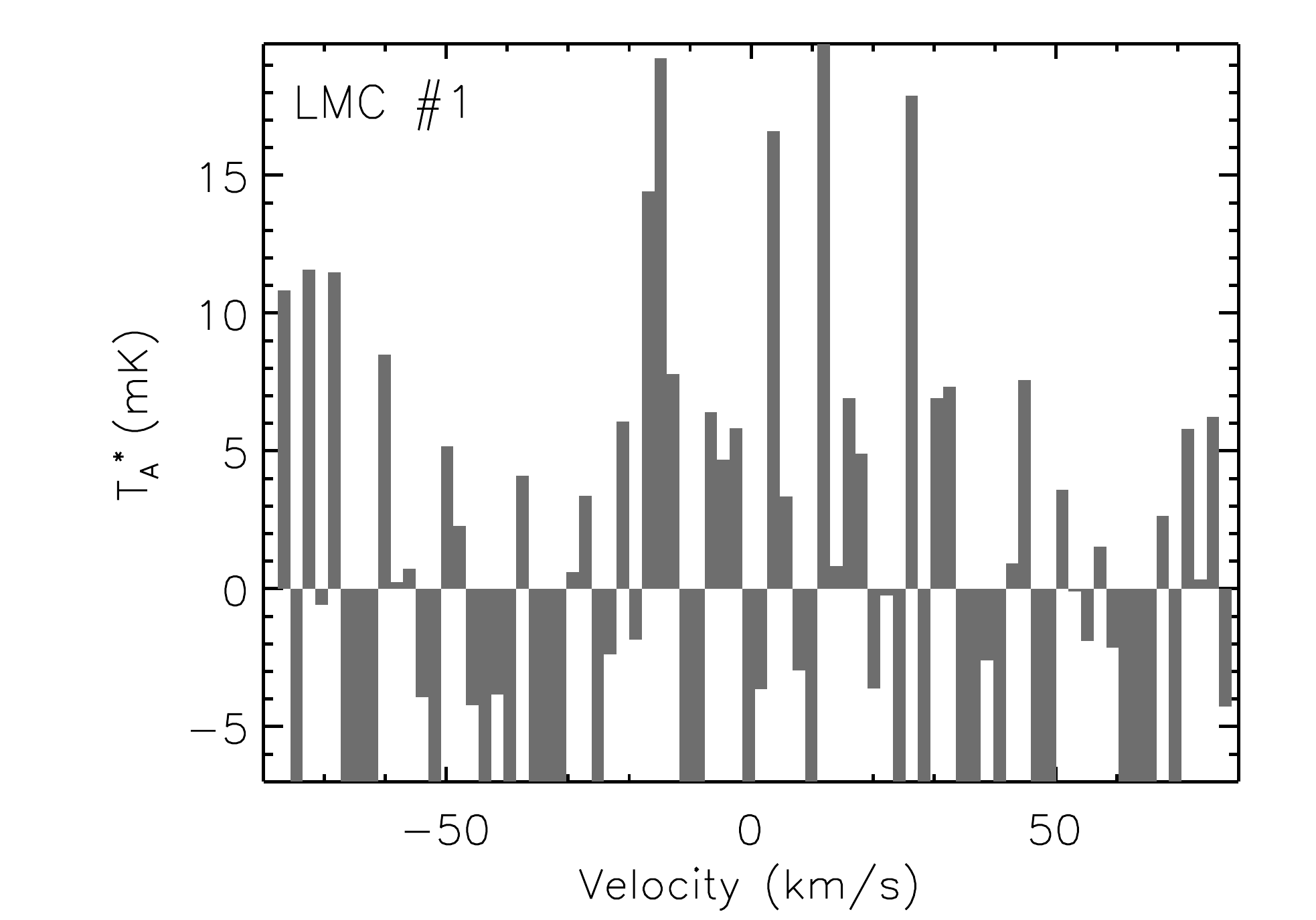}  & 
\hspace{-0.4cm}\includegraphics[height=4cm]{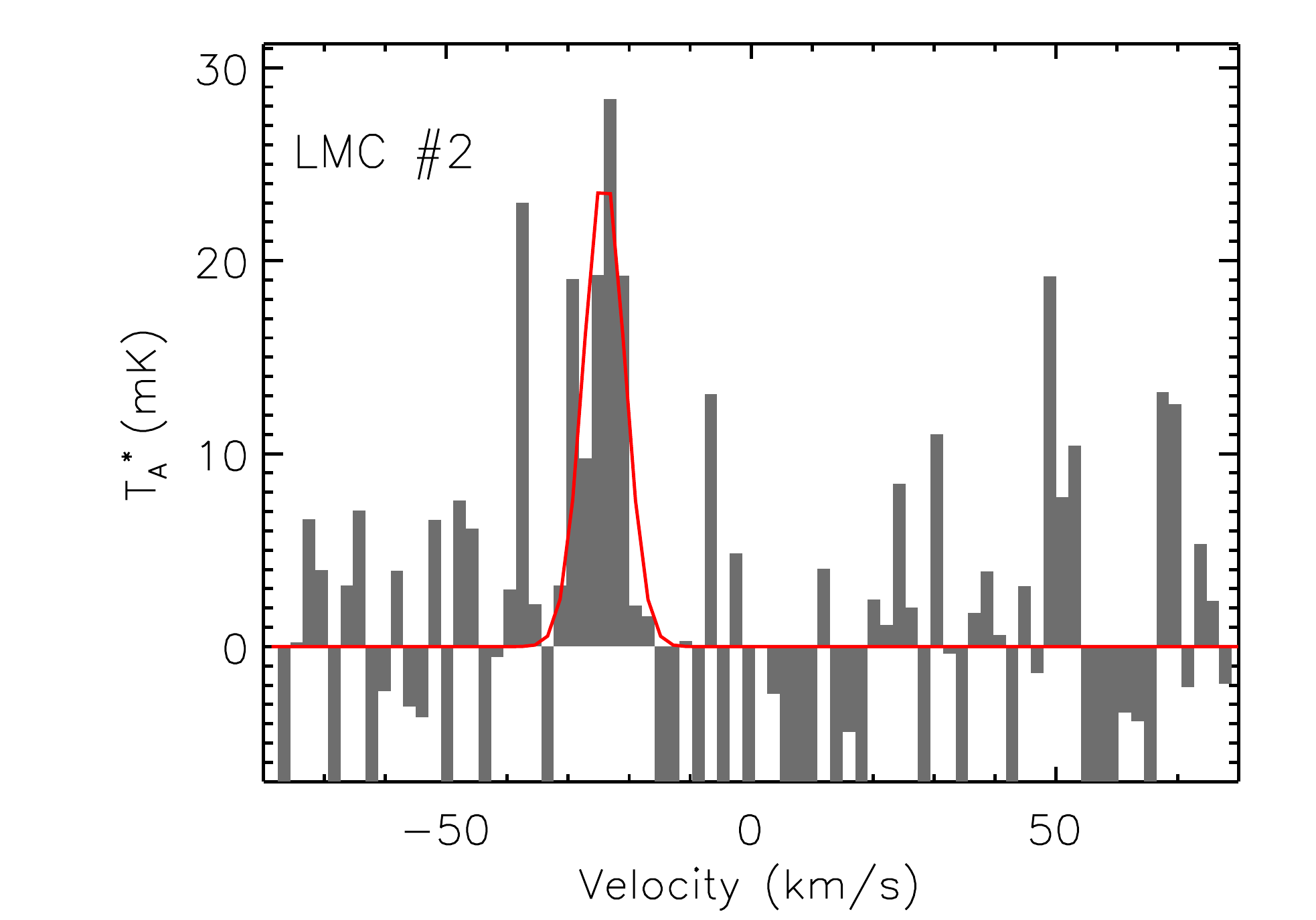}  &
\hspace{-0.4cm}   \includegraphics[height=4cm]{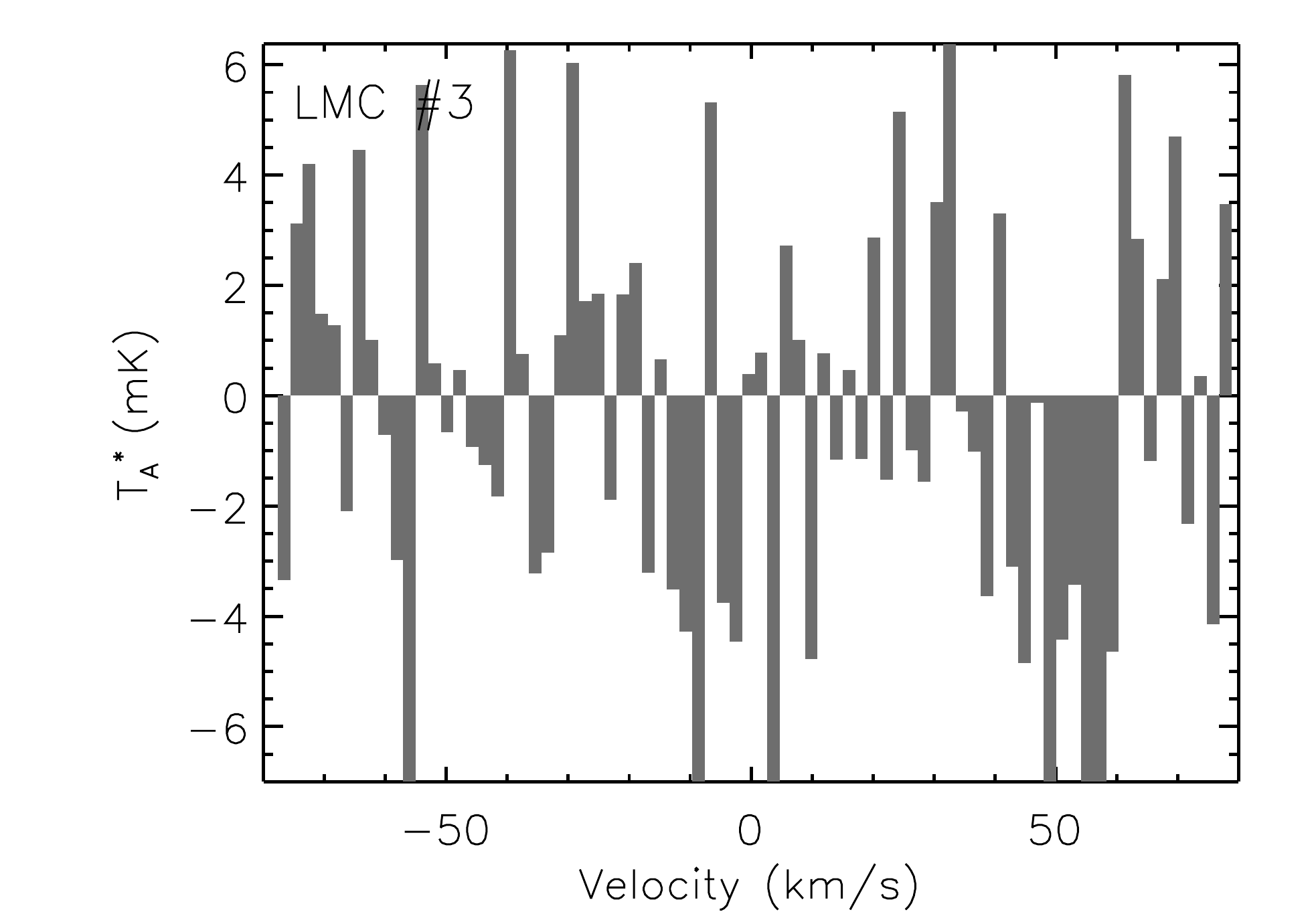} \\
\includegraphics[height=4cm]{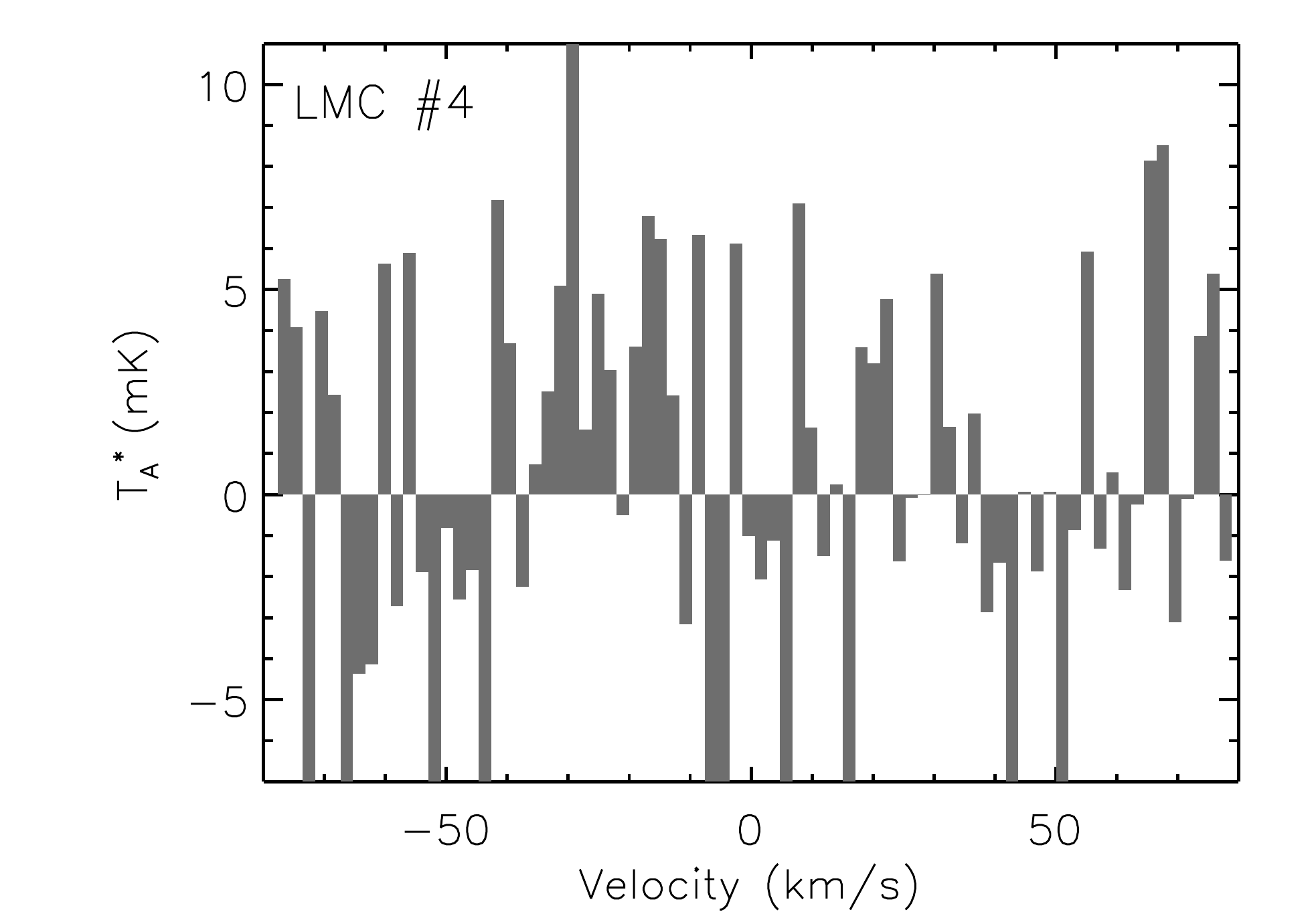}  & 
\hspace{-0.4cm}\includegraphics[height=4cm]{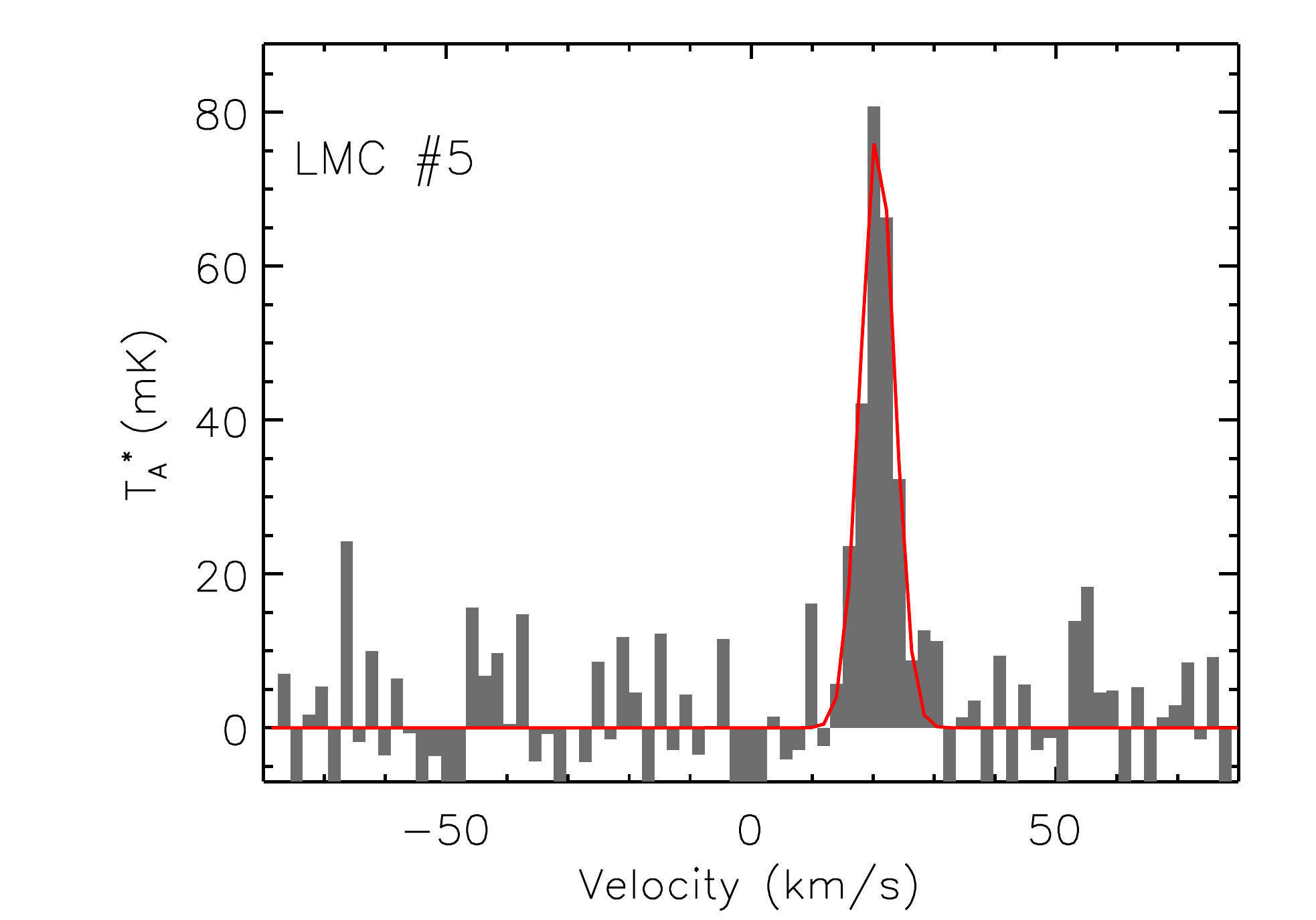}  &
\hspace{-0.4cm}  \includegraphics[height=4cm]{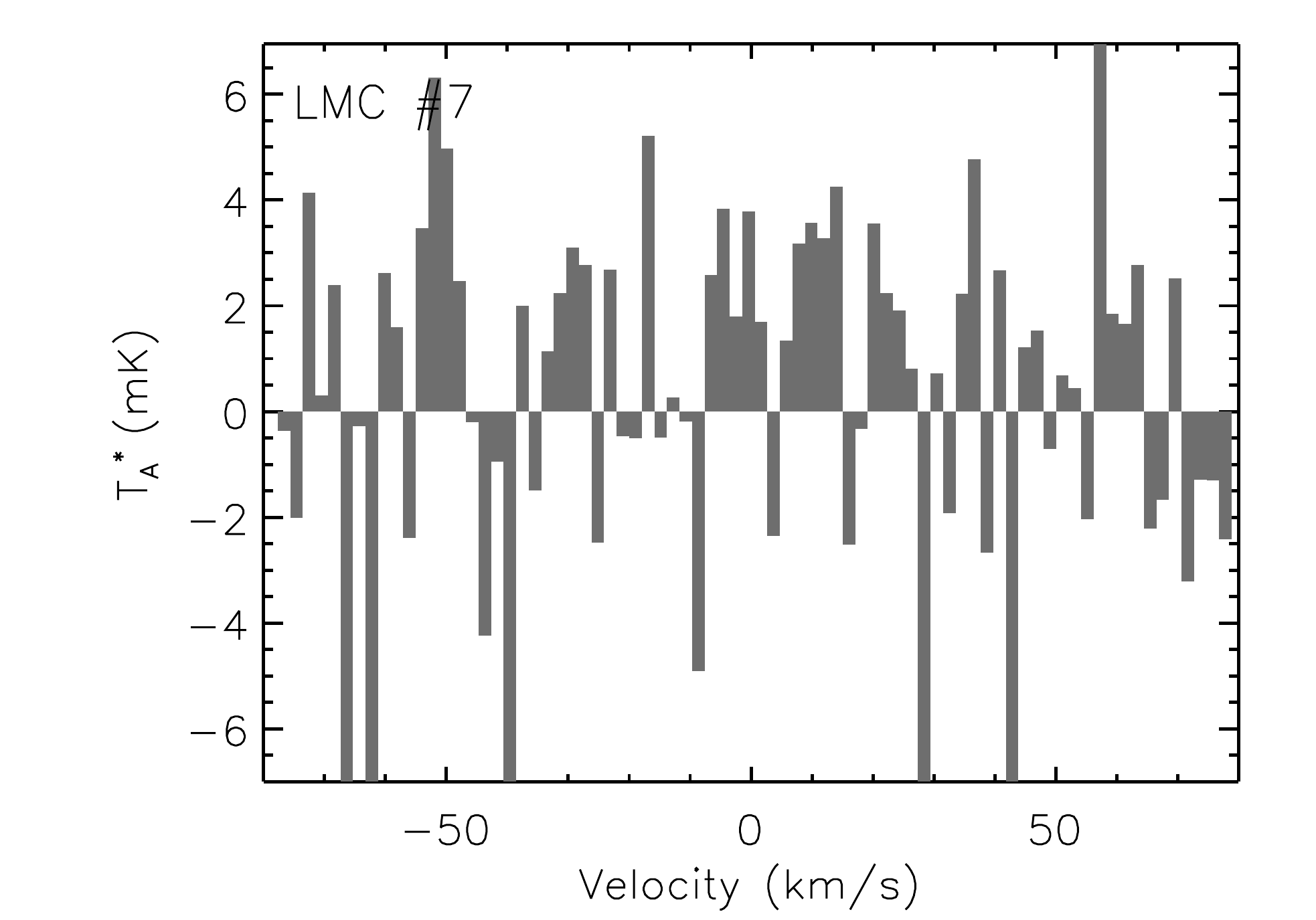} \\
\includegraphics[height=4cm]{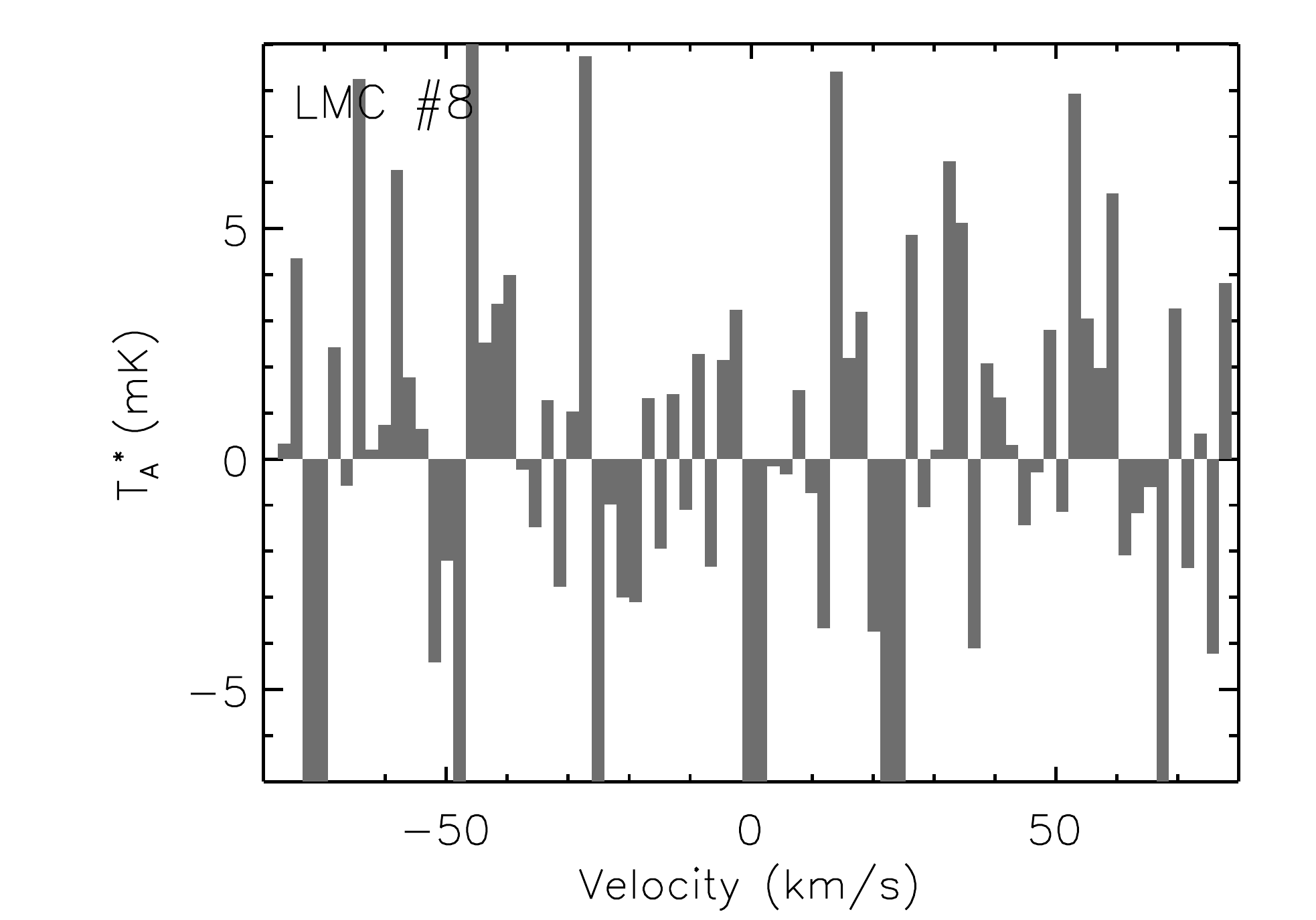}  & 
\hspace{-0.4cm}\includegraphics[height=4cm]{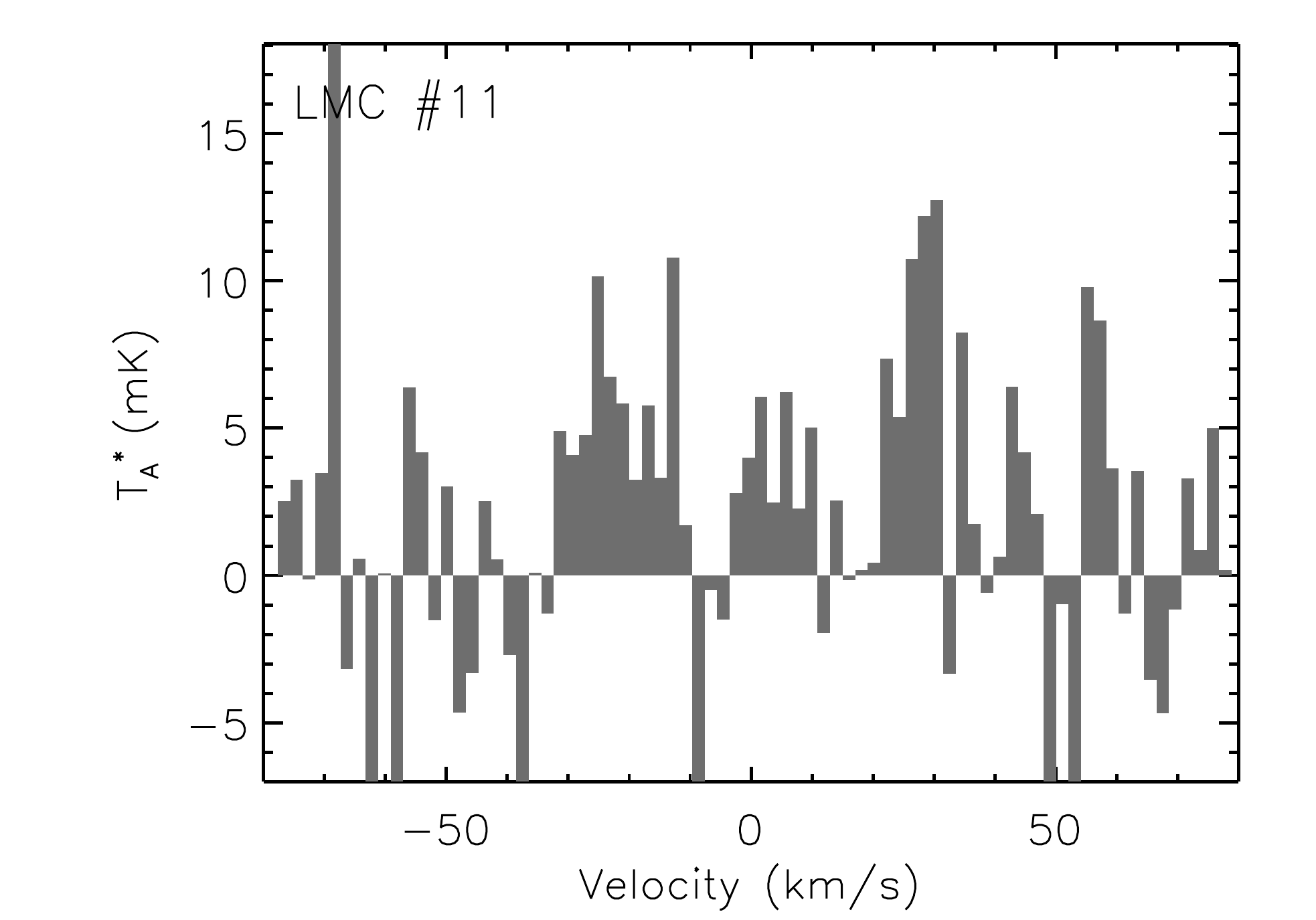}  &
\hspace{-0.4cm}  \includegraphics[height=4cm]{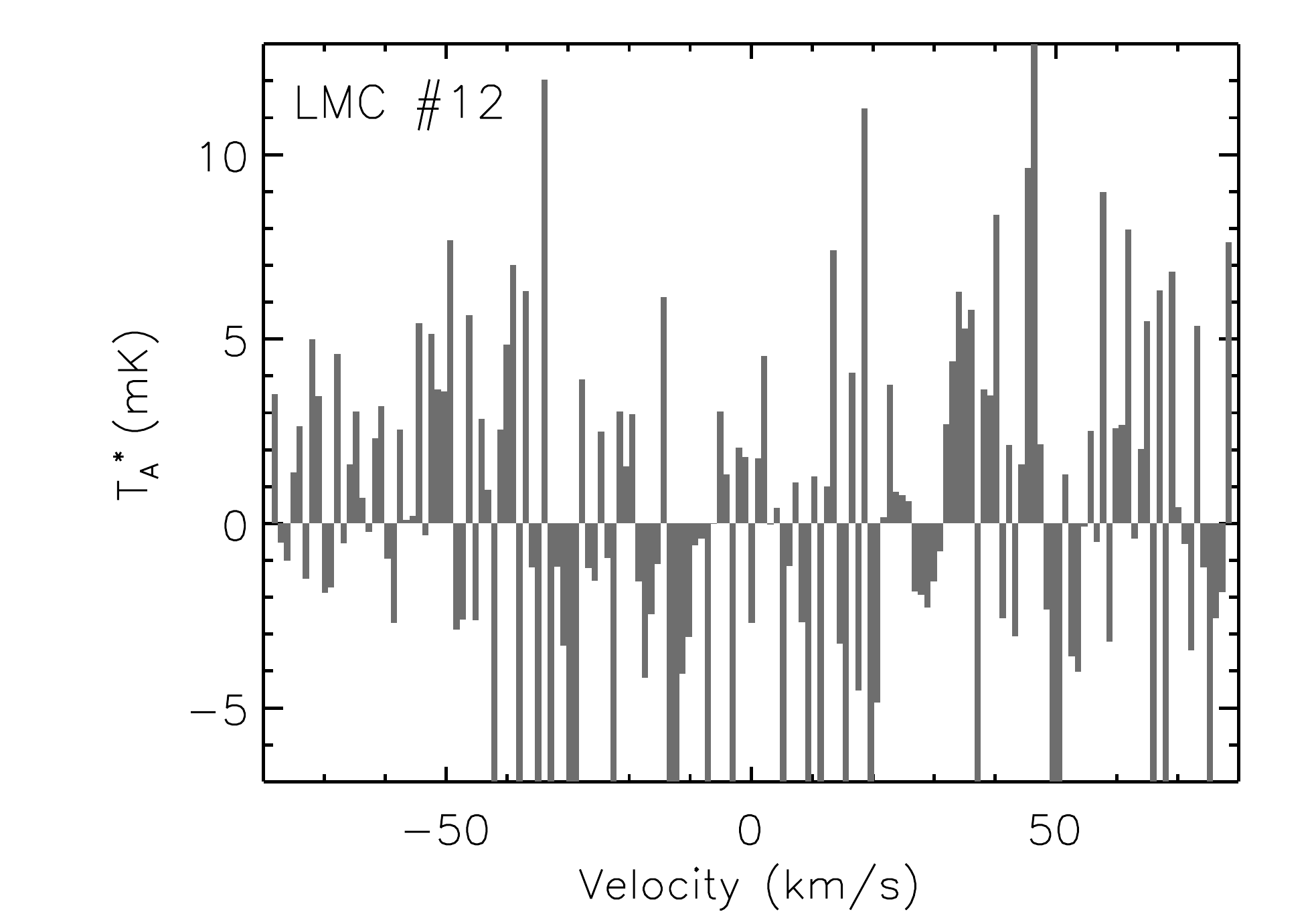} \\ 
\includegraphics[height=4cm]{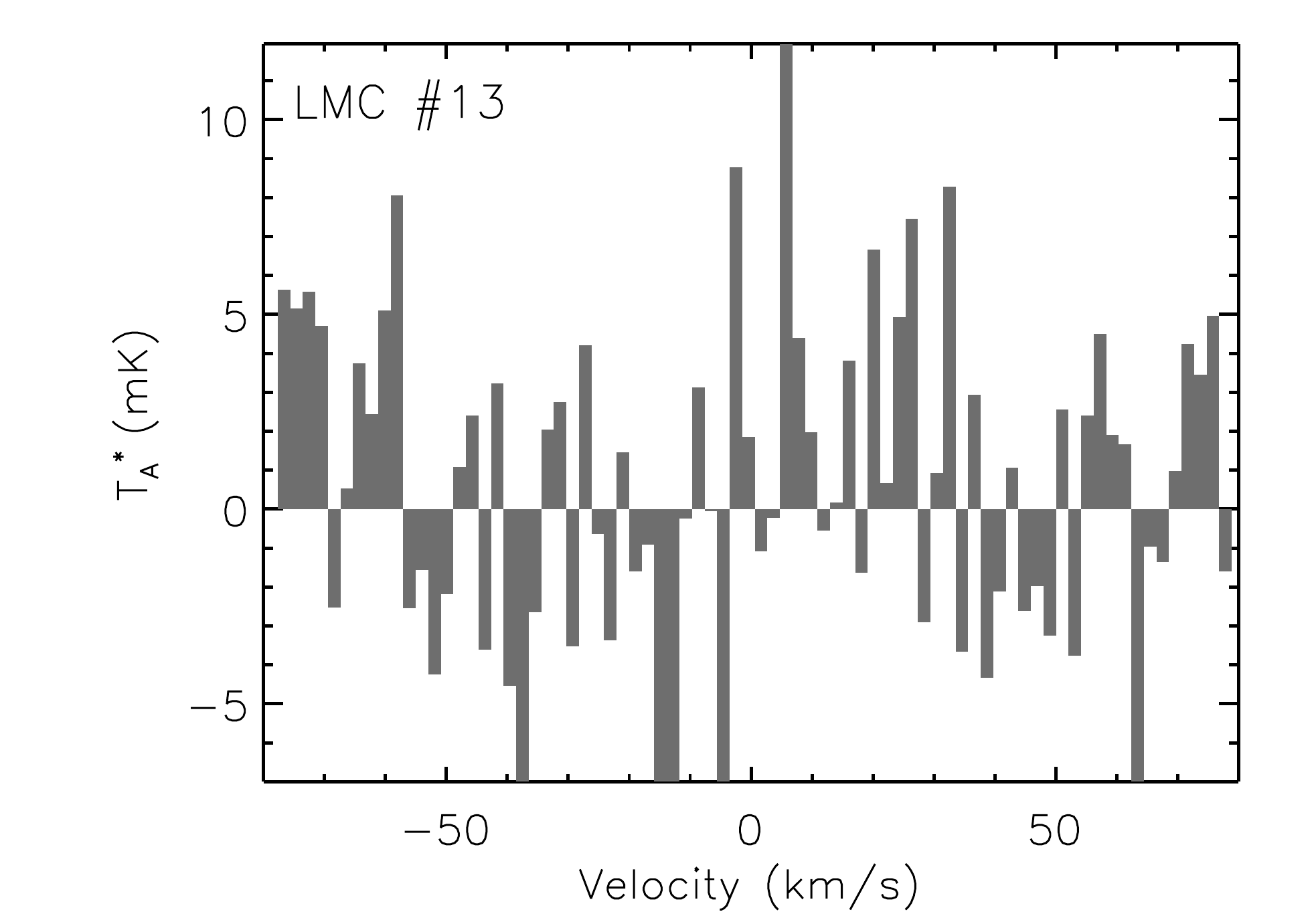}  & 
\hspace{-0.4cm}\includegraphics[height=4cm]{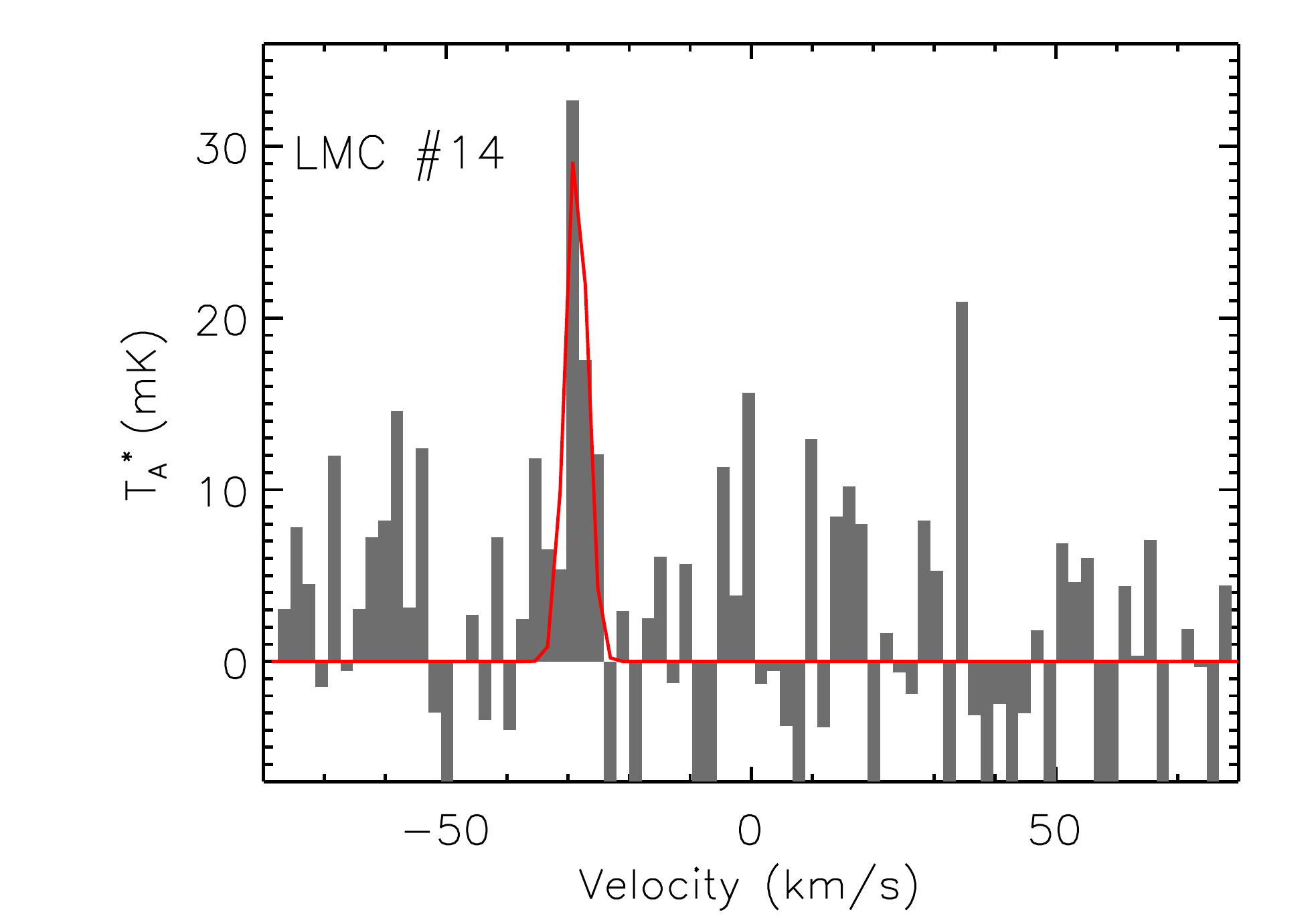}  &
\hspace{-0.4cm}  \includegraphics[height=4cm]{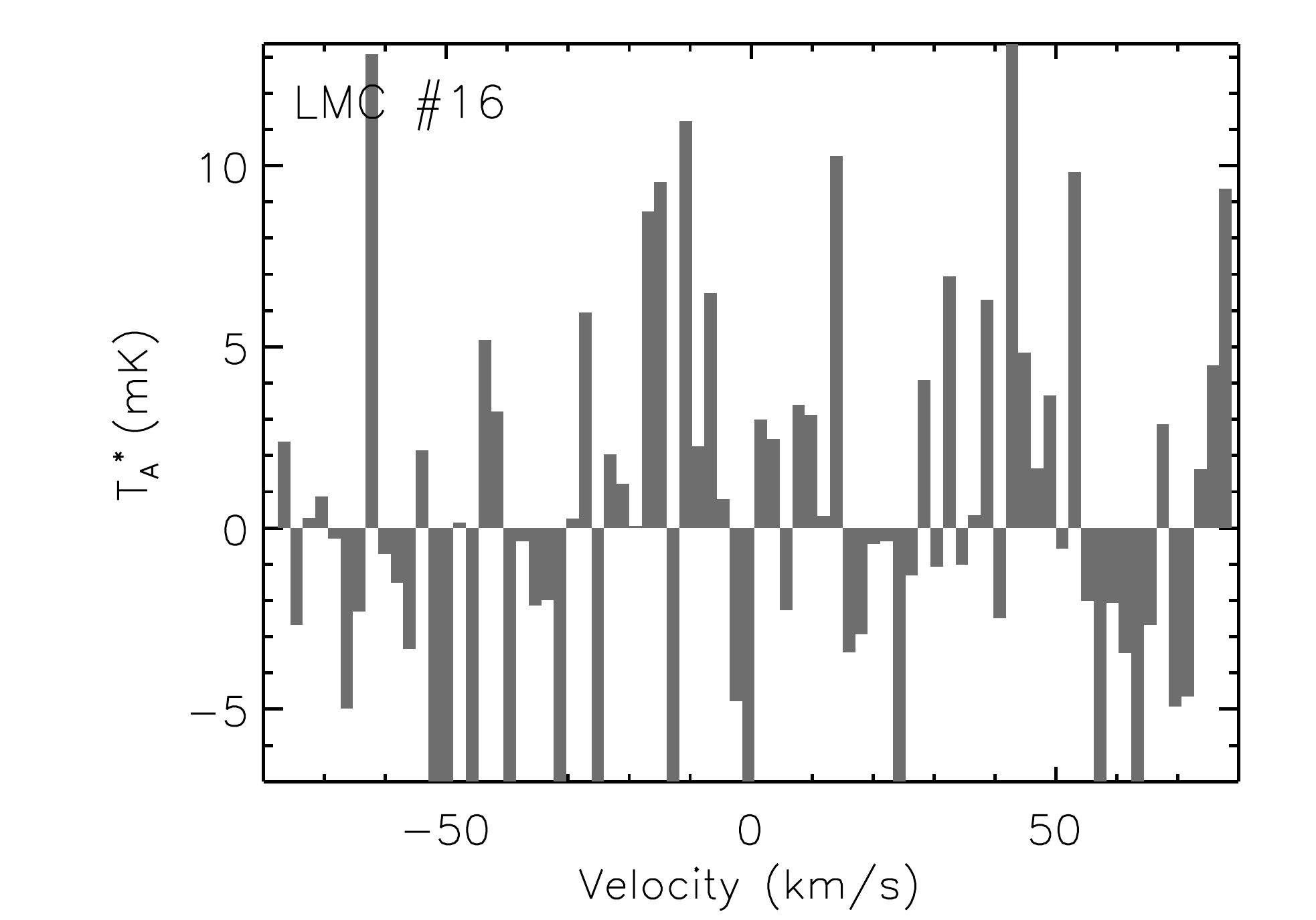} \\ 
\includegraphics[height=4cm]{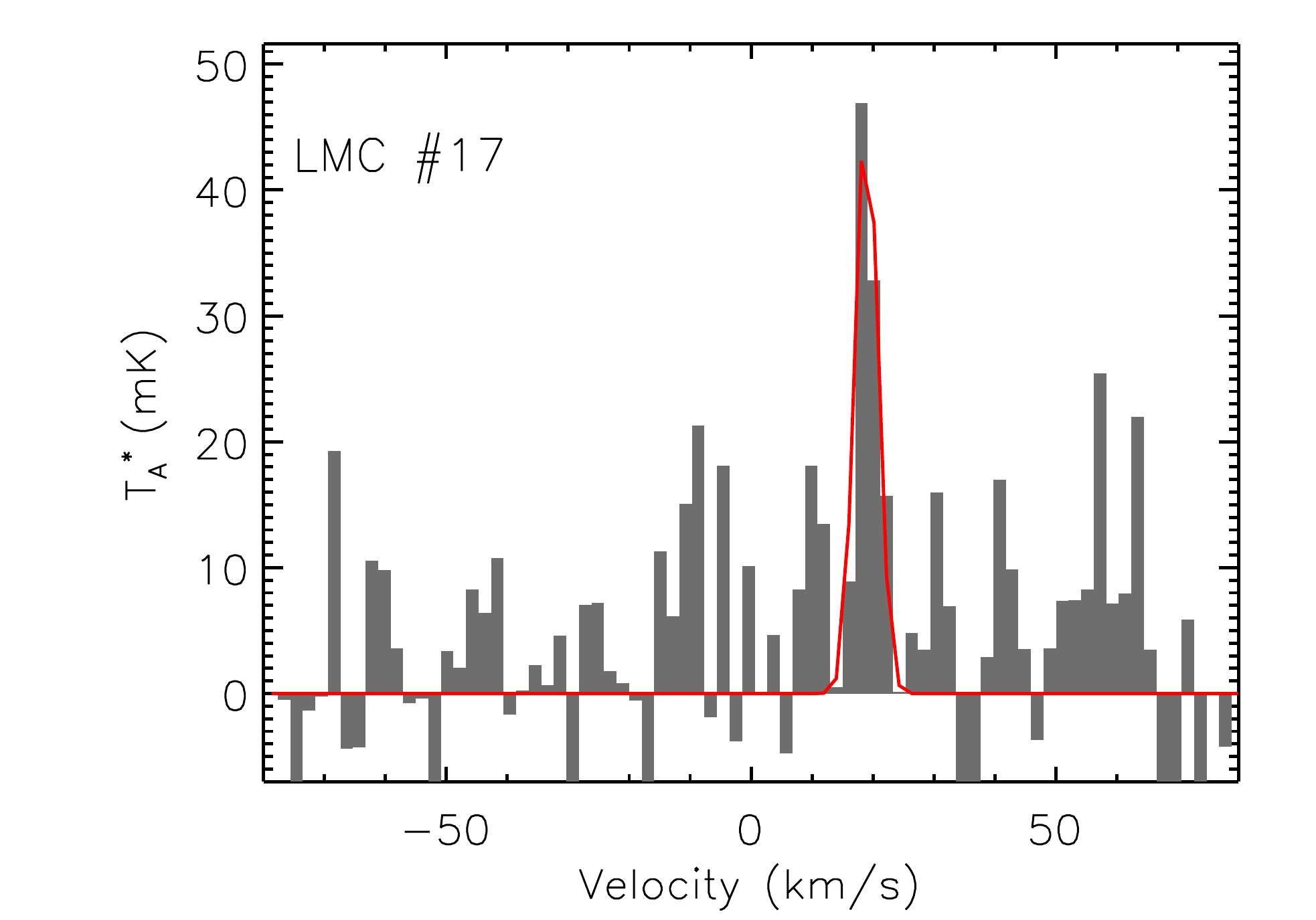}  & 
\hspace{-0.4cm}\includegraphics[height=4cm]{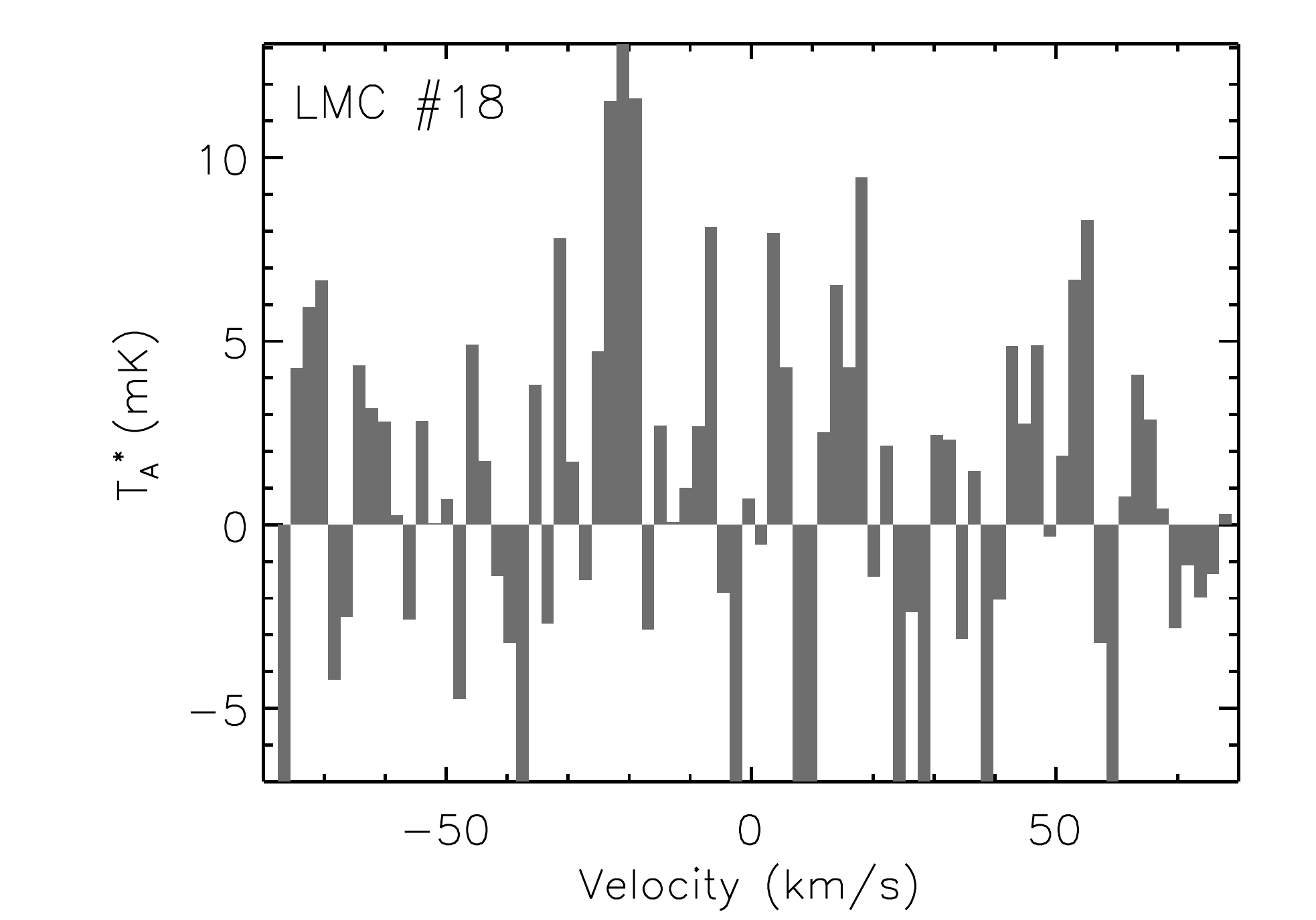}  &
\hspace{-0.4cm}   \includegraphics[height=4cm]{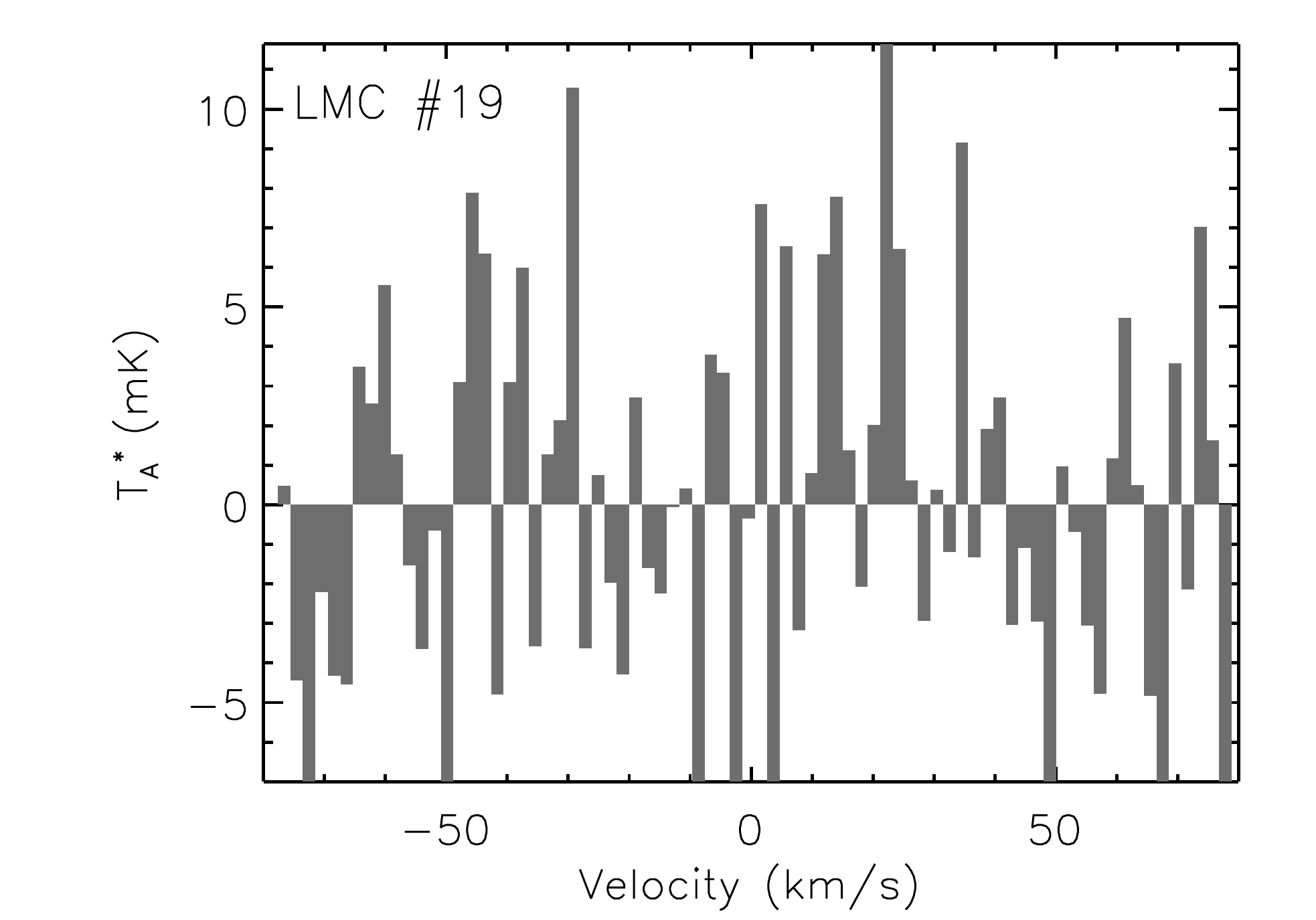}\\ 
\end{tabular}
   \caption{HCN(2$-$1) observations toward the LMC and SMC (2 km~s$^{-1}$ spectral resolution). 
   The bottom x-axis is expressed in velocity with respect to the systemic velocity (v=0 corresponding to 262.2 km~s$^{-1}$ for the 
   LMC and 158 km~s$^{-1}$ for the SMC). We indicate with a red line the Gaussian
   fit on the lines detected at a 3-$\sigma$ level.}
   \label{LMCSMC_HCN}
\end{figure*}

\addtocounter {figure}{-1}
\begin{figure*}
\centering
\vspace{30pt}
\begin{tabular}{ccc}
\includegraphics[height=4cm]{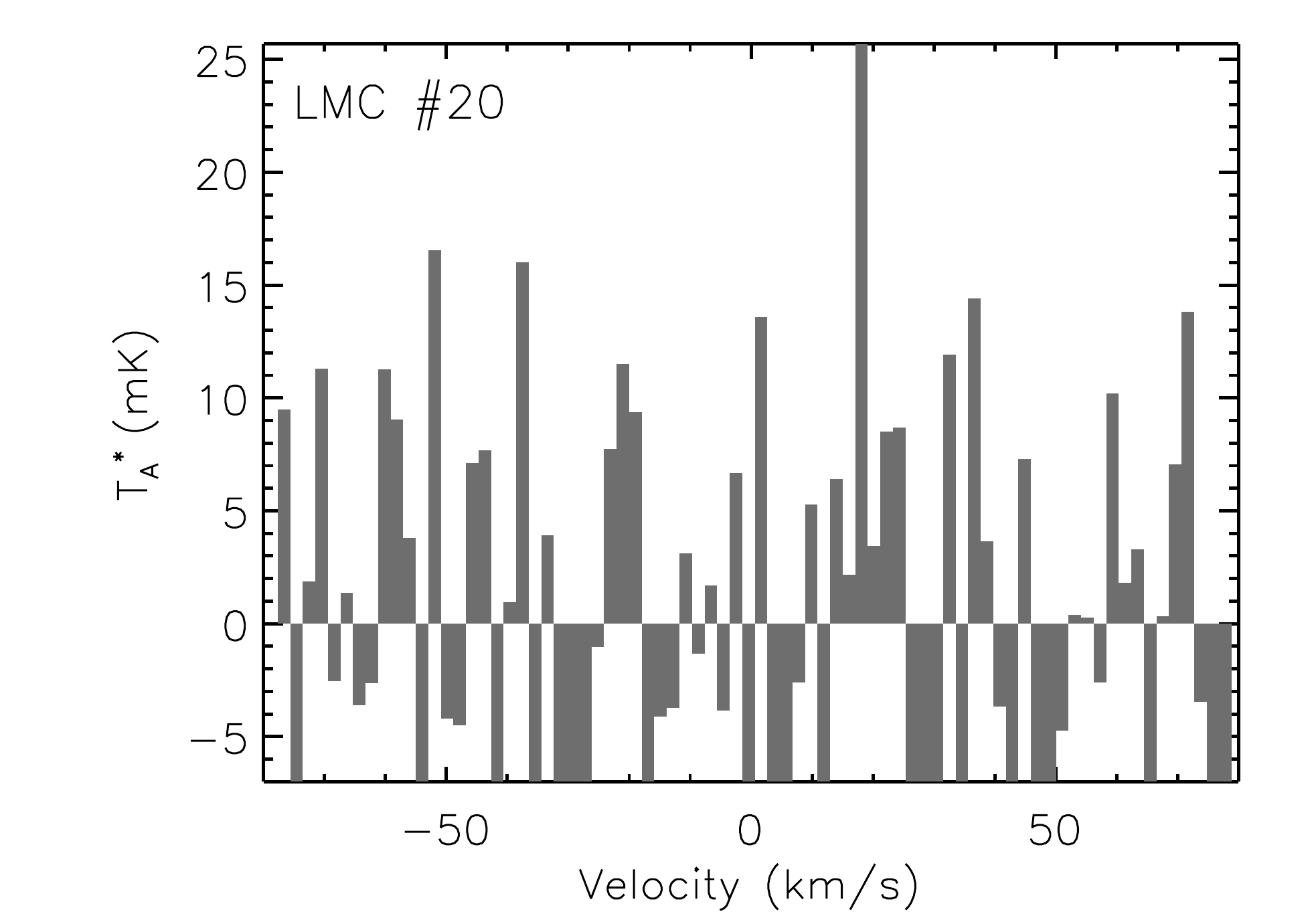}  & 
\hspace{-0.4cm}\includegraphics[height=4cm]{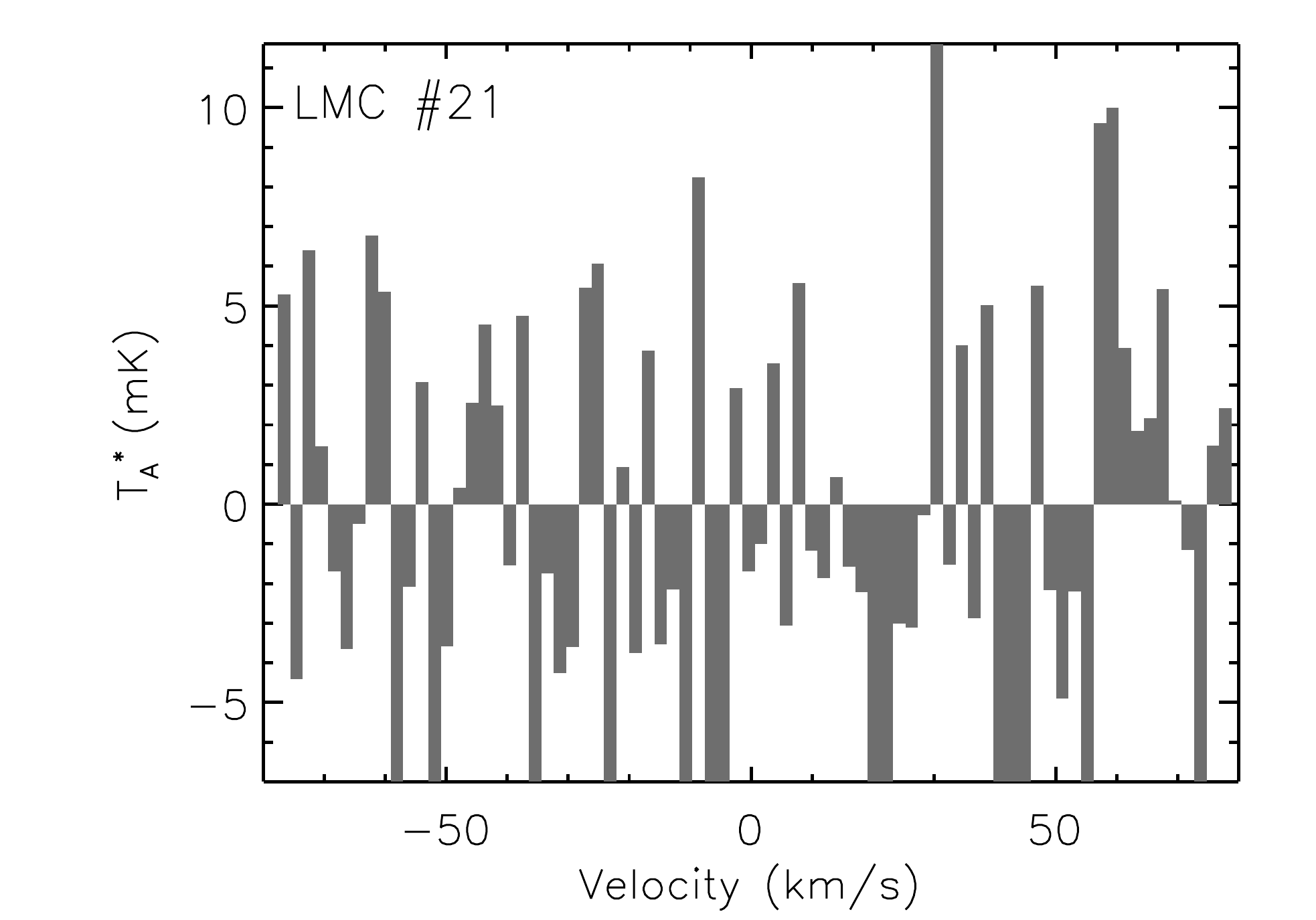}  &
\hspace{-0.4cm}   \includegraphics[height=4cm]{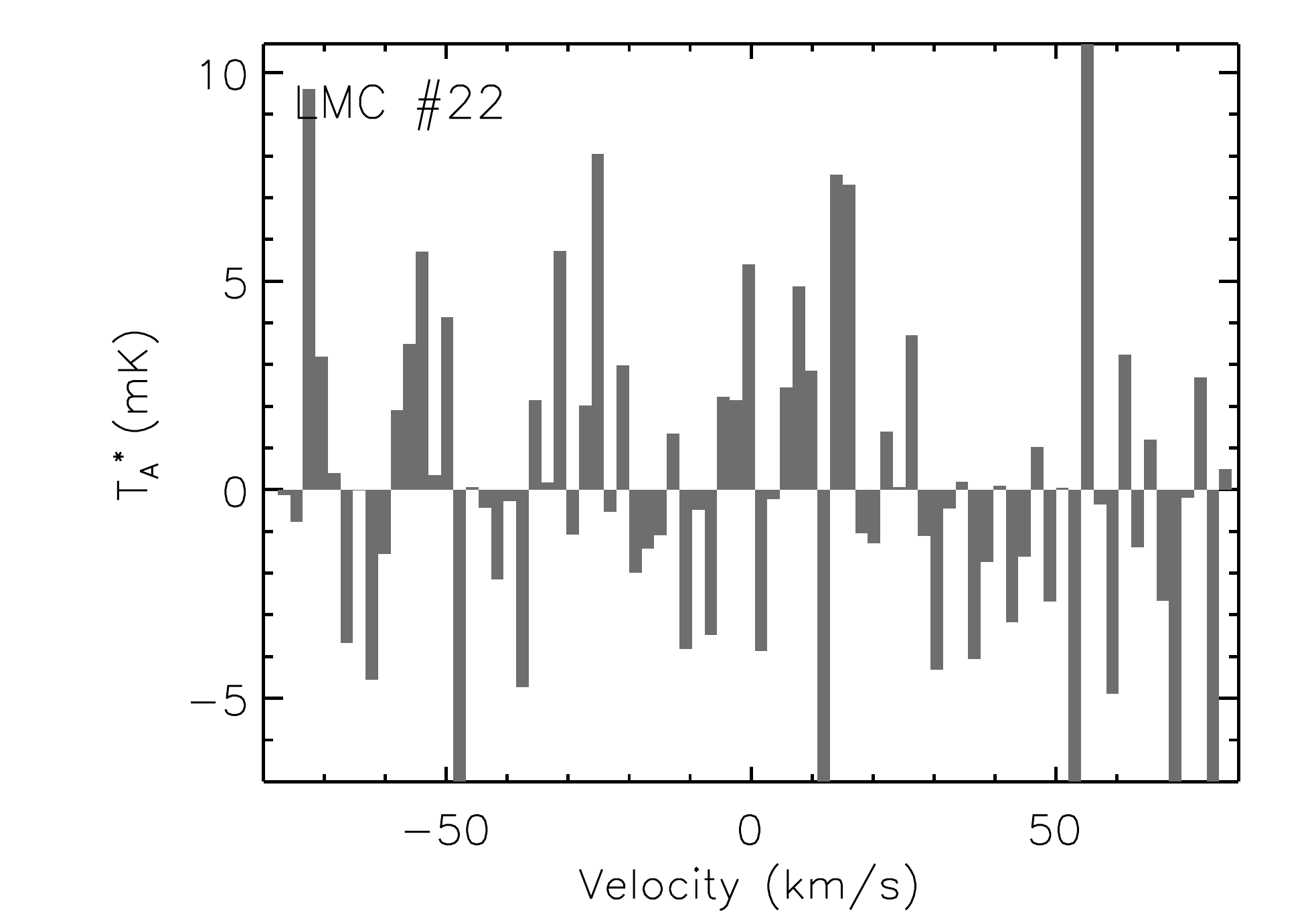} \\
\includegraphics[height=4cm]{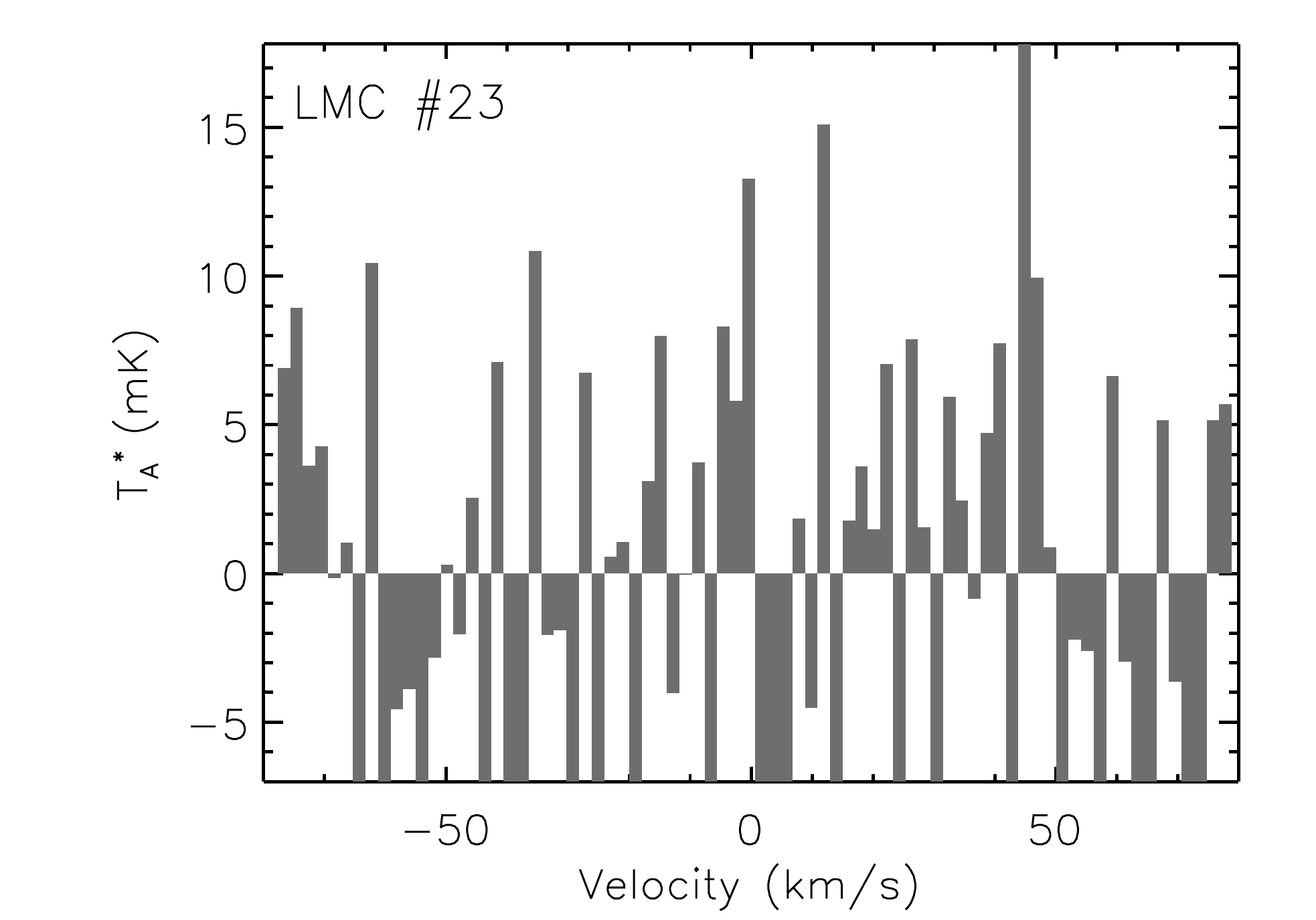}  & 
\hspace{-0.4cm}\includegraphics[height=4cm]{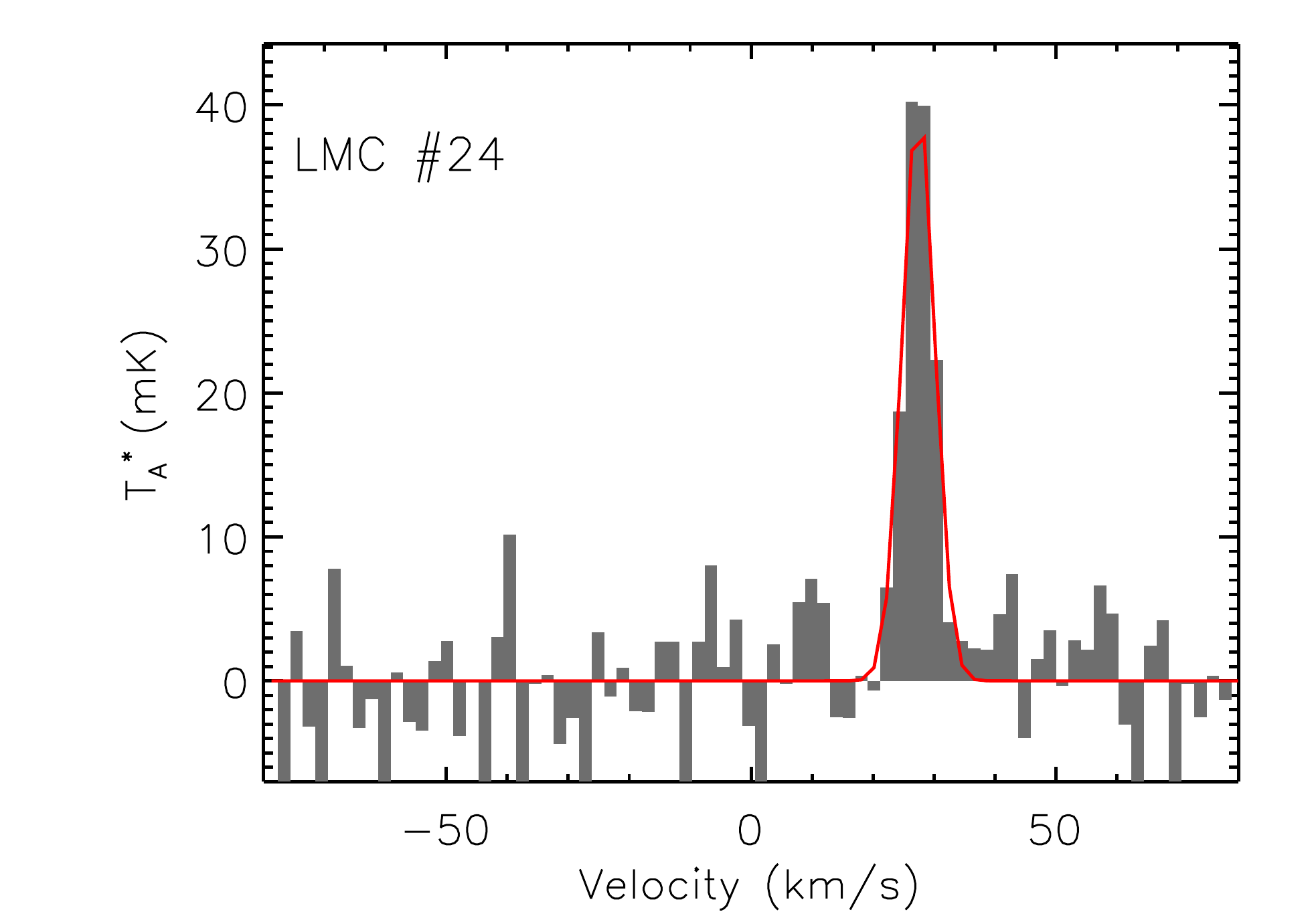}  &
\hspace{-0.4cm}  \includegraphics[height=4cm]{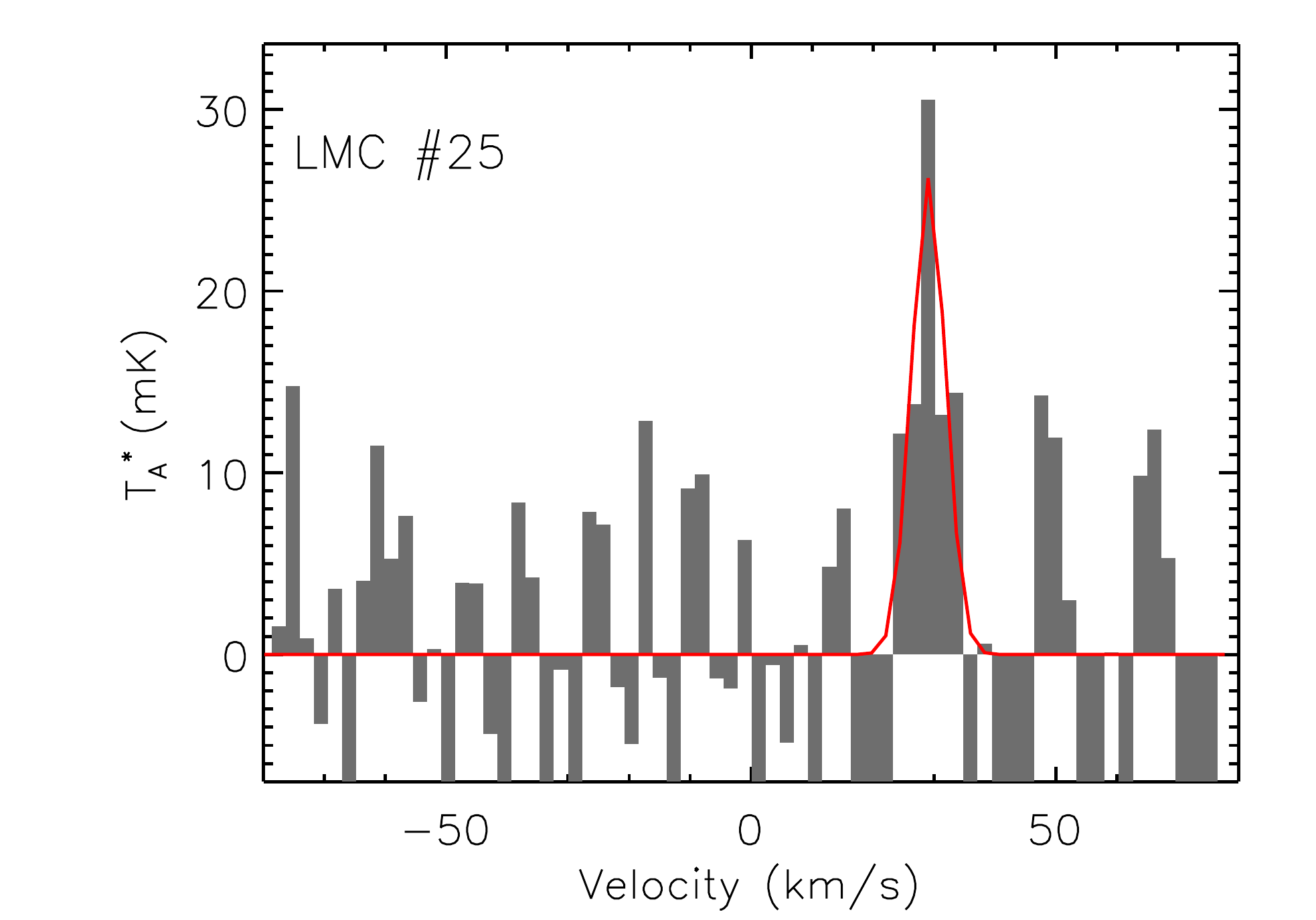} \\
\includegraphics[height=4cm]{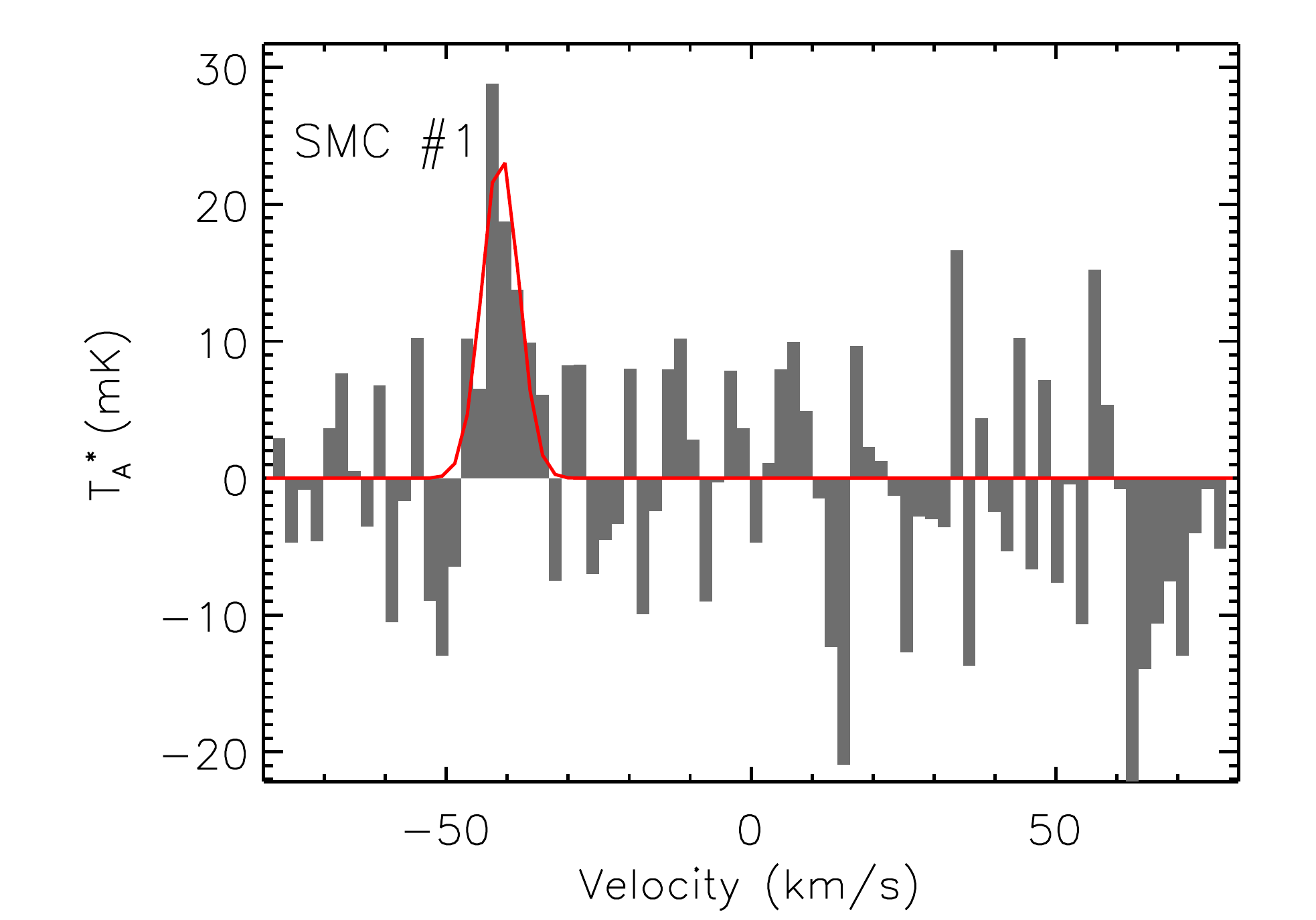}  & 
\hspace{-0.4cm}\includegraphics[height=4cm]{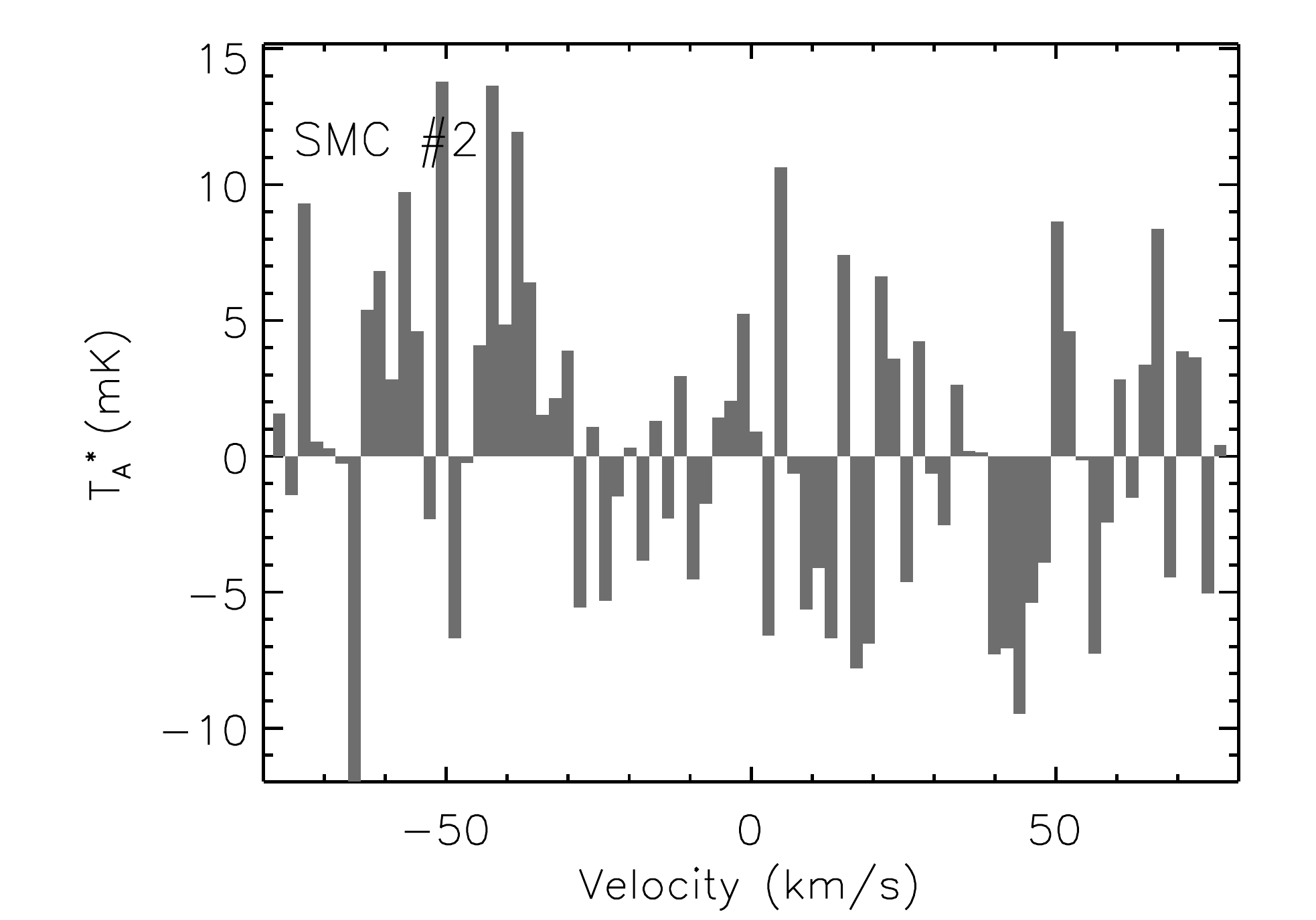}  & 
\hspace{-0.4cm}\includegraphics[height=4cm]{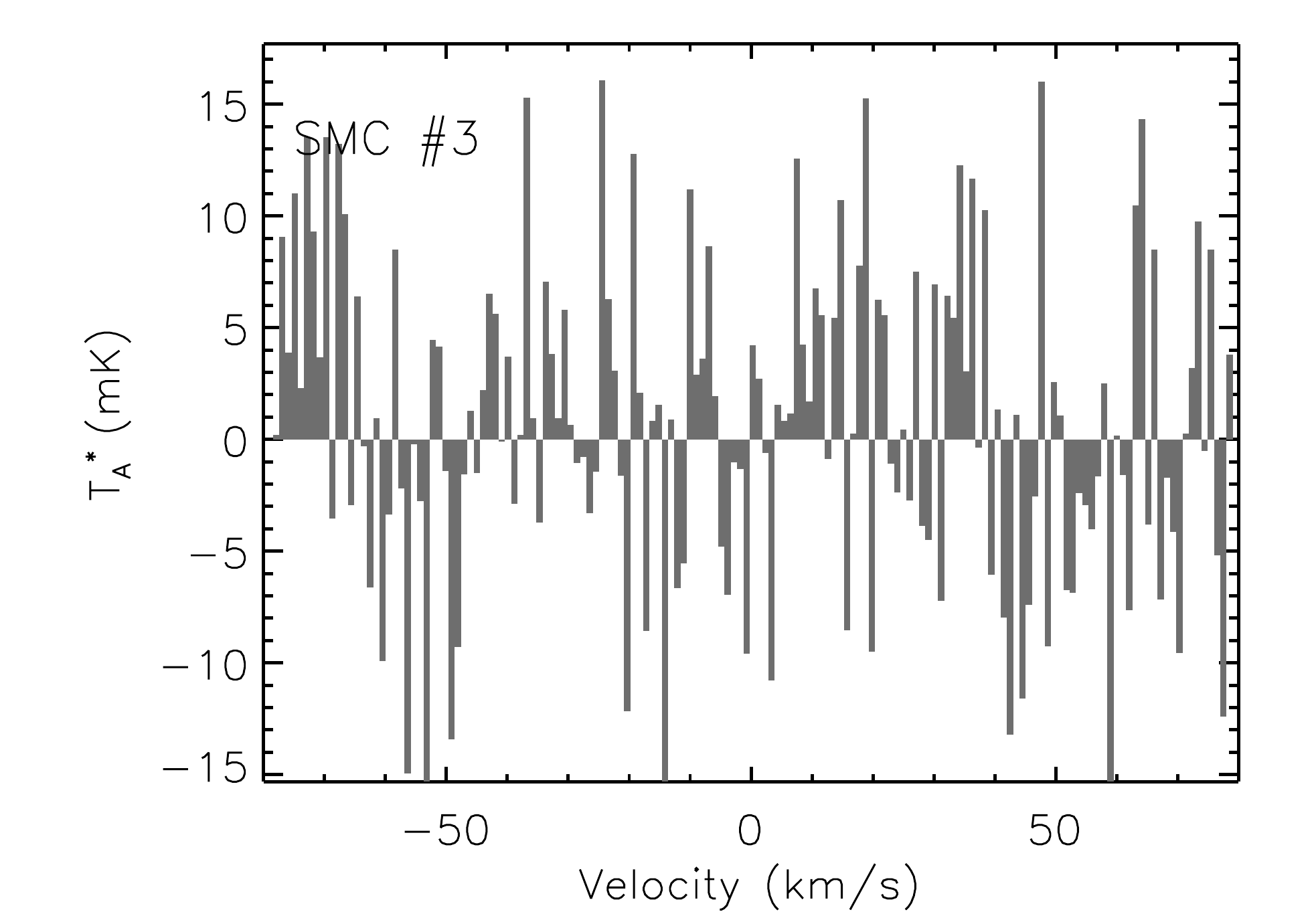} \\
\includegraphics[height=4cm]{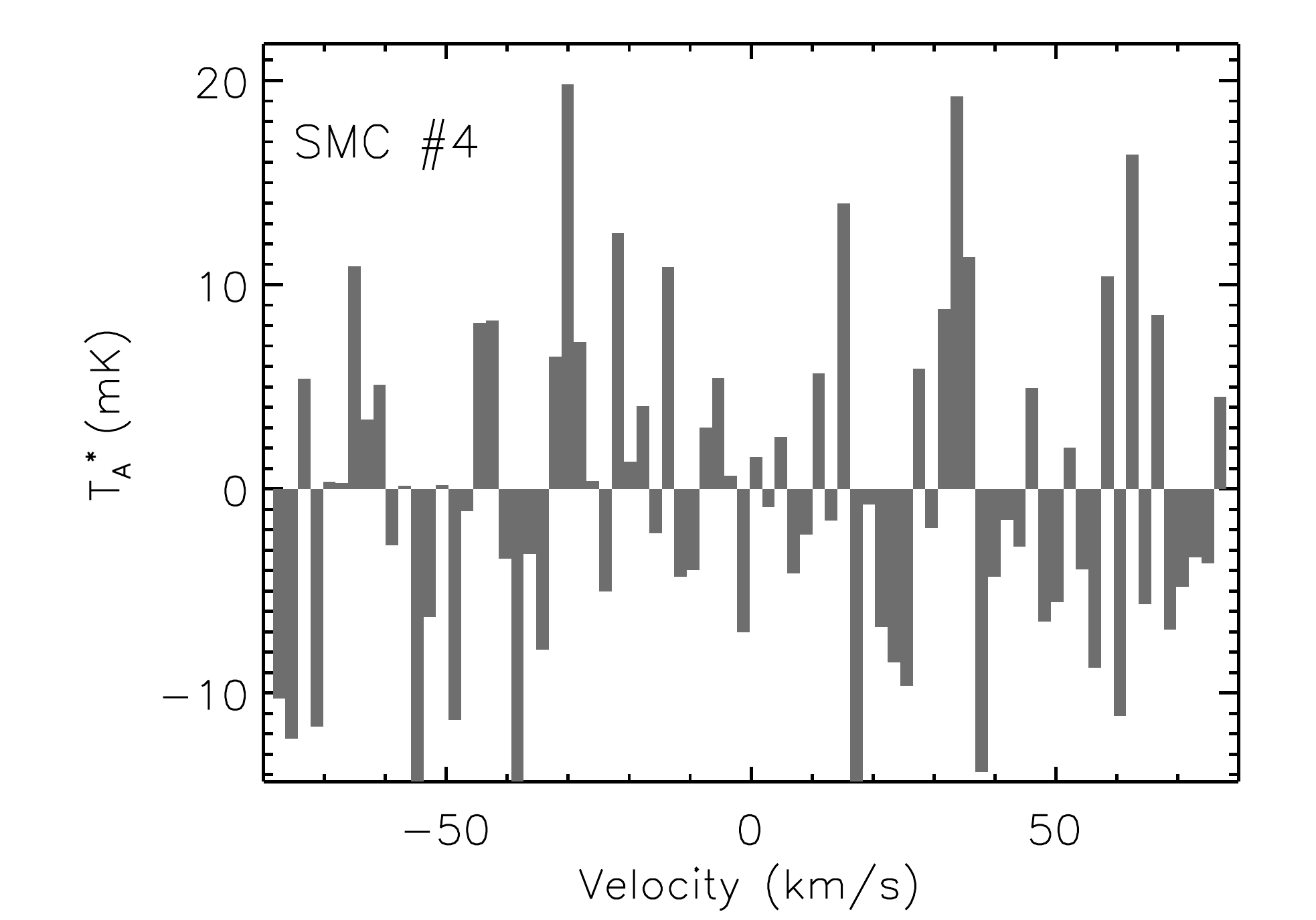}  & 
\hspace{-0.4cm}\includegraphics[height=4cm]{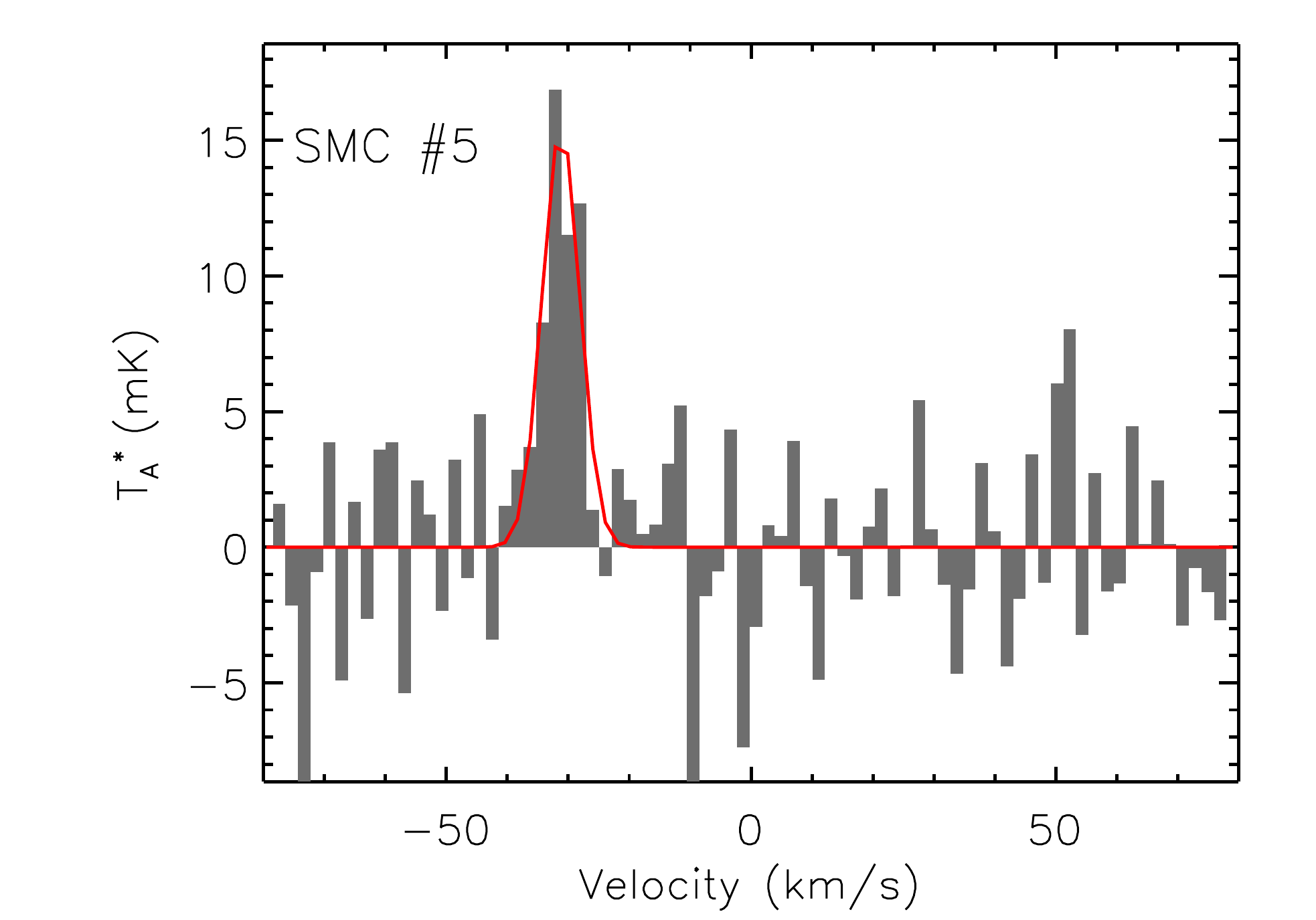}  & 
\hspace{-0.4cm}\includegraphics[height=4cm]{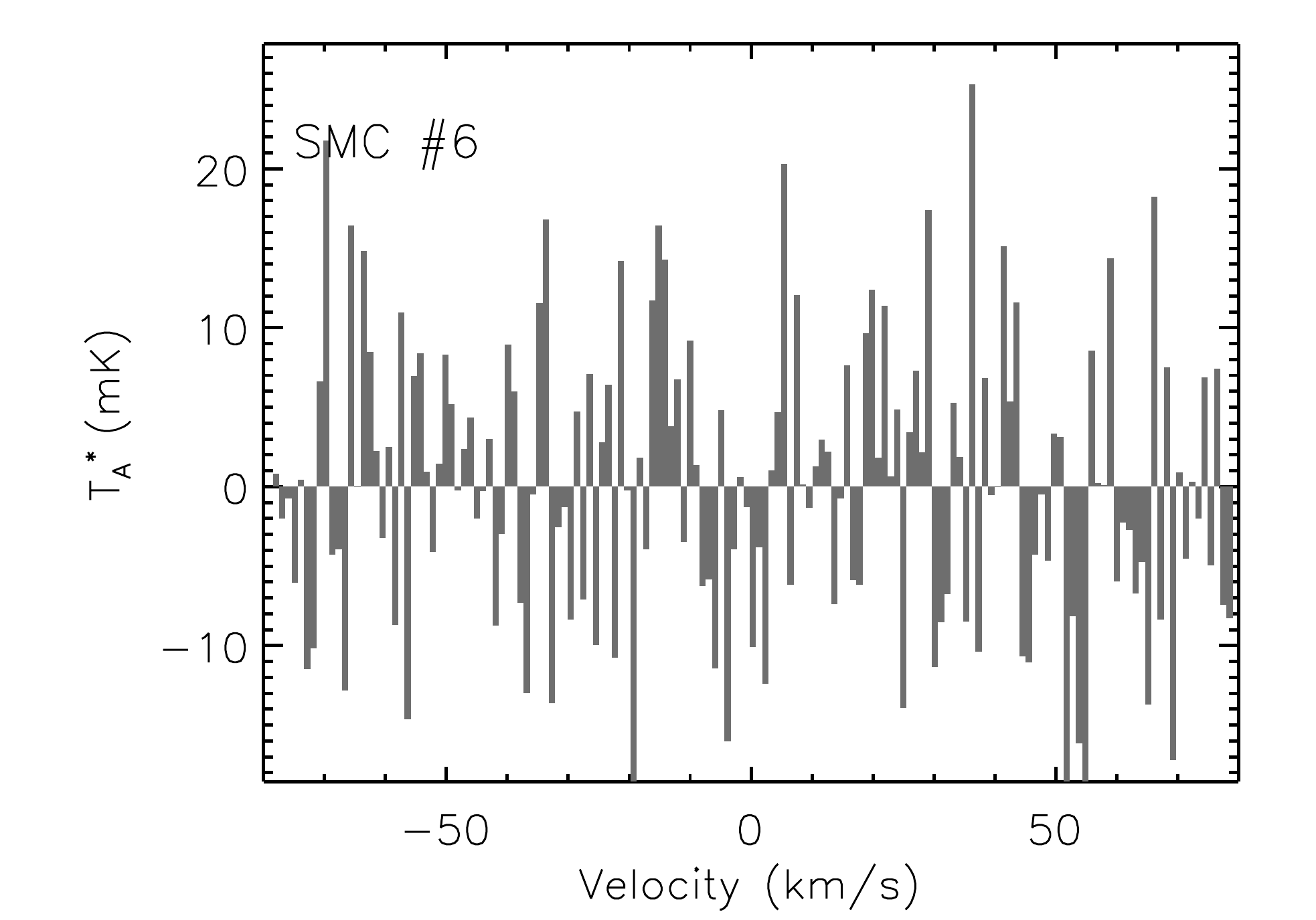} \\
\end{tabular}
\begin{tabular}{cc}
\includegraphics[height=4cm]{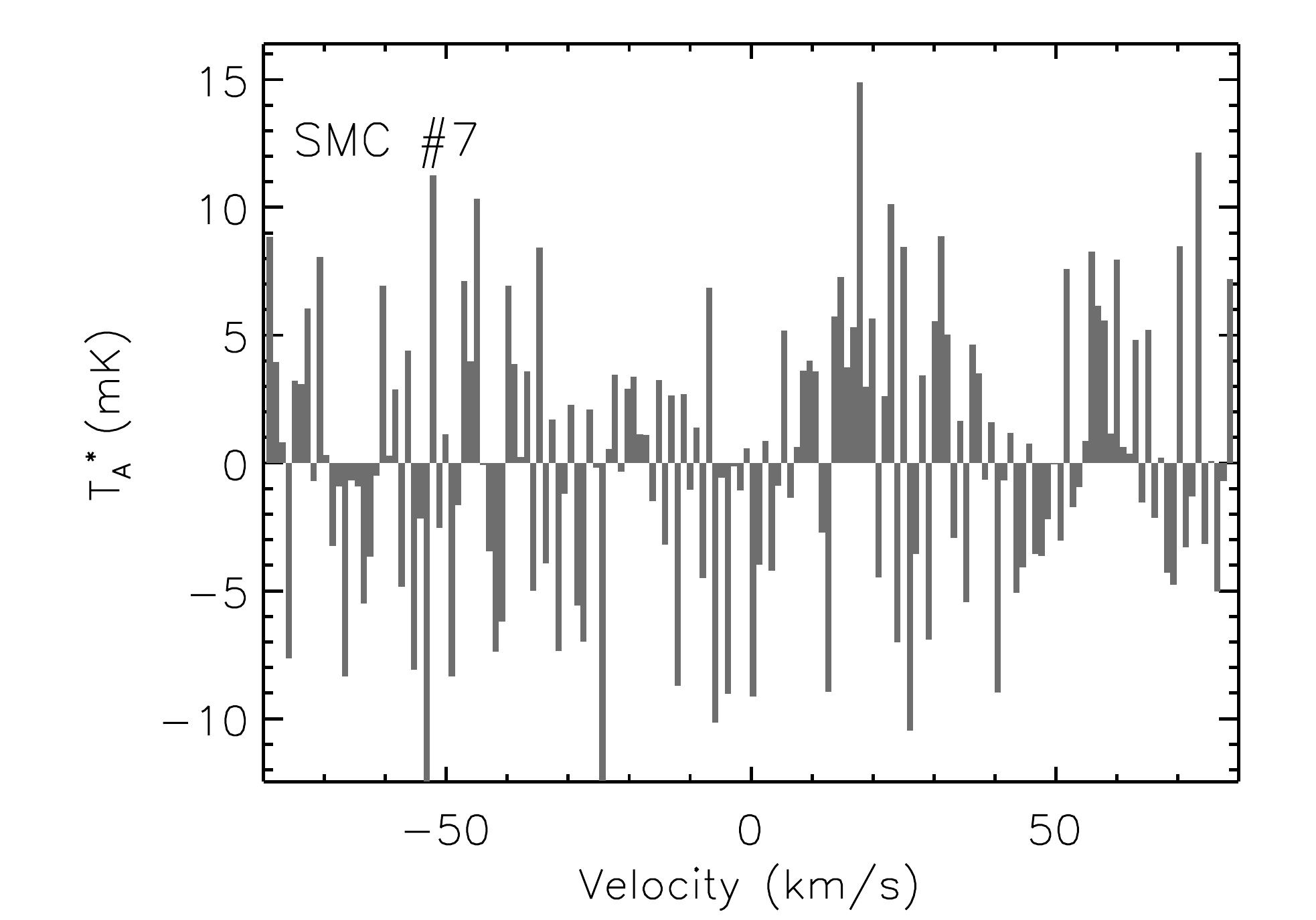}  &
\includegraphics[height=4cm]{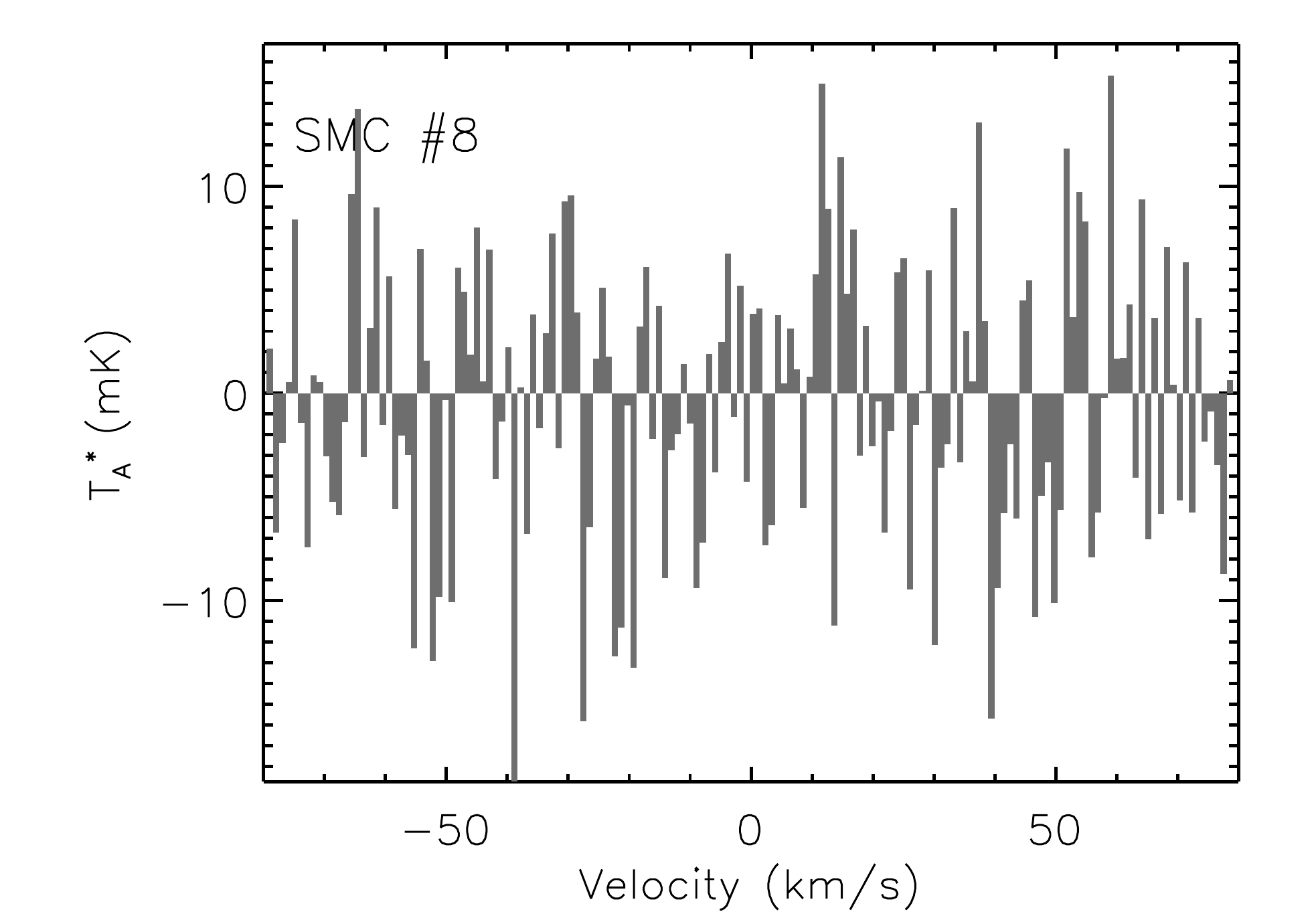}  \\
\end{tabular}
\caption{continued.}
\end{figure*}

\clearpage
\newpage
\section{Mapping campaign}


\begin{table*}
\caption{Characteristics of the mapping campaign sample.}
\label{LineCharacteristics2}
\centering
\begin{tabular}{cccccccccc}
\hline
\hline
\vspace{-5pt}&\\
        &&&&&  \multicolumn{4}{c}{HCO$^+$(2$-$1)}                                   \\
\vspace{-5pt}&\\
\cline{5-10}
\vspace{-5pt}&\\
Name    & Map center   & YSOs  &&  Map & \multicolumn{4}{c}{ $@$ Peak emission $^b$}                         \\
\vspace{-5pt}&\\
\cline{6-10}
\vspace{-5pt}&\\
&       $\alpha$ \& $\delta$ (J2000)    & &&  rms $^c$   &  Position & Vel. Offset $^d$      & FWHM          & $\mathrm{\int T_{mb}~dv}$  & SFR    \\
&($^h$,$^m$,$^s$)~~~($^{\circ}$,$\arcmin$,$\arcsec$)&&&  (K)      & ($^h$,$^m$,$^s$)~~~($^{\circ}$,$\arcmin$,$\arcsec$) & (km~s$^{-1}$)   & (km~s$^{-1}$) & (K~km~s$^{-1}$)         & (\msun~yr$^{-1}$)       \\
\vspace{-5pt}&\\
\hline
\vspace{-5pt}&\\
{\bf LMC}  \\
30Dor      & 05:38:42.4~-69:06:03 & 17 $^a$ & &        3.4$\times$$10^{-2}$            
        & 05:38:49.0~-69:04:39 & -15.1$\pm$0.6 & 6.9$\pm$1.2  & 2.55$\pm$0.07 &  5.4$\times$10$^{-5}$           \\
N11B       & 04:56:48.5~-66:24:18 & 16& &      9.2$\times$$10^{-2}$    
        & 04:56:48.5~-66:24:18& 19.8$\pm$0.3  & 3.7$\pm$0.5  & 1.42$\pm$0.12 & 6.4$\times$10$^{-6}$     \\
N11C       & 04:57:41.3~-66:27:36 & 12& &      5.0$\times$$10^{-2}$    
        & 04:57:56.0~-66:26:28& 17.4$\pm$0.6  & 4.0$\pm$1.1  & -         & -                \\
N44         & 05:22:08.8~-67:58:20 & 26& &    1.1$\times$$10^{-1}$         
        & 05:22:02.7~-67:58:23& 16.9$\pm$0.2  & 4.8$\pm$0.4  & 2.88$\pm$0.28  & 5.4$\times$10$^{-6}$   \\
N55         & 05:32:18.1~-66:26:01 & 5& &      8.2$\times$$10^{-2}$    
        & 05:32:09.3~-66:26:04& 12.0$\pm$0.4 &  3.0$\pm$0.5  & 0.79$\pm$0.09 & 1.4$\times$10$^{-6}$    \\
N105       & 05:09:57.1~-68:53:31 & 19& &     8.8$\times$$10^{-2}$    
        & 05:09:54.0~-68:53:34& -24.6$\pm$0.5  & 6.5$\pm$1.1  & 2.24$\pm$0.21 & 1.6$\times$10$^{-5}$  \\
N113       & 05:13:21.7~-69:21:33 & 19& &     3.7$\times$$10^{-2}$    
        &  05:13:18.3~-69:22:46& -31.6$\pm$0.1  & 5.4$\pm$0.1  & 3.40$\pm$0.06  &  1.4$\times$10$^{-5}$        \\
N159E    & 05:40:07.7~-69:44:53 & 18& &        4.4$\times$$10^{-2}$    
        & 05:40:04.3~-69:44:39& -32.6$\pm$0.1  & 5.0$\pm$0.1  & 2.16$\pm$0.07  & 1.7$\times$10$^{-5}$          \\
N159W   & 05:39:37.2~-69:45:57 & 20& &         5.9$\times$$10^{-2}$    
        & 05:39:37.2~-69:45:22& -28.1$\pm$0.1  & 6.5$\pm$0.1  & 3.44$\pm$0.12 & 1.3$\times$10$^{-5}$           \\
N214       & 05:39:54.7~-71:10:04 & 11& &      4.2$\times$$10^{-2}$    
        & 05:39:54.7~-71:10:06& -37.4$\pm$0.1  & 4.6$\pm$0.2  & 1.48$\pm$0.06 & 5.8$\times$10$^{-6}$   \\
\vspace{-5pt}&\\
\hline
\vspace{-5pt}&\\
{\bf SMC}  \\
N13        & 00:45:23~-73:22:38 &-& &   3.8$\times$$10^{-2}$    & -  & - & - & -  & -                                                      \\
N19        & 00:48:23~-73:05:54 &-& &   3.3$\times$$10^{-2}$    & -  & - & - & -  & -                                                     \\
N66        & 00:59:06~-72:10:44 &-& &   6.1$\times$$10^{-2}$    & -  & - & - & -         & -                                             \\
\vspace{-5pt}&\\
\hline
\end{tabular}
\begin{list}{}{}
\item[$^a$] Total number of YSOs per field. 
\item[$^b$] The various line characteristics are derived via a Gaussian fit to the line. Values are provided if the peak is detected at 3-$\sigma$.
\item[$^c$] The rms is estimated using the {\it go noise} function in CLASS.
\item[$^d$] We use a systemic velocity v$_{\rm sys}$=262.2 km~s$^{-1}$ for the LMC. 
\end{list}
\end{table*} 

\addtocounter {table}{-1}
\begin{table*}
\vspace{20pt}
\caption{continued for HCN(2$-$1).}
\centering
\begin{tabular}{cccccccccccc}
\hline
\hline
\vspace{-5pt}&\\
        &&&&& \multicolumn{4}{c}{HCN(2$-$1)}      \\
\vspace{-5pt}&\\
\cline{5-10}
\vspace{-5pt}&\\
Name    & Map center   & YSOs  && Map & \multicolumn{4}{c}{$@$ Peak emission}                  \\
\vspace{-5pt}&\\
\cline{6-10}
\vspace{-5pt}&\\
&       $\alpha$ \& $\delta$ (J2000)    & &&    rms  &
        Position & Vel. Offset                 & FWHM          & $\mathrm{\int T_{mb}~dv}$ & SFR        \\
&($^h$,$^m$,$^s$)~~~($^{\circ}$,$\arcmin$,$\arcsec$)& &&     (K)  &  
($^h$,$^m$,$^s$)~~~($^{\circ}$,$\arcmin$,$\arcsec$) & (km~s$^{-1}$) & (km~s$^{-1}$)         & (K~km~s$^{-1}$) & (\msun~yr$^{-1}$)      \\
\vspace{-5pt}&\\
\hline
\vspace{-5pt}&\\
{\bf LMC} \\
30Dor      & 05:38:42.4~-69:06:03 & 17 && 2.5$\times$$10^{-2}$ 
        & 05:38:32.4~-69:06:25 & -21.2$\pm$0.1   & 7.2$\pm$0.3           & 0.96$\pm$0.06 & 3.3$\times$10$^{-5}$    \\
N11B       & 04:56:48.5~-66:24:18 & 16 && 6.0$\times$$10^{-2}$  
        & 04:57:03.3~-66:24:00 & 18.9$\pm$2.4    & 9.9$\pm$4.9           & 1.82$\pm$0.31 & 3.8$\times$10$^{-6}$    \\
N11C       & 04:57:41.3~-66:27:36 & 12 && 3.4$\times$$10^{-2}$  
        & 04:57:47.3~-66:28:14 & 13.6$\pm$2.4    & 10.3$\pm$4.8  & 1.17$\pm$0.18  & 3.5$\times$10$^{-6}$   \\
N44         & 05:22:08.8~-67:58:20 & 26 && 7.8$\times$$10^{-2}$  
        & 05:21:59.6~-67:56:54 & 22.1$\pm$0.9    & 5.0$\pm$1.8           & 2.03$\pm$0.23 &  4.1$\times$10$^{-6}$  \\
N55         & 05:32:18.1~-66:26:01 & 5 && 6.3$\times$$10^{-2}$  
        & - & - & - & - & -  \\
N105       & 05:09:57.1~-68:53:31 & 19 && 7.3$\times$$10^{-2}$  
        & 05:09:54.0~-68:53:34 & -26.0$\pm$0.8   & 4.9$\pm$1.5   & 1.08$\pm$0.15 & 1.6$\times$10$^{-5}$ \\
N113       & 05:13:21.7~-69:21:33 & 19 && 2.5$\times$$10^{-2}$  
        & 05:13:18.3~-69:22:29 & -33.2$\pm$0.1   & 6.3$\pm$0.1   & 1.58$\pm$0.05 & 1.5$\times$10$^{-5}$ \\
N159E    & 05:40:07.7~-69:44:53 & 18&& 3.3$\times$$10^{-2}$  
        & 05:40:04.3~-69:44:21 & -32.0$\pm$0.2   & 6.1$\pm$0.3   & 1.07$\pm$0.07 &  9.2$\times$10$^{-6}$\\
N159W   & 05:39:37.2~-69:45:57 & 20&& 4.4$\times$$10^{-2}$  
        & 05:39:37.2~-69:45:39 & -30.6$\pm$0.2   & 7.0$\pm$0.4   & 1.35$\pm$0.10 & 1.8$\times$10$^{-5}$ \\
N214       & 05:39:54.7~-71:10:04 &  11&& 3.0$\times$$10^{-2}$  
        & 05:39:54.7~ -71:10:06 & -38.5$\pm$0.2  & 3.7$\pm$0.4   & 0.52$\pm$0.04  & 5.8$\times$10$^{-6}$ \\
\vspace{-5pt}&\\
\hline
\vspace{-5pt}&\\
{\bf SMC} \\
N13             & 00:45:23~-73:22:38 &-&& 2.8$\times$$10^{-2}$  & - & - & - & - & - \\
N19        & 00:48:23~-73:05:54 &-&& 2.4$\times$$10^{-2}$  & - & - & - & - & -\\
N66        & 00:59:06~-72:10:44 &-&& 4.5$\times$$10^{-2}$  & - & - & - & - & -\\
\vspace{-5pt}&\\
\hline
\end{tabular}
\end{table*} 

\end{document}